\documentclass{siamonline1116}

\usepackage{lipsum}
\usepackage{amsfonts}
\usepackage{graphicx}
\usepackage{epstopdf}
\usepackage{algorithmic}
\ifpdf
  \DeclareGraphicsExtensions{.eps,.pdf,.png,.jpg}
\else
  \DeclareGraphicsExtensions{.eps}
\fi

\numberwithin{theorem}{section}

\newcommand{\TheTitle}{Bayesian inversion in resin transfer molding} 
\newcommand{\TheAuthors}{M. A. Iglesias, M. Park and M.V. Tretyakov}

\headers{\TheTitle}{\TheAuthors}

\title{{\TheTitle}\thanks{Submitted to the editors DATE.}}

\author{
Marco A. Iglesias\thanks{School of Mathematical Sciences, University of Nottingham, University Park,
Nottingham, NG7 2RD, UK.
    (\email{marco.iglesias@nottingham.ac.uk}).}
  \and
Minho Park \thanks{School of Mathematical Sciences, University of Nottingham, University Park,
Nottingham, NG7 2RD, UK.}
\and
M.V. Tretyakov \thanks{School of Mathematical Sciences, University of Nottingham, University Park,
Nottingham, NG7 2RD, UK (\email{Michael.Tretyakov@nottingham.ac.uk}).}
}

\usepackage{amsopn}

\ifpdf
\hypersetup{
  pdftitle={\TheTitle},
  pdfauthor={\TheAuthors}
}
\fi

\usepackage{float}
\usepackage{enumitem}

\usepackage[toc,page,header]{appendix}
\usepackage[top=1.0in,left=0.80in,bottom=1.0in,right=0.80in]{geometry}
\usepackage{amssymb}
 \usepackage{amsmath}
\usepackage{psfrag}

\usepackage{pifont}

\newcommand{\cG}{\mathcal G}
\newcommand{\Ga}{\Upsilon}
\newcommand{\cC}{\mathcal C}
\newtheorem{assumption}{Assumption}

\newtheorem{remark}{Remark}

\numberwithin{equation}{section}
\numberwithin{remark}{section}
\numberwithin{algorithm}{section}
\numberwithin{assumption}{section}

\def\thefigure{\arabic{figure}}
\def\thetable{\arabic{table}}
\numberwithin{table}{section}
\numberwithin{table}{section}
\numberwithin{figure}{section}

\newcommand{\mt}[1]{{\color{black}  #1}}

\begin{document}
\maketitle


\begin{abstract}

We study the Bayesian inverse problem of inferring the permeability of a porous medium within the context of a moving boundary framework motivated by Resin Transfer Molding (RTM), one of the most commonly used processes for manufacturing fiber-reinforced composite materials. During the injection of resin in RTM, our aim is to update our probabilistic knowledge of the permeability of the material by inverting pressure measurements as well as observations of the resin moving domain. We consider both one-dimensional and two-dimensional forward models for RTM. Based on the analytical solution for the one-dimensional case, we prove existence of the sequence of posteriors that arise from a sequential Bayesian formulation within the infinite-dimensional framework. For the numerical characterisation of the Bayesian posteriors in the one-dimensional case, we investigate the application of a fully-Bayesian Sequential Monte Carlo method (SMC) for high-dimensional inverse problems. By means of SMC we construct a benchmark against which we compare performance of a novel regularizing ensemble Kalman algorithm (REnKA) that we propose to approximate the posteriors in a computationally efficient manner under practical scenarios. We investigate the robustness of the proposed REnKA with respect to tuneable parameters and computational cost.
 \mt{We demonstrate} advantages of REnKA compared with SMC with a small number of particles. We further \mt{investigate}, in both the one-dimensional and two-dimensional settings, practical aspects \mt{of REnKA} relevant to RTM, which include the effect of pressure sensors configuration and the observational noise level in the uncertainty in the log-permeability quantified via the sequence of Bayesian posteriors.


\end{abstract}


\section{Introduction}\label{Intro}

In this paper we study the Bayesian inverse problem within the moving boundary setting motivated by applications in manufacturing of fiber-reinforced composite materials. Due to their light weight, high strength, as well as their flexibility to fit mechanical requirements and complex designs, such materials are playing a major role in automotive, marine and aerospace industries \cite{Ast97,AS10,Long05}. The moving boundary problem under consideration arises from Resin Transfer Molding (RTM) process, one of the most commonly used processes for manufacturing composite materials. RTM consists of the injection of resin into a cavity mold with the shape of the intended composite part according to design and enclosing a reinforced-fiber preform previously fabricated. The next stage of RTM is curing of the resin-impregnated preform, which may start during or after the resin injection. Once curing has taken place, the solidified part is demolded from the cavity mold. In the present work we are concerned with the resin injection stage of RTM under the reasonable assumption that curing starts after resin has filled the preform. Though the current study is motivated by RTM, the results can be also used for other applications where a moving boundary problem is a suitable model.

\begin{figure}[htbp]
\begin{center}
\includegraphics[scale=0.4]{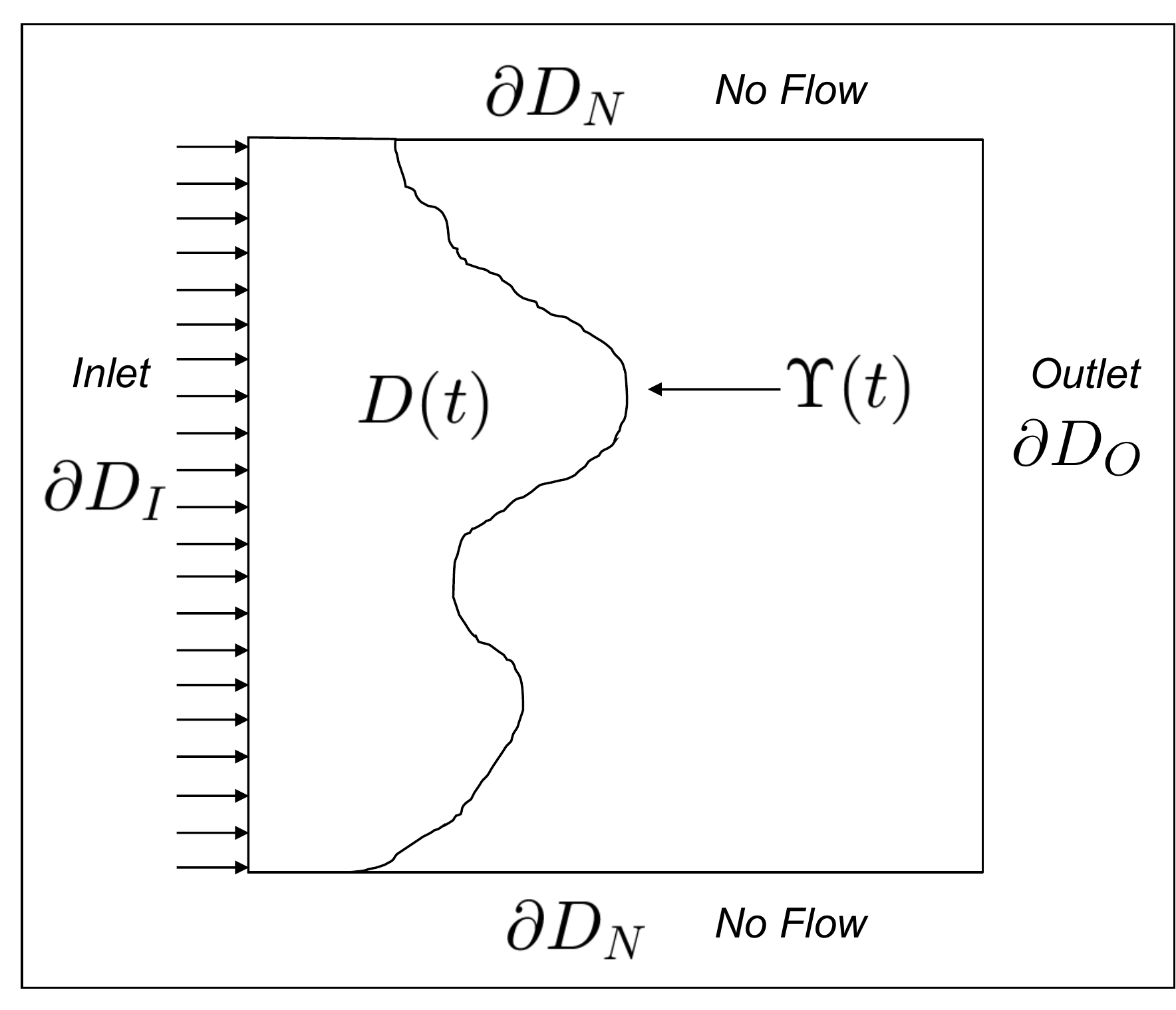}
 \caption{\mt{An example of the} physical configuration of the moving boundary problem.} \label{Fig12A}
\end{center}
\end{figure}

We now describe the (from the inverse problem prospective, forward) model (see further details in  \cite{AS10,TW01,SRTM16}). Let  $D^{\ast}\subset \mathbb{R}^{d}$, $d\in \{1,2\}$, be an open domain representing a physical domain of a porous medium with the permeability $\kappa(x)$ and porosity $\varphi$. The boundary of the domain  $D^{\ast}$ is $\partial D^{\ast}=\partial D_{I}\cup \partial D_{N}\cup
\partial D_{O}$, where $\partial D_{I}$ is the inlet, $\partial D_{N}$ is the perfectly sealed boundary, and $\partial D_{O}$ is the outlet. The domain $D^{\ast}$ is initially filled with air at a pressure $p_0$. This medium is infused with a fluid (resin) with viscosity $\mu$ through an inlet boundary $\partial D_{I}$ at a pressure $p_{I}$ and moves through $D^{\ast}$ occupying a time-dependent domain $D(t)\subset D^{\ast}$, which is bounded by the moving boundary $\Ga(t)$
and the appropriate parts of $\partial D$.
 \mt{An example of the} physical configuration of this problem in 2D is illustrated in Figure \ref{Fig12A}.

The forward problem for the pressure of resin $p(t,x)$ consists of the conservation of mass
\begin{eqnarray}\label{eq:2000}
\nabla \cdot \mathbf{v}   = 0, ~~~~~x\in D(t),\ t>0,
\end{eqnarray}
where the flux $\mathbf{v}(x,t)$ is given by Darcy's law
\begin{eqnarray}\label{eq:2001}
\mathbf{v}(x,t)=- \frac{\kappa(x)}{\mu} \nabla p(x,t)
 \end{eqnarray}%
with the following initial and boundary conditions
\begin{eqnarray}\label{eq:2002}
p(x,t)& =&p_{I},\ x\in \partial D_{I},\ t\geq 0,  \\
 \nabla p(x,t) \cdot \mathbf{n(x)} &=&0,\ x\in \partial D_{N},\ t\geq 0,  \label{eq:2003}\\
V(x,t)&= & - \frac{\kappa(x)}{\mu \varphi}  \nabla p(x,t) \cdot \mathbf{n(x,t)}, ~~x\in \Ga (t),\ t\geq 0,  \label{eq:2004} \\
p(x,t) &=&p_0,\ x\in \Ga (t),\ t>0,    \label{eq:2005}\\
p(x,t) &=&p_0, x \in \partial D_{O}, \ t>0, \label{eq:2005n}\\
p(x,0) &=&p_0,\ x\in D^{\ast}, \label{eq:2006}\\
\Ga (0) &=&\partial D_{I}.   \label{eq:2008}
\end{eqnarray}%
Here $V(t)$ is the velocity of the moving boundary $\Ga(t)$ in the normal direction,
$\mathbf{n(x)}$ and $\mathbf{n(x,t)}$ are the unit outer normals to the corresponding boundaries.
We note that in the considered one and two dimensional cases of this problem, we can view the velocity of the moving boundary as the following derivative:
\begin{eqnarray}\label{eq:2007}
V(x,t) &=& \frac{d \Ga(t)}{dt}\cdot \mathbf{n(x,t)}.
\end{eqnarray}
We remark that for definiteness we have assumed that at the initial time the moving boundary $\Ga(0)$ coincides with the inlet boundary $\partial D_{I}$ and that the constant pressure condition is imposed at the inlet. It is not difficult to carry over the inverse problem methodology considered in this paper to other geometries and other conditions on the inlet (e.g. constant rate). Further, in two (three) dimensional RTM settings one usually models permeability via a second (third)-order permeability tensor to take into account anisotropic structure of the media \cite{AS10,SRTM16} but here for simplicity of the exposition the permeability $\kappa(x)$ is a scalar function. Again, the developed methodology is easy to generalize to the tensor case.

Let us note that in the one-dimensional case the nonlinear problem (\ref{eq:2000})-(\ref{eq:2008}) is analytically simple and admits a closed form solution (see Section~\ref{sec:1d} and \cite{SRTM16}) but the two and three dimensional cases are much more complicated and analytical solution is in general not available.
We remark that in two and three dimensional cases the resin can race around low permeability regions and the front $\Ga$ can become discontinuous creating macroscopic voids behind the main front (see further details in \cite{AS10,SRTM16}) but in this paper we ignore such effects which deserve further study.

It has been extensively recognized \cite{EL06,EL062,MAL14,Matveev,PP99,SC09}
that imperfections in a preform that arise during its fabrication and packing in the molding cavity can lead to variability in fiber placement which results in a heterogenous highly-uncertain preform permeability.  In turn, these unknown heterogeneities in permeability of the preform give rise to inhomogeneous resin flow patterns which can have profound detrimental effect on the quality of the produced part, reducing its mechanical properties and ultimately leading to scrap. To limit these undesirable effects arising due to uncertainties, conservative designs are used which lead to heavier, thicker and, consequently, more expensive materials aimed at avoiding performance \mt{being} compromised. Clearly, the uncertainty quantification of material properties is essential for making RTM more cost-effective. One of the key elements in tackling this problem
is to be able to quantify the uncertain permeability.

In this work we assume that $D^{\ast}$, $\partial D_{O}$, $\partial D_{I}$, $\partial D_{N}$, $p_{I}$, $p_0$,
$\mu$ and $\varphi$ are known deterministic parameters while the permeability $\kappa(x)$ is unknown.
Our objective is within the Bayesian framework to infer $\kappa(x)$ or, more precisely, its natural logarithm $u(x)=\log{\kappa(x)}$ from measurements of pressure $p(x,t)$ at some sensor locations as well as measurements of the front $\Ga(t)$, or alternatively, of the time-dependent domain $D(t)$ at a given time $t>0$. We put special emphasis on computational efficiency of the inference, which is crucial from the applicable point of view.

\subsection{Practical approaches for permeability estimation in fiber-reinforced composites}

While the estimation of preform permeability during resin injection in RTM is clearly an inverse problem constrained by a moving boundary PDE such as (\ref{eq:2000})-(\ref{eq:2008}), most existing practical approaches pose the estimation of permeability in neither a deterministic nor stochastic inverse problems framework. For example, the very extensive review published in 2010 \cite{Review} reveals that most conventional methods for measuring permeability assume that (i) the material permeability tensor is homogenous and (ii) the flow is one-dimensional (including 2D radial flow configurations). Under these assumptions the resin injection in RTM can be described analytically, via expressions derived from Darcy's law, which enable a direct computation of the permeability in terms of quantities that can be measured before or during resin injection. These conventional methods suffer from two substantial practical limitations. First, they do not account for the heterogenous structure of the preform permeability, and although they provide an estimate of an effective permeability, this does not enable the prediction of the potential formation of voids and dry spots. Second, those conventional methods compute the permeability in an off-line fashion (i.e before RTM) with specific mold designs that satisfy the aforementioned assumptions intrinsic to those methods (e.g. rectangular flat molds). This second limitation is not only detrimental to the operational efficiency of RTM but also neglects the potential changes in permeability that can results from encapsulating the preform in cavities with complex designs.

Some practical methodologies for online (i.e. during resin injection) estimation of heterogenous permeability have been proposed in \cite{PC:PC10504,PC:PC23290}.  While these approaches seem to address the aforementioned limitations of conventional methods, they also use a direct approach for the estimation of permeability which faces unresolved challenges. As an example, let us consider the recent work of \cite{PC:PC23290} which uses an experimental configuration similar to the one described in Figure \ref{Fig12A} and which, by using pressure measurements from sensors located within the domain occupied by the preform, computes a finite-difference approximation of the normal flux to the front $ \nabla p(\Ga(t),t)\cdot \mathbf{n}$. In addition, by means of images from CCT cameras, seepage velocity of the resin front is computed in \cite{PC:PC23290}; this velocity is nothing but $V(x,t)$ defined by (\ref{eq:2004}) in the context of the moving boundary problem (\ref{eq:2000})-(\ref{eq:2008}). Under the assumption that $\mu$ and $\varphi$ are known, the approach proposed in \cite{PC:PC23290} consists of finding
\begin{eqnarray}\label{eq:AA2}
\kappa(\Ga(t))=\arg\min_{\theta}\Big\vert\Big\vert  V(x,t)+\frac{\theta}{\varphi \mu}  \nabla p(\Ga(t),t)  \cdot \mathbf{n}\Big\vert \Big\vert
 \end{eqnarray}%
with $V(x,t)$  and $\nabla p(\Ga(t),t)  \cdot \mathbf{n}$ computed from measurements as described above. This approach offers a practical technique to \mt{estimating} $\kappa$ on the moving front and can then potentially infer the whole permeability field during the resin injection in RTM. However, from the mathematical inverse problems perspective, this ad-hoc approach is not recommended as it involves differentiating observations of pressure data for the computation of $\nabla p(\Ga(t),t)  \cdot \mathbf{n}$. Indeed, it is well-known \cite{Kirsch} that differentiation of data is an ill-posed problem that requires regularization. In addition, rather than an inverse problem, the least-squares formulation in (\ref{eq:AA2}) is a data fitting exercise that excludes the underlying constraint given by the moving boundary problem and which entails a global effect induced by $\kappa$. As a result, the estimate of permeability obtained via (\ref{eq:AA2}) has no spatial correlation and thus fails to provide \mt{an accurate global} estimate of the permeability field.

The recent work of \cite{EnKF_RTM} demonstrates \mt{considerable} advantages of using systematic data assimilation approaches to infer permeability during the resin injection of RTM. By means of a standard ensemble Kalman methodology for data assimilation, the approach of \cite{EnKF_RTM} uses measurements from visual observations of the front location to produce updates of the preform permeability within the context of a discrete approximation of the moving boundary problem (\ref{eq:2000})-(\ref{eq:2008}). While the methodology used in \cite{EnKF_RTM} is focused in producing deterministic estimates, the standard Kalman methodology can be potentially used to quantify uncertainty in preform permeability. However, it has been shown that standard Kalman methodologies, such as the one used in \cite{EnKF_RTM}, could result in unstable estimates unless further regularisation to the algorithm is applied \cite{Iglesias2014}.

In addition to the lack of an inverse problem framework that can lead to unstable and ultimately inaccurate estimates of the permeability in resin injection of RTM, most existing approaches (i) do not incorporate the uncertainty in the observed variables and (ii) do not quantify uncertainty in the estimates of the permeability of preform. It is indeed clear from our literature review that the estimation of permeability of preform during resin injection deserves substantial attention from an inverse problems perspective capable of quantifying uncertainty inherent to the fabrication and packing of the preform.

\subsection{The Bayesian approach to inverse problems}

In this paper we propose the application of the Bayesian approach to inverse problems \cite{Andrew} in order to infer the logarithm of the permeability $u(x)=\log{\kappa(x)}$, from observations $\{y_{n}\}_{n=1}^{N}$ collected at some prescribed measurement/observation times $\{t_{n}\}_{n=1}^{N}$ during the resin injection in RTM. At each time $t_{n}$ we observe a vector, $y_{n}$, that contains noisy measurements of resin pressure from sensors as well as some information of the moving domain (or alternatively front location) observed, for example, via CCT cameras or dielectric sensors \cite{Andy_control07}. In the Bayesian approach, the unknown $u(x)$ is a random function that belongs to a space of inputs $X$. A prior probability measure $\mu_{0}(u)=\mathbb{P}(u)$ on $u$ must be specified before the data are collected; this enables us to incorporate prior knowledge which may include design parameters as well as the uncertainty that arises from preform fabrication (i.e. prior to resin injection). In our work we consider Gaussian priors which have been identified as adequate for characterizing the aforementioned uncertainty in log-permeability from the preform fabrication
\mt{\cite{Zhang20111478,MAL14,Matveev} (see also references therein)}.

At each observation time $t_{n}$ during the infusion of resin in RTM, we then pose the inverse problem in terms of computing, $\mu_{n}(u)=\mathbb{P}(u\vert y_{1},\dots,y_{n})$, the (posterior) probability measure of the log-permeability conditioned \mt{on} measurements $y_{1}\dots,y_{n}$. Each posterior $\mu_{n}$ then provides a rigorous quantification of the uncertainty in the log-permeability field given all available measurements up to the time $t_{n}$. Knowledge of each of these posteriors during RTM can then be used to compute statistical moments of the log-permeability under $\mu_{n}$ (e.g. mean, variance) as well as expectations of quantities of interest that may be needed for the optimization of controls (e.g. pressure injection) in RTM.

Although the proposed application of the Bayesian formulation assumes Gaussian priors, the nonlinear structure of the PDE problem, that describes resin injection in RTM, gives rise to a sequence of non-Gaussian Bayesian posteriors $\{\mu_{n}\}_{n=1}^{N}$ which cannot be characterized in a closed form. A sampling approach is then required to compute approximations of these posteriors. Among existing sampling methodologies, Sequential Monte Carlo (SMC) samplers \cite{delmoral,Chopin01082002,BJMS15,Kantas} are particularly relevant for the formulation of the above described inverse problem as they provide a recursive mechanism to approximate the sequence of Bayesian posteriors $\{\mu_{n}\}_{n=1}^{N}$.

Starting with $J$ samples from the prior $u_{0}^{(j)}\sim \mu_{0}$, $j=1,\dots,J$ (i.i.d.), the idea behind SMC is to transform a system of weighted particles $\{W_{n-1}^{(j)},u_{n-1}^{(j)}\}_{j=1}^{J}$ that define $\mu_{n-1}^{J}$ to an updated set $\{W_{n}^{(j)},u_{n}^{(j)}\}_{j=1}^{J}$ that approximates $\mu_{n}$ as the new data $y_{n}$ collected at time $t_{n}$ become available. The weights $\{W_{n}^{(j)}\}_{j=1}^{J}$ are normalised (i.e. $\sum_{j=1}^{J}W_{n}^{(j)} =1$, $W_{n}^{(j)}>0$) and the empirical measure
\begin{eqnarray}\label{00}
\mu_{n}^{J}(u)\equiv \sum_{j=1}^{J} W_{n}^{(j)}\delta_{u_{n}^{(j)}}(u)
\end{eqnarray}
converges to $\mu_{n}$ as $J\to \infty$ ($\delta_{w}$ denotes the Dirac measure concentrated at $w$). Moreover, if $f(u)$ denotes a quantity of interest of the unknown log-permeability $u(x)$, the weighted particles $\{W_{n}^{(j)},u_{n}^{(j)}\}_{j=1}^{J}$ can be easily used to compute the sample mean
\begin{eqnarray}\label{eq35A1}
\mathbb{E}^{\mu_{n}^{J}}(f(u))\equiv   \int_{X}f(u)\mu_{n}^{J}(du) = \sum_{j=1}^{J}W_{n}^{(j)}f(u^{(j)}) ,
\end{eqnarray}
which converges (see for example \cite{delmoral}) to the expectation (under $\mu_{ n}$) of the quantity of interest $\mathbb{E}^{\mu_{n}}(f(u))$.

The recursive computation of the weighted particles in SMC is suitable for the proposed application in RTM as it allows us to update, potentially in real time, our knowledge of the uncertainty in the log-permeability. However, producing accurate approximations of the Bayesian posteriors $\{\mu_{n}\}_{n=1}^{J}$ in the context of the inference of preform log-permeability in RTM represents a substantial computational challenge that arises from the fact that these posterior measures are defined on a (infinite-dimensional) functional space. Upon discretization, these posteriors could be potentially defined on a very high-dimensional space. Unfortunately, it has been shown \cite{David,Failure} that standard Bayesian sampling methodologies such as standard SMC do not scale well with the dimension of the (discretized) unknown; this leads to unstable and ultimately inaccurate algorithms.


The recent works \cite{David,Kantas} developed scalable (dimension independent) sampling algorithms for the approximation of the Bayesian posterior that arises from high-dimensional inverse problems. While these algorithms have a solid theoretical background that ensures their stability and convergence properties, achieving a desirable level of accuracy often comes at extremely high computational cost. More specifically, Bayesian methodologies, that provide approximation of the form (\ref{00}) and that converge asymptotically to the underlying posterior measure $\mu_{n}$, often involve solving the forward model thousands or even millions of times. In the context of the inverse problem for RTM, the numerical solution of the moving boundary (forward) problem in 2D or 3D settings is computationally very intensive. Therefore, the sequential approximation of the Bayesian posteriors of preform's log-permeability must be conducted with scalable computational efficiency so that it can be realistically used within a near real-time optimization loop for RTM. In the proposed work we develop a computational inverse framework that possess such computational efficiency with the ultimate aim of the real-time uncertainty quantification of the reinforced preform's log-permeability.


\subsection{Contributions of this work}

The contributions of this article are the following:
\begin{itemize}
\item[(A)] A Bayesian formulation of the inverse problem to infer log-permeability from sequential data collected during resin injection in RTM.  Both the 2D forward model described by (\ref{eq:2000})-(\ref{eq:2008}) as well as the corresponding 1D version are considered. For the 1D case, we show that application of the infinite-dimensional Bayesian framework of \cite{Andrew} leads to well-posedness of the sequence of Bayesian posteriors.
\item[(B)] Application of a state-of-the-art SMC framework \cite{Kantas} for the approximation of the sequence of Bayesian posteriors $\{\mu_{n}\}_{n=1}^{J}$ that arises from the Bayesian formulation. From this SMC framework, we motivate a novel regularizing ensemble Kalman algorithm (REnKA) that aims at approximating this sequence of posteriors in a computationally efficient manner, thus suitable for its implementation in a practical setting of RTM.
\item[(D)] Numerical investigation of the accuracy and robustness of the proposed REnKA scheme in the 1D case; this involves constructing, via the SMC sampler of \cite{Kantas}, accurate approximations of the posteriors that we use as Benchmark against which we compare the proposed REnKA. The advantages of REnKA in terms of accuracy vs computational cost are showcased by comparing it with the implementation of a low-resolution SMC whose computational cost is comparable to REnKA's.
\item[(E)] Application of REnKA for further investigations of the Bayesian inverse problem in both 1D and 2D. In particular for the 1D case we conduct a numerical investigation of the added value of assimilating the front location relative to the number of pressure sensors. Since the number of pressure sensors that can be physically deployed for preform permeability monitoring in RTM is usually limited, this investigation aims at providing practitioners with guidelines for the number of sensors that can accurately infer preform permeability alongside with its uncertainty. In addition, for the 1D case we study the effect of the frequency of the observations, as well as the observational noise level on the inference of the log-permeability. We further apply REnKA to the 2D forward model and, analogous to the 1D case, we study the effect that the number of pressure sensors have on the inferred log-permeability.
\end{itemize}

The rest of the paper is organized as follows. In Section \ref{sec:1d} we introduce the Bayesian inverse problem of inferring the permeability of a porous media in a 1D moving boundary problem for resin injection in RTM. In Section \ref{SMC_approach} we discuss and apply SMC to approximate the Bayesian posteriors that arise from the Bayesian approach. In Section \ref{ensemble} we introduce REnKA and conduct a numerical investigation of its approximation properties relative to its computational cost. In Section \ref{REnKA_application} we apply REnKA to further investigate relevant practical aspects of the inverse problem in both 1D and 2D; this includes the study of the effect of the number of pressure sensors as well as the noise level \mt{on} accuracy of the inferred log-permeability and its uncertainty. Some conclusions are presented in Section~\ref{Conclusions}.

\section{Bayesian inversion of a one-dimensional RTM model} \label{sec:1d}

In this section we apply the Bayesian approach to infer log-permeability in the context of the one-dimensional version of the forward problem defined in (\ref{eq:2000})-(\ref{eq:2008}). The corresponding 1D moving boundary problem induces a sequence of forward maps that we define in Section~\ref{1D} and that we aim at inverting with the Bayesian formalism that we introduce in Section~\ref{BIP_1D}. This sequence of 1D forward maps admits a closed-form solution that can be numerically approximated at a very low computational cost. This will enable us in Section~\ref{SMC_approach} to obtained accurate numerical approximations of the solution to the Bayesian inverse problem; we use this accurate approximations as a benchmark for assessing the approximation properties of the ensemble Kalman algorithm that we introduce in Section~\ref{ensemble} .

\subsection{The Forward 1D RTM model}\label{1D}
Let us consider a one-dimensional porous media with physical domain $D^{\ast}\equiv [0,x^{\ast}]\subset \mathbb{R}$. As before, we denote by $\kappa(x)$ ($x\in D^{\ast}$) and $\phi>0$ the permeability and porosity of the porous medium, respectively. Resin with viscosity $\mu$ is injected at $x=0$ at a pressure $p_{I}$. The pressure at the moving front (outlet) $\Ga(t)$ is prescribed and equal to $p_0$. The initial pressure distribution before injection is also set to $p_{0}$. For convenience of the subsequent analysis, we parameterize the permeability in terms of its natural logarithm $u(x)\equiv \log\kappa(x)$. The pressure $p(x,t)$ and the moving front $\Ga(t)$ are given by the solution to the following model
\begin{eqnarray}
\frac{d}{dx}\Bigg[  \frac{1}{\mu}e^{u(x)}\frac{d}{dx}p(x,t)\Bigg] &=&0,~~~~x\in (0,\Ga(t)),~~ t>0,  \label{eq1} \\
p(x,0) &=&p_{0},~~~~x\in (0,x^{\ast }],  \label{eq3} \\
p(0,t) &=&p_{I},~~~~t\geq 0,  \label{eq4} \\
\frac{d}{dt}\Ga(t) +  \frac{1}{\phi\mu}e^{u(\Ga(t))}\frac{d}{dx}p(\Ga(t),t)&=&0,~~t>0, ~~~\ \ \Ga(0)=0, \label{eq5} \\
p(\Ga(t),t) &=&p_{0},\ ~~~~~~~ t>0.  \label{eq6}
\end{eqnarray}
The solution to (\ref{eq1})-(\ref{eq6}) can be obtained analytically by the following proposition (see \cite{AS10,TW01,SRTM16}).
\begin{proposition}\label{Prop1}
Given $u\in X\equiv C[0,x^{\ast}]$, let us define
\begin{equation}\label{eq7}
F_{u}(x):=\int_{0}^{x}e^{-u(z)} dz,\qquad \textrm{and}\qquad W_{u}(x):=\int_{0}^{x}F_{u}(\xi)d\xi.
\end{equation}
The unique solution $\Ga(t)$, $p(x,t)$ of (\ref{eq1})-(\ref{eq6}) is given for $t\geq 0$ by
\begin{eqnarray}
\Ga(t) &=&W_{u}^{-1}\bigg((p_{I}-p_{0})\frac{t}{\mu\phi}\bigg),  \label{eq8}
\end{eqnarray}
\begin{eqnarray}\label{eq10}
p(x,t) = \left\{ \begin{array}{cc}
p_{I}-(p_{I}-p_{0})\frac{F_u(x)}{F_u(\Ga(t))}, & \  x\in D(t)\equiv (0,\Ga(t)),\\
p_{0},&  \  x\in D^{\ast}\setminus D(t), \end{array}\right.
\end{eqnarray}
\end{proposition}
 The quantity of interest arising from the RTM injection model is the so-called filling time: the time it takes the front $\Ga(t)$ to reach the right boundary of the domain of interest $[0,x^{\ast}]$. Filling time, denoted by $\tau^{\ast}$, is defined by $\Ga(\tau^{\ast})=x^{\ast}$. From (\ref{eq8}) and the definition in (\ref{eq7}) it follows \cite{SRTM16} that $\tau^*$ is given by
\begin{eqnarray} \label{eq10B}
\tau^{*}=\frac{\mu\phi}{(p_{I}-p_{0})}\int_{0}^{x^{*}}F_{u}(\xi)d\xi.
\end{eqnarray}

Note that the parameters $p_{0}$ and $p_{I}$ are prescribed control variables and thus known. In addition, we assume that $\mu$ and $\phi$ are known constants. As stated earlier, we are interested in the inverse problem of estimating the permeability, or more precisely its natural logarithm $u(x)=\log{\kappa}(x)$ given time-discrete measurements of the front location as well as the pressure from $M$ sensors located at $\{x_{m}\}_{m=1}^{M}\subset [0,x^{\ast}]$. We denote by $\{t_{n}\}_{n=1}^{N}$ the set of $N$ observation times. For fixed (assumed known) parameters $p_{I}$, $p_0$, $\phi$ and $\mu$, the solution to the PDE model (\ref{eq1})-(\ref{eq6}) induces the $n$th forward map $\cG_{n}:C[0,x^{\ast}]\to \mathbb{R}^{M+1}$ defined by
\begin{eqnarray}\label{eq11}
\cG_{n}(u) \equiv \Big[\cG_{n}^{\Ga}(u) ,\cG_{n}^{p}(u) \Big]^{T}=\Big[ \Ga(t_{n}\wedge \tau^{\ast}),\bigg\{p(x_{m},t_{n}\wedge \tau^{\ast}) \bigg\}_{m=1}^{M}  \Big]^{T}.
\end{eqnarray}
Given $u(x)=\log{\kappa(x)}\in X$, the evaluation of the forward map $\cG_{n}(u)$ predicts the location of the front and the pressure at the sensor locations at the time $t=t_{n}$. Since observation times are prescribed before the experiment, there is no assurance that for a given $u$, the corresponding filling time satisfies $t_{n}\leq \tau^{\ast}$ for all $n=1,\dots, N$. In other words, the front could reach the right end of the domain before we observe it at time $t_{n}$. In a real experimental setting, the process stops at time $\tau^{\ast}$. However, in the inverse problem of interest here, observation times are selected beforehand, and the search of optimal $u$'s within the Bayesian calibration of the $n$th forward map can lead to filling times greater than some observation times. In this case $(t_{n}>\tau^{\ast}$), the definition \eqref{eq11} yields $\cG_{n}(u) =[ \Ga(\tau^{\ast}),\{p(x_{m},\tau^{\ast})\}_{m=1}^{M}]^{T}$.

The following theorem ensures the continuity of the forward map, which is necessary for justifying the application of the Bayesian framework in Section \ref{Pos}.
\begin{theorem}\label{Theo1}
The forward map $\cG_{n}:C[0,x^{*}]\to \mathbb{R}^{M+1}$ is continuous.
\end{theorem}
For the proof of this theorem, see Appendix~\ref{Ape_Theo}.

In the following subsection we apply the Bayesian framework for inverse problems in order to invert observations of $\cG_{n}(u)$.
\begin{remark}\label{rem0}
We note that for the present work the porosity $\varphi$ is an assumed known constant; our objective is to infer the log-permeability $u(x)=\log{\kappa(x)}$. However, the Bayesian methodology that we apply can be extended to the case where the unknown is not only $\log{\kappa}(x)$ but also $\varphi$, and can include the case where $\varphi=\varphi(x)$ is a spatial function defined on the physical domain $D^{\ast}$.
\end{remark}

\subsection{The Bayesian Inverse Problem}\label{BIP_1D}

Suppose that, at each observation time $t=t_{n}$, we collect noisy measurements of the front location as well as pressure measurements from sensors. We denoted these measurement by $y_{n}^{\Ga}\in \mathbb{R}^{+}$ and $y_{n}^{p}\in \mathbb{R}^{M}$, respectively. Our aim is to solve the inverse problem of estimating the log permeability $u(x)=\log{\kappa(x)}$ given all the data $y_{1}^{p},y_{1}^{\Ga}, \dots, y_{n}^{p},y_{n}^{\Ga}$ up to time $t=t_{n}$. We assume that the aforementioned observations are related to the unknown $u(x)$, via the forward map (\ref{eq11}), in terms of
\begin{eqnarray}\label{eq12}
y_{n}^{p} &=\cG_{n}^{p}(u)+\eta_{n}^{p},\\
y_{n}^{\Ga} &=\cG_{n}^{\Ga}(u)+\eta_{n}^{\Ga},\label{eq12B}
\end{eqnarray}
where $\eta_{n}^{\Ga}$ and $\eta_{n}^{p}$ are realizations of Gaussian noise with zero mean and covariance $\Gamma_{n}^{\Ga}$ and $\Gamma_{n}^{p}$, respectively, i.e. $\eta_{n}^{\Ga}\sim N(0, \Gamma_{n}^{\Ga})$ and $\eta_{n}^{p}\sim N(0, \Gamma_{n}^{p})$ (i.i.d.). For simplicity we assume that both measurements of the front location and pressures are uncorrelated in time. We additionally assume that $\eta_{n}^{\Ga}$ and $\eta_{n}^{p}$ are uncorrelated for all $n=1,\dots, N$.

Note that (\ref{eq12})-(\ref{eq12B}) can be written as
\begin{eqnarray}\label{eq12C}
y_{n}=\cG_{n}(u)+\eta_{n},\qquad \eta_{n}\sim N(0,\Gamma_{n}),
\end{eqnarray}
with
\begin{eqnarray}\label{eq12D}
y_{n}\equiv\left[ \begin{array}{c}
y_{n}^{\Ga}  \\
y_{n}^{p} \end{array}\right]
,\qquad  \eta_{n}\equiv \left[ \begin{array}{c}
\eta_{n}^{\Ga} \\
\eta_{n}^{p} \end{array}\right],\qquad  \Gamma_{n}\equiv \left[ \begin{array}{cc}
\Gamma_{n}^{\Ga} &0\\
0 &\Gamma_{n}^{p}  \end{array}\right].
\end{eqnarray}
\begin{remark}\label{MT1}
Due to the nature of the RTM problem, we have that the pressure $p(x_{m},t)$ at each sensor $x_{m}$ should increase with time as well as the fact that $\cG_{n+1}^{\Ga}(u)\ge\cG_{n}^{\Ga}(u)$. However, the Gaussian noise in (\ref{eq12})-(\ref{eq12B}) can make the observations $y_{n}^{p}$ and $y_{n}^{\Ga}$ ``unphysical''. In practice, observations need to be post-processed before using them for the Bayesian inverse problem and unphysical $y_{n}^{p},y_{n}^{\Ga}$ should be excluded. We leave the question of how to incorporate such a post-processing framework for future study. Here we follow the traditional point of view on data modeled via (\ref{eq12})-(\ref{eq12B}) and choose sufficiently small $\Gamma_{n}^{\Ga}$ and $\Gamma_{n}^{p}$ so that the probability of $y_{n}^{p},y_{n}^{\Ga}$ being unphysical is very low.
\end{remark}

We adopt the Bayesian framework for inverse problems where the unknown $u(x)=\log{\kappa(x)}$ is a random field and our objective is to characterize the sequence of distributions of $u$ conditioned \mt{on} the observations which we express as $u\vert y_{1},\dots, y_{n}$. In other words, at each observation time $t=t_{n}$ we aim at computing the Bayesian posterior $\mu_{n}(u)=\mathbb{P}(u\vert y_{1},\dots, y_{n})$. From this distribution we can obtain point estimates of the unknown that can be used in real time to, for example, modify controls (\mt{e.g.. $p_{I}$}). More importantly, as we stated in the Introduction, the aforementioned distribution enables us to quantify uncertainty not only of the unknown but also of quantities of interest that may be relevant to an optimization of resin injection in RTM.

Even though for the \mt{illustrative} purposes the \mt{model} presented in this section \mt{is} discretized on a relatively low dimensional space (e.g. 60 cells), our aim is to introduce a general computational framework independent of the size of the discretized domain. We therefore consider an infinite-dimensional formulation of the Bayesian inverse problem for which the unknown $u$ belongs to a functional space $X$. The discretization of the Bayesian inverse problem will be conducted at the last stage of the computational algorithm, when the posteriors are sampled/approximated. Thus, we are aiming at robust mesh-invariant computational algorithms.

\subsubsection{The Prior}\label{Prior}

For the Bayesian approach that we adopt in this work, we require to specify a prior distribution $\mu_{0}(u)=\mathbb{P}(u)$ of the unknown, before the data are collected. This distribution comprises all our prior knowledge of the unknown and may include, for example, the regularity of the space of admissible solutions to the inverse problem. For the present work we consider Gaussian priors which have been used to characterize the uncertainty in the (log) permeability that arises from the preform fabrication
\mt{\cite{Zhang20111478,MAL14,Matveev} (see also references therein)}. In particular, here we consider stationary Gaussian priors $\mu_{0}=N(\overline{u},\cC)$ with covariance operator $\cC$ that arises from the Wittle-Matern correlation function defined by \cite{matern1,whittlematern,Rasmussen,Stein}:
  \begin{equation}\label{eq13}
c(x,y) =\sigma_{0}^{2}\frac{2^{1-\nu}}{\Gamma(\nu)}\Bigg(\frac{\vert x-y\vert }{l} \Bigg)^{\nu}K_{\nu}\Bigg(\frac{\vert x-y\vert }{l}\Bigg),
\end{equation}
where $\Gamma$ is the gamma function, $l$ is the characteristic length scale, $\sigma_{0}^2$ is an amplitude scale and $K_{\nu}$ is the modified Bessel function of the second kind of order $\nu$. The parameter $\nu$ controls the regularity of the samples. It can be shown \cite{Matt,Stein} that, for any $\nu>0$, if $u\sim \mu_{0}$, then $u\in C[0,x^{\ast}]$ almost-surely, i.e. $\mu_{0}([0,x^{\ast}])=1$. This requirement, together with the continuity of the forward map ensures the well-posedness of the Bayesian inverse problems as we discuss in the next subsection. In the context of composite preform's  permeability, it is natural to choose the mean $\overline{u}$ according to the log-permeability intended by the design of the composite part \cite{SRTM16}.

For computational purposes we use the prior to parametrize the unknown $u$ in terms of its Karhunen-Loeve (KL) expansion \cite{KL}:
\begin{eqnarray}\label{eqKL}
u(x)=\overline{u}(x)+\sum_{k=1}^{\infty}\lambda_{k}^{1/2} v_{k}(x) u_{k}
\end{eqnarray}
with coefficients $u_{k}$ and where $\lambda_{k}$ and $v_{k}$ are the eigenvectors and eigenfunctions of $\mathcal{C}$, respectively. A random draw from the prior $u\sim N(\overline{u},\mathcal{C})$ can then be obtained from (\ref{eqKL}) with drawing $u_{k}\sim N(0,1)$ i.i.d.

\subsubsection{The Posterior}\label{Pos}

From (\ref{eq12C}) and our Gaussian assumptions on the observational noise, it follows that for a fixed $u\in X$,
we have $y_{n}=\cG_{n}(u)+\eta_{n} \sim N(\cG_{n}(u),\Gamma_{n})$. Therefore, the likelihood of $y_{n} \vert u$ is given by
\begin{equation}\label{eq14}
 l_{n}(u,y_{n}) \propto \exp\Big[-\frac{1}{2}\vert\vert \Gamma_{n}^{-1/2}(y_{n}-\cG_{n}(u))\vert\vert^{2} \Big].
\end{equation}
At a given time $t=t_{n}$, the Bayesian posterior $\mu_{n}(u)=\mathbb{P}(u\vert y_{1},y_{2},\dots,y_{n})$ is defined by the following infinite-dimensional version of Bayes's rule.
\begin{theorem}[Bayes Theorem \cite{Andrew}]\label{Bayes}
Let $\{\cG_{s}\}_{s=1}^{N}$ be the sequence of forward maps defined by (\ref{eq11}) and let $\{l_{s}(u;y_{s})\}_{s=1}^{N}$ be the corresponding likelihood functions (\ref{eq14}). Let $\mu_0=N(\overline{u},\mathcal{C})$ be the prior distribution with correlation function (\ref{eq13}). Then, for each $n\in \{1,\dots,N\}$, the conditional distribution of $u|y_{1},\cdots,y_{n}$, denoted by $\mu_{n}$, exists. Moreover, $\mu_{n}\ll\mu_0$ with the Radon-Nikodym derivative
\begin{eqnarray}\label{eq17}
\frac{d \mu_{n}}{d\mu_{0}}(u) =\frac{1}{Z_{n}} \prod\limits_{s=1}^{n}l_{s}(u,y_{s}),
\end{eqnarray}
where
\begin{eqnarray}\label{eq18}
Z_{n}=\int_{X} \prod\limits_{s=1}^{n} l_{s}(u,y_{s})\mu_{0}(u)du>0.
\end{eqnarray}
\end{theorem}
\textbf{Proof:}
The proof follows from the application of Theorem 6.31 in \cite{Andrew} and the continuity of the forward maps (Theorem \ref{Theo1}) on a full $\mu_0$-measure set $X$.
$\Box$

Note that from our assumption of independence of $\eta_{1},\dots,\eta_{n}$, the right hand side of (\ref{eq17}) is the likelihood of $y_{1},\dots,y_{n}\vert u$.

\begin{remark}\label{rem1}
Due to the assumption of independence between front location and pressure measurements, the likelihood (\ref{eq14}) can be expressed as
\begin{equation}\label{eq14A}
 l_{n}(u,y_{n}) \propto  l_{n}^{p}(u,y_{n}^p)l_{n}^{\Ga}(u,y_{n}^{\Ga}),
\end{equation}
where
\begin{eqnarray}\label{eq14B}
 l_{n}^{\beta}(u,y_{n}^{\beta}) \propto  \exp\Big[-\frac{1}{2}\vert\vert [\Gamma_{n}^{\beta}]^{-1/2}(y_{n}^{\beta}-\cG_{n}^{\beta}(u))\vert\vert^{2} \Big],\qquad \beta\in\{p,\Ga\}.
\end{eqnarray}
This enables us to define two particular cases of the inverse problem. The first case corresponds to the assimilation of only pressure measurements $y_{n}^{p}$, while in the second case only front \mt{location measurements}  $y_{n}^{\Ga}$ are assimilated. Similar arguments to those that led to Theorem \ref{Bayes} can be applied (with $l_{s}^{\beta}(u,y_{n}^{\beta})$ instead of $l_{s}(u,y_{n})$) to define the Bayesian posteriors $\mu_{n}^{p}$ and $\mu_{n}^{\Ga}$ associated to these two Bayesian inverse problems. In Section \ref{meas} we will study these two particular cases with an eye towards understanding the added value of assimilating observations of the front location with respect to assimilating only pressure measurements.
\end{remark}

\section{Approximating the posteriors via Sequential Monte Carlo method}\label{SMC_approach}

In the previous section we have established the well-posedness of the Bayesian inverse problem associated to inferring the log-permeability in the one-dimensional moving boundary problem (\ref{eq1})-(\ref{eq6}). The solution of this inverse problem is the sequence of posterior measures $\{\mu_{n}\}_{n=1}^{N}$ defined by Theorem \ref{Bayes}. As we discussed in Section \ref{Intro}, these posteriors cannot be expressed analytically and so a sampling approach is then required to compute the corresponding approximations. Note that the sampling of each posterior $\mu_{n}$ ($n=1,\dots,N$) can be performed independently by, for example, Markov chain Monte Carlo (MCMC) methods. However, we reiterate that, for the present application SMC samplers are rather convenient as they exploit the sequential nature of the \mt{considered} inverse problem by enabling a recursive approximation of the posterior measures as new data (in time) become available. Such recursive approximations of the posterior could enable practitioners to update their probabilistic knowledge of preform's log-permeability which is, in turn, essential to develop real-time optimal control strategies for RTM under the presence of uncertainty.

Recognizing that the inverse problem under consideration involves inferring a function potentially discretized on a very fine grid, it is vital to consider the application of SMC samplers such as the one introduced in \cite{Kantas}, carefully designed for approximating measures defined on a high-dimensional space. In this section we review and apply this scheme for the approximation of the Bayesian posteriors $\{\mu_{n}\}_{n=1}^{N}$ that we defined in the previous section. The aims of this section are to (i) provide a deeper quantitative understanding of the accuracy of the fully-Bayesian methodology of \cite{Kantas} with respect to its computational cost under practical computational conditions; (ii) provide a motivation for the proposed REnKA that we propose from this SMC sampler in Section \ref{ensemble}; and (iii) define accurate approximations of $\{\mu_{n}\}_{n=1}^{N}$ which we use as a benchmark for testing our REnKA scheme. 

In Section \ref{standard_SMC} we briefly discuss the essence of the standard SMC  that we then use in \mt{Sections~\ref{Kantas_SMC}-\ref{note}} to review methodological aspects of the adaptive-tempering SMC sampler for high-dimensional inverse problems of \cite{Kantas}. We then apply this SMC in Section \ref{SMC_1d_ex} for the solution of the Bayesian inverse problem in the 1D case defined in the previous section.
\mt{In Section~\ref{SMC_small} we assess practical limitations of the SMC.}

%

\subsection{Standard SMC for Bayesian inference}\label{standard_SMC}

As we discussed in the Introduction, starting with the prior $\mu_{0}$, the objective of SMC is to recursively compute an approximation of the sequence of Bayesian posteriors $\{\mu_{n}\}_{n=1}^{N}$ in terms of weighted particles. More specifically, assume that at the observation time $t_{n}$, we have a set of $J$ particles $\{u_{n-1}^{(j)}\}_{j=1}^{J}$ with, for simplicity, equal weights ($W_{n}^{(j)}=1/J$, $j=1,\dots,J$), which provides the following particle approximation of $\mu_{n-1}(u)=\mathbb{P}(u\vert y_{1},\dots, y_{n-1})$:
\begin{eqnarray}\label{eq35A0}
\mu_{n-1}^{J}(u)\equiv \frac{1}{J}\sum_{j=1}^{J} \delta_{u_{n-1}^{(j)}}(u)\approx \mu_{n-1}(u).
\end{eqnarray}
The objective now is to construct a particle approximation of $\mu_{n}(u)=\mathbb{P}(u\vert y_{1},\dots, y_{n})$, which includes the new data $y_{n}$ collected at time $t_{n}$. In a standard SMC framework \cite{del2004feynman,Chopin01082002,Doucet}, this particle approximation is constructed by means of an importance sampling step with proposal distribution $\mu_{n-1}$. To illustrate this methodology, let us first note formally that
\begin{eqnarray}\label{eq35A2}
\mathbb{E}^{\mu_{n}}(f(u))\equiv \int_{X}f(u)\mu_{n}(du) =\frac{Z_{n-1}}{Z_{n}}\int_{X}f(u) l_{n}(u,y_{n})\mu_{n-1}(du)\nonumber\\
 = \Big[\int_{X} l_{n}(u,y_{n})\mu_{n-1}(du)\Big]^{-1}\int_{X}f(u) l_{n}(u,y_{n})\mu_{n-1}(du),
\end{eqnarray}
where we have used
\begin{eqnarray}\label{eq35A3}
 \frac{d \mu_{n}}{d\mu_{n-1}}(u) =\frac{Z_{n-1}}{Z_{n}} l_{n}(u,y_{n})\quad \textrm{and} \quad  \frac{Z_{n}}{Z_{n-1}}=\int_{X} l_{n}(u,y_{n})\mu_{n-1}(du),
\end{eqnarray}
which can be obtained directly from Theorem \ref{Bayes}. An approximation of (\ref{eq35A2}) can be obtained by
\begin{eqnarray}\label{eq35A4}
\mathbb{E}^{\mu_{n}^{J}}(f(u)) = c_{n}^{-1}\sum_{j=1}^{J}f(u^{(j)}) l_{n}(u^{(j)},y_{n})=\sum_{j=1}^{J}W_{n}^{(j)}f(u^{(j)}),
\end{eqnarray}
where
\begin{eqnarray}\label{eq35A5}
W_{n}^{(j)}\equiv c_{n}^{-1}  l_{n}(u_{n}^{(j)},y_{n}), \qquad c_{n}\equiv  \sum_{j=1}^{J}l_{n}(u_{n-1}^{(j)},y_{n}).
\end{eqnarray}
From (\ref{eq35A4}) we see that the importance (normalized) weights $W_{n}^{(j)}$ assigned to each particle $u_{n-1}^{(j)}$ define the following empirical (particle) approximation of $\mu_{n}$:
\begin{eqnarray}\label{eq35A6}
\mu_{n}^{J}(u)\equiv \sum_{j=1}^{J}W_{n}^{(j)} \delta_{u_{n-1}^{(j)}}(u).
\end{eqnarray}
However, the accuracy of such empirical approximation relies on $\mu_{n-1}$ being sufficiently close to $\mu_{n}$; when this is not the case, after a few iterations (observation times) the algorithm may produce only a few particles with nonzero weights. This is a well-known issue of weight degeneracy that often arises from the application of empirical (importance sampling) approximations within the context SMC samplers \cite{Failure}. Weight degeneracy is routinely measured in terms of the Effective Sample Size (ESS) statistic \cite{Kong}:
\begin{eqnarray}\label{eq37}
\textrm{ESS}\equiv \Bigg[\sum_{j=1}^{J}(W_{n}^{(j)})^2\Bigg]^{-1},
\end{eqnarray}
which takes a value between 1 and $J$; $ESS=J$ when all weights are equal and $ESS=1$ when the distribution is concentrated at one single particle. A common approach to alleviate weight degeneracy is, for example, to specify a threshold for the ESS below which resampling  (often multinomially) according to the weights $\{W_{n}^{(j)}\}_{j=1}^{J}$ is performed. Resampling discards particles with low weights by replacing them with several copies of particles with higher weights. The approximation of a sequence of measures via the combination of the importance sampling step followed with resampling leads to the Sequential Importance Resampling (SIR) scheme \cite{Doucet}.

It is important to note that the aforementioned resampling step in SIR can clearly lead to the lack of diversity in the population of resampled particles. This is, in turn, detrimental to the approximation of the sequence of posteriors. The general aim of the standard SMC approach is to diversify these particles by a mutation step with involves replacing them with samples from a Markov kernel $\mathcal{K}_{n}$ with invariant distribution $\mu_{n}$. In the following subsection we provide a discussion of the aforementioned mutation in the context of the SMC sampler for high-dimensional inverse problem \cite{Kantas}. We refer the reader to \cite{delmoral,del2004feynman,Chopin01082002,Doucet} for a thorough treatment of more standard SMC samplers.

\subsection{SMC for high-dimensional inverse problems}\label{Kantas_SMC}

The weight degeneracy in the importance sampling step described above is more pronounced when the two consecutive measures $\mu_{n-1}$ and $\mu_{n}$ differ substantially from each other. This has been particularly associated with complex (e.g. multimodal) measures defined in high-dimensional spaces. When the change from $\mu_{n-1}$ to $\mu_{n}$ is abrupt, the importance sampling step can result in a sharp failure, whereby the approximation of $\mu_{n}$ is concentrated on a single particle \cite{Failure}. Recent work for high-dimensional inference problems has suggested \cite{Kantas,BJMS15} that further stabilization of the importance  weights is needed by defining a smooth transition between $\mu_{n-1}$ and $\mu_{n}$. For the present work, we consider the annealing approach of \cite{Neal2001,delmoral}, where $q_{n}$ intermediate artificial measures $\{ \mu_{n,r}\}_{r=0}^{q_{n}}$ are defined such that $\mu_{n,0}=\mu_{n-1}$ and $\mu_{n,q_{n}}=\mu_{n}$. These measures can be bridged by introducing a set of $q_{n}$ tempering parameters denoted by $\{\phi_{n,r}\}_{r=1}^{q_{n}}$ that satisfy $0=\phi_{n,0}<\phi_{n,1}<\cdots< \phi_{n,q_{n}}=1$ and defining each $\mu_{n,r}$ as the probability measure with density proportional to $l_{n}(u,y_{n})^{\phi_{n,r}}$ with respect to $\mu_{n-1}$. More specifically, $\mu_{n,r}$ satisfies
\begin{eqnarray}\label{eq37A1}
\frac{d \mu_{n,r}}{d\mu_{n-1}} \propto  \Big[ l_{n}(u,y_{n})\Big]^{\phi_{n,r}},
\end{eqnarray}
which, formally, implies
\begin{eqnarray}\label{eq37A2}
 \frac{d \mu_{n,r}}{d\mu_{n,r-1}} \propto \Big[l_{n}(u,y_{n})\Big]^{\phi_{n,r}-\phi_{n,r-1}}.
\end{eqnarray}
Note that when $q_{n}=1$, $\phi_{n,1}-\phi_{n,0}=1$ and so expression (\ref{eq37A2}) reduces to (\ref{eq35A3}). We now follow the SMC algorithm for high-dimensional inverse problems as described in \cite{Kantas}.

\subsubsection{Selection Step}

The first stage of the SMC approach of \cite{Kantas} is a selection step which consists of careful selection of the tempering parameters which define the intermediate measures $\{\mu_{n,r}\}_{r=0}^{q_{n}}$; these are in turn approximated by the application of the SIR scheme described above. Let us then assume that at an observation time $t_{n}$ and iteration level $r-1$, the tempering parameter $\phi_{n,r-1}$ has been specified, and that a set of particles  $u_{n,r-1}^{(j)}$ provides the following approximation (with equal weights) of the intermediate measure $\mu_{n,r-1}$:
\begin{equation}\label{eq37B}
\mu_{n,r-1}^{J}(u)\equiv \frac{1}{J}\sum_{j=1}^{J} \delta_{u_{n,r-1}^{(j)}}(u)\approx \mu_{n,r-1}(u).
\end{equation}
From (\ref{eq37A2}) we can see that the new tempering parameter $\phi_{n,r}$ must be selected to ensure that $\phi_{n,r}-\phi_{n,r-1}$ is sufficiently small, so that the subsequent measure $\mu_{n,r}$ is close to $\mu_{n,r-1}$ thus preventing a sharp failure of the empirical approximation of $\mu_{n,r}$ (\ref{eq35A6}). In particular, once the next tempering parameter $\phi_{n,r}$ is specified, we note from expression (\ref{eq37A2}) that the importance weights for the approximation of $\mu_{n,r}$ are given by
\begin{eqnarray}\label{eq36}
W_{n,r}^{(j)} =\mathcal{W}_{n,r-1}^{(j)}[\phi_{n,r}] = \frac{\big[l_{n}(u_{n,r-1}^{(j)},y_{n})\big]^{\phi_{n,r}-\phi_{n,r-1}}}{\sum_{s=1}^{J}\big[l_{s}(u_{n,r-1}^{(s)},y_{n})\big]^{\phi_{n,r}-\phi_{n,r-1}}}.
\end{eqnarray}
Recognizing that the ESS in (\ref{eq37}) quantifies weight degeneracy in SIR, the approach of \cite{Kantas} (see also \cite{Ajay}) proposes to define on-the-fly the next tempering parameter $\phi_{n,r}$ by imposing a fixed, user-defined value $J_{thres}$ \mt{on} the ESS. More specifically, $\phi_{n,r}$ is defined by the solution to the following equation:
\begin{eqnarray}\label{eq38}
ESS_{n,r}(\phi)\equiv  \Bigg[\sum_{j=1}^{J}(\mathcal{W}_{n,r-1}^{(j)}[\phi])^2\Bigg]^{-1}= J_{thres},
\end{eqnarray}
which may, in turn, be solved by a simple bisection algorithm \mt{on} the interval $(\phi_{n,r-1},1]$. An approximation of $\mu_{n,r}$ is then given by the weighted particle set $\{u_{n,r-1}^{(j)},W_{n,r}^{(j)}\}_{j=1}^{J}$. If at the $r-1$ level, we find that $ESS_{n,r}(1)>J_{thresh}$, \mt{it} implies that no further tempering is required and thus one can simply define $\phi_{n,r}=1$. We note that the number of tempering steps $q_n$ is random.

While the tempering approach described above is aimed at preventing ESS from falling below a specified threshold $J_{thres}$ and thus avoiding a sharp failure of the empirical approximation of $\mu_{n,r}$, resampling is still required to discard particles with very low weights. Let us then denote by $\hat{u}_{n,r}^{(j)}$ ($j=1,\dots, J$) the particles, with equal weights, that result from resampling with replacement \mt{of} the set of particles $u_{n,r-1}^{(j)}$ according to the weights $W_{n,r}^{(j)}$.

\subsubsection{Mutation Phase}

As stated in the preceding subsection, at the core of the SMC methodology is a mutation phase that adds diversity to the population of the resampled particles $\hat{u}_{n,r}^{(j)}$. In the context of the tempering approach described above, this mutation is conducted by means of sampling from a Markov kernel $\mathcal{K}_{n,r}$ with invariant distribution $\mu_{n,r}$. Similar to the approach of \cite{Kantas}, here we consider mutations given by running $N_{\mu}$ steps of an MCMC algorithm with $\mu_{n,r}$ as its target distribution. More specifically,
we consider the preconditioned Crank-Nicolson (pcn)-MCMC method from \cite{David} with target distribution $\mu_{n,r}$ and reference measure $\mu_{0}$. Formally, these two measures are related by
 \begin{eqnarray}\label{eq34C}
\frac{d \mu_{n,r}}{d\mu_{0}} \propto  l_{n,r}(u,y_{n})\equiv \big[l_{n}(u,y_{n})\big]^{\phi_{n,r}} \prod\limits_{s=1}^{n-1}l_{s}(u,y_{s}).
\end{eqnarray}
The pcn-MCMC method for sapling $\mu_{n,r}$ is summarised in Algorithm \ref{MCMC_al} (see Appendix~\ref{Ape_SMC}). Under reasonable assumptions this algorithm produces a $\mu_{n,r}$-invariant Markov kernel \cite{Kantas}. The resulting particles denoted by $u_{n,r}^{(j)}$ ($u_{n,r}^{(j)}\sim \mathcal{K}_{n,r}(\hat{u}_{n,r}^{(j)},\cdot)$) then provide the following particle approximation of $\mu_{n,r}$:
\begin{eqnarray}\label{eq39}
\mu_{n,r}^{J}\equiv \frac{1}{J}\sum_{j=1}^{J} \delta_{u_{n,r}^{(j)}} \to \mu_{n,r}\qquad \textrm{as}~~J\to\infty,  \end{eqnarray}
where the convergence is proven in a suitable metric for measures \cite{BJMS15}. Note that at the end of the iteration $r=q_{n}$, the corresponding particle approximation $\mu_{n}^{J}=\mu_{n,q_{n}}^{J}\equiv \frac{1}{J}\sum_{j=1}^{J} \delta_{u_{n,q_{n}}^{(j)}}$ provides the desired approximation of the posterior that arises from the Bayesian inverse problem of interest. This SMC sampler is summarized in Algorithm~\ref{SMC_al} (see Appendix~\ref{Ape_SMC}).
\begin{remark}\label{resampling}
For simplicity, here we \mt{use the} resampling step at every iteration of the SMC sampler. However, whenever $ESS_{n,r}(1)>J_{thresh}$ (and so $\phi_{n,r}=1$)
\mt{the resampling step} can be \mt{skipped}; this involves using the corresponding weighted particle approximation and modifying the formula for the incremental weights as discussed in \cite[Section 4.3]{Kantas}.
\end{remark}

\subsection{A note on tempering}\label{note}
Let us define the following inverse of the increment in tempering parameters:
\begin{eqnarray}\label{reg_param}
\alpha_{n,r}=\frac{1}{ \phi_{n,r}-\phi_{n,r-1}},
\end{eqnarray}
and note that $0\leq \phi_{n,r}\leq 1$ implies $\alpha_{n,r}\ge 1$. In addition, expression (\ref{eq37A2}) can be written as
\begin{eqnarray}\label{reg_param2}
 \frac{d \mu_{n,r}}{d\mu_{n,r-1}} \propto \big[l_{n}(u,y_{n})\big]^{\alpha_{n,r}^{-1}}  \propto \exp \Bigg[ -\frac{1}{2\alpha_{n,r}}\vert\vert (\Gamma_{n})^{-1/2}(y_{n}-\cG_{n}(u))\vert\vert^2  \Bigg]\nonumber\\
=\exp\Bigg[ -\frac{1}{2}\vert\vert (\alpha_{n,r}\Gamma_{n})^{-1/2}(y_{n}-\cG_{n}(u))\vert\vert^2  \Bigg],
\end{eqnarray}
where we have used the definition of the likelihood in (\ref{eq14}). Informally, we can then interpret each iteration of the SMC sampler (at a given observation time $t_{n}$) as the solution of a Bayesian inverse problem that consists of finding $\mu_{n,r}$ given the prior $\mu_{n,r-1}$ and the data:
\begin{eqnarray}\label{eq42}
 y_{n}=\cG_{n}(u)+\tilde{\eta}_{n},\qquad \tilde{\eta}_{n,r}\sim N(0,\alpha_{n,r}\Gamma_{n}).
\end{eqnarray}
From (\ref{reg_param}) and the fact that $0\leq \phi_{n,r}\leq 1$, it follows that $ \alpha_{n,r}\ge 1$. Therefore, (\ref{eq42}) is nothing but the original problem (\ref{eq12C}) albeit with a noise $\tilde{\eta}_{n,r}$ that has an inflated covariance $\alpha_{n,r}\Gamma_{n}$.

We also note that $\alpha_{n,r}$ plays the role of a regularization parameter in the sense that it controls the transition between $\mu_{n,r-1}$ and $\mu_{n,r}$. The larger the $\alpha_{n,r}$ the smoother this transition. Alternatively, we can see that $\alpha_{n,r}$ can be interpreted as a ``temperature'' in the tempering scheme which, in turn, flattens out the likelihood function at the observation time $t_{n}$. Clearly, more tempering will be required whenever $\vert\vert (\Gamma_{n})^{-1/2}(y_{n}-\cG_{n}(u))\vert\vert^2$ is large; this can for example happen if the observational data are accurate (i.e small $\Gamma_{n}$) and/or many observations are available.

The amount of tempering is controlled by the number of parameters obtained via (\ref{eq38}). The greater the number of tempering parameters, the larger the $\alpha_{n,r}$'s  which in turn indicates that more regularization is needed to ensure a stable transition between those measures. This has also, in turn, an associated increase in iterations and thus in computational cost.

\subsubsection{Computational aspects of SMC}

The main computational cost of the SMC sampler previously discussed is attributed to the mutation step for which $N_{\mu}$ steps of the pcn-MCMC algorithm are performed. At each observation time $t_{n}$ and iteration $r$, the SMC sampler then requires $J\ N_{\mu}$ evaluations of the $n$th forward map $\cG_{n}$. Therefore, the computational cost of computing $\mu_{n}$ is $ q_{n}  g_{n}J \ N_{\mu} $, where $g_{n}$ denotes the computational cost of evaluating $\cG_{n}$ which, in turn, corresponds to solving the moving boundary
\mt{problem} from time $t=0$ up to time $t_{n}$. The total computational cost of computing the full sequence of posteriors $\{\mu_{n}\}_{n=1}^{N}$ is then
\begin{eqnarray}\label{cost_SMC}
\mathbf{C}_{SMC}\equiv J \ N_{\mu}  \sum_{n=1}^{N} q_{n}  \frac{g_{n}}{g_{N}},
\end{eqnarray}
which is expressed in terms of $g_{N}$, the cost of evaluating $\cG_{N}$ (i.e. solving the forward model from time zero up to the final observation time).

The work of  \cite{Kantas} has suggested that accurate approximations of the posterior via SMC samplers require, for example, values of $N_{\mu}=20$ and $J=10^4$. If we assume for a moment that only one observation time $N=1$ is considered and that only one tempering step $q_{1}=1$ \mt{is} required to compute $\mu_{1}$, the computational cost in this case would be approximately $10^5$ times the cost of solving the forward model from time $t=0$ up to time $t_{1}$. Such cost would be clearly computationally prohibitive for practical applications, where the aforementioned forward simulation may take several minutes of CPU time. In particular, for the 2D or 3D version of the RTM process, the high computational cost of the SMC sampler becomes impractical.  While reducing the values of $J$ and $N_{\mu}$ may result in a more affordable computational cost, this is substantially detrimental to the level of accuracy of the SMC sampler as we show via numerical experiments in Section \ref{SMC_small}.

\subsection{Numerical examples with SMC}\label{SMC_1d_ex}
In this subsection we report the results from the numerical application of the SMC sampler discussed in the previous subsection. The objective is to approximate the sequence of Bayesian posteriors that arise from the 1D moving boundary problem defined in Section~\ref{sec:1d} for the experimental set-up described in Section \ref{set_up}. In Section~\ref{benchmark}
we discuss the numerical results obtained via the SMC sampler with a very high number of particles which results in accurate approximations of the Bayesian posteriors. These approximations are then used in Section~\ref{SMC_small} to assess the practical limitations of the scheme under certain choices of tunable parameters and number of particles. These limitations motivate the approximate methods that we propose in Section \ref{ensemble}.

\subsubsection{Experimental set-up}\label{set_up}

\begin{figure}[h]
\begin{center}

\includegraphics[scale=0.35]{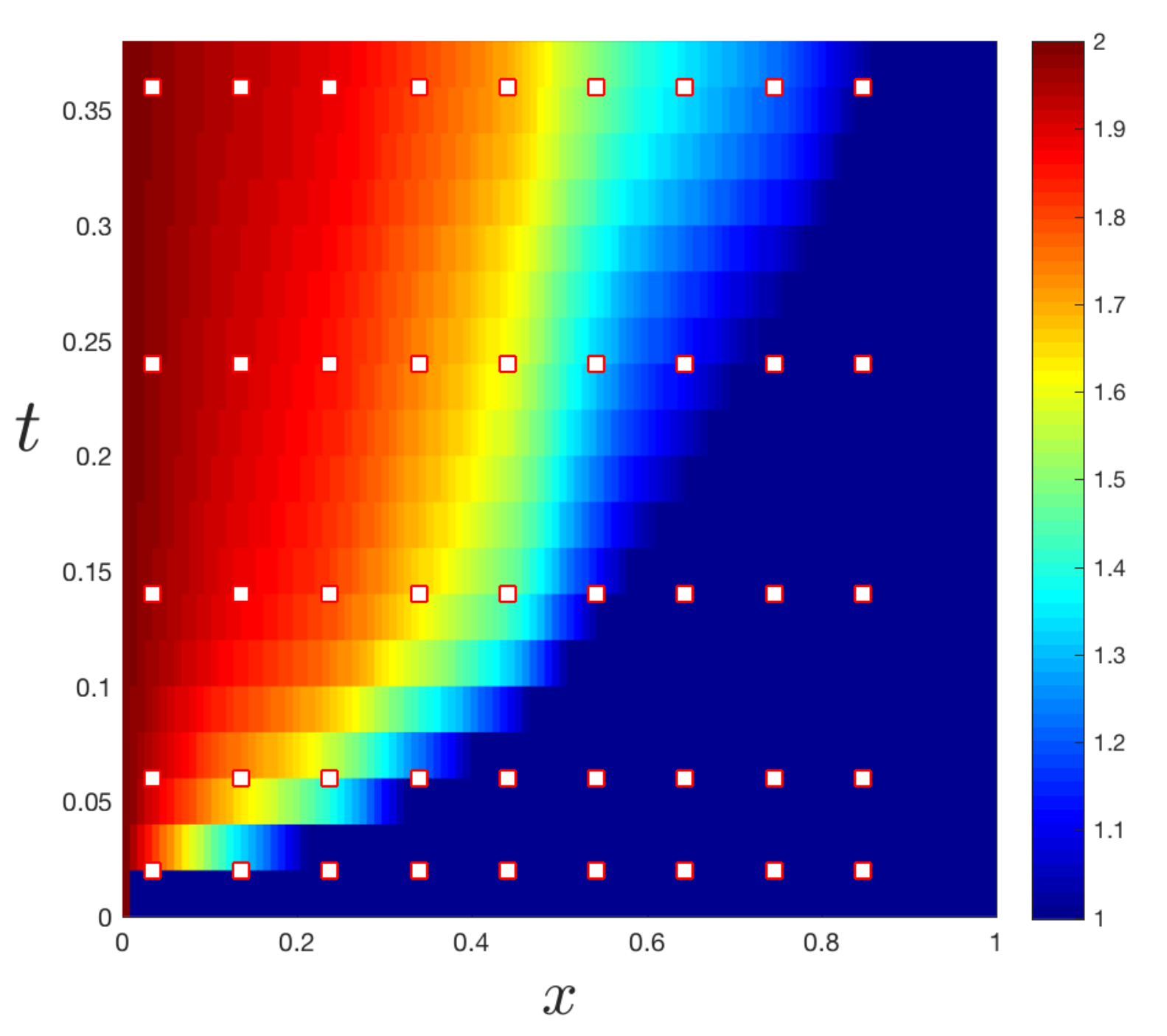}
\includegraphics[scale=0.35]{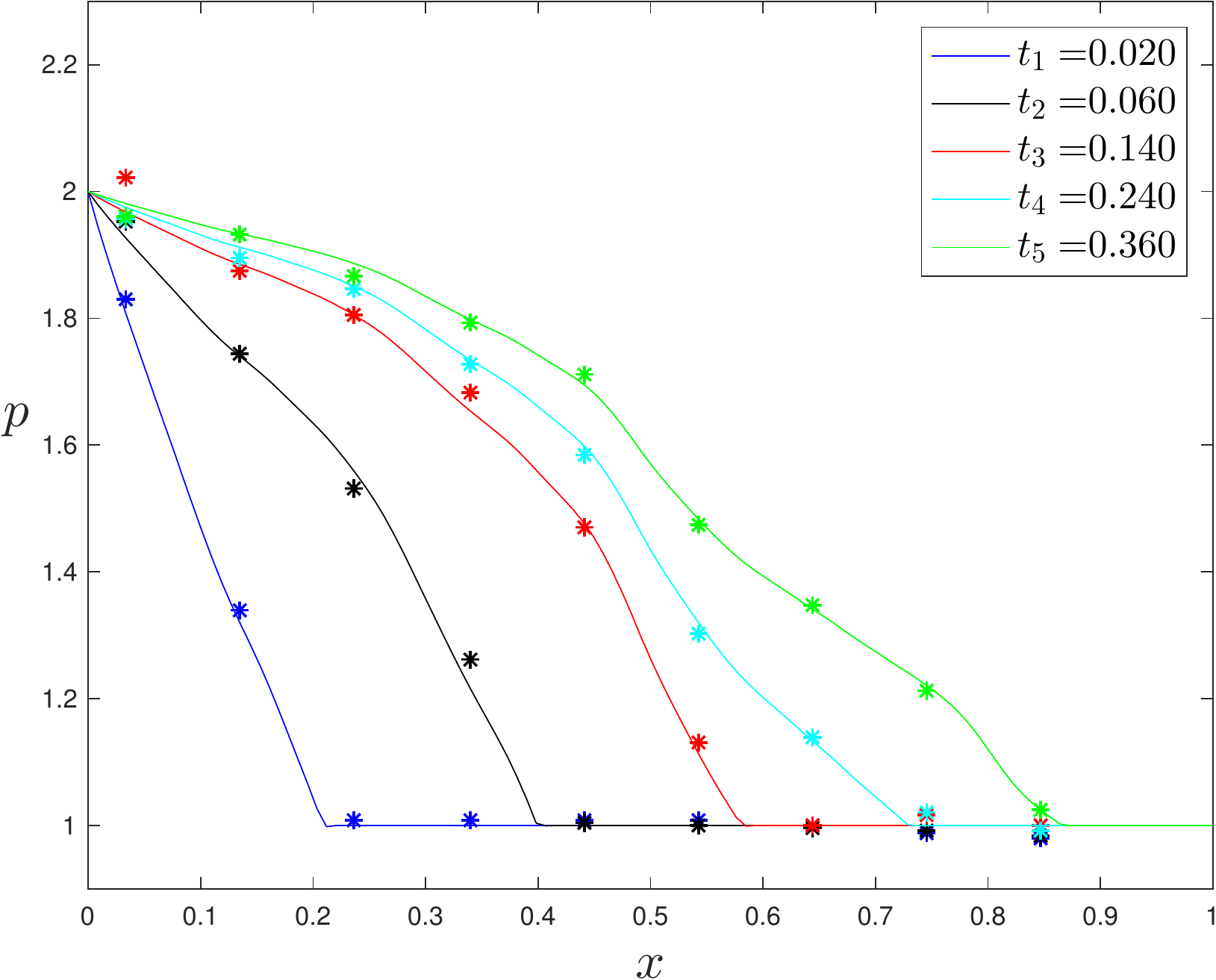}
\includegraphics[scale=0.35]{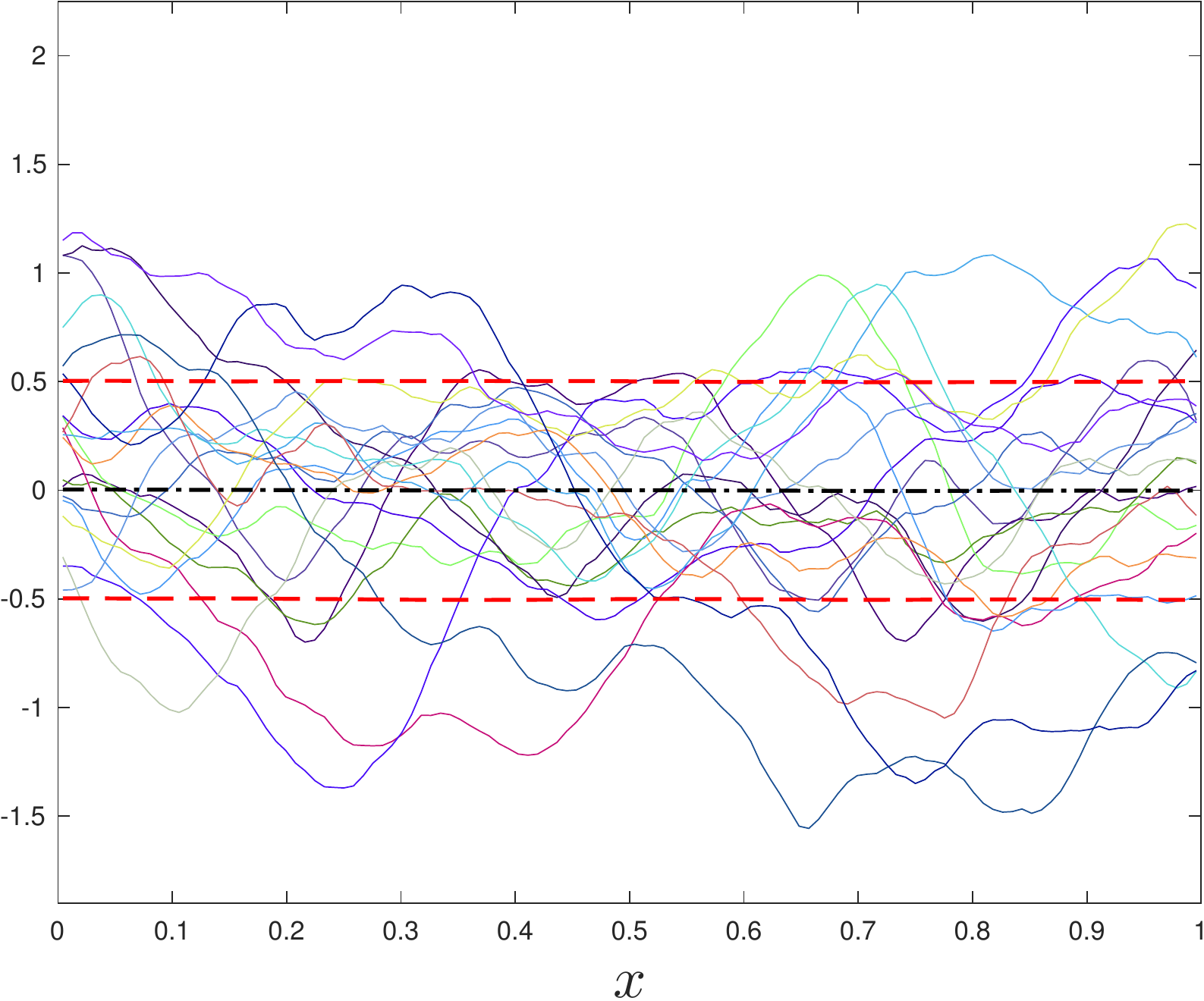}
 \caption{Left: True pressure field $p^{\dagger}(x,t)$ and space-time measurement configuration with $M=9$ sensors and $N=5$ observation times. Middle: True pressure at observation times $\{p^{\dagger}(x,t_{n})\}_{n=1}^{5}$ (curves) together with the corresponding synthetic data $y_{n}^{p}$ (dots). Right: Samples from the prior.} \label{Fig1}
\end{center}
\end{figure}

We consider a dimensionless version of the one-dimensional model (\ref{eq1})-(\ref{eq6}) \mt{which together with its numerical approximation} is described in Appendix~\ref{num_imple}. The dimensionless values for the control variables are $p_{0}=1$ and $p_{I}=2$. We use a Gaussian prior distribution $\mu_{0}=N(\overline{u},\mathcal{C})$ with the covariance operator $\mathcal{C}$ that arises from the covariance function defined in (\ref{eq13}).  We numerically solve (off-line) the eigenvalue problem associated to the matrix that results from discretizing $\mathcal{C}$; the corresponding eigenvector/eigenvalues are then stored for subsequent use in the parameterization of the log-permeability in the SMC sampler. The KL expansion (\ref{eqKL}) becomes a truncated sum with a number of elements equal to the the total number of eigenvalues of this matrix; these are, in turn, equal to the number of cells used for the discretization of the domain $D^{\ast}=[0,1]$. No further truncation to this KL expansion is carried out. A few samples from the prior are displayed in Figure \ref{Fig1} (right). Pointwise percentiles (0.02, 0.25, 0.5, 0.75 and 0.98) of the prior are displayed in Figure \ref{Fig2} (top-left). Tuneable parameters of the prior for the present experiments are $\sigma_{0}^{2}=0.5$, $\nu=1.5$, $l=0.05$ and $\overline{u}(x)= 0.0$ for all $x\in D^{\ast}$.

In order to generate synthetic data, we define the ``true/reference'' log permeability field  $u^{\dagger}$ whose graph (red curve) is displayed in Figure \ref{Fig2} (top-left); this function is a random draw from the prior described above. We use $u=u^{\dagger}$ in the numerical implementation of (\ref{eq1})-(\ref{eq6}) in order to compute the true pressure field $p^{\dagger}(x,t)$ as well as the true front location $\Ga^{\dagger}(t)$. The plot of $p^{\dagger}(x,t)$  is shown in Figure \ref{Fig1} (left) together with the space-time configuration of $M=9$ pressure sensors and $N=5$ observation times. The graphs of $\{p(x,t_{n})\}_{n=1}^{5}$ are shown in Figure~\ref{Fig1} (middle). The true locations of the front $\{\Ga^{\dagger}(t_{n})\}_{n=1}^{5}$ are 0.21 ,0.40, 0.58, 0.73 and 0.87. Synthetic data are then generated by means of $y_{n}^{p} =\{p^{\dagger}(x_{m},t_{n})\}_{m=1}^{9}+\eta_{n}^{p}$ and $y_{n}^{\Ga}=\Ga^{\dagger}(t_{n})+\eta_{n}^{\Ga}$, where $\eta_{n}^{p}$ and $\eta_{n}^{\Gamma}$ are Gaussian noise (see subsection \ref{BIP_1D}) with standard deviations equal to $1.5\%$ of the size of the noise-free observations. Synthetic pressure data $\{y_{n}^{p}\}_{n=1}^{5}$ are superimposed on the graphs of $\{p(x,t_{n})\}_{n=1}^{5}$ in Figure \ref{Fig1} (middle). Synthetic front locations $\{y_{n}^{\Ga}\}_{n=1}^{5}$ are  0.21,  0.39, 0.59, 0.74, 0.86. In order to avoid inverse crimes, synthetic data are generated by using a finer discretization (with 120 cells) than the one used to approximate the posteriors (with 60 cells).

\subsubsection{Application of SMC}\label{benchmark}

\begin{figure}[htbp]
\begin{center}

\includegraphics[scale=0.3]{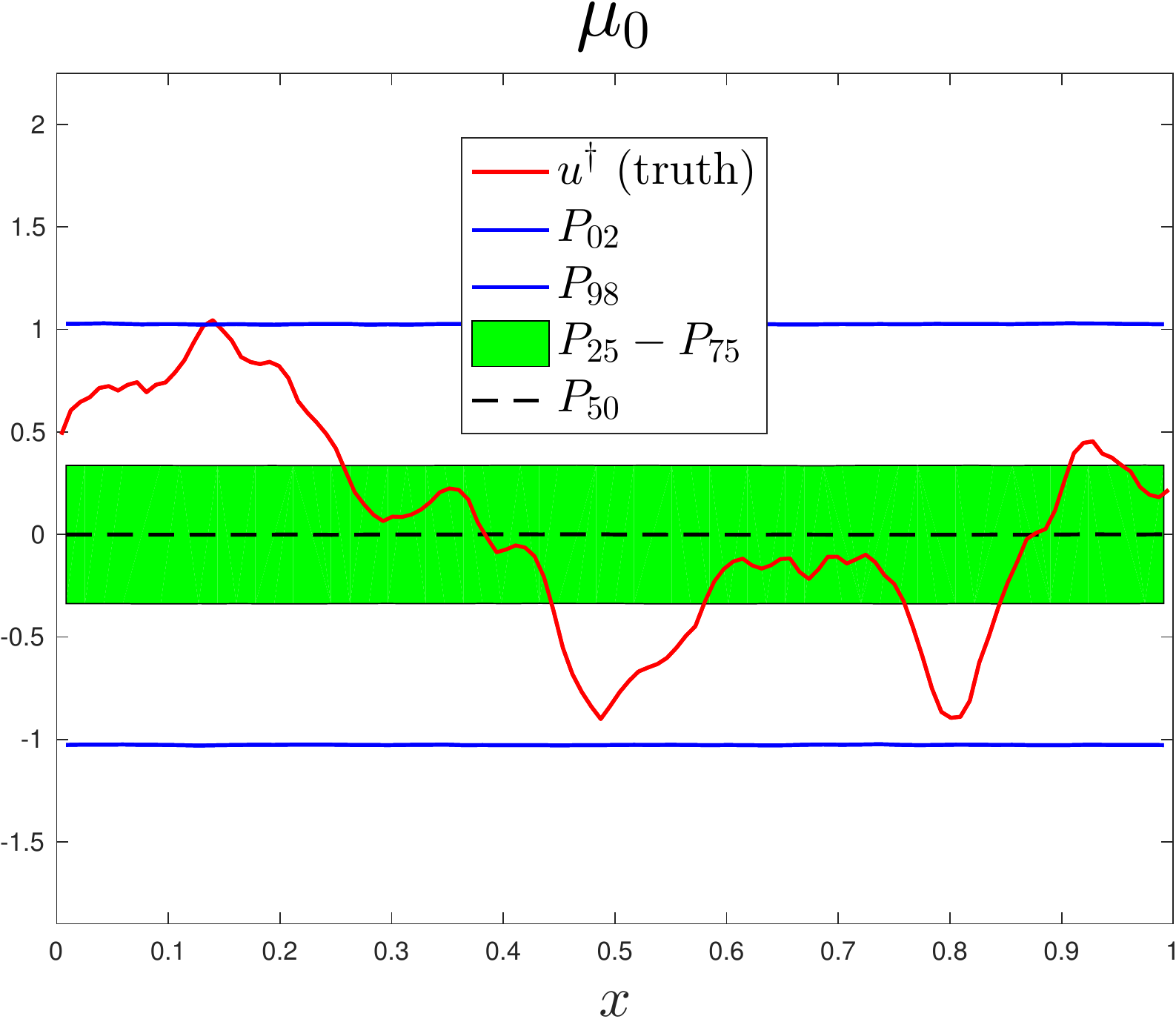}
\includegraphics[scale=0.3]{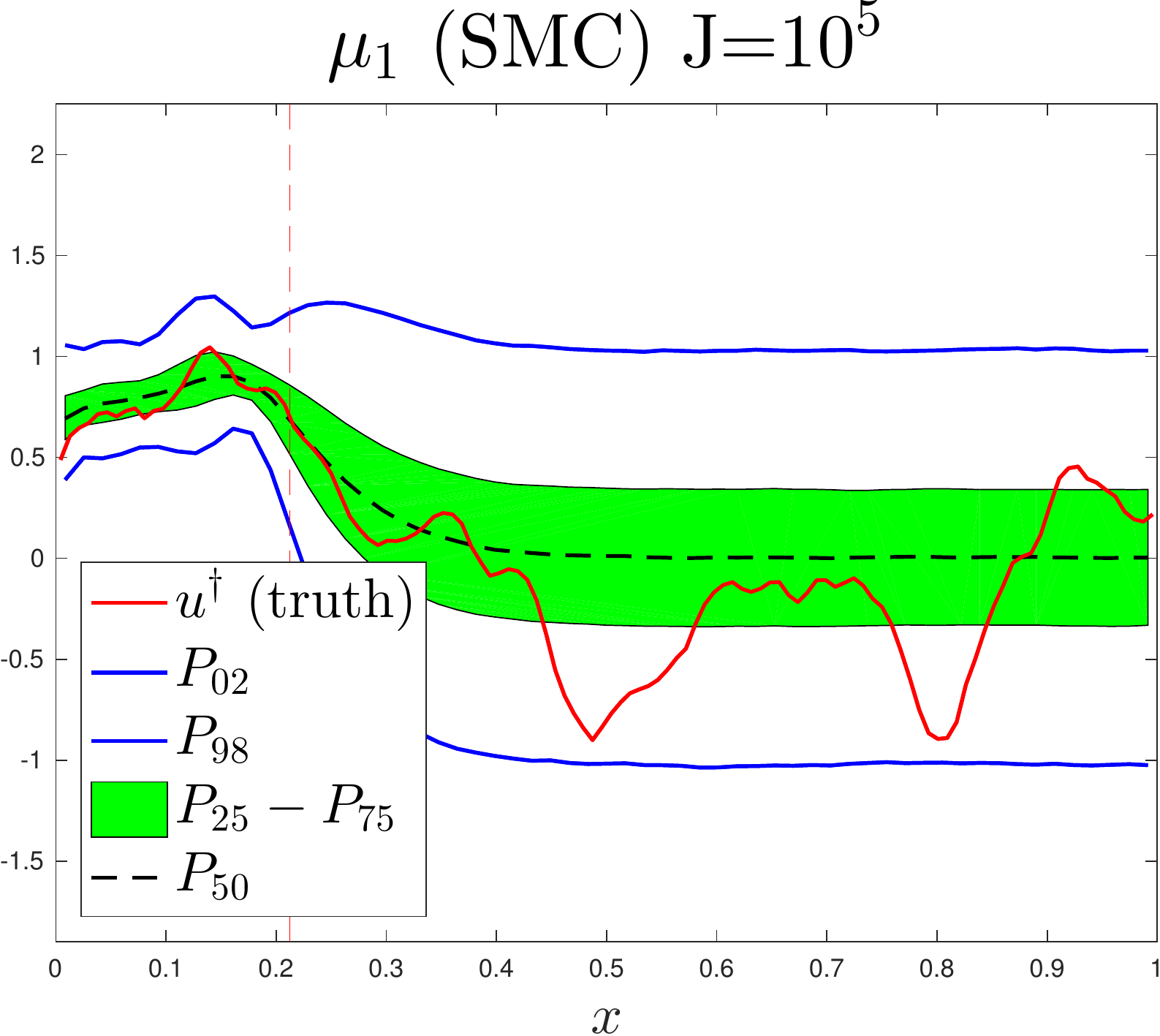}
\includegraphics[scale=0.3]{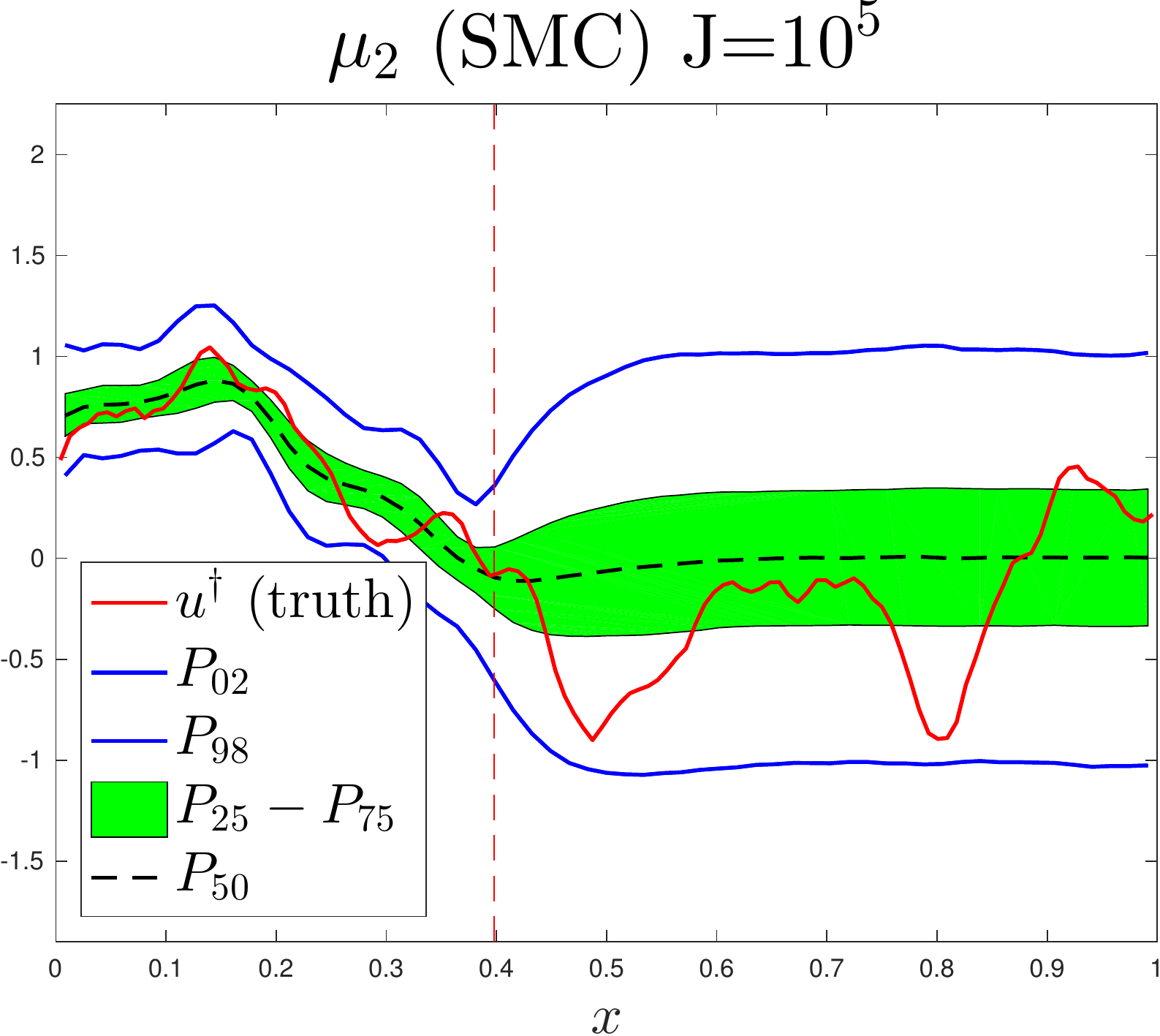}
\includegraphics[scale=0.3]{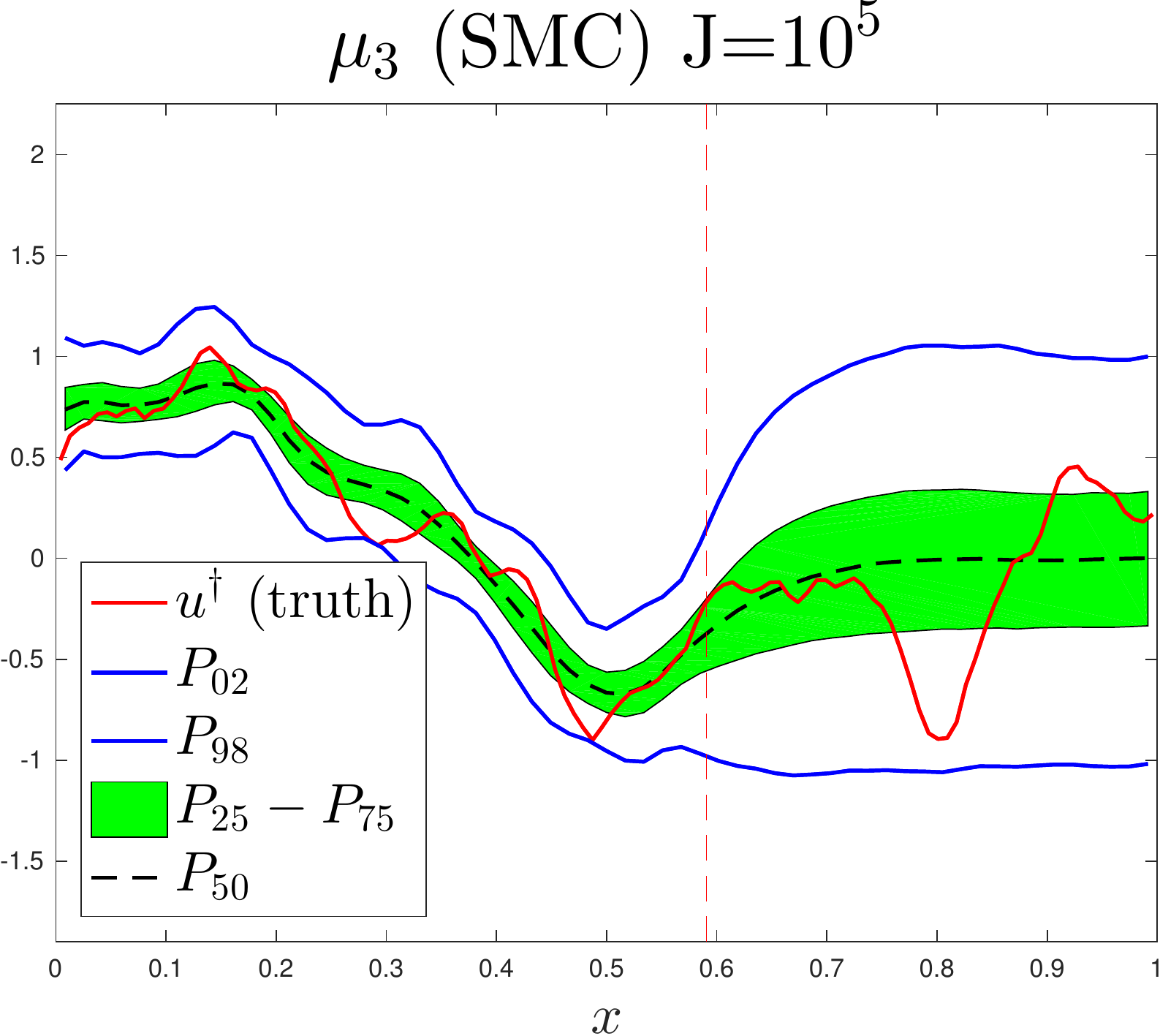}
\includegraphics[scale=0.3]{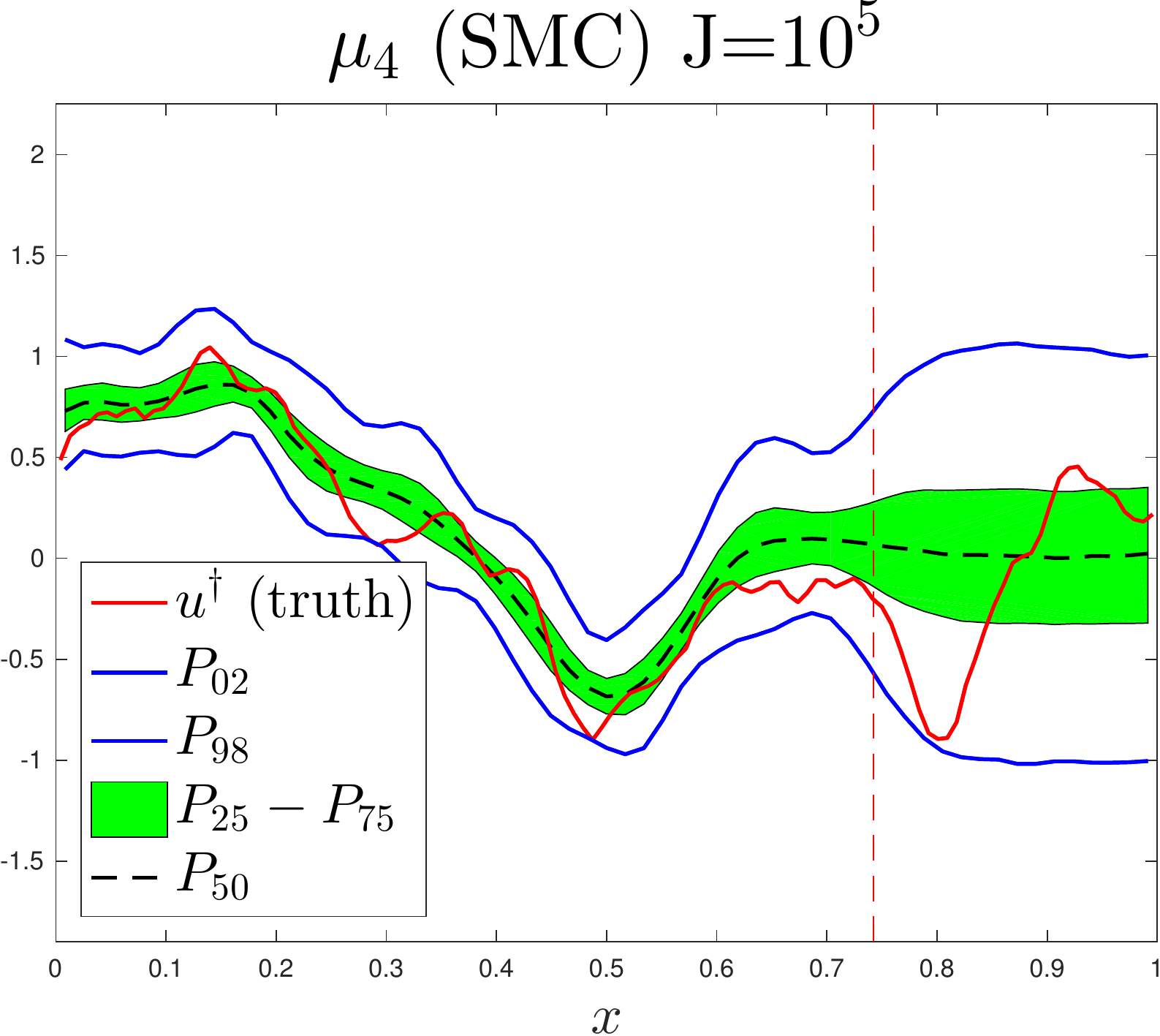}
\includegraphics[scale=0.3]{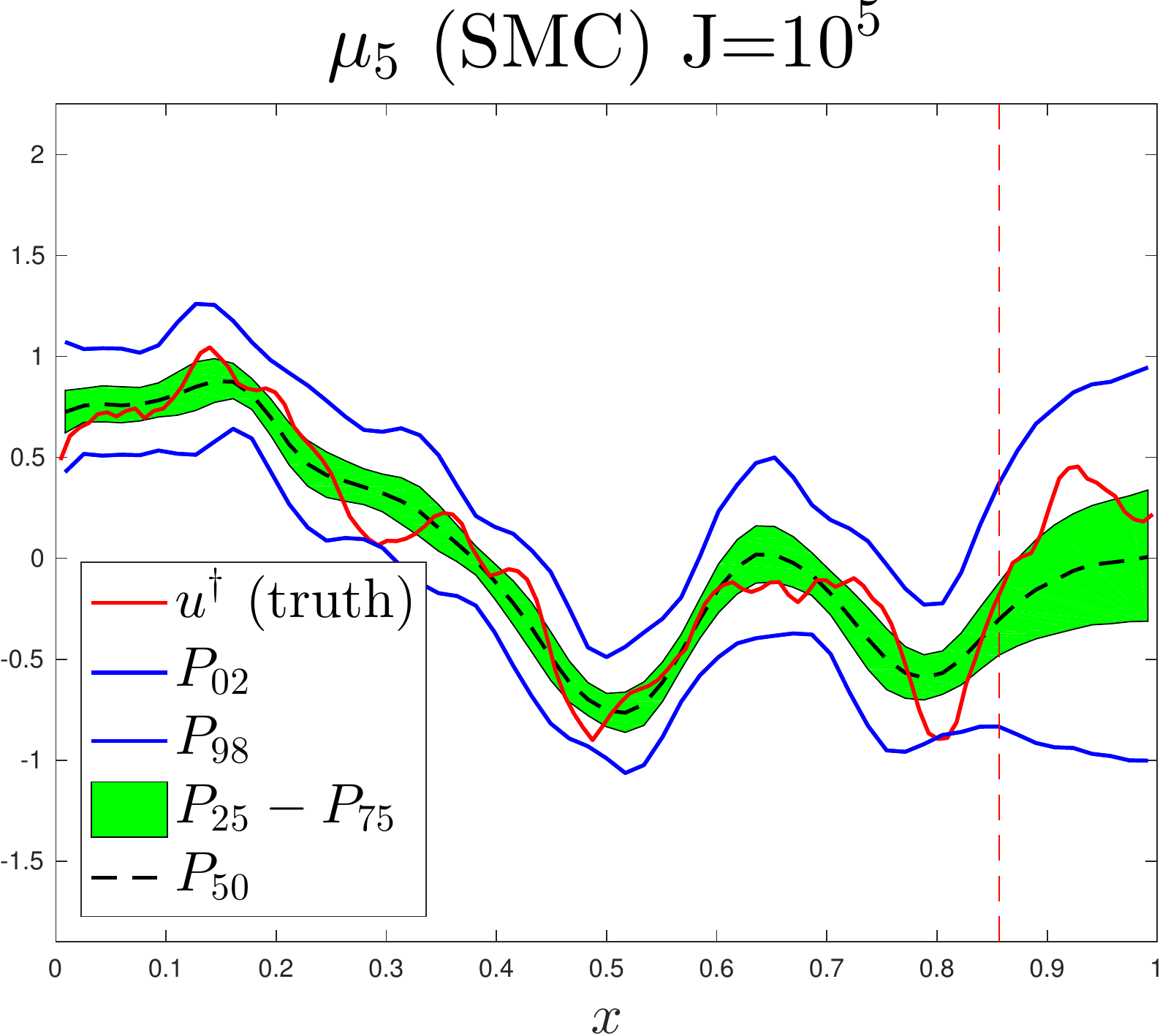}

 \caption{Top-left: Percentiles of the prior log-permeability $\mu_{0}$. Top-Middle to bottom-right: Percentiles of the posteriors $\{\mu_{n}\}_{n=1}^{5}$ obtained via SMC with large number of samples $J=10^5$. Solid red line corresponds to the graph of the true log-permeability $u^{\dagger}$. Vertical dotted line indicates the location ot the true front $\Ga^{\dagger}(t_{n})$.} \label{Fig2}
\end{center}
\end{figure}

In this subsection we report the application of the SMC sampler of \cite{Kantas} (see Algorithm \ref{SMC_al} in Appendix~\ref{Ape_SMC}) which, as described in the preceding section, provides a particle approximation of each posterior that converges to the exact posterior measure $\mu_{n}$ as the number of particles $J$ goes to infinity. In order to achieve a high-level of accuracy we use $J=10^5$ number of particles which is substantially larger compared to the number of particles (e.g. $10^3$ to $10^4$) often used in existing applications of SMC for high-dimensional inverse problems \cite{Chopin01082002,Kantas}. In addition, we consider the selection of tunable parameters $N_{\mu}=20$ and $J_{thresh}=J/3$ similar to the ones suggested in \cite{Kantas}. For each observation time $t_{n}$, we store the ensemble of particles $\{u_{n}^{(j)}\}_{j=1}^{10^5}$ that approximates the corresponding posterior $\mu_{n}$.  From this ensemble, we compute the 0.02, 0.25, 0.5, 0.75, 0.98 posterior percentiles displayed in Figure \ref{Fig2} (top-middle to bottom-right), where we also include the graph of the true log-permeability (red curve). The vertical line in these figures indicate the true location of the front $\Ga^{\dagger}(t_{n})$ at each observation time $t_{n}$. We can clearly appreciate that the uncertainty band defined by these percentiles is substantially reduced as more observations (in time) are assimilated. In fact, the main reduction of the uncertainty is observed in the region of the moving domain $D^{\dagger}(t_{n})=[0,\Ga^{\dagger}(t_{n})]$ at the corresponding observation time $t_{n}$. It is then clear that at each observation time $t_{n}$, measurements collected from pressure sensors with $x_{m}\in D^{\ast}\setminus D^{\dagger}(t_{n})$ are not very informative of the log-permeability field. This comes as no surprise when we recognize that the pressure field given by (\ref{eq10}) depends on the permeability field only in the region of the moving domain $D(t)$. In other words, the values of the permeability in the region defined by $D^{\ast}\setminus D(t_{n})$ have no effect on $p(x,t_{n})$; hence the $n$th likelihood function is independent of $u$ in this region. We can indeed observe from Figure \ref{Fig2} that the percentiles of the log-permeability in this region (see domain to the right of the vertical lines) is similar to those from the prior. However, due to the regularity of the log-permeability enforced in the prior $\mu_{0}$, there is a smooth transition in the uncertainty band at the interface defined by the front location $\Ga(t_{n})$.

The number of intermediate tempering distributions that SMC adaptively computed to approximate the sequence of posteriors $\{\mu_{n}\}_{n=1}^{5}$ were the following:
\begin{eqnarray}\label{it}
q_{1}=4,\quad q_{2}=3, \quad q_{3}=3, \quad q_{4}=2,\quad q_{5}=3.
\end{eqnarray}
We use these numbers in (\ref{cost_SMC}) to compute the total computational cost of approximating the sequence $\{\mu_{n}\}_{n=1}^{5}$.  The values of $g_{n}$ (i.e. cost of evaluating each $\cG_{n}$) are estimated by the average CPU time from 1000 simulations computed with different log-permeabilities sampled from the prior. We obtain that total cost is approximately $1.5\times 10^7$ times the cost of evaluation the 5th forward map $\cG_{5}$ (i.e. at the final observation time). Clearly, this computational cost is prohibitive for the two and three dimensional problems where, as stated earlier, evaluating the forward map can take several minutes of CPU time.  For the present one-dimensional case we are able to afford this cost due to the relatively low cost associated with solving the 1D moving boundary problem.

\subsection{Reducing the cost of SMC by adjusting tunable parameters}\label{SMC_small}

Given the high computational cost of computing accurate approximations of the posteriors with SMC, it is reasonable to ask whether its computational cost can be reduced by adjusting the tunable parameters in (\ref{cost_SMC}). By reducing either the number of particles $J$ and/or the number of MCMC steps $N_{\mu}$, we can achieve a substantial decrease in the computational cost. The selection of $J_{thresh}$ also determines the computational cost as it, in turns, defines the number of tempering steps for each posterior.  However, it is essential to understand the effect of decreasing these tunable parameters on accuracy of the SMC sampler. In this subsection we aim at understanding this effect by comparing the application of the SMC sampler with smaller number of particles $J$ and different choices of the tunable parameters $N_{\mu}$ and $J_{thresh}$. This requires creating a Benchmark against which we can compare performance of SMC. The Benchmark is obtained by the highly-resolved characterization of the posteriors that we computed in the preceding section by using the SMC sampler with large number of particles $(J=10^5)$. In Appendix~\ref{Ape_SMC} we provide further discussions of the performance and diagnostics of the SMC sampler applied to approximate these posterior measures. These diagnostics offer evidence that the SMC sampler has been successfully applied, thereby providing accurate characterization of the posterior that we may use as a Benchmark to compare against the posteriors computed via algorithms with lower resolution/accuracy. The numerical investigation below is aimed at assessing SMC with different selections of ensemble size $J$  as well as the tunable parameters $N_{\mu}$ and $J_{thresh}$.

For the reasons stated above, through the rest of the this and the following sections, we refer to the aforementioned highly-resolved SMC particle approximations (with $J=10^5$) as the ``exact''  sequence of posteriors $\{\mu_{n}\}_{n=1}^{5}$ that we use for subsequent comparisons purposes. Moreover, for these comparisons we assume that the sample mean and variance of these SMC samples are exact approximations of the mean $\mathbb{E}^{\mu_{n}}$ and variance $\mathbb{V}^{\mu_{n}}$ of the posterior $\mu_{n}$. In other words, we assume
\begin{eqnarray}\label{eq:1001}
\mathbb{E}^{\mu_{n}} =\overline{u}_{n,10^5},\qquad \mathbb{V}^{\mu_{n}}=\sigma_{n,10^5}^{2},
\end{eqnarray}
where
\begin{eqnarray}\label{eq:1001B}
\overline{u}_{n,J} \equiv \frac{1}{J} \sum_{j=1}^{J}u_{n}^{(j)},\qquad \sigma_{n,J}^{2}\equiv \frac{1}{J-1} \sum_{j=1}^{J-1}(u_{n}^{(j)}-\overline{u}_{n,J})^2.
\end{eqnarray}

\begin{figure}[htbp]
\begin{center}
\includegraphics[scale=0.75]{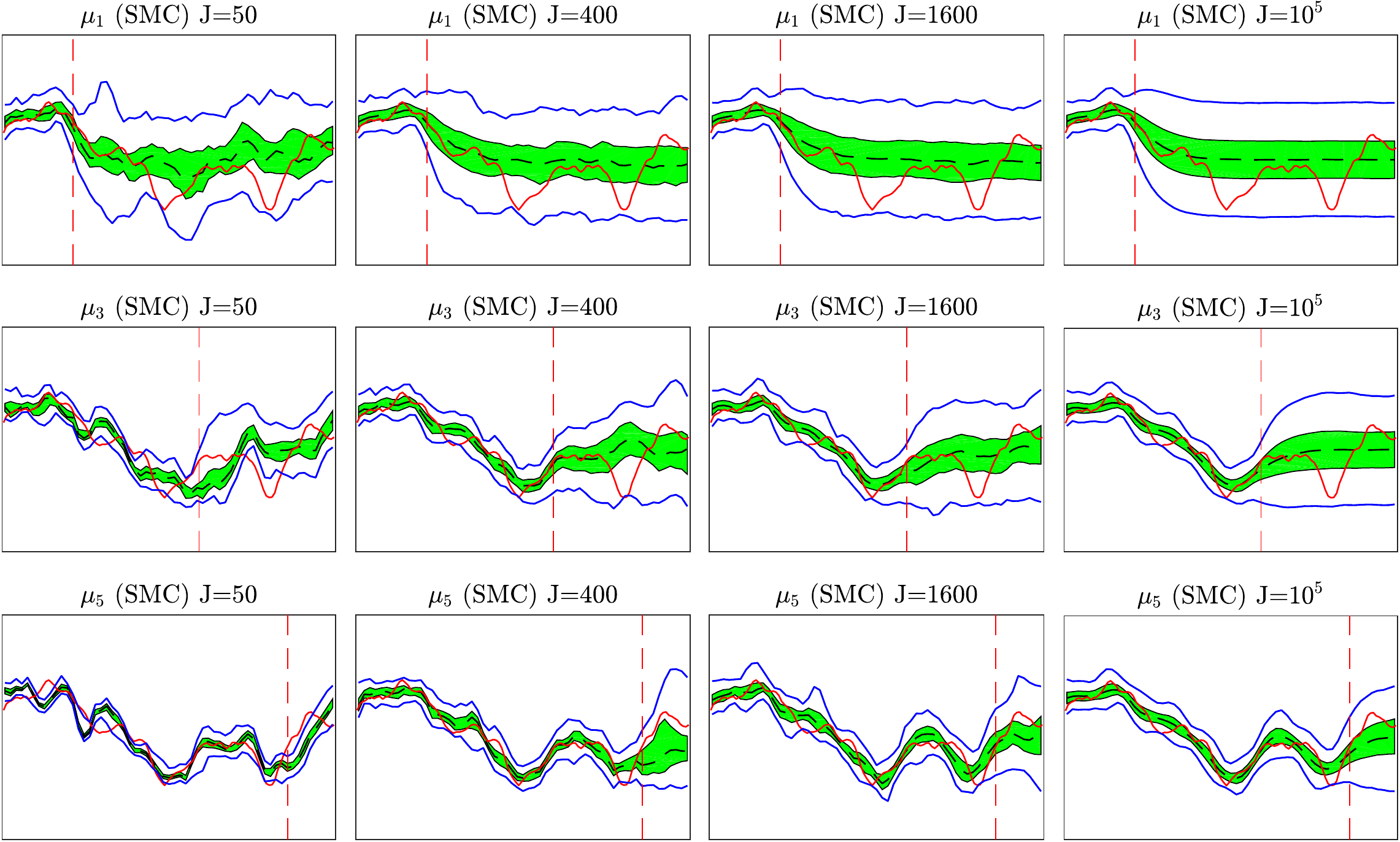}
 \caption{Percentiles of the posteriors $\mu_{n}$'s ($n=1,3,5$) obtained via SMC with (from left to right) $J=50, 400, 1600, 10^5$. Solid red line corresponds to the graph of the true log-permeability $u^{\dagger}$. Vertical dotted line indicates the location ot the true front $\Ga^{\dagger}(t_{n})$.} \label{Fig3A}
\end{center}
\end{figure}

Let us now consider the application of the SMC sampler for the following choices of small number of particles: $J=50, 100,200,400,800,1600,3200,6400$. We also consider three choices of the tunable parameter $J_{thresh}$ ($J_{thresh}=J/3,J/2,2J/3$) and two choices of $N_{\mu}$ ($N_{\mu}=5,20$). In Figure \ref{Fig3A} we show percentiles of the log-permeability posteriors $\mu_{n}$ (for $n=1,3$ and $5$) obtained using the aforementioned SMC sampler for some of those choices of the number particles $J$, and with the same selection of tunable parameters $N_{\mu}=20$, $J_{thresh}=J/3$ that we used for the highly-resolved SMC with large particles; percentiles from the latter are included in the right column of Figure \ref{Fig3A} for comparison purposes. We can see that as the ensemble of particle increases, the approximation of SMC improves when compared to the one provided by the highly-resolved SMC. Note that very small number of particles results in very poor approximations of these percentiles.

\begin{figure}[htbp]
\begin{center}

\includegraphics[scale=0.34]{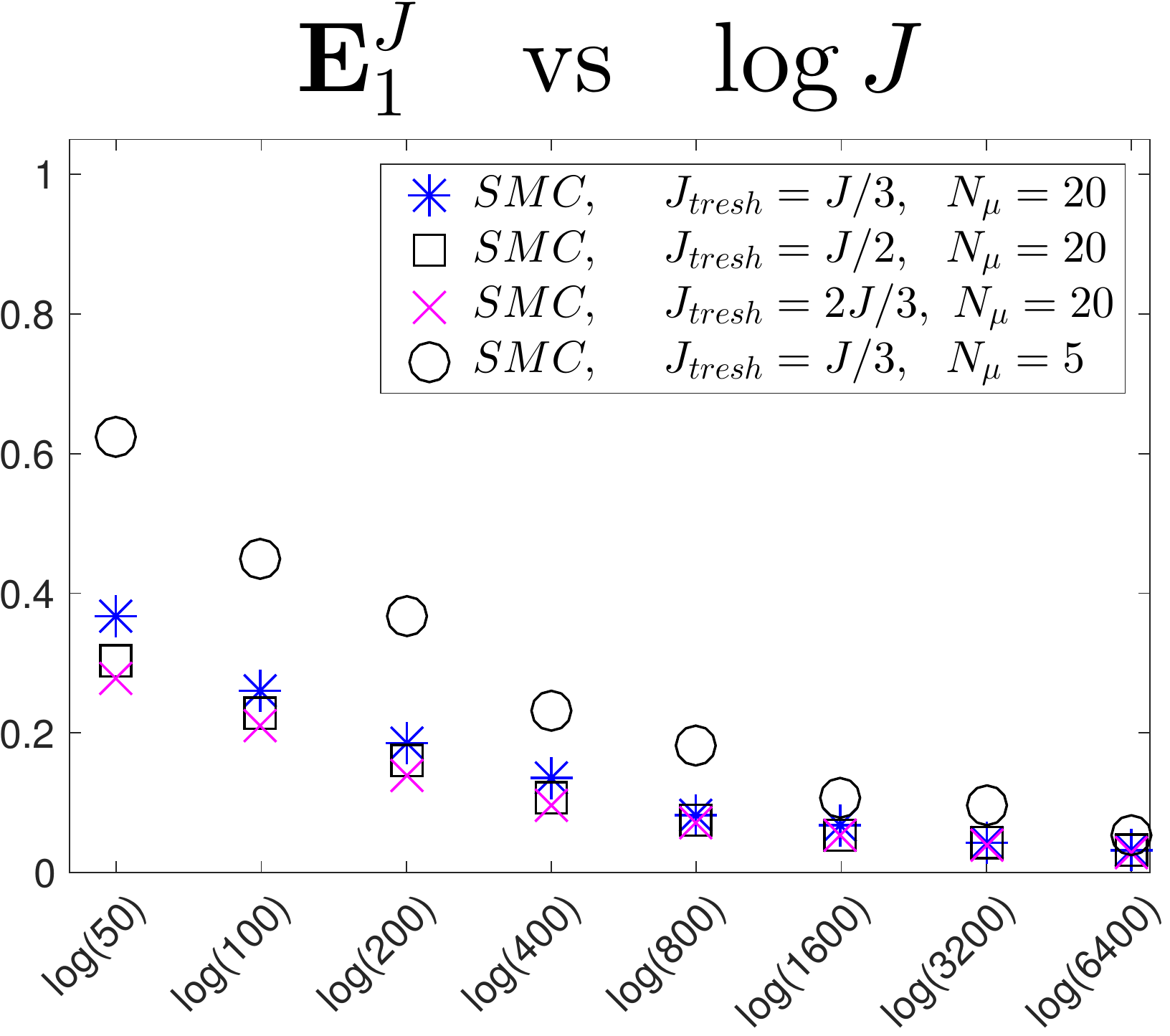}~\includegraphics[scale=0.34]{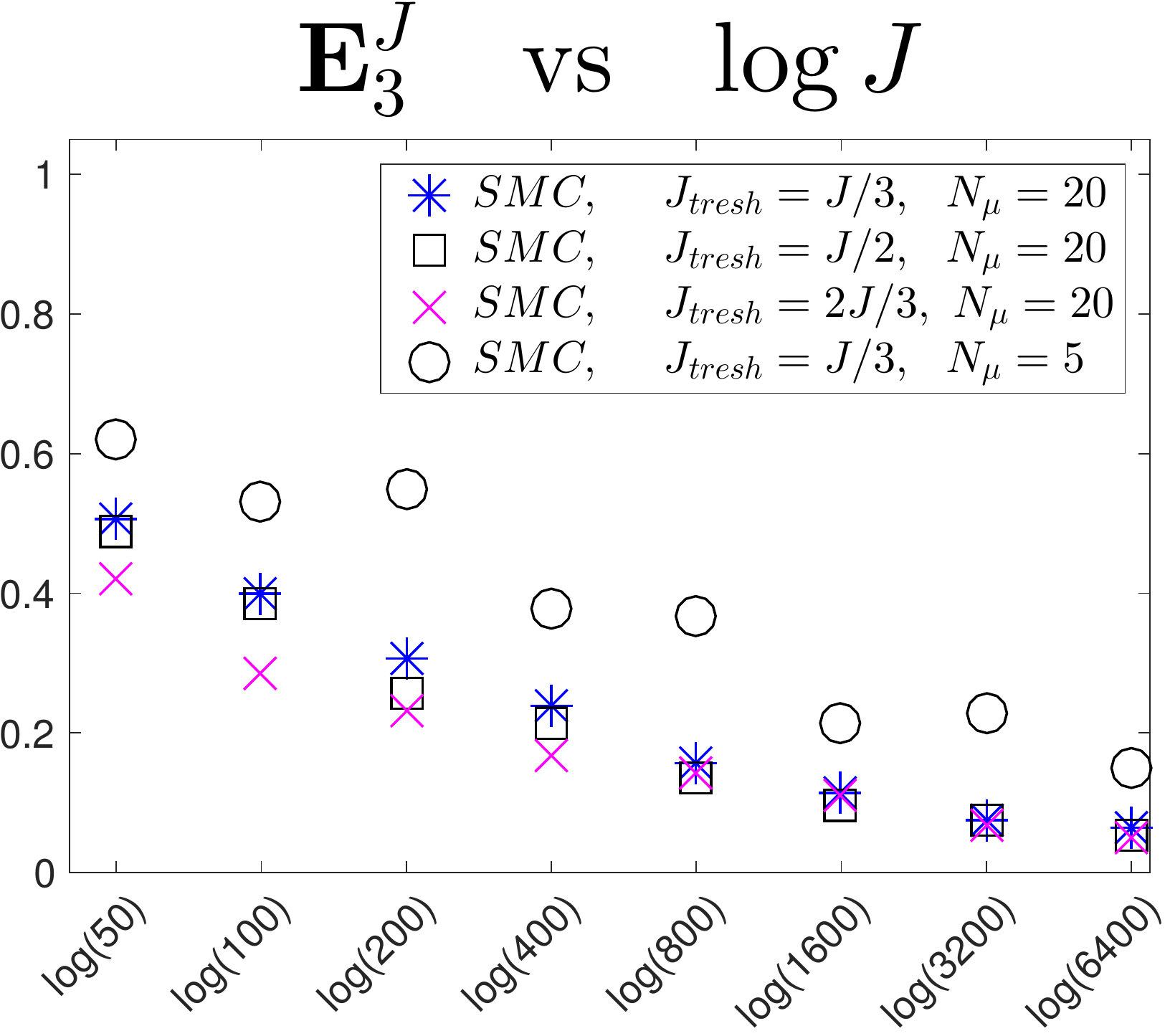}~\includegraphics[scale=0.34]{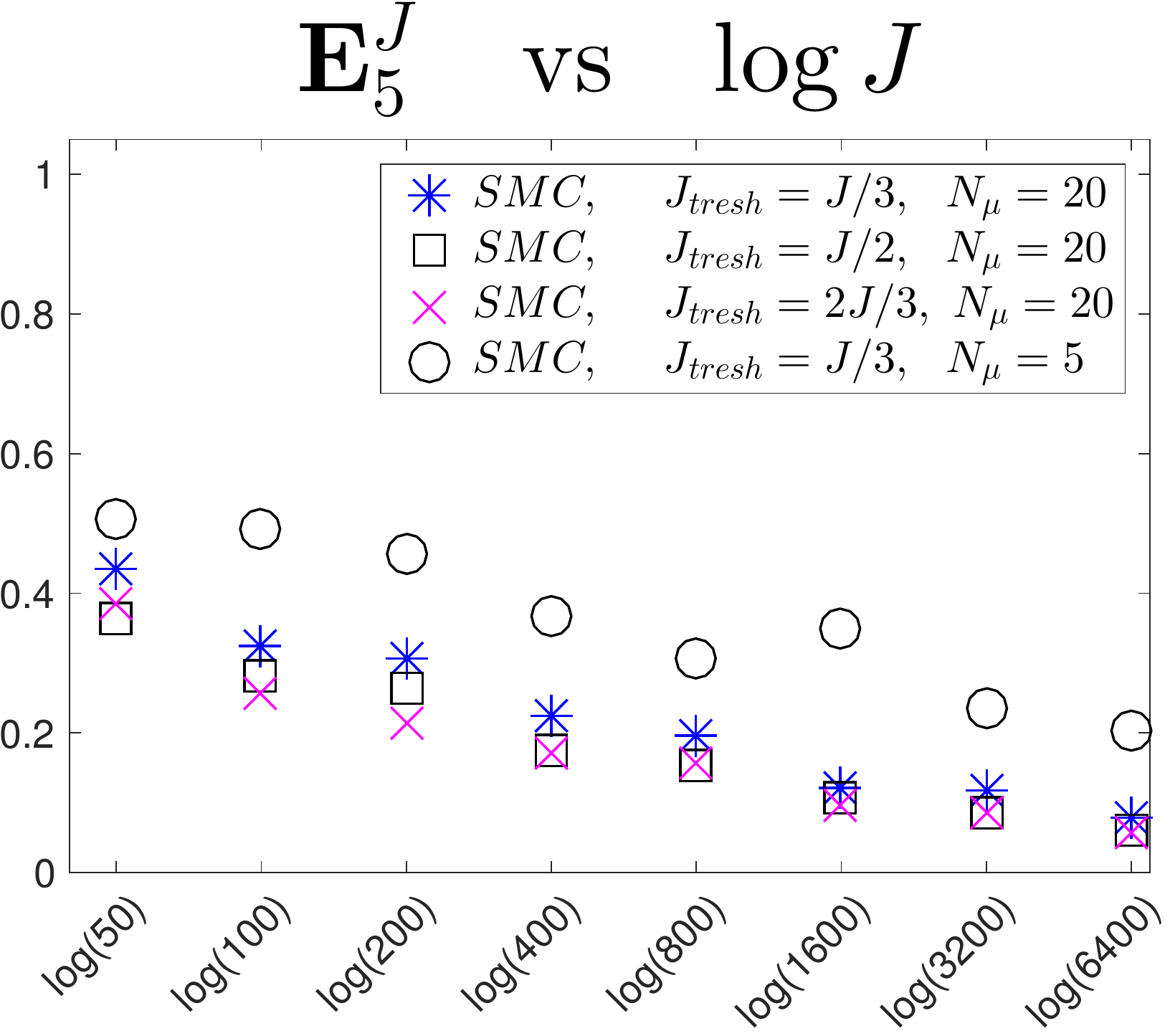}\\
\includegraphics[scale=0.34]{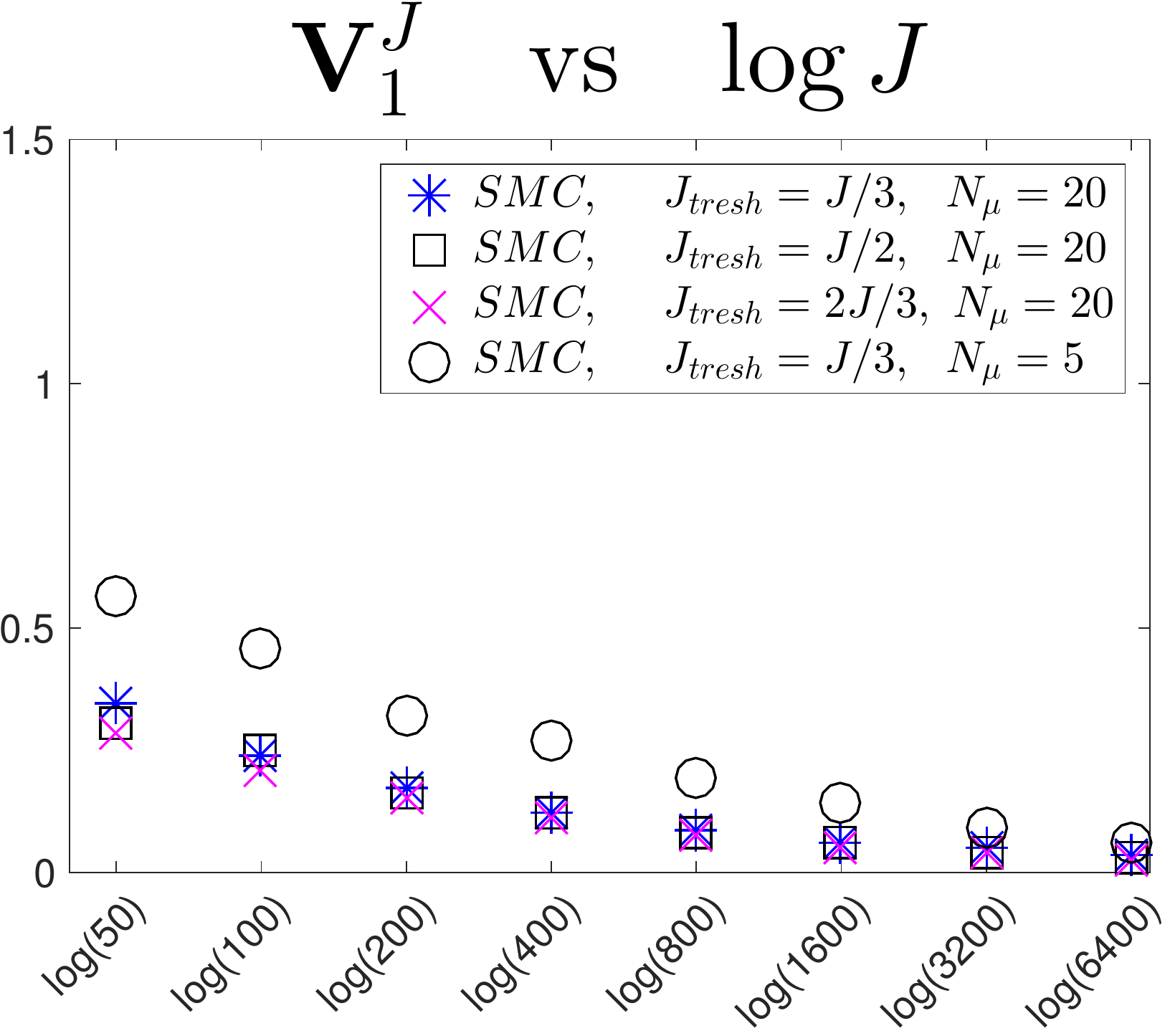}~\includegraphics[scale=0.34]{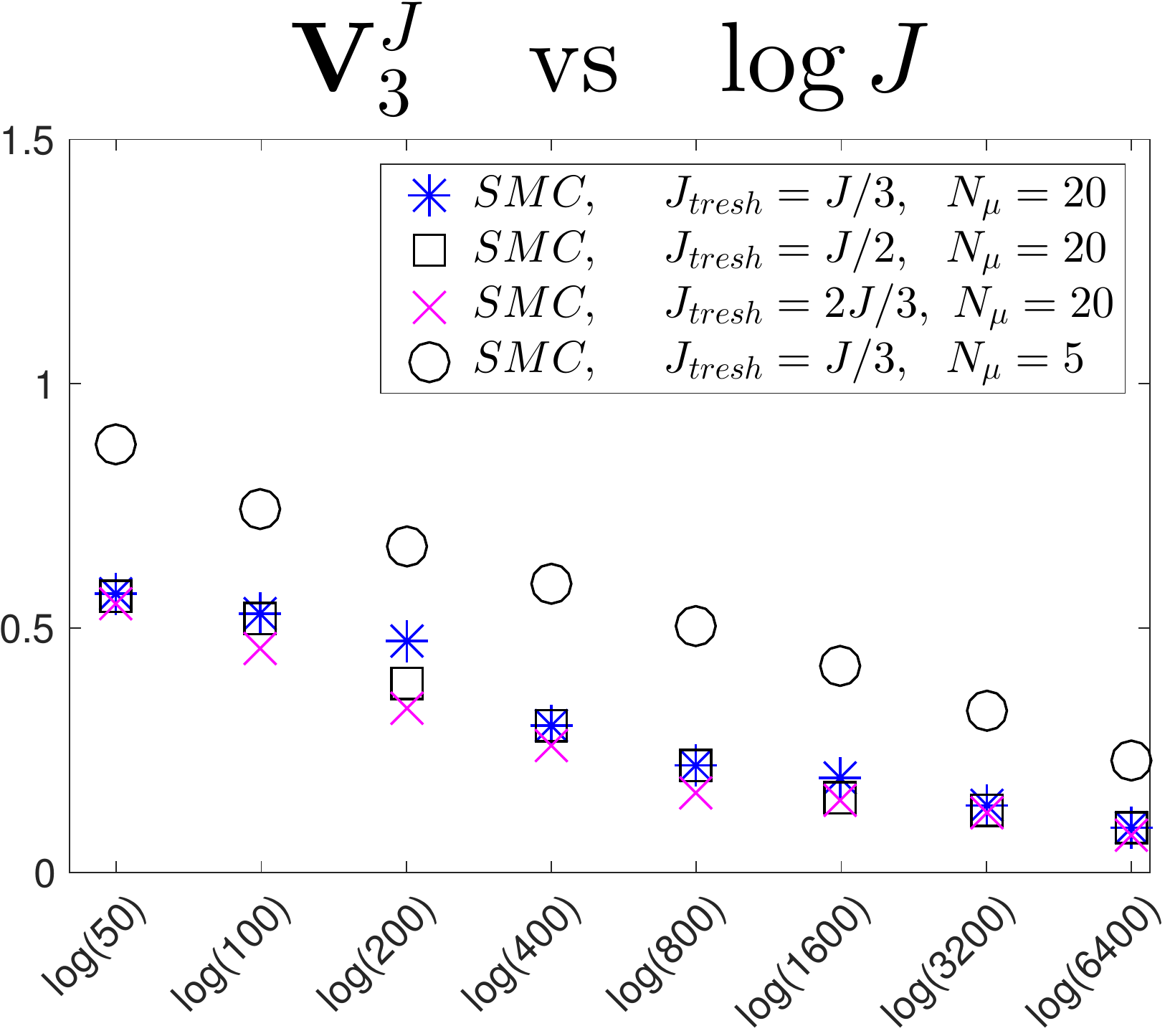}~\includegraphics[scale=0.34]{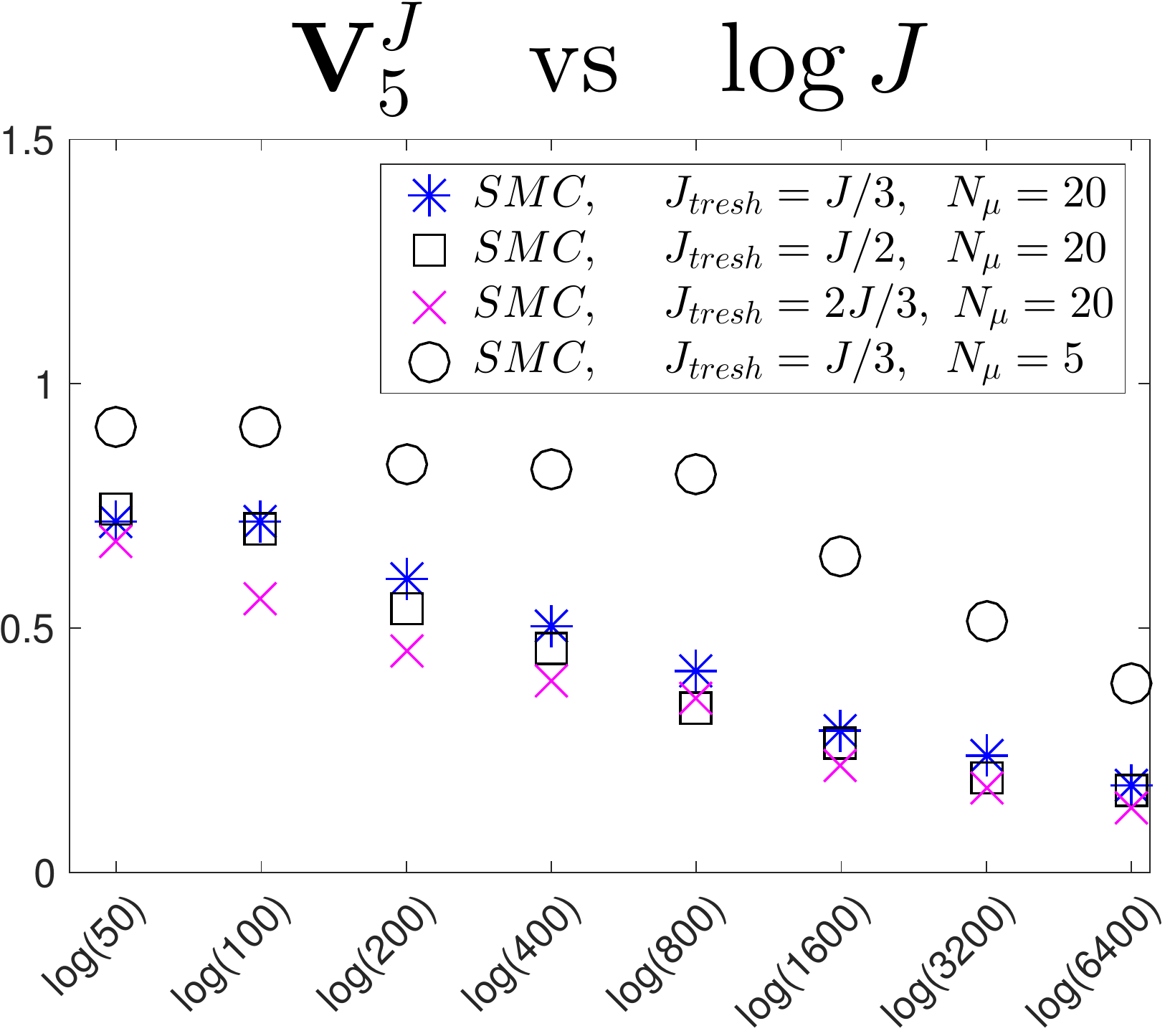}\\
\includegraphics[scale=0.34]{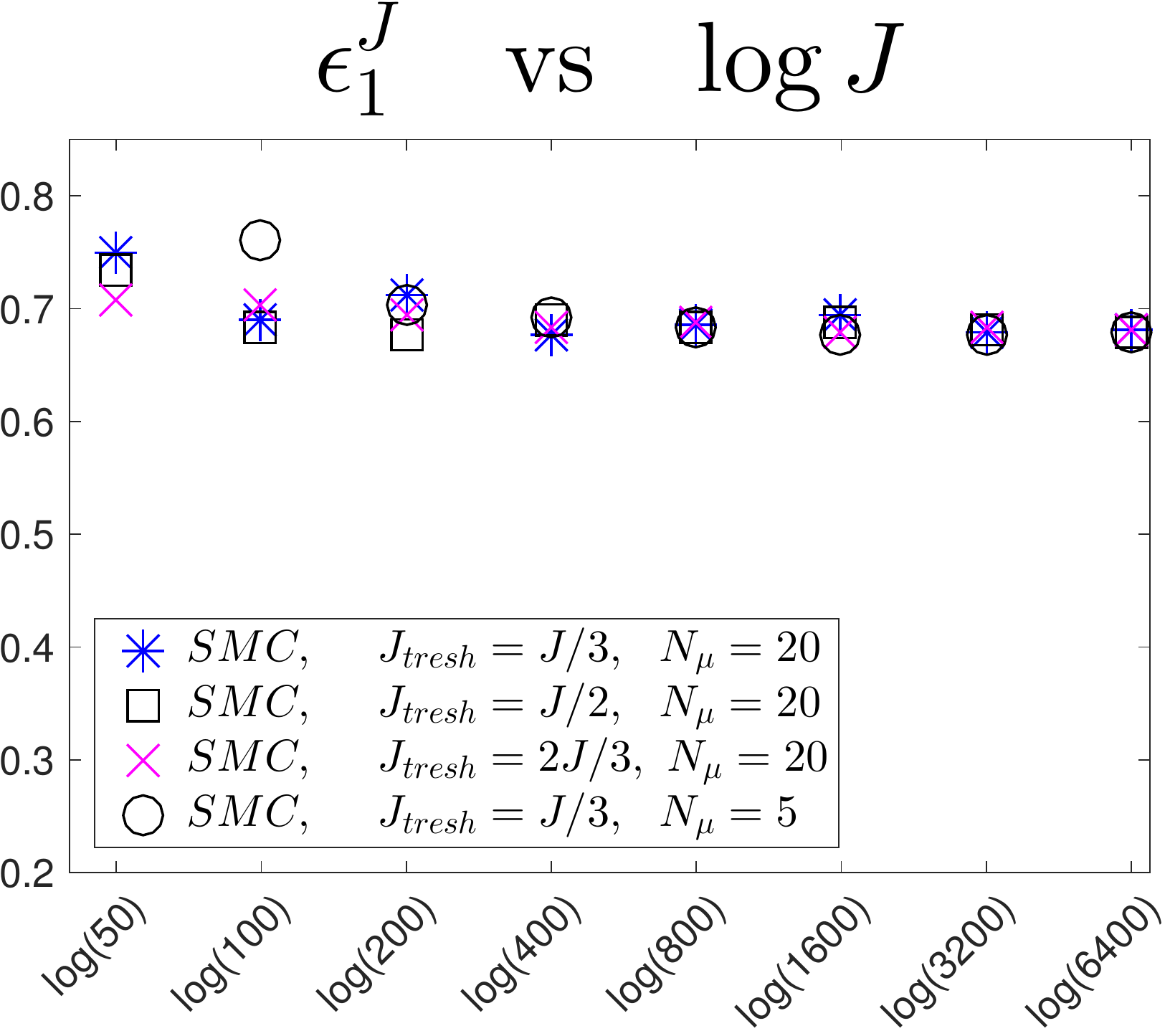}~\includegraphics[scale=0.34]{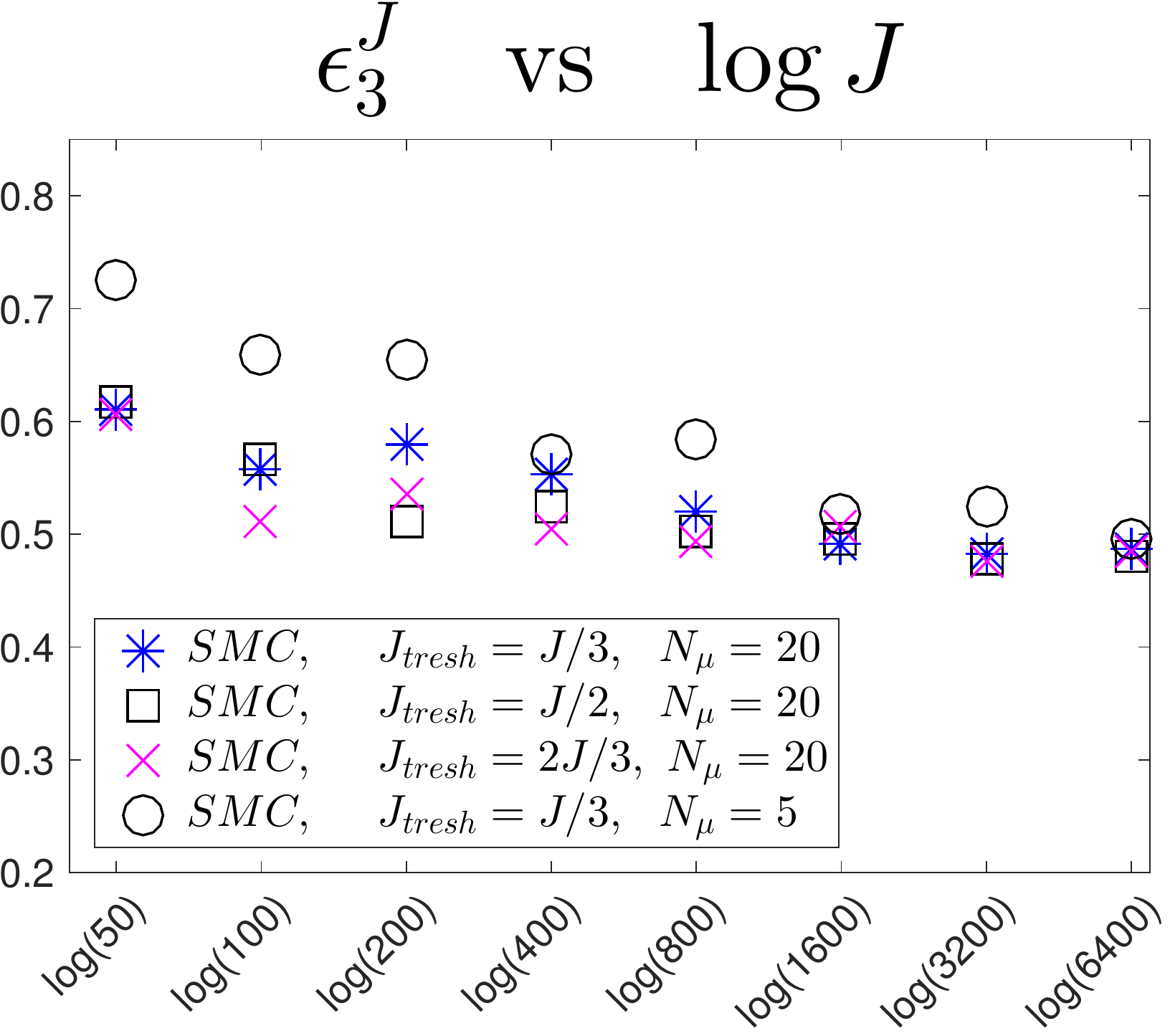}~\includegraphics[scale=0.34]{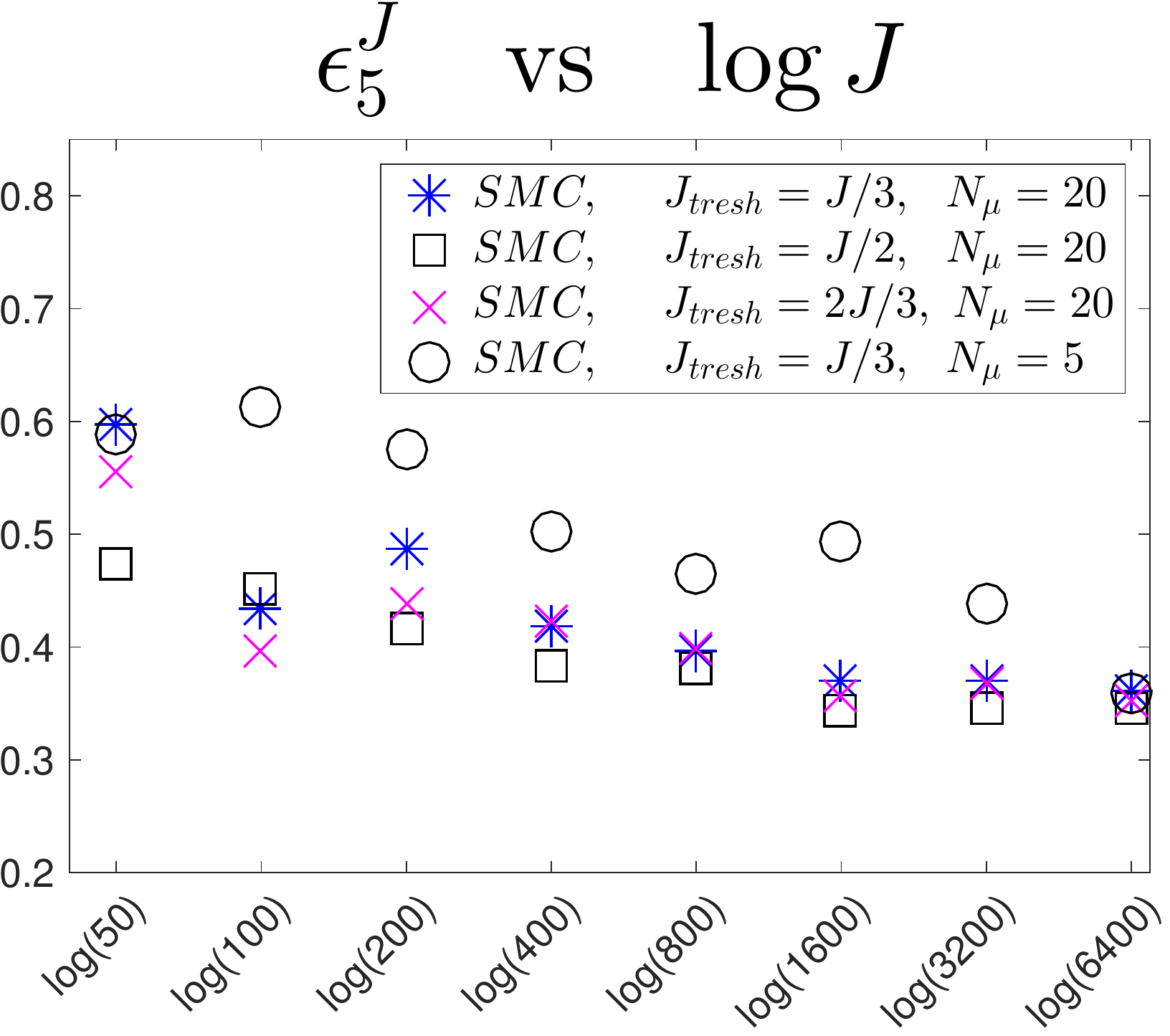}
 \caption{SMC approximations. Top and middle: Relative errors of mean (top) and variance (middle) of the posteriors $\mu_{n}$, (from left to right) $n=1,3,5$, obtained with SMC  with different choices of small (log) ensemble size $\log(J)$ and tunable (SMC) parameters $J_{thresh}$ and $N_{\mu}$. Bottom: Relative errors with respect to the truth $u^{\dagger}$ of the ensemble mean.} \label{Fig3B}
\end{center}
\end{figure}

\begin{figure}[h]
\begin{center}

\includegraphics[scale=0.475]{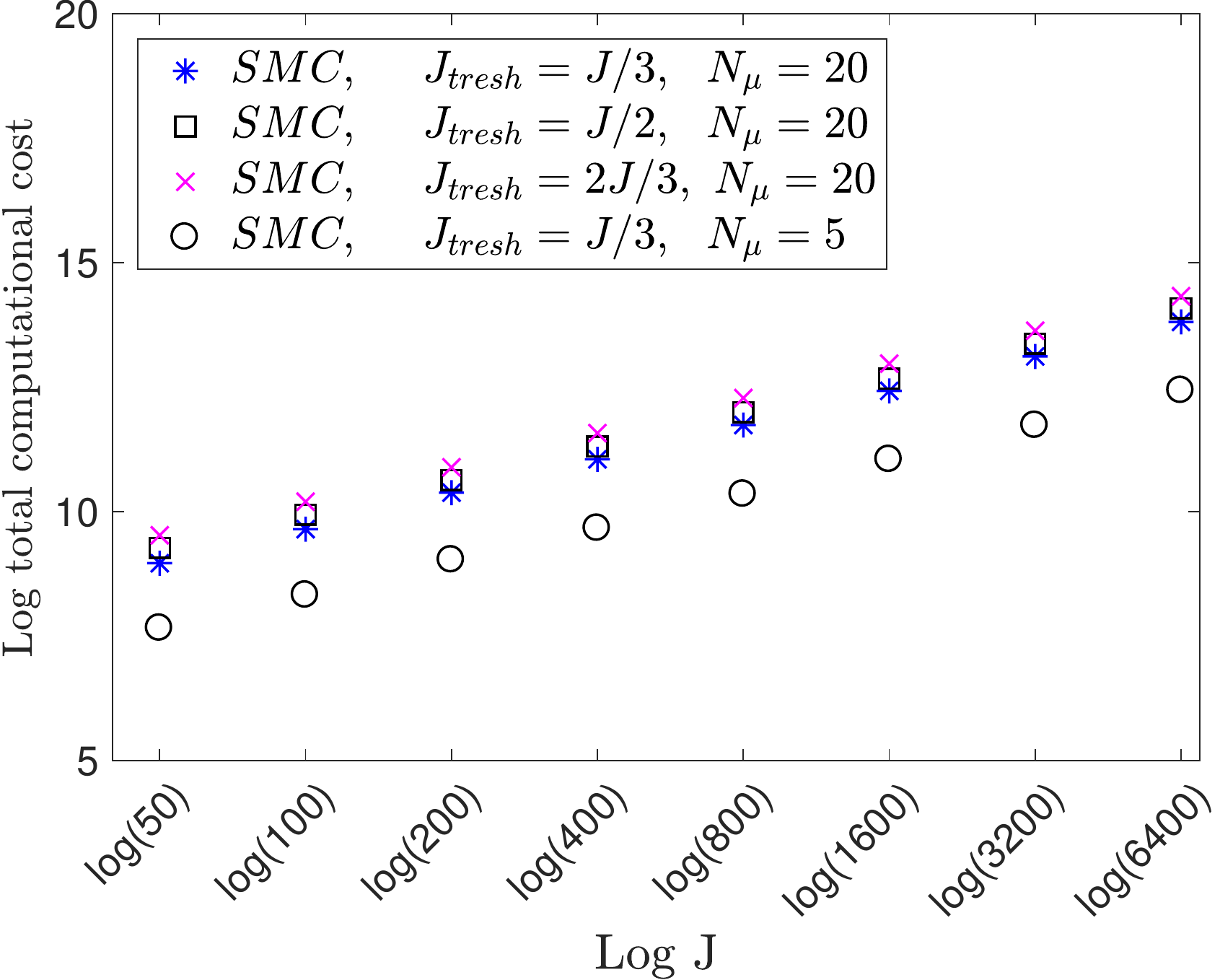}
\includegraphics[scale=0.475]{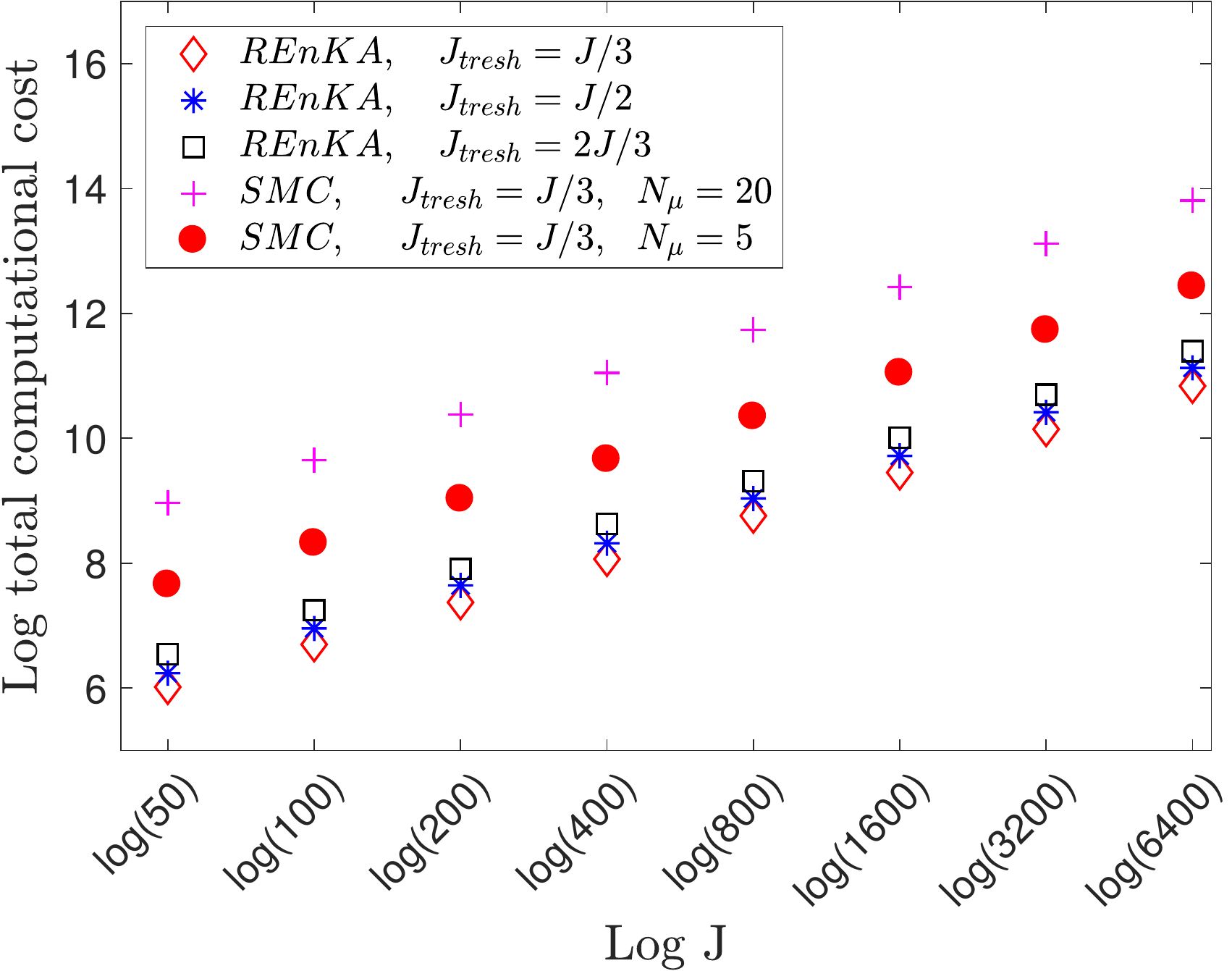}

 \caption{Total computational cost in terms of  $\cG_{5}$-forward model evaluations. Left: Total computational cost obtained via SMC with different choices of $N_{\mu}$ and $J_{thresh}$. Right: Comparison of total computational cost obtained via REnKA with different choices of $J_{thresh}=J/3$ against the cost of SMC with different selection of tunable parameters $N_{\mu}$ and $J_{thresh}$. } \label{Fig3C}
\end{center}
\end{figure}

In order to quantify the level of approximation obtained with SMC with the aforementioned selections of parameters, we compute the $L^{2}(D^{\ast})$-relative errors of the mean and variance with respect to the posterior measure approximated with the highly-resolved SMC computed as described in the preceding subsection. More precisely, let us define
\begin{eqnarray}\label{mean_var}
\mathbf{E}_{n}^{J}\equiv \frac{\vert\vert\mathbb{E}^{\mu_{n}}-\overline{u}_{n,J}\vert\vert_{L^{2}(D^{\ast})}}{\vert\vert \mathbb{E}^{\mu_{n}} \vert\vert_{L^{2}(D^{\ast})}},\qquad \mathbf{V}_{n}^{J}\equiv \frac{\vert\vert \mathbb{V}^{\mu_{n}}-\sigma_{n,J}^{2}\vert\vert_{L^{2}(D^{\ast})}}{\vert\vert \mathbb{V}^{\mu_{n}}\vert\vert_{L^{2}(D^{\ast})}},
\end{eqnarray}
where $\mathbb{E}^{\mu_{n}}$ and $\mathbb{V}^{\mu_n}$ are the $\mu_{n}$-posterior mean and variance characterized via SMC with large $J$ from (\ref{eq:1001})-(\ref{eq:1001B}). In the previous expressions $\overline{u}_{n,J}$ and  $\sigma_{n,J}^{2}$ are the sample mean and variance defined in (\ref{eq:1001}) obtained from the ensemble $\{u_{n}^{(j)}\}_{j=1}^{J}$ computed via SMC for the choices of small $J$ stated above and with the aforementioned selections of tunable parameters. In addition, we consider  the estimator of the \mt{true} log-permeability defined by the ensemble mean $\overline{u}_{n,J}$ and thus we monitor the corresponding $L^{2}(D^{\ast})$-relative error defined by
\begin{eqnarray}\label{truth_err}
\epsilon_{n}^{J}\equiv \frac{\vert\vert u^{\dagger}-\overline{u}_{n,J}\vert\vert_{L^{2}(D^{\ast})}}{\vert\vert u^{\dagger} \vert\vert_{L^{2}(D^{\ast})}}.
\end{eqnarray}
Quantities $\mathbf{E}_{n}^{J}$, $ \mathbf{V}_{n}^{J}$ and $\epsilon_{J}^{n}$ are random variables that depend on the initial ensemble of particles that we generate from the prior $\mu_{0}$. We thus report these quantities (for each $n=1,3,5$) averaged over 15 experiments corresponding to different selections of the initial ensemble of particles. In Figure \ref{Fig3B} we display $\mathbf{E}_{n}^{J}$ (top), $ \mathbf{V}_{n}^{J}$ (middle) and $\epsilon_{n}^{J}$ (bottom) for (from left to right) $n=1,3,5$ as a function of the aforementioned selections of ensemble size $J$. For brevity we omit the results for $n=2,4$ as they display similar behaviour. The total computational cost of computing the full sequence of posteriors (i.e. $\mathbf{C}_{SMC}$ from (\ref{cost_SMC})) is shown in Figure \ref{Fig3C} (left). We reiterate that this cost is expressed in terms of the number of evaluations of the 5th forward map $\cG_{5}$.

While the numerical analysis of the convergence of the SMC sampler is beyond the scope of this work, the results presented in this section are aimed at understanding the level of accuracy of SMC with relatively small number of particles and for a selection of tunable parameters which may enable the use of this method in more practical scenarios. From these results it is clear that the selections of $J_{thresh}$ have no substantial effect on the accuracy of the scheme in terms of approximating the mean and variance of each posterior. Similarly, the computational cost with respect to our selections of $J_{thresh}$ does not seem to vary significantly. It is evident that the main effect in terms of accuracy is the choice of MCMC steps (i.e. parameter $N_{\mu}$). Indeed, note that the error obtained with $N_{\mu}=5$ is considerably larger than the one with $N_{\mu}=20$ although the computational cost of the former is one quarter of the computational cost of the latter. We conclude that even though decreasing $N_{\mu}$ can offer computational affordability, it is detrimental to the approximation properties of the scheme. This comes as no surprise as it is well known that the mutation step that involves running MCMC is crucial for the accuracy of any SMC methodology.

The behavior of the SMC sampler with respect to the number of particles $J$ is as expected. On the one hand, an increase in $J$ corresponds to a decrease in the error with respect to the mean and variance. On the other hand, the computational cost, $\mathbf{C}_{SMC}$, increases with $J$. Note that there is a clear linear relationship between these two variables which is, in turn, obvious from (\ref{cost_SMC}) provided that $q_{n}$ is invariant with respect to $J$. Indeed, for the cases considered here, the number of intermediate tempering distributions (not reported) computed at each observation time, is invariant with respect to our choices of $J$. This is somewhat an expected outcome since our choice of $J_{tresh}$ in (\ref{eq38}) is always a fraction of $J$. It is also worth mentioning that the effect of $J$ is less noticeable when we look at the error with respect to the truth. At each observation time, we notice that the $\epsilon_{n}$ seems to converge to a nonzero value as $J$ increases. Note that convergence to the truth is not ensured due to the limited number of measurements inverted and the potential lack of identifiability of the log-permeability.

The results reported in this subsection suggest that achieving a reduction in the computational cost by reducing $N_{\mu}$ has a severe detrimental effect in the accuracy of the SMC sampler with small number of particles. In addition, $J_{thresh}$ does not seem to have a substantial effect in either the accuracy or computational cost. Clearly, we are only then limited to the number of particles $J$ to control the computational cost of the sampler without severely compromising accuracy of the approximate posteriors.

\section{Approximating the posteriors via a regularizing ensemble Kalman algorithm}\label{ensemble}

In the previous section we have demonstrated, by means of numerical examples, that an accurate approximation of the Bayesian posteriors via the state-of-the art SMC samplers results in a very high computational cost; hence it is unfeasible for practical applications such as the 2D resin injection in RTM introduced in Section \ref{Intro}. In this section we propose a regularizing ensemble Kalman algorithm (REnKA) that aims at providing an accurate approximation of the sequence of Bayesian posteriors at a much lower computational cost. In Section~\ref{REnKA} we introduce REnKA as a Gaussian approximation from the SMC sampler of \cite{Kantas} discussed in the preceding section. The proposed REnKA in the context of existing ensemble Kalman methods is discussed in Section~\ref{REnKA_liet}. A numerical investigation of the convergence properties of REnKA is reported in
Section~\ref{REnKA_num}.

For the subsequent development of the proposed scheme, we extend the domain of definition of the sequence of forward maps $\cG_{n}$ introduced in (\ref{eq11}). More specifically, we assume $\cG_{n}:\mathcal{X}\to \mathbb{R}^{M+1}$ where $\mathcal{X}$ is a Hilbert space such that $X=C[0,x^{*}]\hookrightarrow\mathcal{X}$ (compactly). We denote by $<\cdot ,\cdot>_{\mathcal{X}}$ and $<\cdot ,\cdot>$ the inner products in $\mathcal{X}$ and $\mathbb{R}^{M+1}$, respectively. In addition, we define $\mathcal{Z}\equiv \mathcal{X}\times\mathbb{R}^{M+1}$ with inner product denoted by $<\cdot ,\cdot>_{Z}$.

\subsection{Motivation for REnKA}\label{REnKA}

Motivated by the SMC tempering approach described in the previous section, we now propose an ensemble Kalman algorithm whose aim is to approximate $\{\mu_{n,r}\}_{r=1}^{q_{n}}$ by a sequence of Gaussian measures $\{\nu_{n,r}\}_{r=1}^{q_{n}}$ which are, in turn, characterised by a set of particles with equal weights. Suppose that, at time $t=t_{n}$ we have an ensemble $\{u_{n,r-1}^{(j)}\}_{j=1}^{J}$ of $J$ samples from a Gaussian measure $\nu_{n,r-1}$ that approximates $\mu_{n,r-1}$, and a prescribed tempering parameter $\phi_{n,r-1}$. We may then solve (\ref{eq38}) for the new $\phi_{n,r}$ and define the regularization parameter $\alpha_{n,r}$ in (\ref{reg_param}). We now wish to make a transition from $\nu_{n,r-1}$ to a Gaussian measure $\nu_{n,r}$ that approximates $\mu_{n,r}$. To this end, let us define the augmented variable
\begin{eqnarray}\label{eqE1}
z=(u,\cG_{n}(u))^{T}\in  Z
\end{eqnarray}
and note that, in terms of this variable, we may rewrite (\ref{eq42}) as
\begin{eqnarray}\label{eqE2}
y_{n}=Hz+\tilde{\eta}_{n},\qquad \tilde{\eta}_{n,r}\sim N(0,\alpha_{n,r}\Gamma_{n}),
\end{eqnarray}
where $H=(0,I)$ and $I$ is the identity operator. One can see that by reformulating the inverse problem in terms of the augmented variable, the resulting forward map (i.e. $H$) acting on this variable is linear.

From \eqref{eqE1} we define the following augmented particles
\begin{eqnarray}\label{eqE3}
z_{n,r-1}^{(j)}= (u_{n,r-1}^{(j)}, \cG_{n}(u_{n,r-1}^{(j)}))^{T},
\end{eqnarray}
and construct the empirical Gaussian measure:
\begin{eqnarray}\label{eqE4}
\hat{\nu}_{n,r-1}(z) \equiv  N(\overline{z}_{n,r-1},C_{n,r-1}),
\end{eqnarray}
where
\begin{eqnarray}\label{eqE5}
\overline{z}_{n,r-1}\equiv \frac{1}{J}\sum\limits_{j=1}^{J} z_{n,r-1}^{(j)},
\end{eqnarray}
and
\begin{eqnarray}\label{eqE6}
C_{n,r-1} (\cdot)\equiv \frac{1}{J-1}\sum\limits_{j=1}^{J}(z_{n,r-1}^{(j)}-\overline{z}_{n,r-1})\langle z_{n,r-1}^{(j)}-\overline{z}_{n,r-1},\cdot\rangle_{Z}.
\end{eqnarray}

The Gaussian measure $\hat{\nu}_{n,r-1}(z)$ is used to approximate the measure, denoted by $\hat{\mu}_{n,r-1}(z)$, that arises from pushing forward $\mu_{n,r-1}(u)$ under (\ref{eqE1}). By using this Gaussian approximation of $\hat{\mu}_{n,r-1}(z)$, we then provide a Bayesian formulation of the inverse problem given by \eqref{eqE2}. More specifically, we wish to compute $\hat{\nu}_{n,r}(z)\equiv \mathbb{P}(z\vert y_{n})$ given $\hat{\nu}_{n,r-1}(z)$ and the data from (\ref{eqE2}). A formal application of Bayes theorem yields
\begin{eqnarray}\label{eqE7}
\frac{d\hat{\nu}_{n,r}}{d\hat{\nu}_{n,r-1}}(z)\propto \exp\Big[-\frac{1}{2} \vert\vert  (\alpha_{n,r}\Gamma_{n})^{-1/2}(y_{n}-Hz)\vert\vert^2\Big].
\end{eqnarray}
Moreover, from (\ref{eqE4}) and the linearity of the forward map $H$ (on the augmented variable), it follows by standard arguments \cite{0266-5611-5-4-011} that
\begin{eqnarray}\label{eqE8}
\hat{\nu}_{n,r}(z)= N\big(\overline{z}_{n,r-1}+  \mathcal{K}_{n,r}(y_{n}-H\overline{z}_{n,r-1}),(I-  \mathcal{K}_{n,r}H)C_{n,r-1}\big),
\end{eqnarray}
where
\begin{eqnarray}\label{eqE9}
\mathcal{K}_{n,r}\equiv   C_{n,r-1}H^{T}(HC_{n,r-1} H^{T}+\alpha_{n,r}\Gamma)^{-1}.
\end{eqnarray}
Let us then note that $C_{n,r-1}$ in (\ref{eqE6}) can be written as
\begin{eqnarray}\label{eqE10}
C_{n,r-1}=\left[\begin{array}{cc}
C_{n,r-1}^{uu} & C_{n,r-1}^{uw}\\
(C_{n,r-1}^{uw})^{T} & C_{n,r-1}^{ww}\end{array}\right],
\end{eqnarray}
where
\begin{eqnarray}\label{eqE11}
C_{n,r-1}^{ww}(\cdot)=&\frac{1}{J-1}\sum\limits_{j=1}^{J}(\cG(u_{n,r-1}^{(j)})-\overline{\cG}_{n,r-1})\langle G(u_{n,r-1}^{(j)})-\overline{\cG}_{n,r-1},\cdot\rangle,\\
C_{n,r-1}^{uw}(\cdot) =& \frac{1}{J-1}\sum\limits_{j=1}^{J} (u_{n,r-1}^{(j)}-\overline{u}_{n,r-1})\langle \cG(u_{n,r-1}^{(j)})-\overline{\cG}_{n,r-1},\cdot \rangle,\label{eqE12}\\
C_{n,r-1}^{uu}(\cdot) =& \frac{1}{J-1}\sum\limits_{j=1}^{J} (u_{n,r-1}^{(j)}-\overline{u}_{n,r-1})\langle u_{n,r-1}^{(j)}-\overline{u}_{n,r-1},\cdot \rangle_{\mathcal{X}},\label{eqE13}
\end{eqnarray}
and
$$\overline{\cG}_{n,r-1} \equiv \frac{1}{J}\sum\limits_{j=1}^{J} \cG_{n}(u_{n,r-1}^{(j)}),\qquad \overline{u}_{n,r-1} \equiv \frac{1}{J}\sum\limits_{j=1}^{J}u_{n,r-1}^{(j)}.$$
Informally, we use the block structure of \eqref{eqE10} and define $\nu_{n,r}$, the approximation of $\mu_{n,r}$, as the marginal measure of $\hat{\nu}_{n,r} $ given by
\begin{eqnarray}\label{eqE14}
\nu_{n,r}(u)\equiv N\big(\overline{u}_{n,r-1}+  \mathcal{K}_{n,r}^{u}(y_{n}-\overline{\cG}_{n,r-1}),C_{n,r-1}^{uu}-  \mathcal{K}_{n,r}^{u}(C_{n,r-1}^{uw})^{T}\big),
\end{eqnarray}
where
\begin{eqnarray}\label{eqE15}
\mathcal{K}_{n,r}^{u}=  C_{n,r-1}^{uw}(C_{n,r-1}^{ww}+\alpha_{n,r}\Gamma)^{-1}.
\end{eqnarray}
Although the measure \eqref{eqE14} is fully characterised by its mean and covariance, for the subsequent tempering step we need a particle approximation of $\nu_{n,r}(u)$. We can obtain those particles by updating the current set of particles $u_{n,r-1}^{(j)}$ via the formula
 \begin{eqnarray}\label{eqE16}
u_{n,r}^{(j)} =u_{n,r-1}^{(j)}+C_{n,r-1}^{uw}(C_{n,r-1}^{ww} +\alpha_{n,r}\Gamma   )^{-1}(y_{n,r}^{(j)}-\cG_{n}(u_{n,r-1}^{(j)})),
\end{eqnarray}
where
\begin{eqnarray}\label{eqE17}
y_{n,r}^{(j)}\equiv y_{n}+\eta_{n,r}^{(j)},\qquad \eta_{n,r}^{(j)}\sim N(0,\alpha_{n,r}\Gamma_{n}).
\end{eqnarray}
Indeed, under the standard assumption that the noise $\eta_{n,r}^{(j)}$ is independent from the variables $u_{n,r-1}^{(j)}$ and $\cG_{n}(u_{n,r-1}^{(j)})$, it can be shown \mt{by} the standard arguments in Kalman-based methods (see for example  \cite{Burgers}) that
\begin{eqnarray}\label{eqE18}
\nu_{n,r}^{J}(u)\equiv \frac{1}{J}\sum_{j=1}^{J}\delta_{u_{n,r}^{(j)}}(z)\to \nu_{n,r}(u),\quad J\to \infty.
\end{eqnarray}
Expression \eqref{eqE16} and selection of the regularisation parameter \mt{$\alpha_{n,r}$} based on the adaptive tempering approach discussed in Section \ref{Kantas_SMC} constitutes the proposed scheme summarized in Algorithm \ref{EnKF_al}.

\begin{algorithm}\caption{Regularizing ensemble Kalman algorithm (REnKA)}\label{EnKF_al}
\begin{algorithmic}
\STATE{ Let $\{u_{0,0}^{(j)}\}_{j=1}^{J}\sim \mu_{0}$ be the initial ensemble of $J$ particles.}
\STATE{ Define the tunable parameter $J_{thresh}$.}
\FOR{ $n=1,\dots,N$}
\STATE{ Set $r=0$ and $\phi_{n,0}=0$;}
\WHILE{ $\phi_{n,r}<1$}
\STATE{$r\to r+1$}
\STATE{  \textbf{Compute the $n$th likelihood} (\ref{eq14}) $l_{n}(u_{n,r-1}^{(j)},y_{n})$ for $j=1,\dots, J$.}
\STATE{  (this implies computing $\cG(u_{n,r-1}^{(j)})$ needed below).}
\STATE{    \textbf{Compute the tempering parameter} $\phi_{n,r}$:}

\IF { $\min_{\phi\in (\phi_{n,r-1},1]}\textrm{ESS}_{n,r}(\phi)>J_{thresh}$}
        \STATE   set $\phi_{n,r}=1$.
\ELSE
                \STATE compute $\phi_{n,r}$ such that $\textrm{ESS}_{n,r}(\phi)\approx J_{thresh}$ using a bisection algorithm on $(\phi_{n,r-1},1]$.

\ENDIF

\STATE{  Construct $C_{n,r-1}^{uw}$,  $C_{n,r-1}^{ww}$ defined by expressions (\ref{eqE11})-(\ref{eqE12}).}
\STATE{  Update each ensemble member:}

\FOR {$j=1,\dots,J$}
\STATE{ \begin{eqnarray}\label{eq:m16}
u_{n,r}^{(j)} =u_{n,r-1}^{(j)}+C_{n,r-1}^{uw}(C_{n,r-1}^{ww} +\alpha_{n,r}\Gamma   )^{-1}(y_{n,r}^{(j)}-\cG_{n}(u_{n,r-1}^{(j)})),
\end{eqnarray}}
\STATE{ where
$$\alpha_{n,r}=(\phi_{n,r}-\phi_{n,r-1})^{-1},\qquad y_{n,r}^{(j)}=y_{n}+\eta_{n,r}^{(j)},$$}
\STATE{ with  $\eta_{n,r}^{(j)}\sim N(0,\alpha_{n,r}\Gamma_{n}).$}
\ENDFOR

\ENDWHILE
\STATE{  Set $u_{n+1,0}^{(j)}\equiv u_{n,r-1}^{(j)}$. Approximate $\mu_{n}$ with $\nu_{n}^{J}\equiv \frac{1}{J}\sum_{j=1}^{J} \delta_{u_{n,r}^{(j)}}.$}
\ENDFOR

\end{algorithmic}
\end{algorithm}

\begin{remark}
Note that the key assumption for the proposed scheme is the Gaussian approximation of $\hat{\mu}_{n,r-1}(z)$ provided by (\ref{eqE8}).
It is clear that \mt{the measure $\hat{\mu}_{n,r-1}(z)$} is, \mt{as a rule,} non-Gaussian and the aforementioned assumption will result in a methodology that will, in general, not converge to the posteriors $\mu_{n}$ as the ensemble size $J\to \infty$. Nevertheless, we will show via numerical examples that \mt{this} approximation provides reasonably accurate estimates using only a small number of particles.
\end{remark}
It is not difficult to see that the main computational cost of REnKA, in terms of the cost of evaluating the forward model at the final observation time, is given by
\begin{eqnarray}\label{cost_RENKA}
\mathbf{C}_{REnKA}\equiv J  \sum_{n=1}^{N} q_{n}  \frac{g_{n}}{g_{N}},
\end{eqnarray}
where, as before, $g_{n}$ denotes the computational cost of evaluating the $\cG_{n}$-forward map. As we will demonstrate via numerical experiments, for the moving boundary problem of Section~\ref{1D}, REnKA offers a computationally affordable and thus practical approach to approximate the solution to the Bayesian inverse problem that arises from RTM.

It is important to mention that, at a given observation time $t_{n}$ and iteration level $r$, the value of $\sum_{j=1}^{J} l_{n}(u_{n,r-1}^{(j)},y_{n})^{1-\phi_{n,r-1}}$ may be zero to machine precision. In this case, the tempering parameter $\phi_{n,r}$ is not be computable, via a bisection scheme on $(\phi_{n,r-1},1]$, as stated in Algorithm \ref{EnKF_al}. This computational issue is more likely to arise at the early iterations of the scheme for which the value of $\phi_{n,r-1}$ is not sufficiently close to one. This can be overcome, for example, by simply adapting the bisection algorithm in order to first compute a $\phi_{\ast}$ such that $\sum_{j=1}^{J} l_{n}(u_{n,r-1}^{(j)},y_{n})^{\phi_{\ast}-\phi_{n,r-1}}>0$. If $\min_{\phi\in (\phi_{n,r-1},\phi_{\ast}]}\textrm{ESS}_{n,r}(\phi)>J_{thresh}$ we then set $\phi_{n,r}=\phi_{\ast}$; otherwise, we find $\phi_{n,r}$ by solving (\ref{eq38}) via a bisection algorithm on $(\phi_{n,r-1},\phi_{\ast}]$. For the numerical experiments reported in the present work, zero values to machine precision for $\sum_{j=1}^{J} l_{n}(u_{n,r-1}^{(j)},y_{n})^{1-\phi_{n,r-1}}$ were only encountered where a large number of measurements were inverted in the 2D setting of Section \ref{REnKA_2D_invest}.

\subsection{REnKA in the context of existing ensemble Kalman methods for inverse problems}\label{REnKA_liet}

Ensemble Kalman methods for inverse/calibration problems have been widely used in the last decades \cite{evensen2009data}.  More recently, using iterative Kalman methods with a regularization parameter (e.g. $\alpha_{n,r}$ in (\ref{eq:m16})) have been proposed \mt{for} a wide class of applications. In particular, the proposed REnKA scheme can be related to the recently developed regularizing ensemble Kalman method introduced in \cite{EnsembleYo} for solving classical (deterministic) inverse problems. More precisely, Algorithm \ref{EnKF_al} is nothing but a sequential version of the iterative scheme presented in \cite{EnsembleYo} except for the selection of the regularization parameter $\alpha_{n,r}$. While in the present work we have motivated Algorithm \ref{EnKF_al} from the SMC framework, the algorithm in \cite{EnsembleYo} was obtained as an ensemble approximation of the regularizing Levenberg-Marquardt \mt{scheme} initially developed in \cite{Hanke} for solving ill-posed inverse problems. In the context of the proposed work, REnKA aims at providing a derivative-free approximation to the solution of
\begin{eqnarray}\label{ls_RENKA}
\vert \vert \Gamma_{n}^{-1/2}(y_{n}- \cG_{n}(u))\vert\vert \to \min
\end{eqnarray}
in a stable (regularized) fashion. The regularization parameter in \cite{EnsembleYo} was selected according to the discrepancy principle in order to regularize the inverse problem posed by (\ref{ls_RENKA}). In contrast, the present work uses the adaptive tempering approach of \cite{Kantas} for the selection of this regularization parameter in the context of SMC. It is clear that tempering can be understood as a regularization to the Bayesian inverse problem; the effect of $\alpha_{n,r}$ is to flatten out the posterior and allow for a more controlled/regularized transition between the sequence of measures. Other works highlighting the connection between ensemble Kalman methods and SMC approaches include \cite{GaussianM1,GaussianM2,SS17}. In addition, by noticing from (\ref{reg_param}) that, for each $n=1,\dots, N$, $\sum_{r=1}^{q_{n}}\alpha_{n,r}^{-1}=1$, the proposed REnKA can be also understood as a sequential version of the ensemble smoother with multiple data assimilation proposed by \cite{EMERICK20133}. However, it is important to reiterate that the adaptive selection of $\alpha_{n,r}$ proposed here is inherited from the SMC approach of \cite{Kantas}. This selection differs substantially from the strategy proposed in \cite{EMERICK20133} for which the number of intermediate tempering distributions $q_{n}$ is fixed and selected a priori.

\subsection{Numerical approximating the posterior with REnKA}\label{REnKA_num}

\begin{figure}[htbp]
\begin{center}
\includegraphics[scale=0.7]{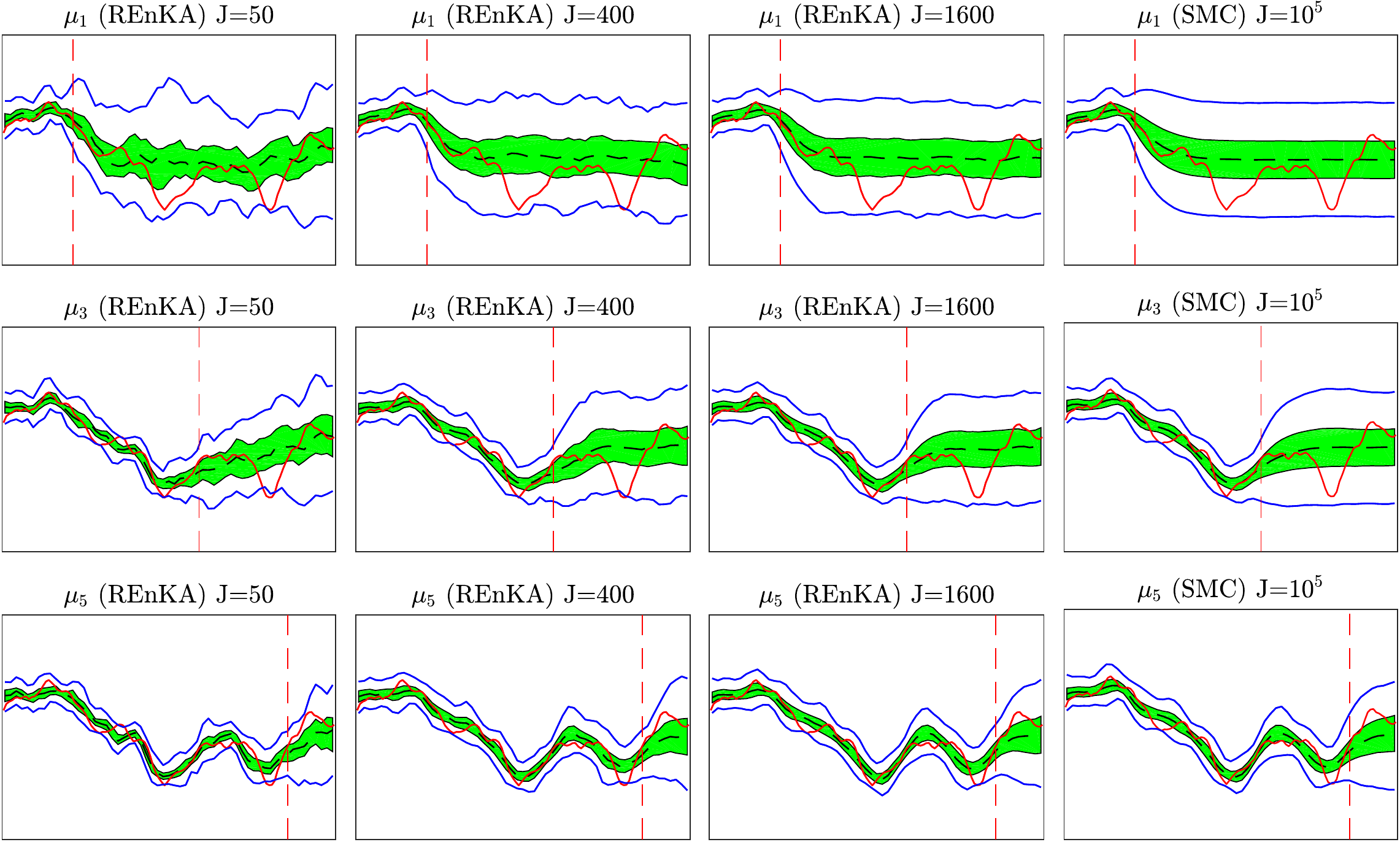}
 \caption{Percentiles of the posteriors $\mu_{1}$, $\mu_{3}$, and $\mu_{5}$ obtained via REnKA with (from left to right) $J=50, 400, 1600$ and SMC (right column) with $J=10^5$. Solid red line corresponds to the graph of the true log-permeability $u^{\dagger}$. Vertical dotted line indicates the location ot the true front $\Ga^{\dagger}(t_{n})$.} \label{Fig4A}
\end{center}
\end{figure}

In this subsection we report the results from applying REnKA proposed in Section~\ref{REnKA} for the approximation of the sequence of posteriors $\{\mu_{n}\}_{n=1}^{5}$ that we introduced in the framework of Section \ref{SMC_1d_ex}. The algorithm is applied with the same selection of ensemble sizes ($J=50$,100, 200, 400, 800, 1600, 3200, 6400) that we use for the SMC sampler of Section~\ref{SMC_small}. In addition, we consider three choices of the tunable parameter $J_{thresh}$ ($J_{thresh}=J/3,J/2,2J/3$). In Figure \ref{Fig4A} we display the percentiles of the log-permeability posteriors $\mu_{1}$, $\mu_{3}$ and $\mu_{5}$ obtained with REnKA, for a fixed set of initial ensembles, and for some of these choices of $J$. For comparison purposes we include the fully resolved posterior (via SMC) in the right column of Figure \ref{Fig4A}. We can clearly observe that the approximations provided by REnKA improves as the ensemble size $J$ increases. More importantly, we can note that the uncertainty band defined by these percentiles provided better approximations than those from SMC with the same number of particles (see Figure \ref{Fig3A}).

\begin{figure}[htbp]
\begin{center}

\includegraphics[scale=0.32]{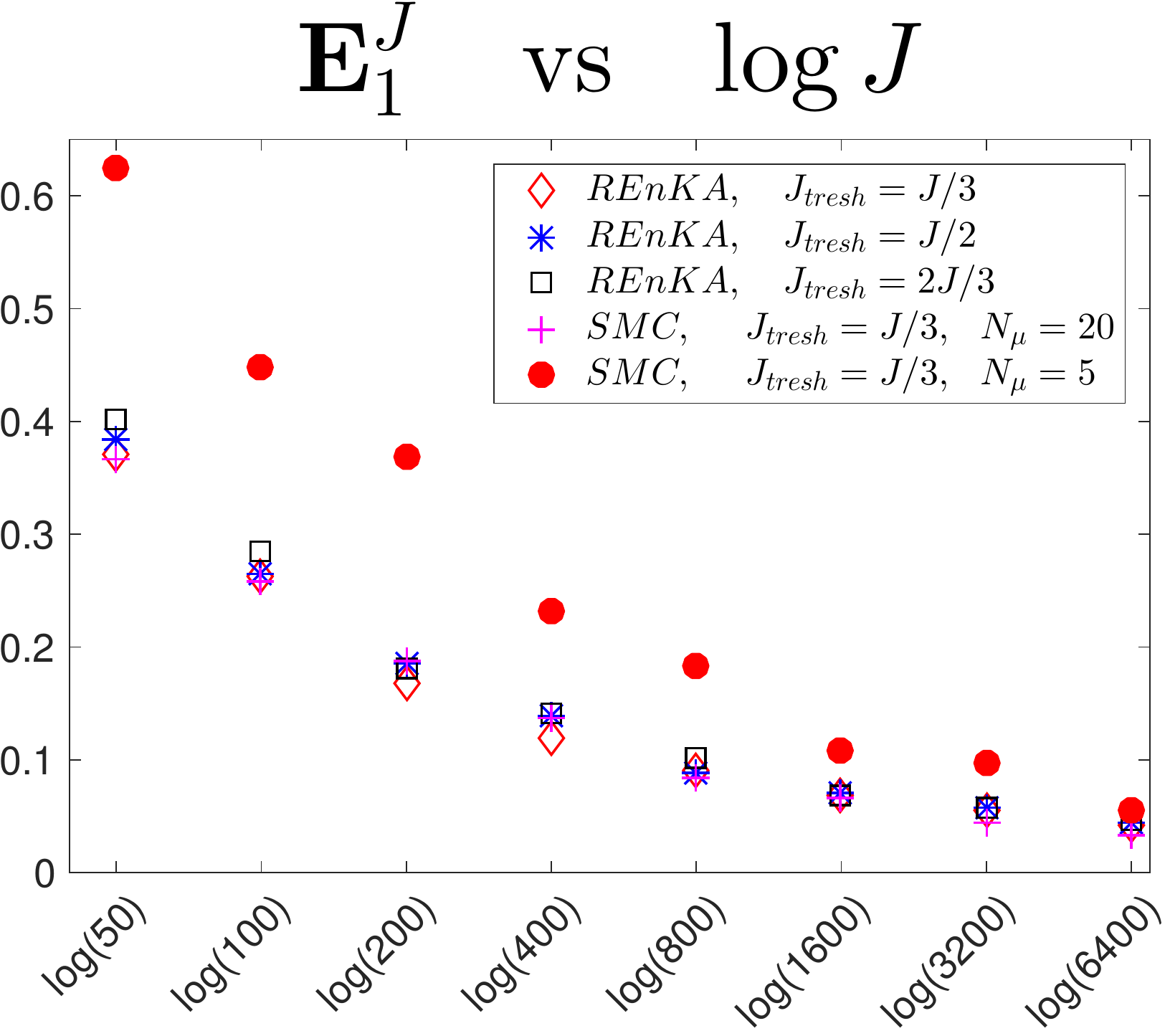}~\includegraphics[scale=0.32]{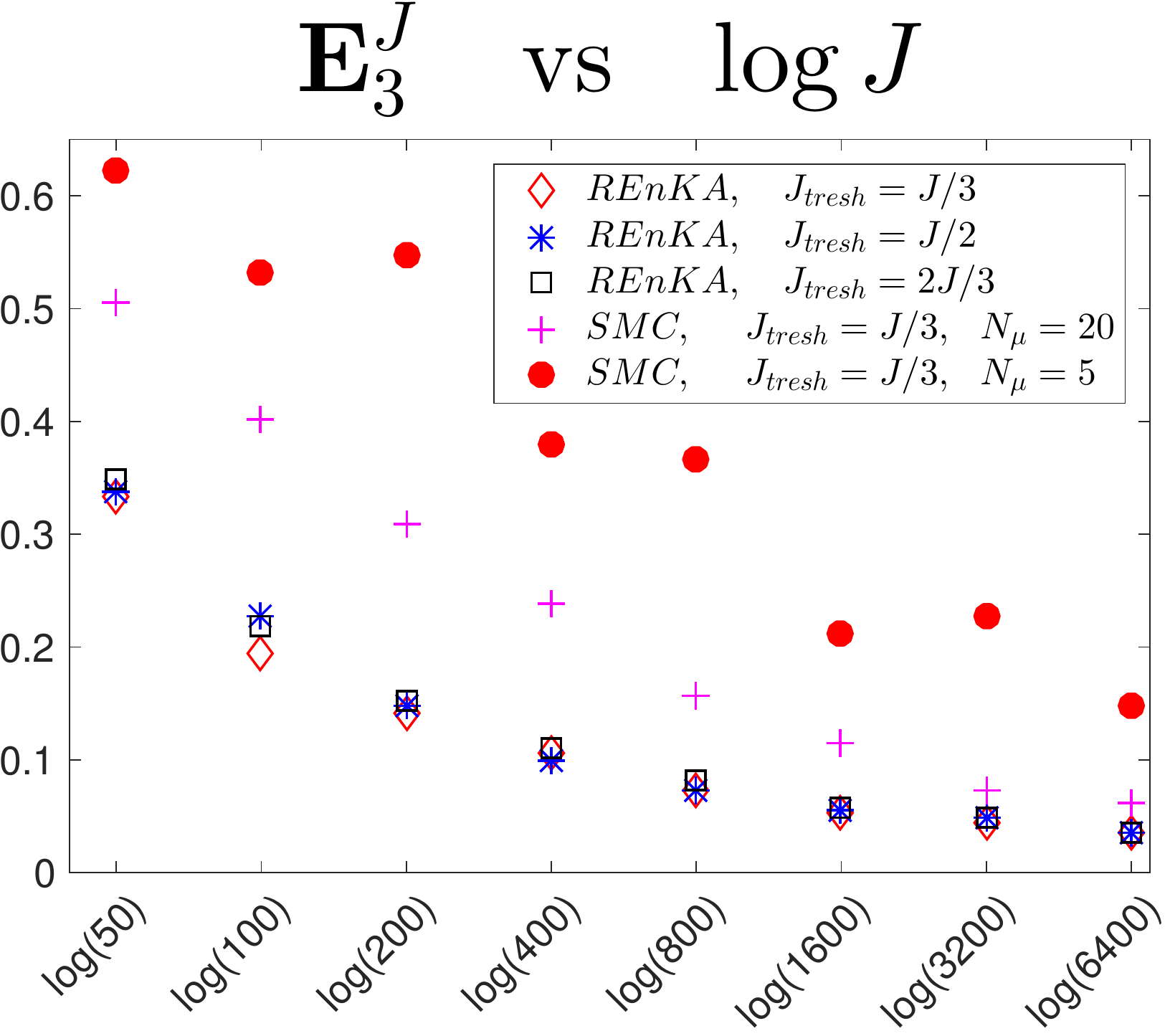}~\includegraphics[scale=0.32]{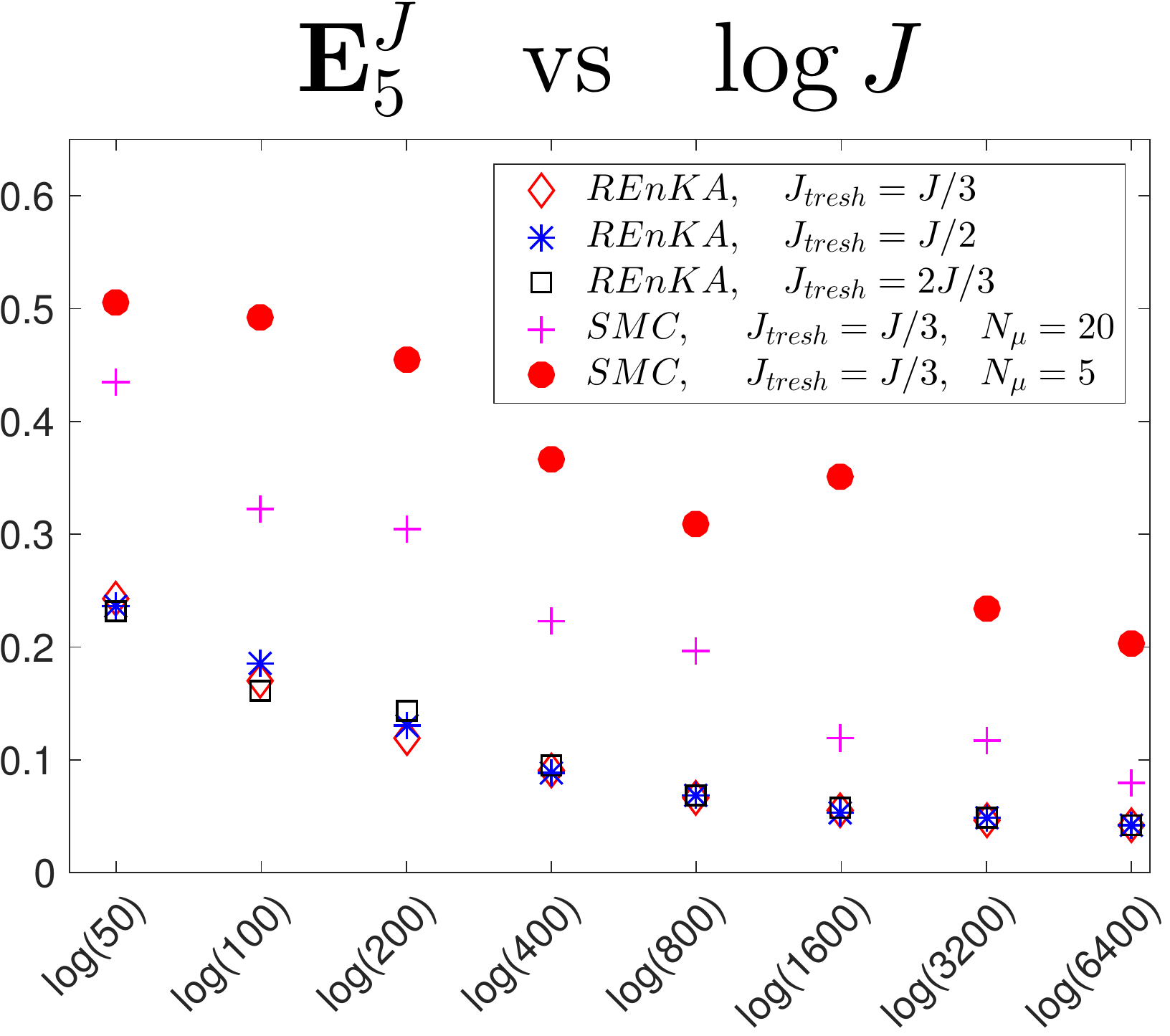}\\
\includegraphics[scale=0.32]{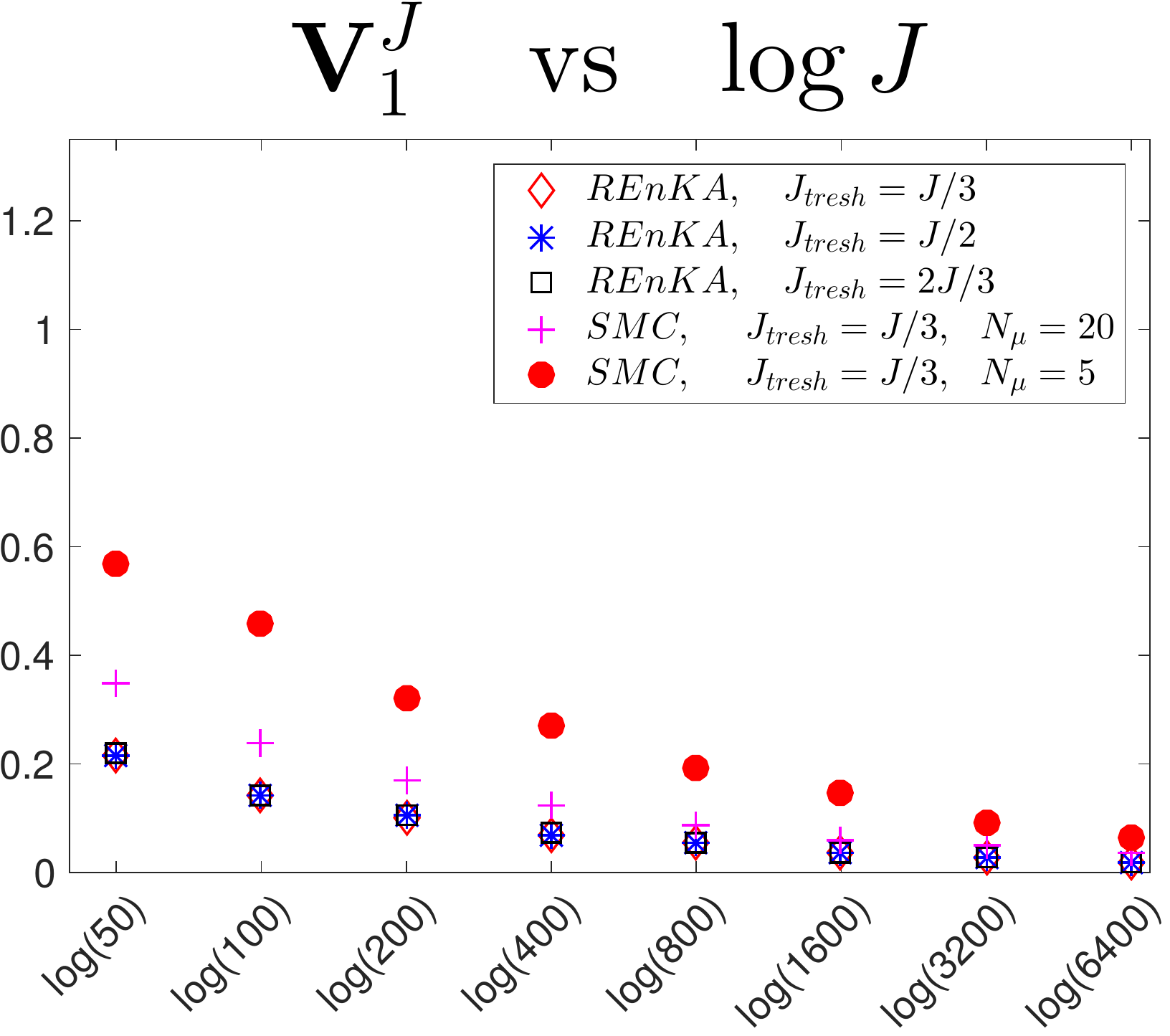}~\includegraphics[scale=0.32]{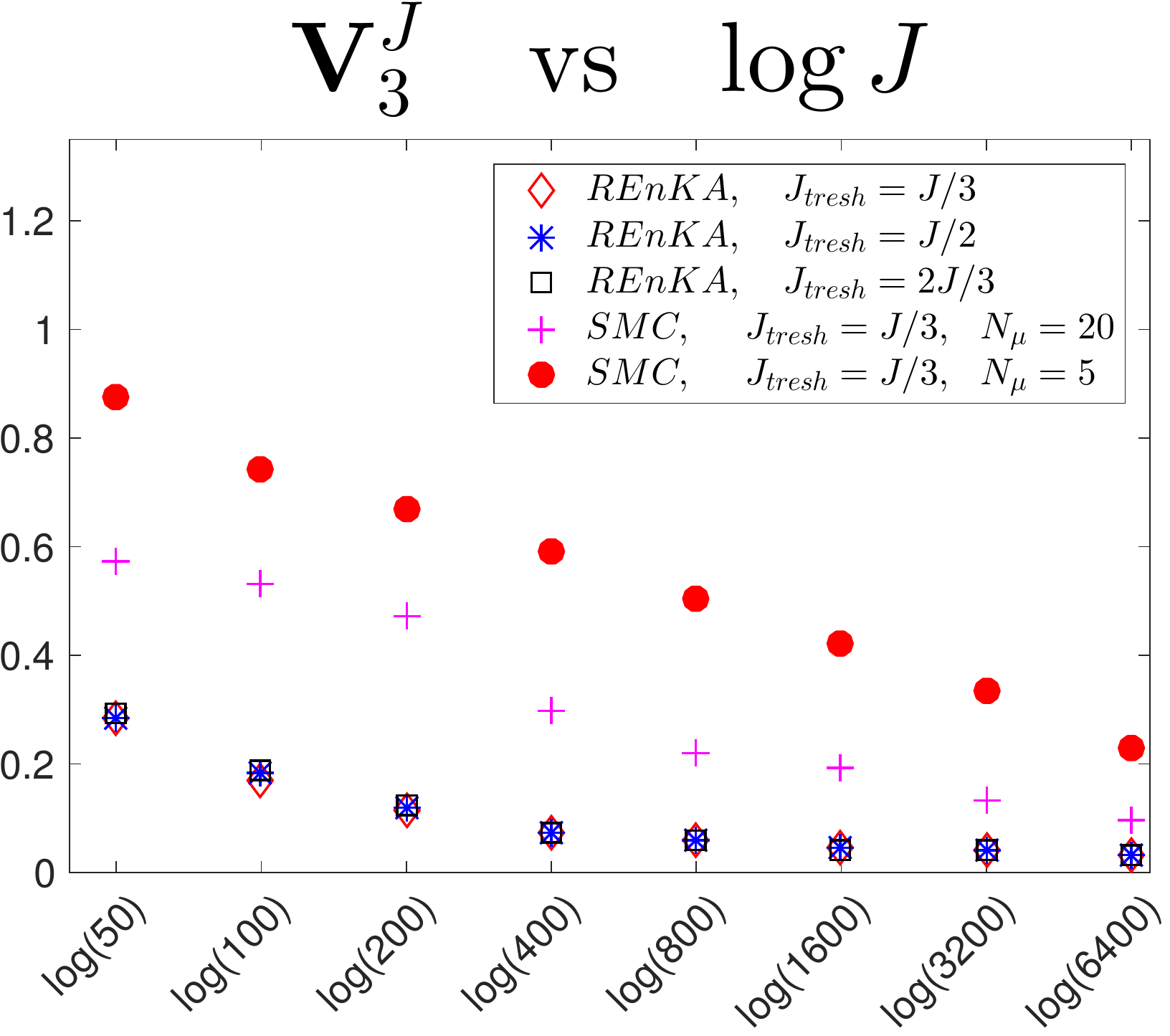}~\includegraphics[scale=0.32]{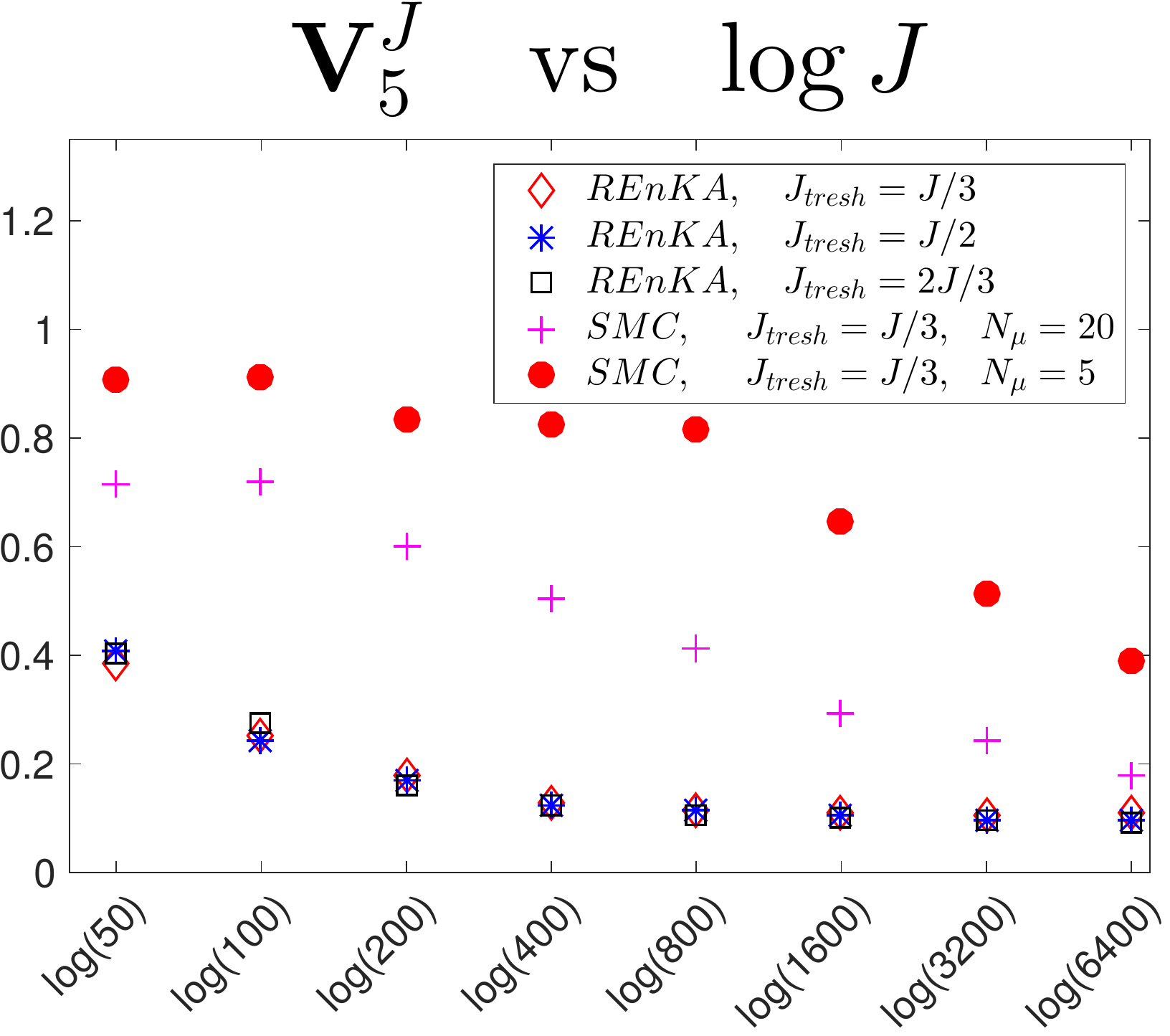}\\
\includegraphics[scale=0.32]{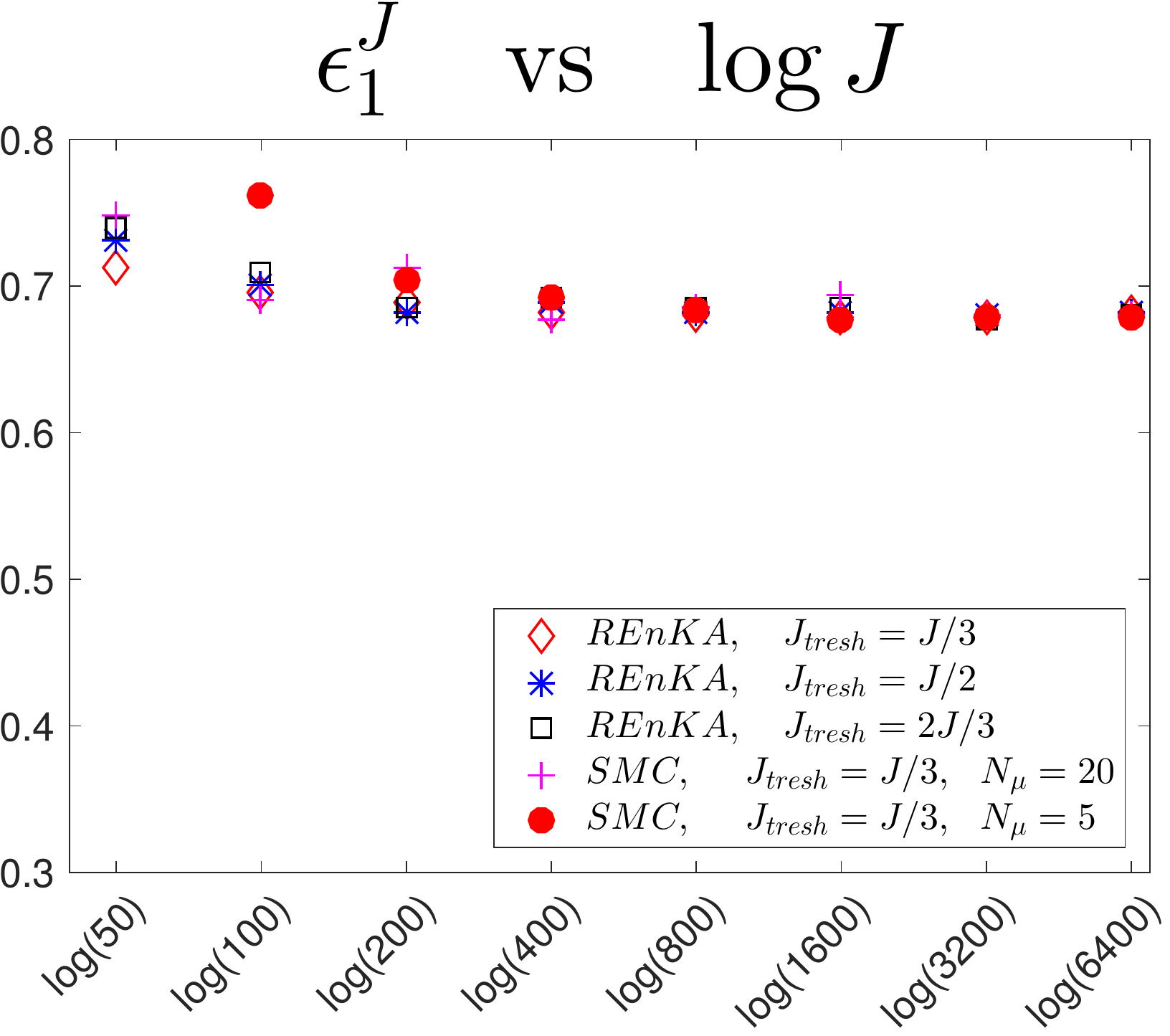}~\includegraphics[scale=0.32]{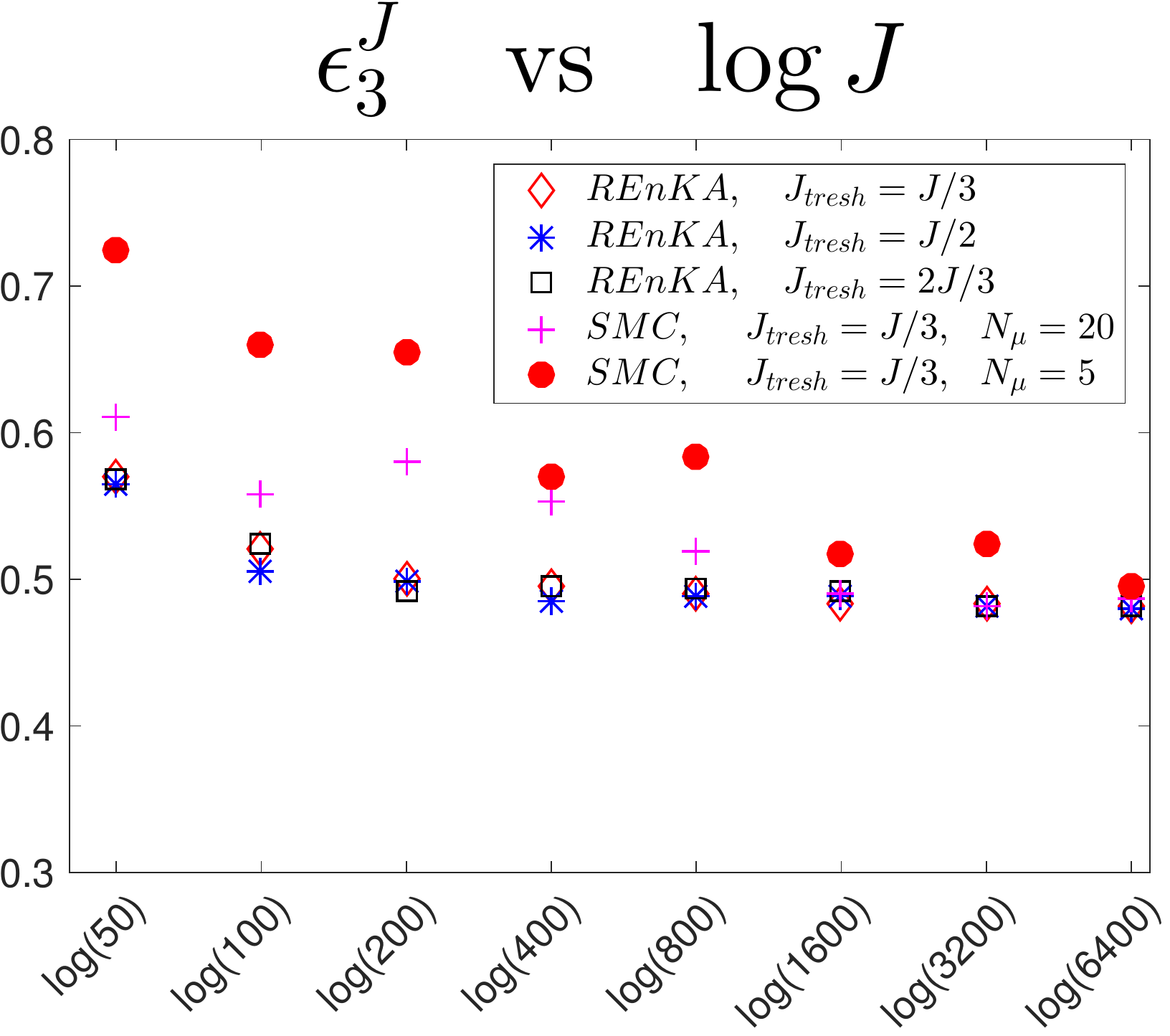}~\includegraphics[scale=0.32]{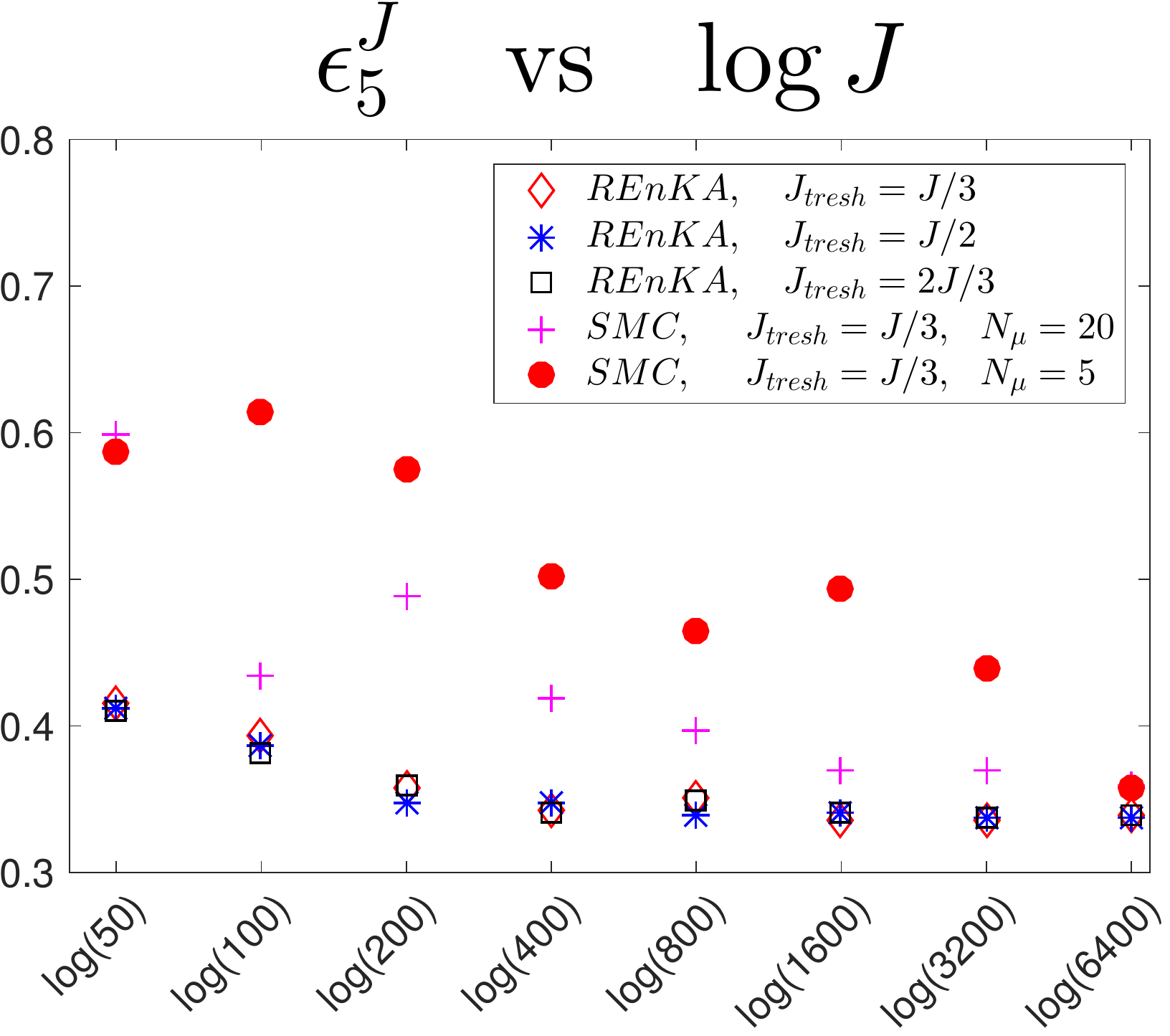}
 \caption{Top and middle: Relative errors of mean (top) and variance (middle) of the posteriors $\mu_{n}$, (from left to right) n=1,3, 5, obtained via REnKA with different choices of (log) ensemble size $\log(J)$ and tunable parameter $J_{thresh}$. Bottom: Relative errors of the ensemble mean obtained via REnKA with respect to the truth $u^{\dagger}$. }\label{Fig4B}
\end{center}
\end{figure}

 We quantify the level of accuracy of REnKA with respect to the Benchmark defined by the highly-resolved SMC sampler reported in Section~\ref{benchmark}. In Figure \ref{Fig4B} we display $\mathbf{E}_{n}^{J}$ (top), $ \mathbf{V}_{n}^{J}$ (middle) and $\epsilon_{n}^{J}$ (bottom) for (from left to right) $n=1,3,5$ computed with the REnKA samples with various ensemble sizes $J$. Similar results (not shown) are obtained for $n=2,4$. The total computational cost ($\mathbf{C}_{REnKA}$ from (\ref{cost_RENKA})) of computing the full sequence of posteriors is shown in Figure \ref{Fig3C} (right). In Figure \ref{Fig4B} and Figure \ref{Fig3C} (right) we also include some of the results obtained with the SMC samplers with the same choice of small number of particles discussed in Section~\ref{SMC_small}. These results speak for themselves; given a small number of particles, REnKA provides a much more accurate approximation of the posterior measures than SMC. For example, note that for the final measure $\mu_{5}$, REnKA (applied with $J_{thresh}=J/3$) with an ensemble of $J=200$ particles yields $\mathbf{E}_{5}^{200}\equiv 12\%$, $\mathbf{V}_{5}^{200}\equiv 18\%$  at a computational cost of $1.6\times 10^3$ $\cG_{5}$-forward model evaluations. In order to obtain a similar level of accuracy ($\mathbf{E}_{5}^{200}\equiv 11\%$, $\mathbf{V}_{5}^{200}\equiv 24\%$), we need to apply the SMC sampler (say with $J_{thresh}=J/3$, and $N_{\mu}=20$) with $J=6400$ particles for which the computational cost is approximately $5\times 10^5$ $\cG_{5}$-forward model evaluations.

The results above not only demonstrate that, when a small number of particles is used, the performance (accuracy vs computational cost) of REnKA outperforms SMC, but also these results show that REnKA is robust for reasonable selections of the tunable parameter $J_{thresh}$. Similar to SMC, this parameter determines the number of tempering distributions at each observation time and thus has an impact on the computational cost of the scheme. It is also important to remark that even though the relative errors of REnKA with respect to the mean and variance decrease as the ensemble size increases, these errors seem to converge to a non-zero value thereby indicating that REnKA does not provide an asymptotic convergence to the posterior measures as $J\to \infty$. Nevertheless, our results clearly showcase the advantages of REnKA for approximating these measures in an accurate and efficient fashion for a limited and realistic computational cost.

\section{Numerical investigations of the Bayesian Inverse problem in RTM}\label{REnKA_application}

Having numerical evidence from Section~\ref{REnKA_num} that REnKA is a robust and computationally feasible approach for addressing the Bayesian inverse problem defined in Section \ref{sec:1d}, in this section we use REnKA to provide further practical insights in the RTM Bayesian inverse problem. Section~\ref{num_1D} is devoted to the 1D case studied earlier which includes the study of Section~\ref{meas}, concerning the effect of the number/type of measurements on the solution to the Bayesian inverse problem. The effect of the number of observation times as well as the observational noise level are investigated in Sections~\ref{meas_time} and~\ref{noise}, respectively. The application of REnKA for the 2D forward model \mt{stated} in Section \ref{Intro} is reported in Section~\ref{REnKA_2D_invest}. For the 1D (resp. 2D) case we consider REnKA with a fixed number of $J=1000$ (resp. $J=150$). In all cases we select the tunable parameter $J_{thresh}=J/3$. For clarity \mt{of} the notation, in the expression for the ensemble mean and variance (\ref{eq:1001B}), we then omit the index $J$ as appropriate. We will focus our numerical investigations in terms of the following quantities defined at each observation time $t_{n}$ ($n=1,\dots, N$):
\begin{itemize}
\item[(A)] The $L^{2}(D^{\ast})$-relative error with respect to the truth, $\epsilon_{n}$ defined in (\ref{truth_err}) in terms of the ensemble mean $\overline{u}_{n}$.
\item[(B)]  A measure of the uncertainty provided by the $L_{2}(D^{\ast})$-norm of the ensemble variance $\sigma _{n}^{2}$ relative to the prior, i.e.
\begin{eqnarray}\label{truth_var0}
\Sigma_{n}\equiv \frac{\vert\vert \sigma_{n}^{2}\vert\vert_{L^{2}(D^{\star})}}{\vert\vert \sigma_{0}^{2}\vert\vert_{L^{2}(D^{\star})}};
\end{eqnarray}
\item[(C)] A normalized $L^{2}(D^{\dagger}(t_n))$-error with respect to the truth defined by
\begin{eqnarray}\label{truth_err2}
\epsilon_{n}^{\Ga}\equiv \frac{1}{\vert D^{\dagger}(t_n)\vert } \vert\vert u^{\dagger}-\overline{u}_{n}\vert\vert_{L^{2}(D^{\dagger}(t_n))};
\end{eqnarray}
\item[(D)] A visual inspection of the percentiles of the approximate posterior characterized with the ensemble.
\end{itemize}
For (A)-(C) we report quantities averaged over 15 experiments corresponding to different selection of the initial ensemble from the prior.

\subsection{Further investigations of the 1D case}\label{num_1D}
In this section we study how the number of observation times and the observational noise level influence the solution to the Bayesian inverse problem in the 1D case.

\subsubsection{Effect of the number and type of measurements}\label{meas}

\begin{figure}[htbp]
\begin{center}

\includegraphics[scale=0.25]{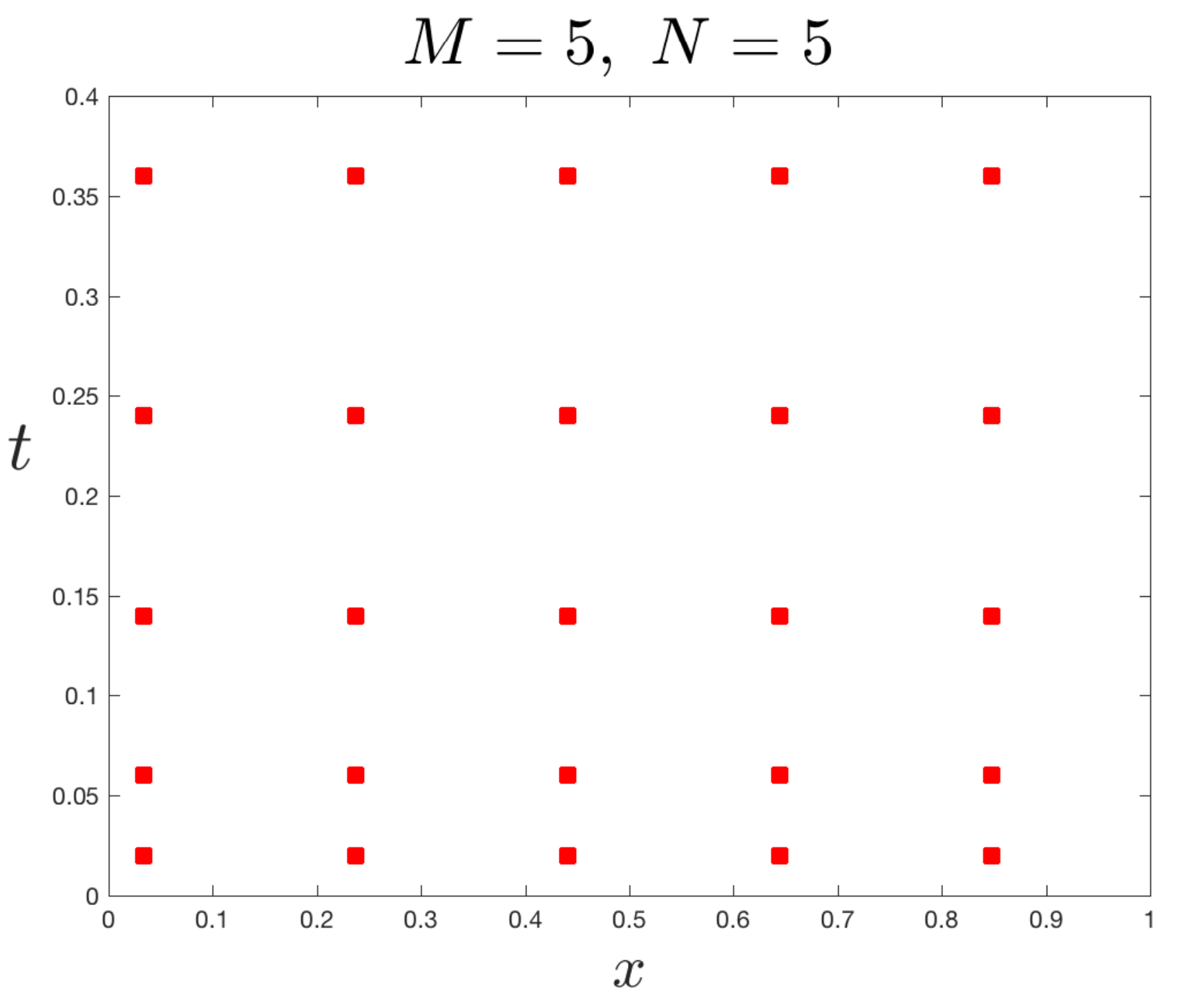}
\includegraphics[scale=0.25]{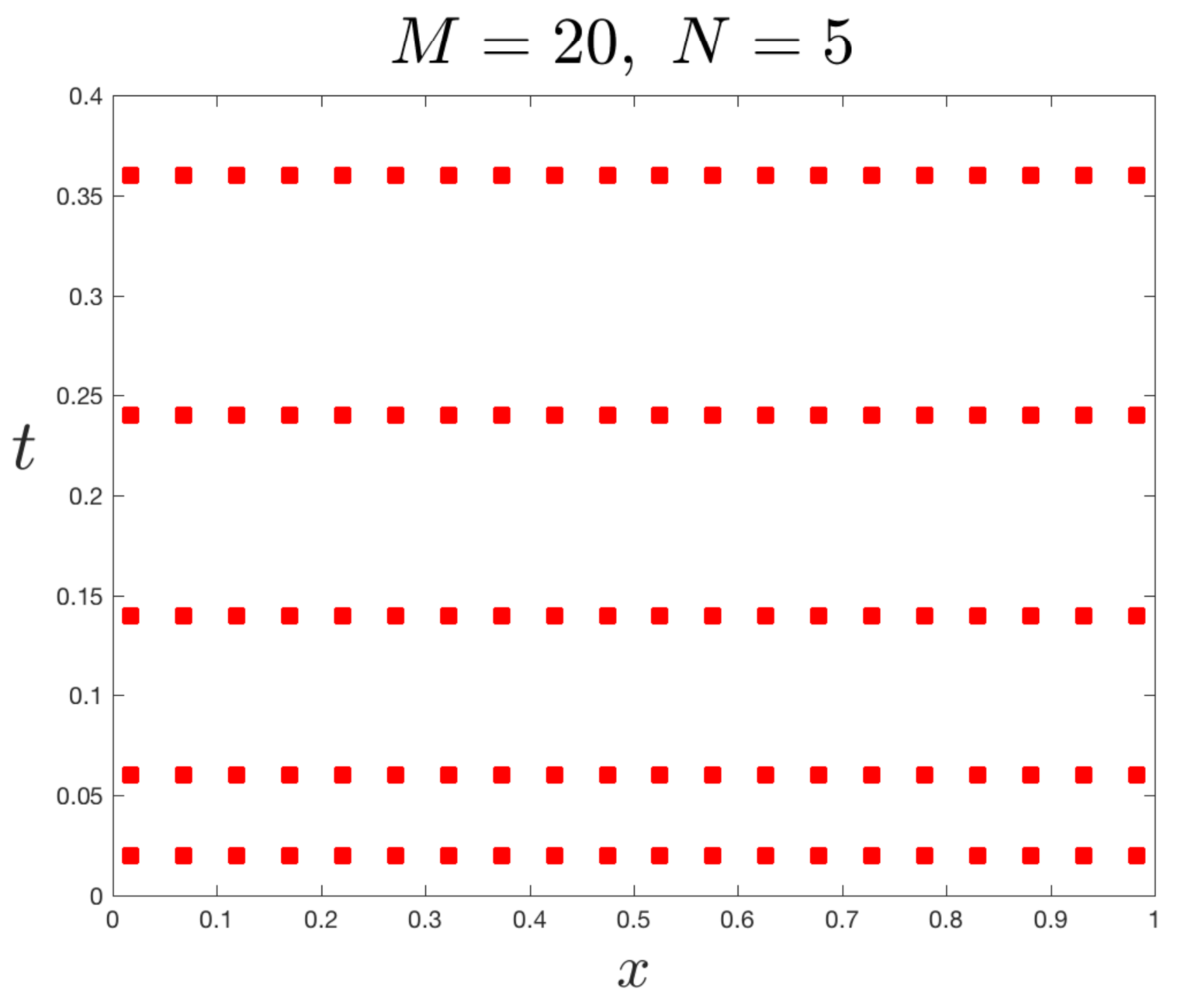}
\includegraphics[scale=0.25]{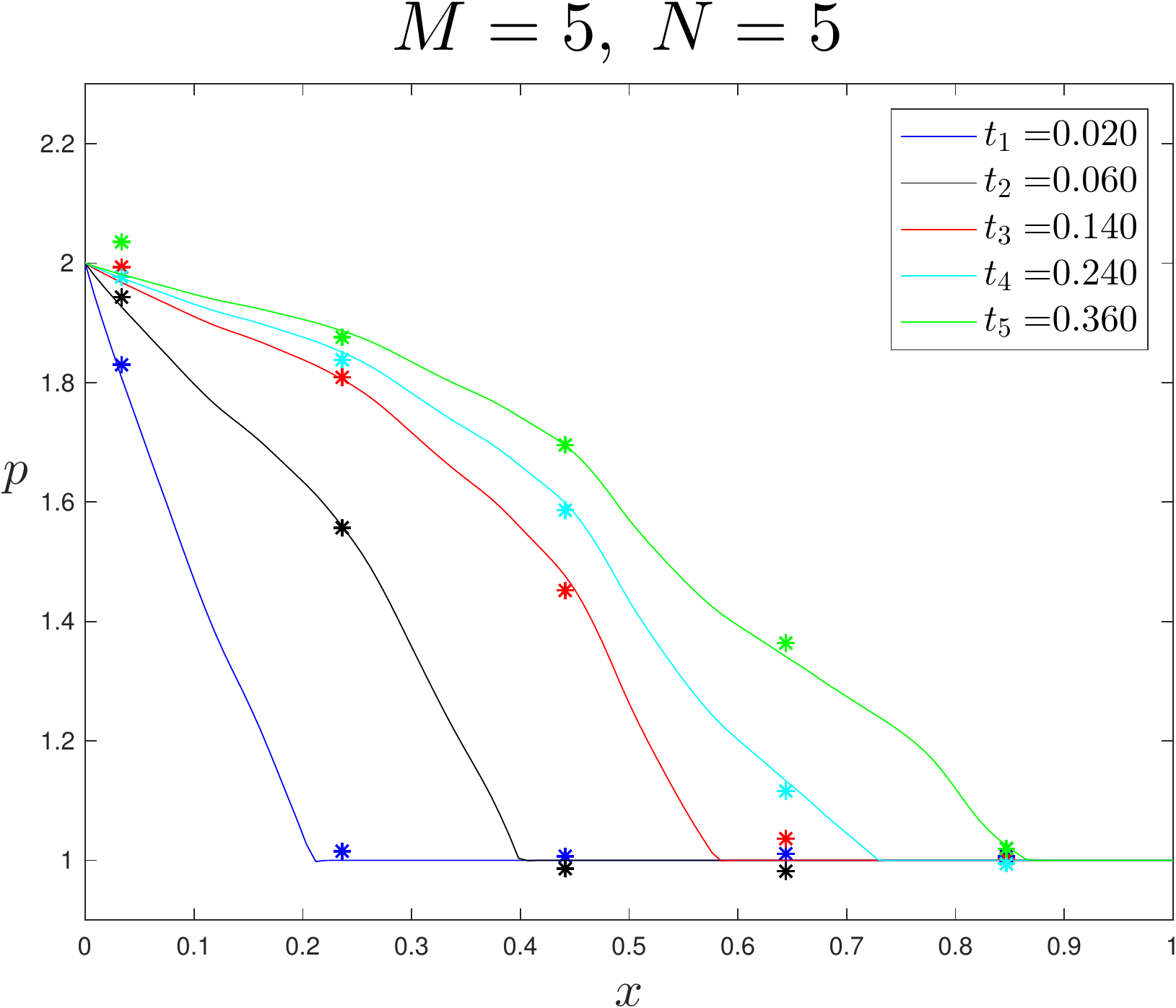}
\includegraphics[scale=0.25]{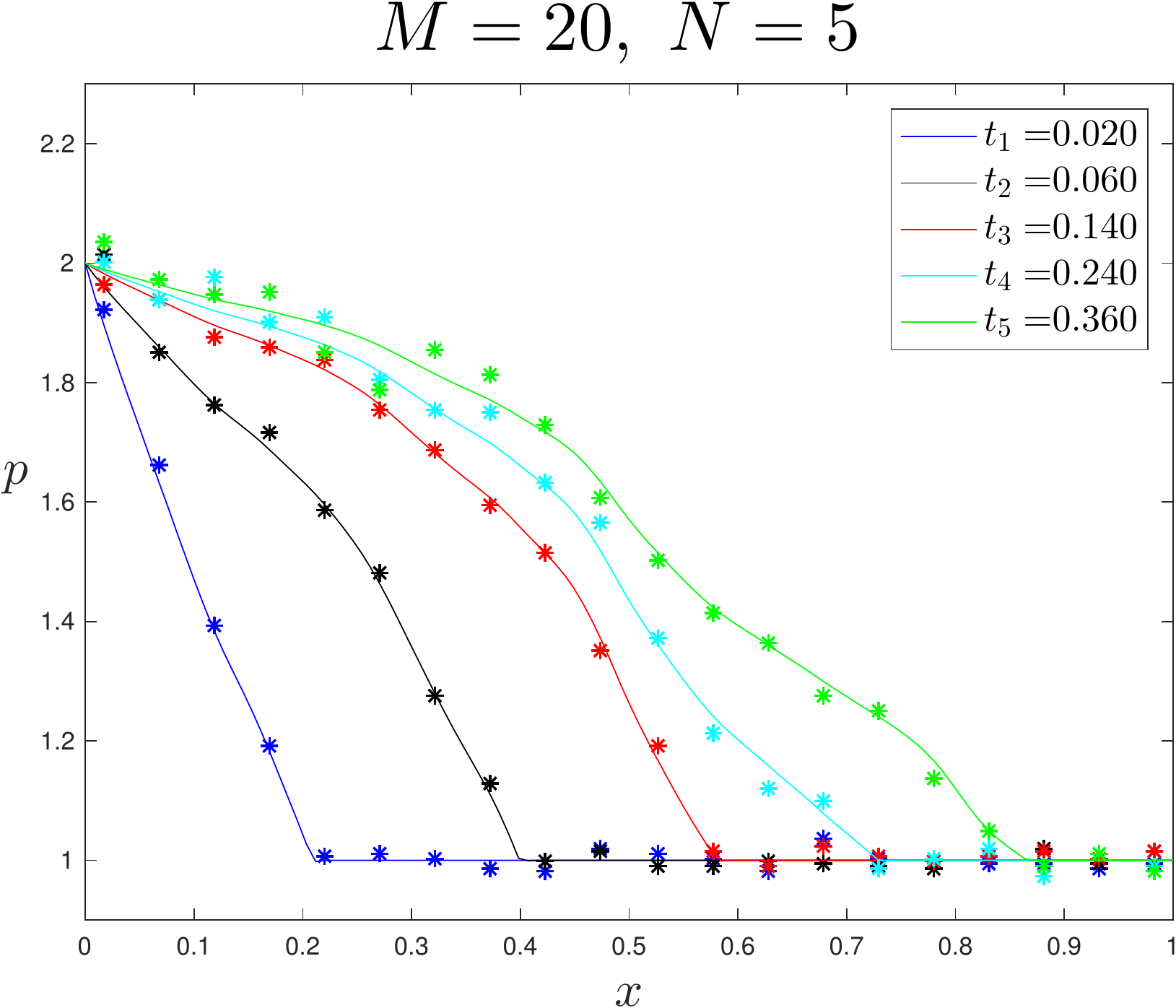}
 \caption{Left and middle-left: Space-time measurement configuration of $M=5$ (left) and $M=20$ (left-middle) pressure sensors with $N=5$ observation times. Middle-right and right: True pressure at observation times $\{p^{\dagger}(x,t_{n})\}_{n=1}^{N}$ together with the corresponding synthetic data $y_{n}^{p}$ (indicated with asterisks) for these measurement configurations.} \label{Fig5}
\end{center}
\end{figure}

Since the number of pressure sensors available for real RTM processes \mt{is usually} limited, in this section we study the effect of the number of pressure measurement locations in the solution to the Bayesian inverse problem.  In addition, we wish to understand the added value of \mt{front location measurements} in reducing uncertainty characterized by the posterior of the log-permeability. Let us then consider the same experimental set-up and relevant model parameters as described in Section~\ref{set_up}. We recall that the pressure measurement configuration consisted of $M=9$ sensor locations (see Figure \ref{Fig1} (left)). Here we additionally consider cases with less ($M=0$, $M=5$) and more ($M=20$) pressure measurement locations. These cases include the two particular \mt{settings} discussed in Remark \ref{rem1}; the case $M=0$ corresponds to the \mt{setting} where we only assimilate \mt{the front location measurement} at each observation time. The space-time pressure measurement configuration with $M=5$ and $M=20$ pressure sensors are displayed in the left and left-middle panels of Figure \ref{Fig5}. For consistency, the same synthetic front-location data generated \mt{in Section~\ref{set_up}} for the case with $M=9$ sensors were also used for these cases ($M=0,5,20$). However, for the new pressure measurement configurations (with $M=5,20$), synthetic pressure data were generated as described in Section~\ref{set_up}, for the same observation times $\{t_{n}\}_{n=1}^{5}$ and with the same noise level of $1.5\%$. The graphs of $\{p^{\dagger}(x,t_{n})\}_{n=1}^{5}$ are displayed in Figure~\ref{Fig5} (middle-right and right panels) together with the corresponding synthetic pressure data.

We apply REnKA for each of these new pressure measurement configurations ($M=5$, $M=20$) with tunable parameter and ensemble size as described earlier. REnKA is then also applied to each set of pressure data collected for all the pressure measurement configurations under investigation ($M=5$, $M=9$, $M=20$) but now with \mt{front location measurements} excluded from the inversion (see Remark \ref{rem1} with $\beta=p$). In addition, we apply REnKA for the case where the measurements of pressure are excluded $(M=0)$ and only \mt{front location measurements} are inverted (see Remark \ref{rem1} with $\beta=\Ga$). The seven measurement configurations are summarized below:
\small
\begin{eqnarray}
(M=0, \textrm{Front:}\checkmark),&(M=5, \textrm{Front:}\checkmark),&\quad (M=9, \textrm{Front:}\checkmark),\quad (M=20, \textrm{Front:}\checkmark), \nonumber\\
(M=5, \textrm{Front:} \mathcal{X}),& (M=9, \textrm{Front:} \mathcal{X}),&\quad (M=20, \textrm{Front:} \mathcal{X}), \label{combo}
\end{eqnarray}
\normalsize
where $\checkmark$ (resp. $\mathcal{X}$) indicates that measurements of front are included (resp. excluded) in the application of REnKA.

\begin{figure}[htbp]
\begin{center}
\includegraphics[scale=0.45]{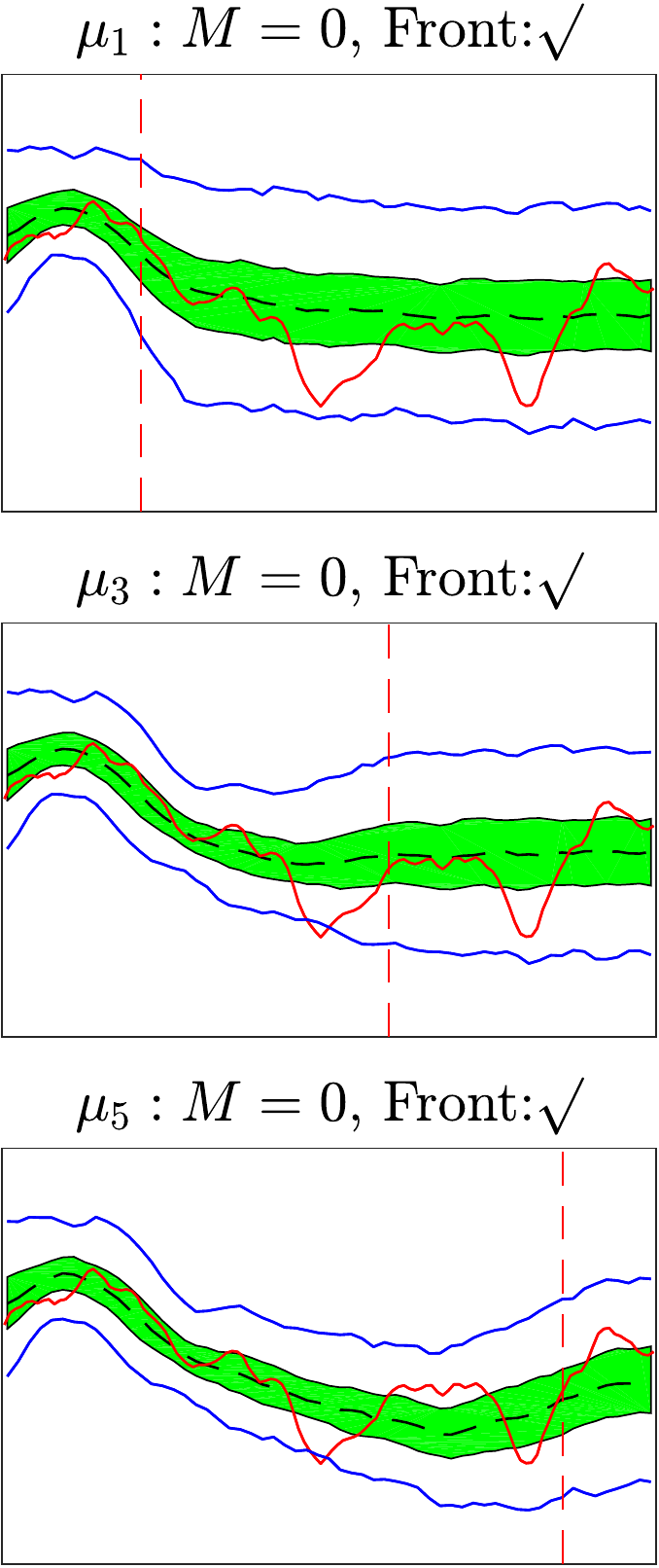}
\includegraphics[scale=0.45]{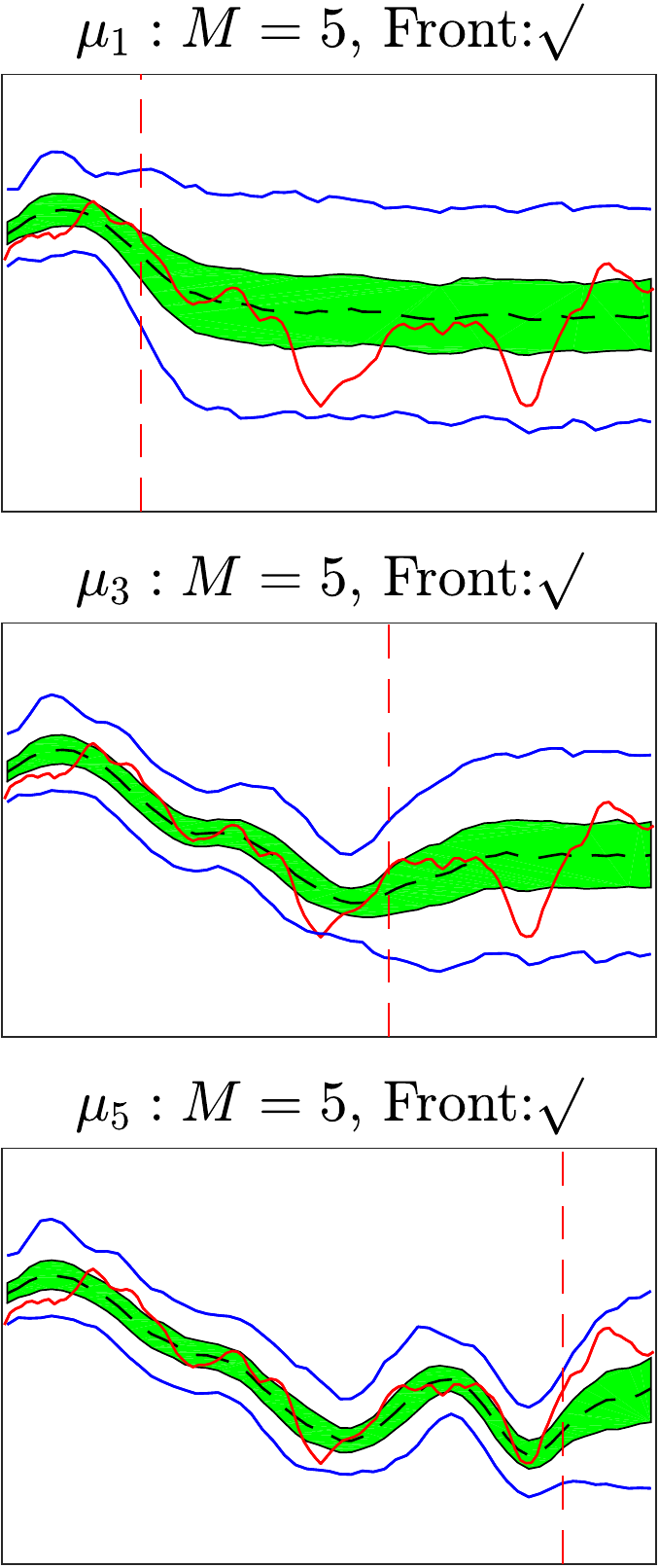}
\includegraphics[scale=0.45]{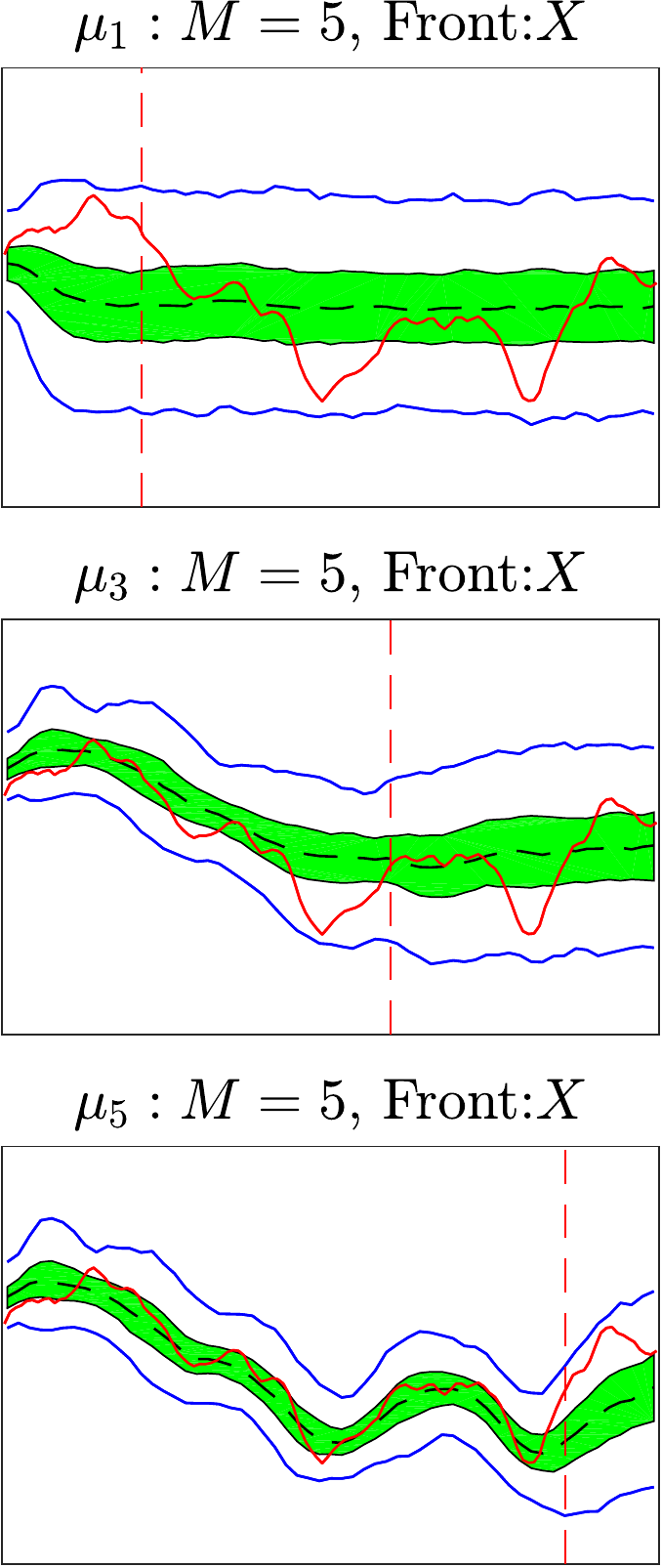}
\includegraphics[scale=0.45]{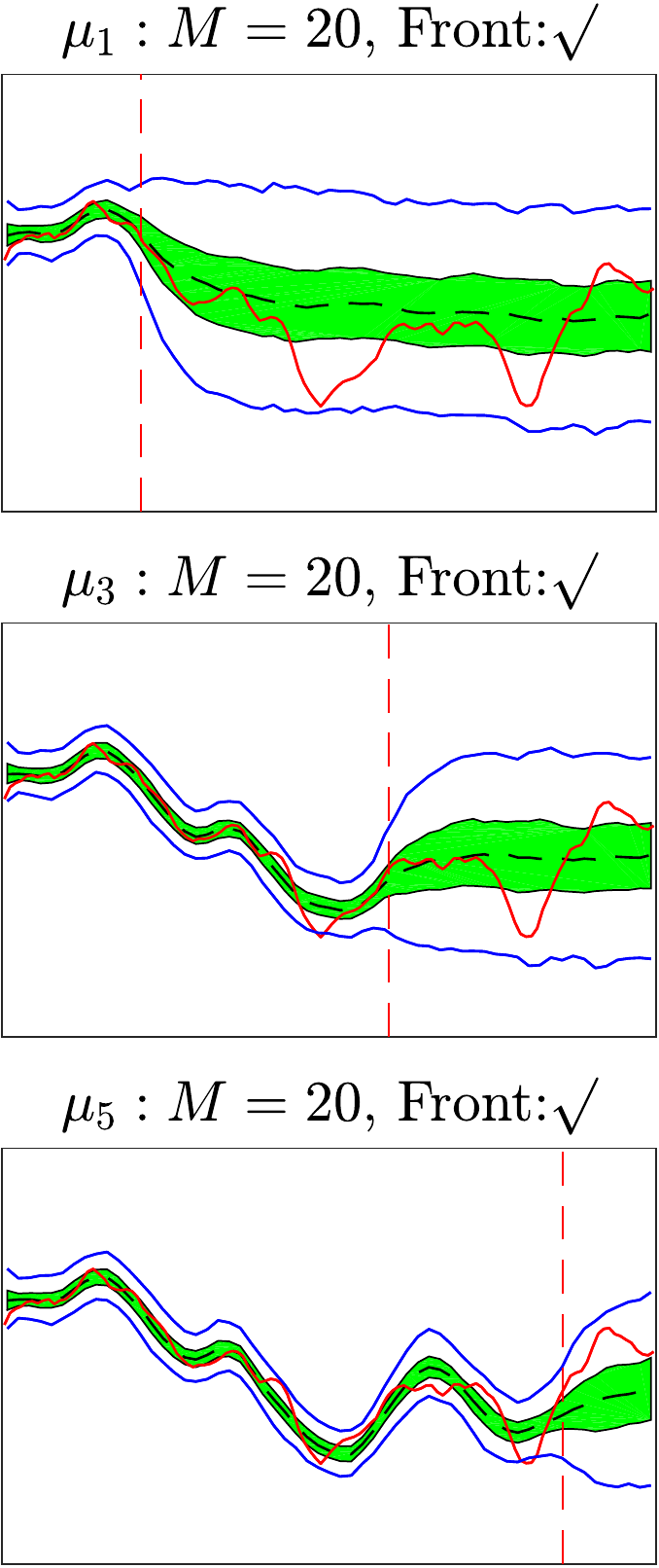}
\includegraphics[scale=0.45]{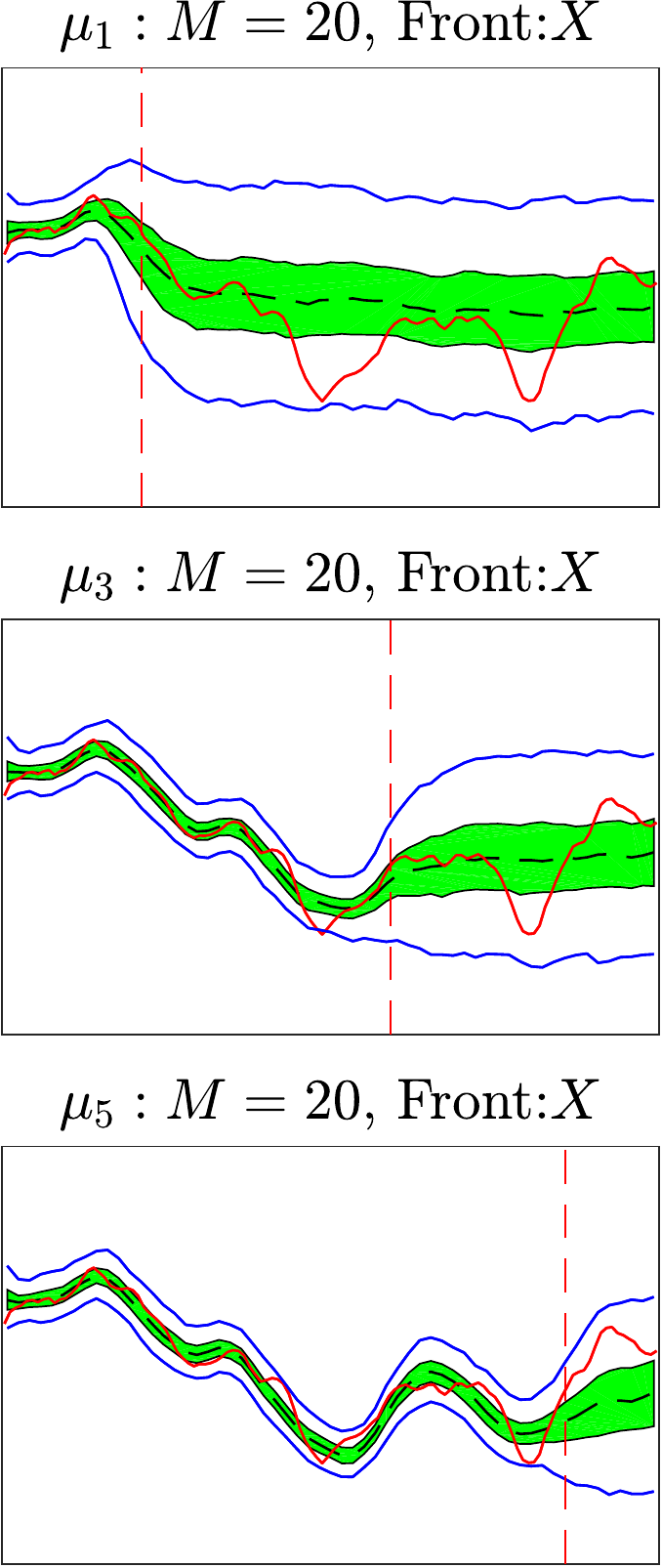}
 \caption{Percentiles of the posteriors $\mu_{1}$ (top), $\mu_{3}$ (middle) and $\mu_{5}$ (bottom) obtained via REnKA for different pressure measurement configurations with (left to right) $M=0,5,20$ sensor locations. For each of these pressure configurations we compare the percentiles obtained with ($\checkmark$) and without ($\mathcal{X}$) inverting measurements of the front location.}  \label{Fig6}
\end{center}
\end{figure}

\begin{figure}[htbp]
\begin{center}

\includegraphics[scale=0.35]{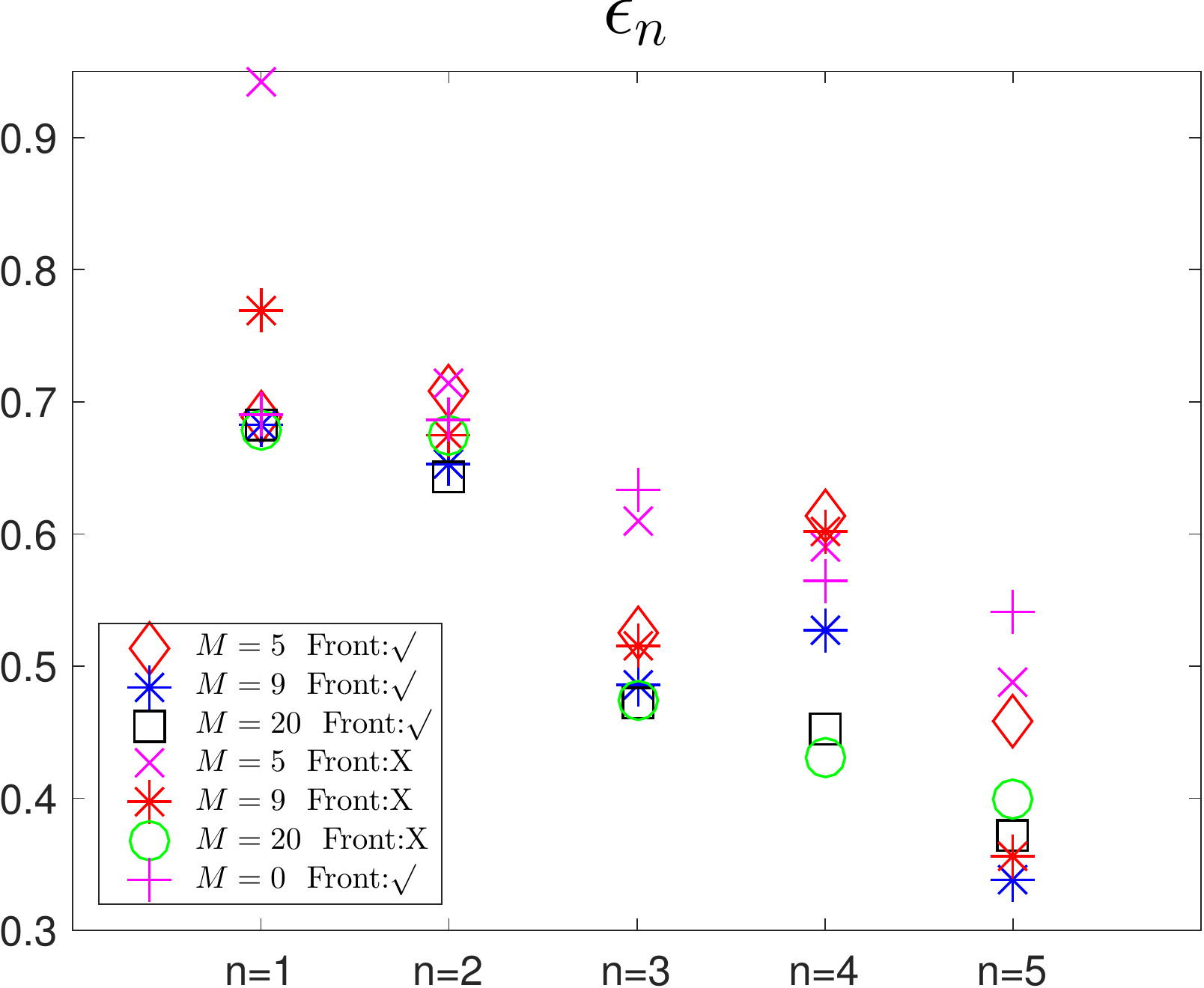}
\includegraphics[scale=0.35]{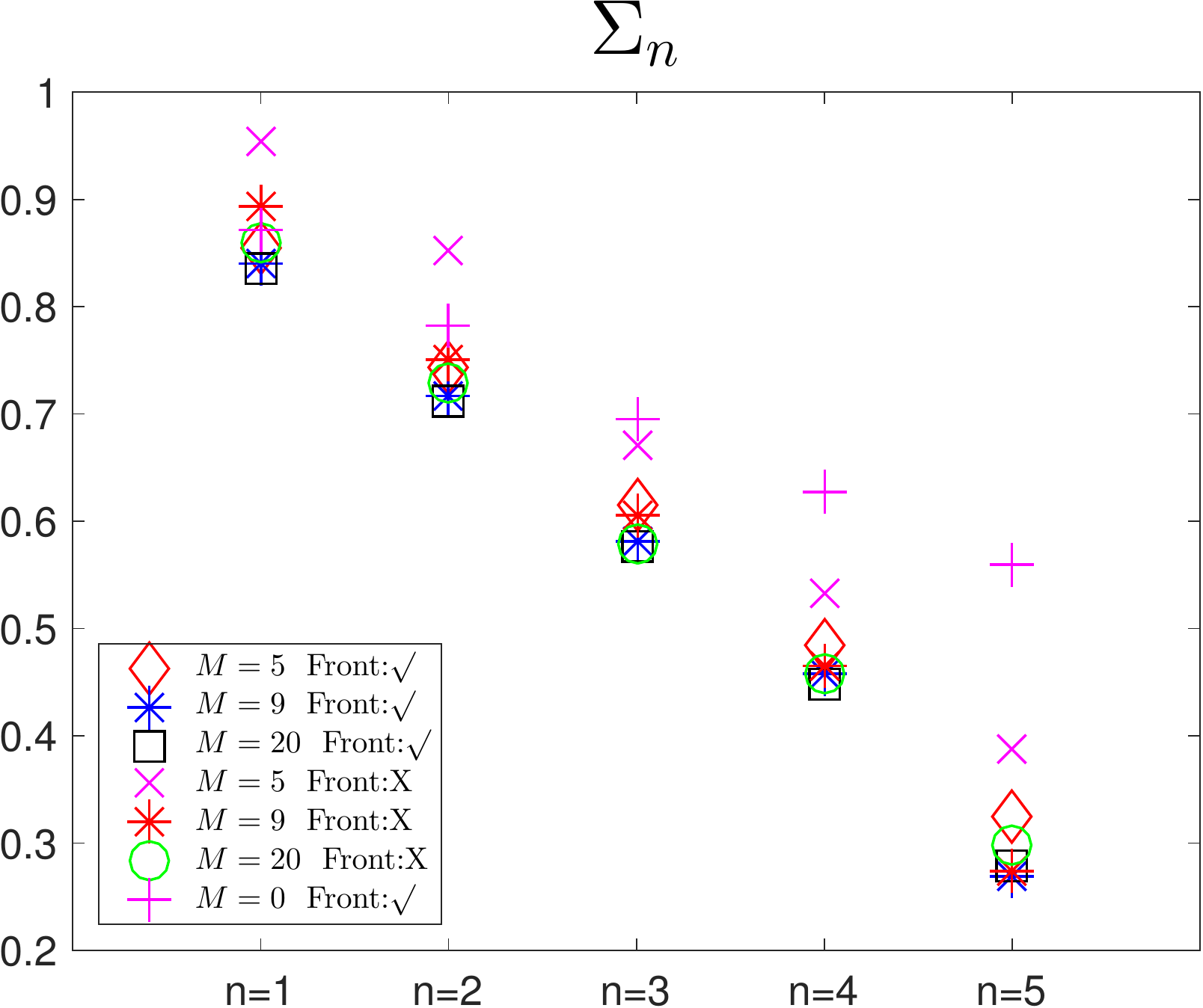}
\includegraphics[scale=0.35]{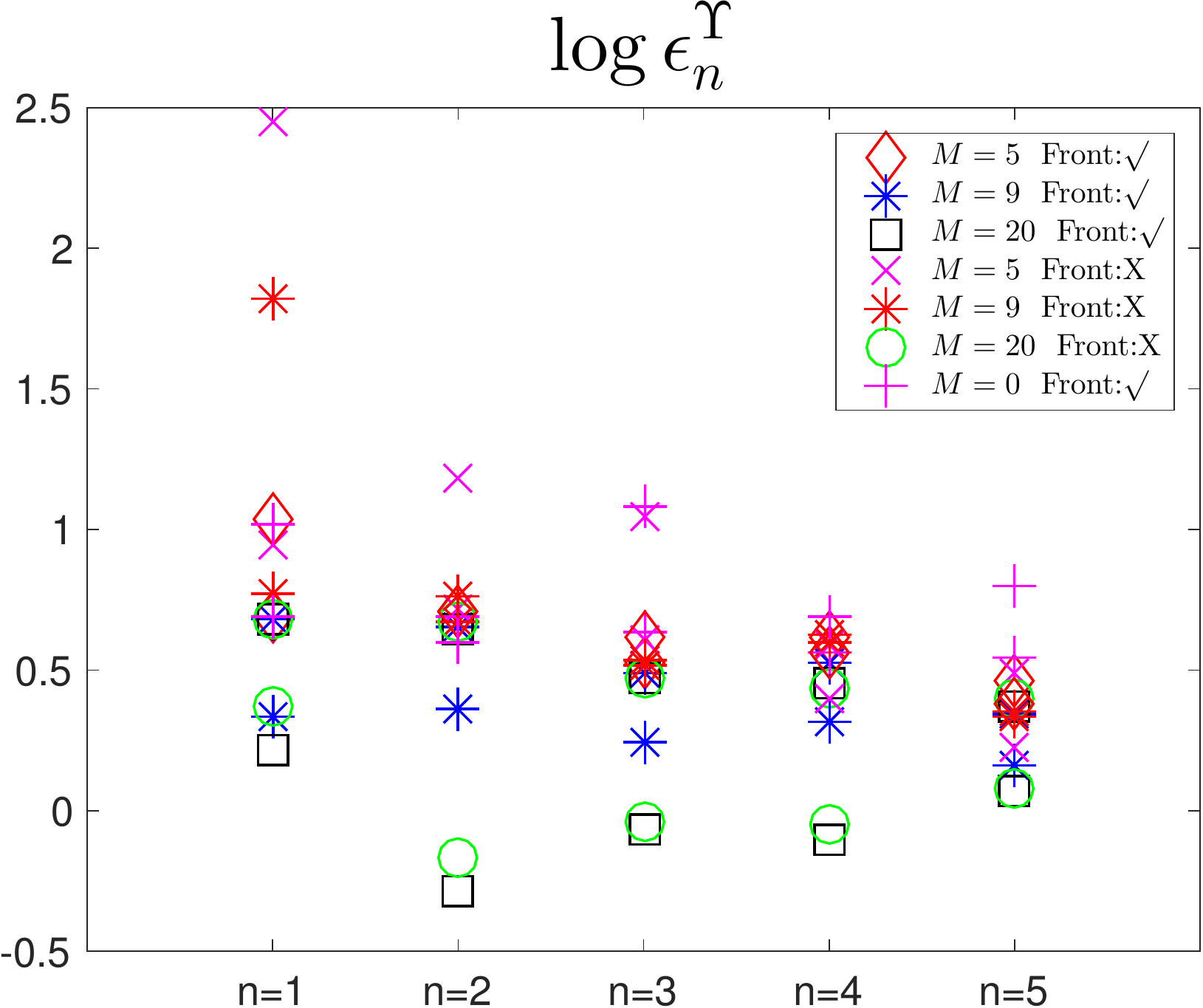}

 \caption{Experiments with the measurement configurations (\ref{combo}) and fixed $N=5$ observation times. Left: Relative error w.r.t the truth $\epsilon_{n}$ (on the domain $D^{\ast}$). Middle: Uncertainty in terms of $\Sigma_{n}$. Right: (log) error with respect to the truth $\epsilon^{\Upsilon}$ on the moving domain $D^{\dagger}(t_{n})$.}  \label{Fig6B}
\end{center}
\end{figure}

In Figure \ref{Fig6} we show posterior percentiles of the log-permeability for some of the combinations described in (\ref{combo}). These posteriors were obtained using the same fixed initial ensemble. We note that the effect of inverting the front (Front: $\checkmark$) in reducing the uncertainty band defined by the posterior percentiles is mainly noticeable at the first observation time at which the front has not yet reached most 
\mt{of the} pressure sensors. Therefore, pressure measurement from locations that have not been reached by the front provide little information of the log-permeability.  As more observations (in time) are assimilated and more pressure sensors are reached by the front, the effect of inverting the front becomes less significant. We can also observe that, even at earlier observation times, the effect of inverting the front has little effect on reducing this uncertainty band when larger number ($M=20$) of pressure sensors are used for the inversion. These results are confirmed in Figure \ref{Fig6B} (middle) where we display the plot of $\Sigma_{n}$. Indeed, $\Sigma_{n}$ decreases as more (fixed) observations in time are assimilated. Note that at the final observation time $t_{5}$, the added value of the measurement of front location is only noticeable when a small number of pressure sensors $M=5$ are considered. In fact, when no pressure data are inverted $M=0$, we can note that $\Sigma_{n}$ is substantially reduced. Therefore, measurements of front location are indeed informative of the log-permeability when only a limited number pressure sensors are available; this also includes the case where pressure measurements are completely absent. However, increasing the values of $M$ has no significant effect on the reduction of $\Sigma_{n}$, thus suggesting that adding more pressure sensors will no further reduce the posterior uncertainty in the log-permeability.

The plots of $\epsilon_{n}$ are displayed in the left panel of Figure \ref{Fig6B}. Although $\epsilon_{n}$ decreases as more (fixed) observations in time are assimilated, these results neither reveal the added value of the front location nor provide insight in the effect of the number of pressure sensors in terms of improving the accuracy of the infer log-perm (i.e. reducing $\epsilon_{n}$). Indeed, note that for $t=t_{4}$, the results from configuration $(M=20, \textrm{Front:} \mathcal{X})$ yield smaller $\epsilon_{n}$ than $(M=20, \textrm{Front:}\checkmark)$. Furthermore, at $t=t_{5}$ we note that $M=9$ results in smaller errors than $M=20$. It is then clear that, $\epsilon_{n}$, as defined in (\ref{truth_err}) does not offer sufficient evidence that increasing the number of pressure sensors results in more accurate inference of the truth. As we recall that the definition of this error  (see (\ref{truth_err})) involves the norm defined on the whole physical domain $D^{\ast}=[0,1]$, we should also then reemphasize that, at a given observation time $t_{n}$, the reduction of the uncertainty in the log-permeability field mainly occurs in the region of the moving domain $D^{\dagger}(t_{n})=[0,\Ga^{\dagger}(t_{n})]$, where $\Ga^{\dagger}$ is the true front location. Therefore, when applying REnKA (or any Bayesian inversion approach), the estimator of the truth (ensemble mean for REnKA) at time $t_{n}$, may not necessarily results in decrease of the error in the region $D^{\ast}\setminus D^{\dagger}(t_{n})$ as no informative measurements have there yet been collected. In order to further understand this effect, let us then consider, $\epsilon_{n}^{\Ga}$, the error with respect the truth defined by (\ref{truth_err2}). This error accounts for error of the estimator w.r.t the truth only in the region $D^{\dagger}(t_{n})$, where measurements at time $t_{n}$ (as well as earlier measurements) have been collected.  In contrast to $\epsilon_{n}$, the error $\epsilon_{n}^{\Ga}$ is defined on the moving domain $D^{\dagger}(t_{n})$. Therefore, we shall not necessarily expect a decrease of $\epsilon_{n}^{\Ga}$ as function of the assimilation times $t_{n}$.

In Figure \ref{Fig6B} (right) we display the log of $\epsilon_{n}^{\Ga}$ from which we can now fully appreciate that more accurate estimates of the permeability (in the domain $D^{\dagger}(t_{n})$) are obtained by inverting both pressure and front location as opposed to only inverting pressure data. Moreover, Figure \ref{Fig6B} (right) now enable us to see that increasing the number of pressure locations, $M$, yields more accurate estimates of the log-permeability. However, it also reveals that a further increase in the number of pressure measurements locations (e.g. $M=20$) has no significant effect in the accuracy of the estimates when compared with smaller and more realistic 
\mt{(from the applicable point of view)} 
measurement configurations (e.g. $M=9$). The (log) total computational cost together with the final (log) error $\epsilon_{n}^{\Ga}$ is displayed in Figure \ref{Fig6C} (left). Computational cost is expressed in terms of forward model evaluations as discussed in subsection \ref{SMC_small}. Note that higher computational cost is associated \mt{with} larger measurements (i.e. $M=20$) as this requires more tempering for the computation of each posterior $\mu_{n}$ (see discussion in Section~\ref{note}).

\subsubsection{Effect of the number of observation times}\label{meas_time}

In previous examples we have considered a fixed number $N=5$ of observation times $\{t_{n}\}_{n=1}^{5}$, where pressure and front locations were measured. We now study influence of the number of observation times, $N$, in the solution of the Bayesian inverse problem. Let us then fix the number of pressure sensors to $M=9$ and consider a set of experiments with different number of total observation times ($N=1,2,3,4,5,8,10,12,14,16$). For each of these $N$ we select the final time $t_{N} =0.36$. In addition, for \mt{any} two given $N_{1}$ and $N_{2}$ with $N_{1}<N_{2}$ we select the observation times so that $\{t_{n}\}_{n=1}^{N_{1}}\subset \{t_{n}\}_{n=1}^{N_{2}}$. The space-time measurement configuration for some of these cases is displayed in Figure \ref{Fig6BB} (top) as well as the synthetic pressure data (bottom) generated as before; for these examples we consider the general Bayesian inverse problem where both pressure and front location are assimilated.

\begin{figure}[htbp]
\begin{center}
\includegraphics[scale=0.38]{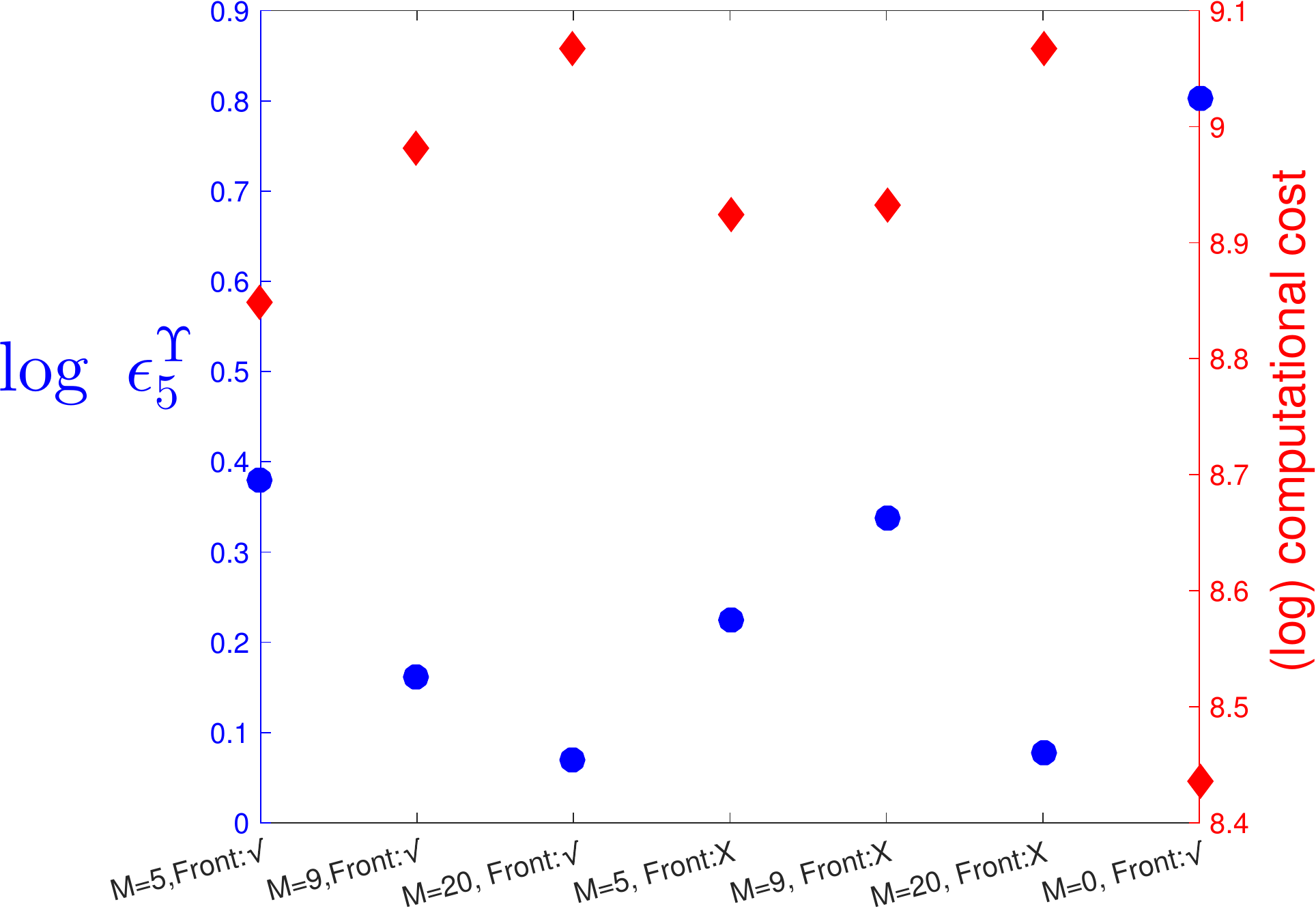},\qquad
\includegraphics[scale=0.38]{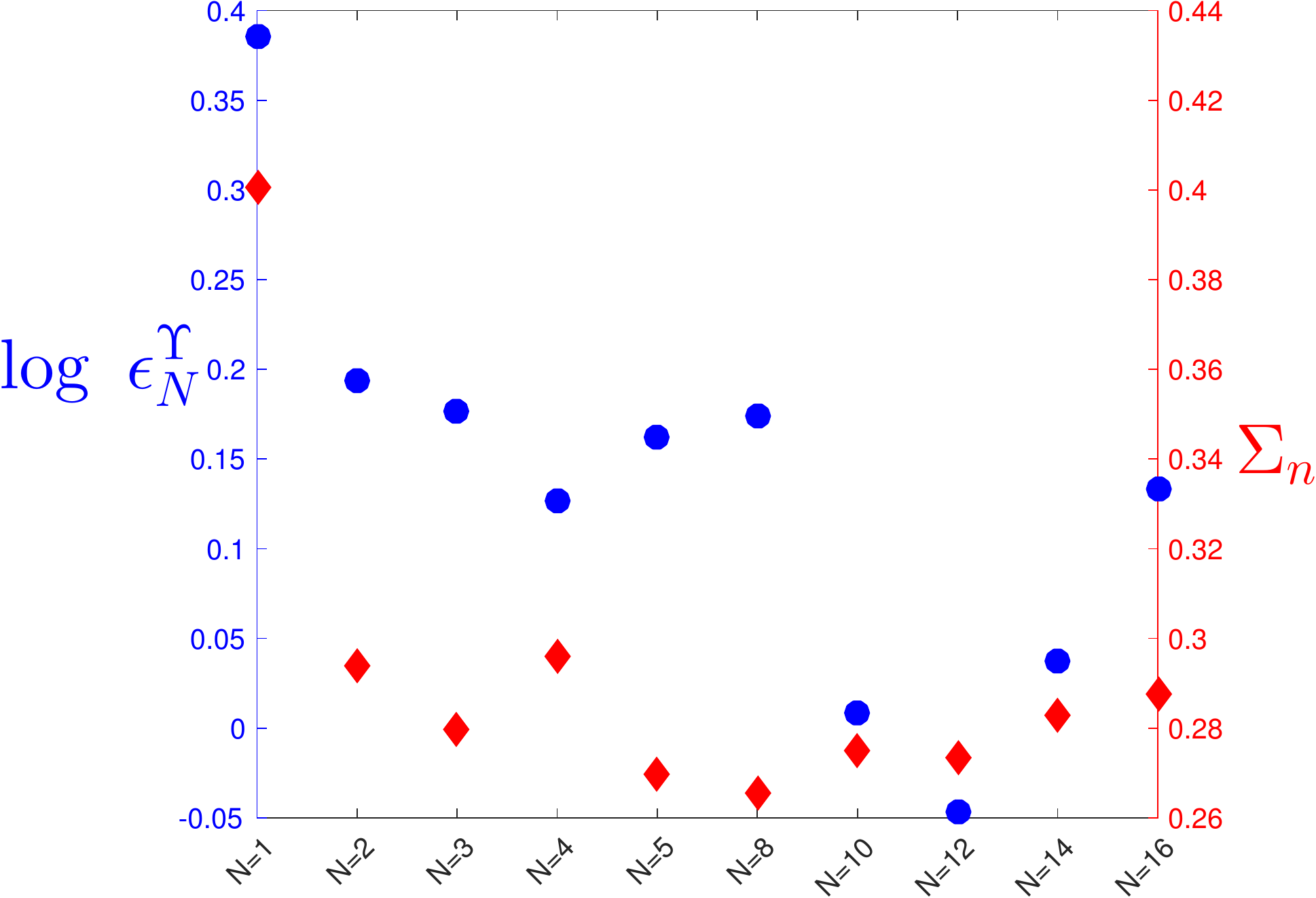}

 \caption{Left: Final $\log \epsilon_{5}^{\Upsilon}$ as well as the total (log) computational cost for the measurement configurations of (\ref{combo}) and fixed $N=5$ observation times. Right: Final $\log \epsilon_{N}^{\Upsilon}$ and $\Sigma_{N}$ for different choices of total observation times $N$ and fixed $M=9$.}  \label{Fig6C}
\end{center}
\end{figure}

\begin{figure}[htbp]
\begin{center}

\includegraphics[scale=0.23]{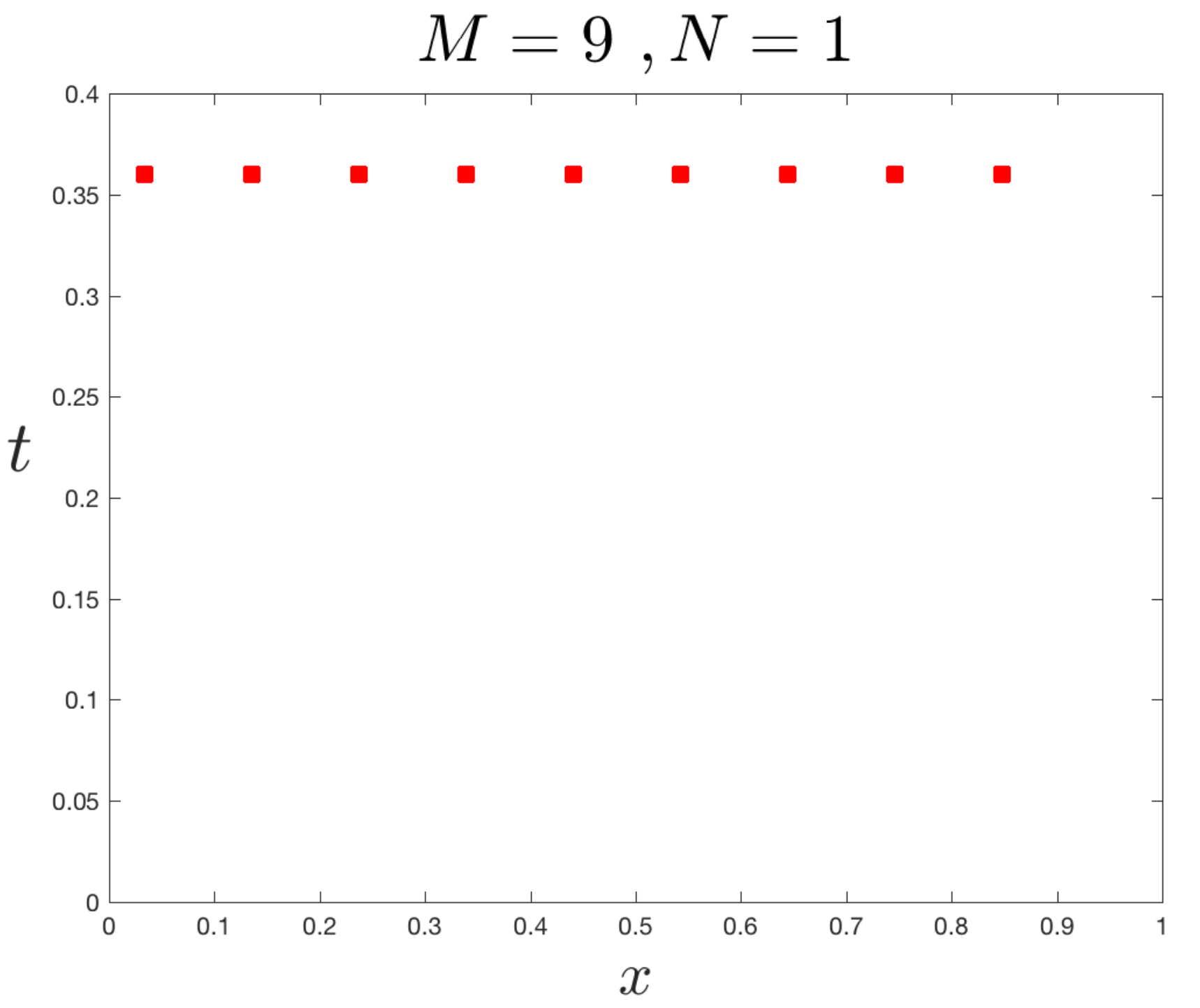}
\includegraphics[scale=0.23]{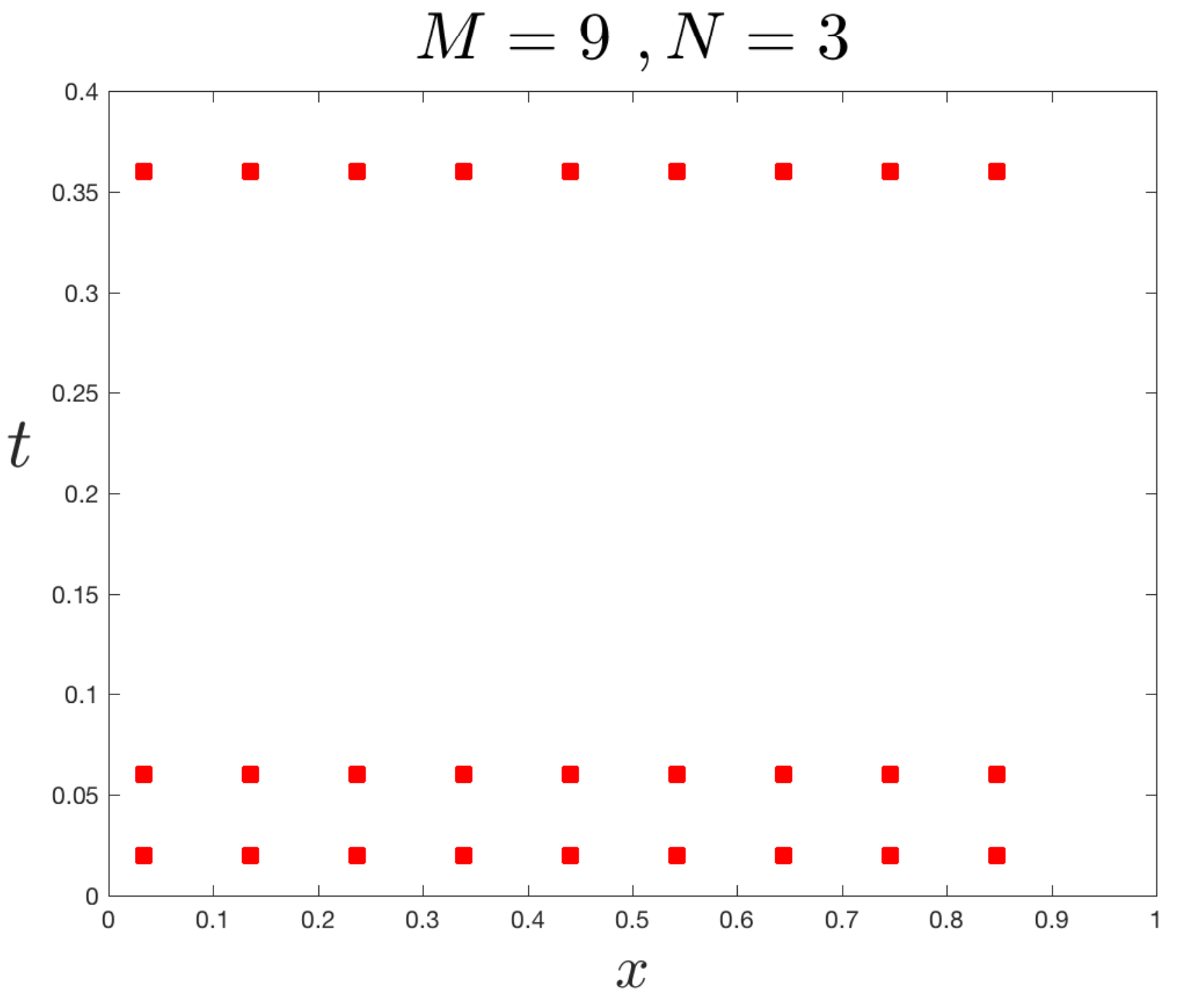}
\includegraphics[scale=0.23]{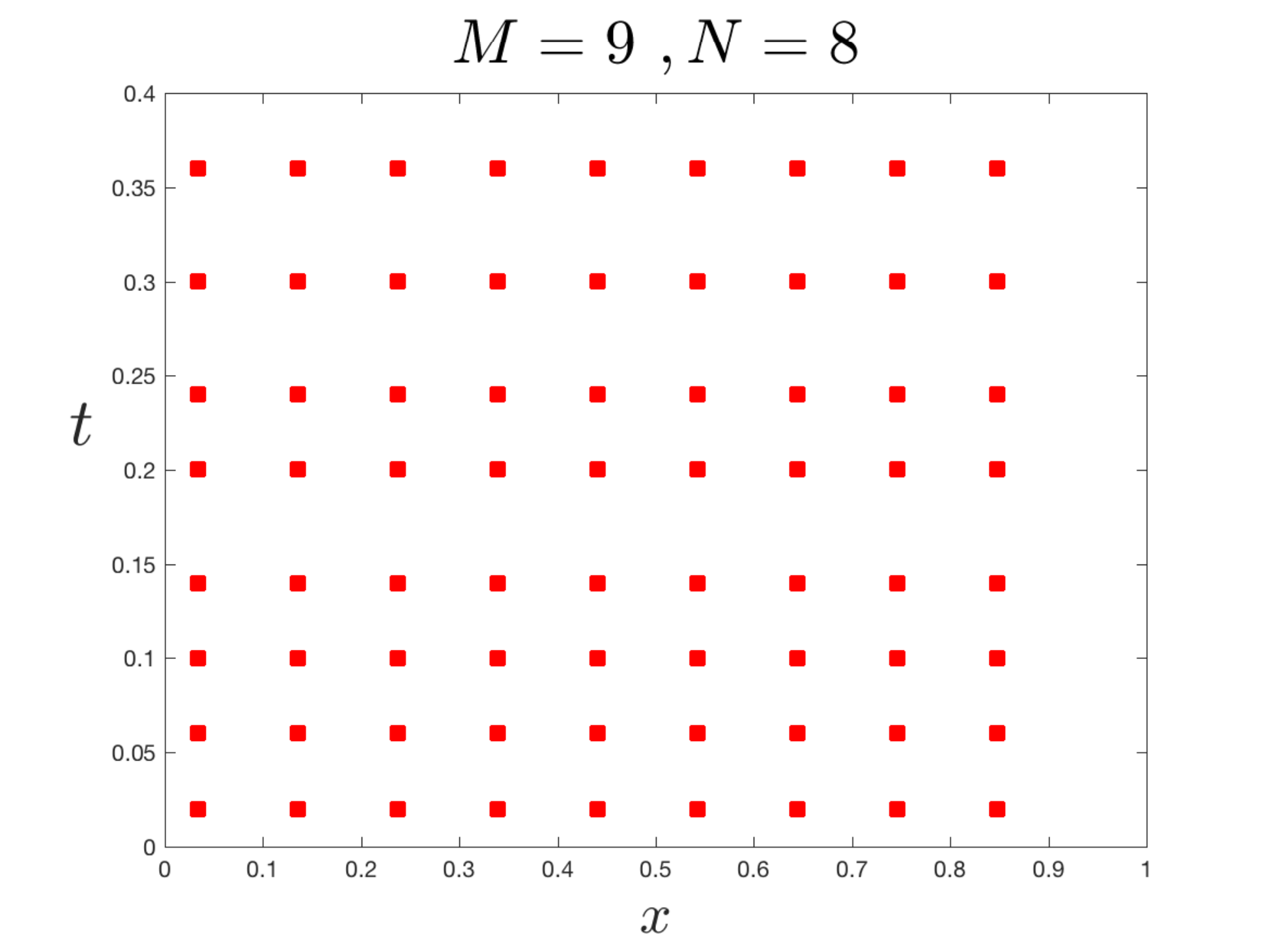}
\includegraphics[scale=0.23]{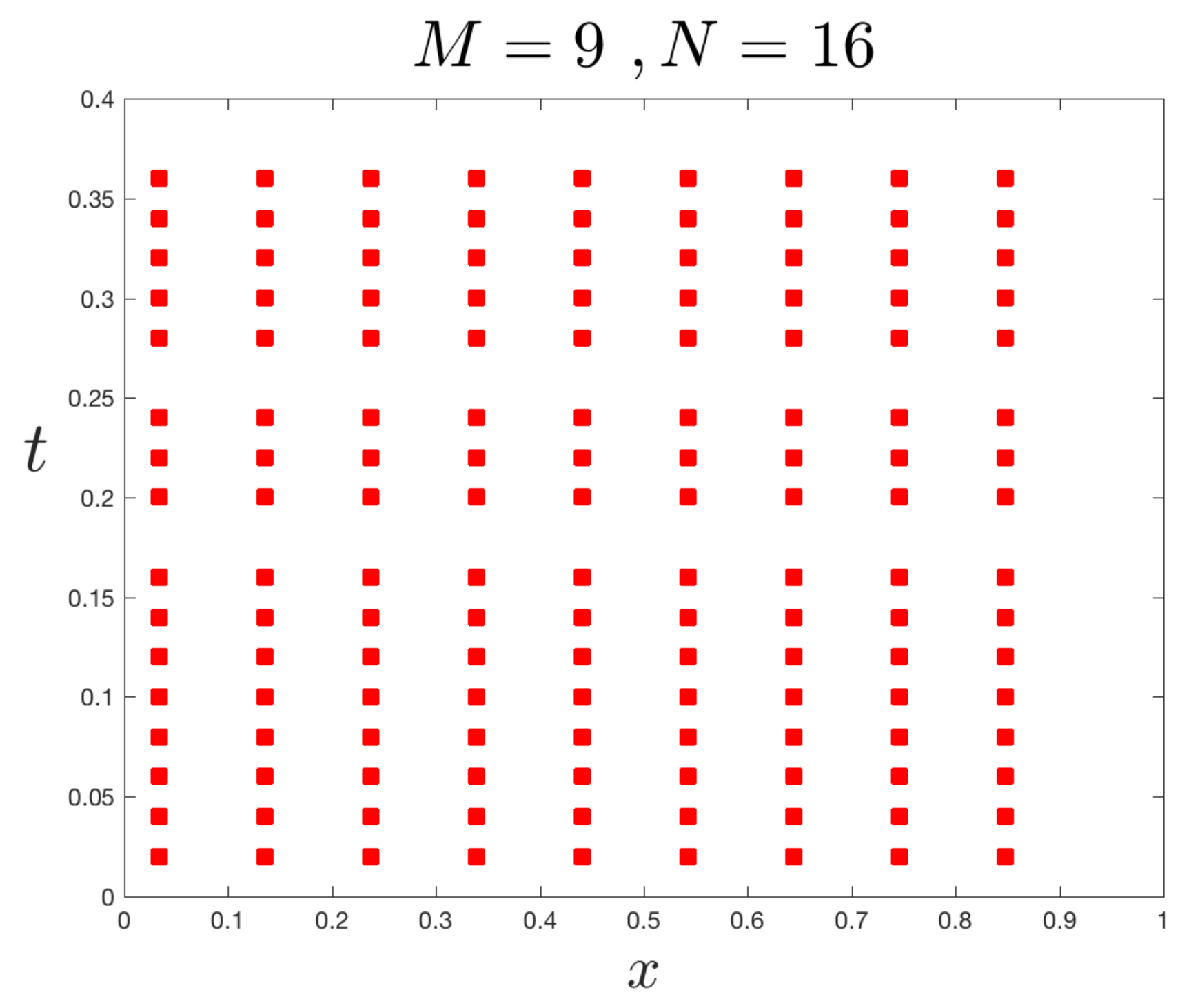}\\
\includegraphics[scale=0.25]{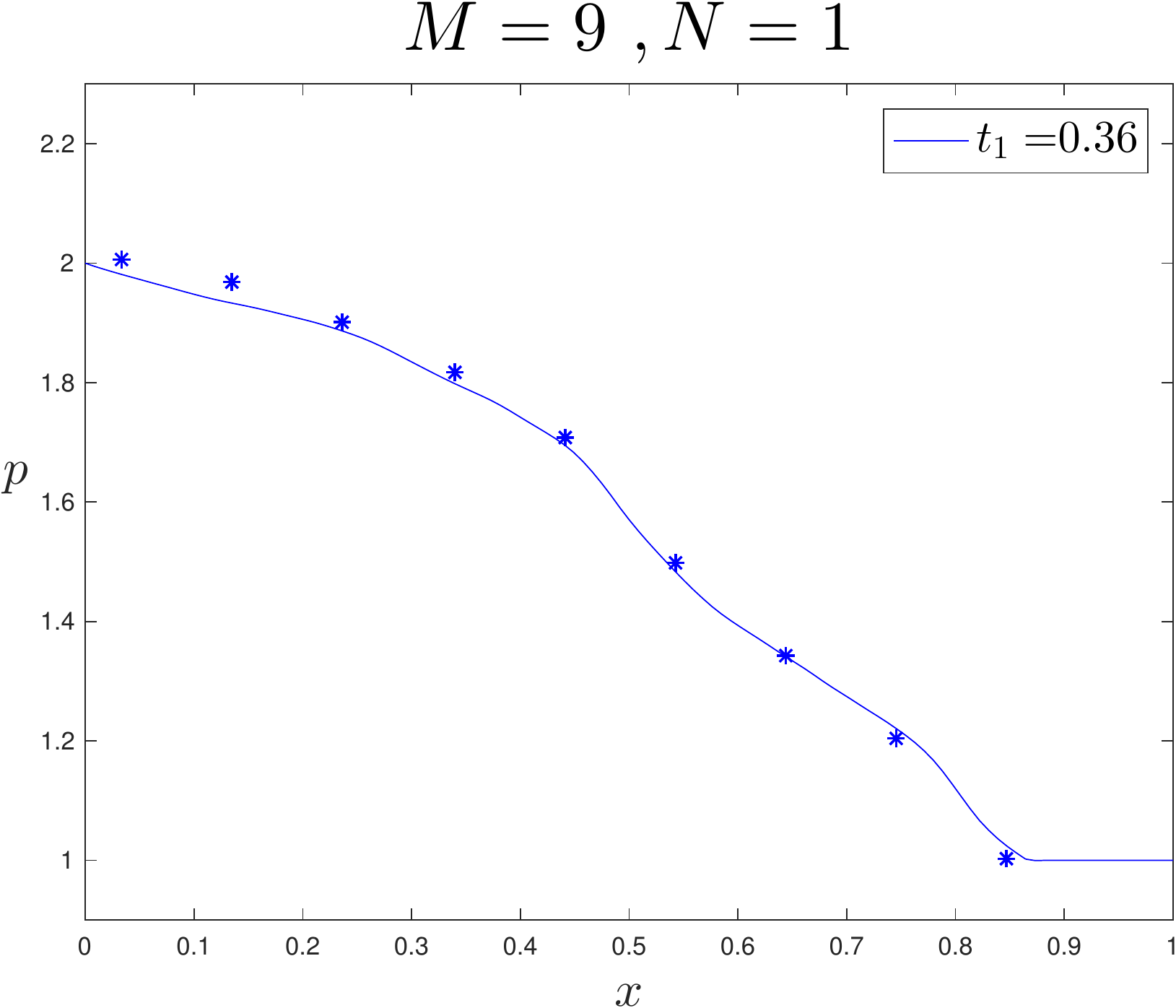}
\includegraphics[scale=0.25]{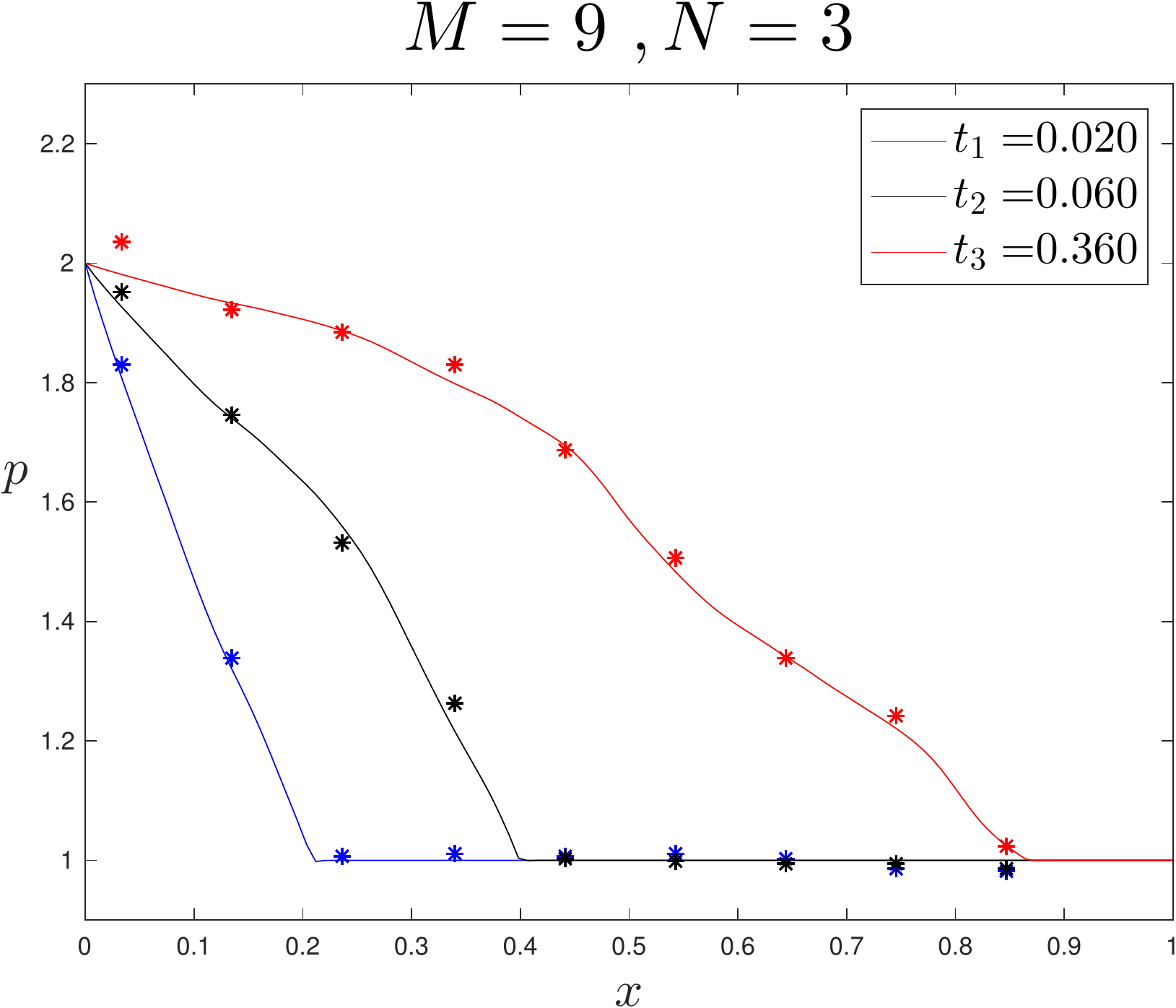}
\includegraphics[scale=0.25]{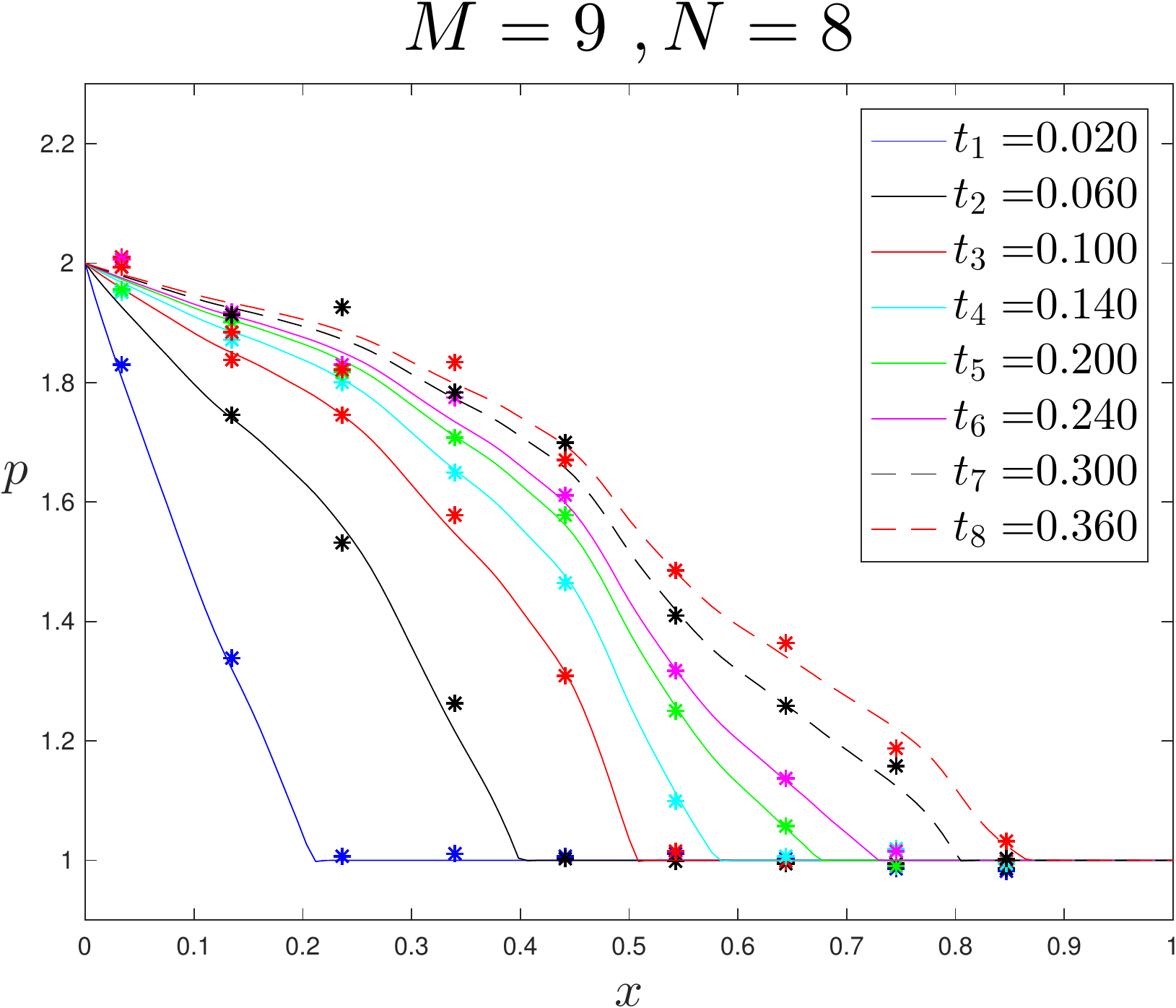}
\includegraphics[scale=0.25]{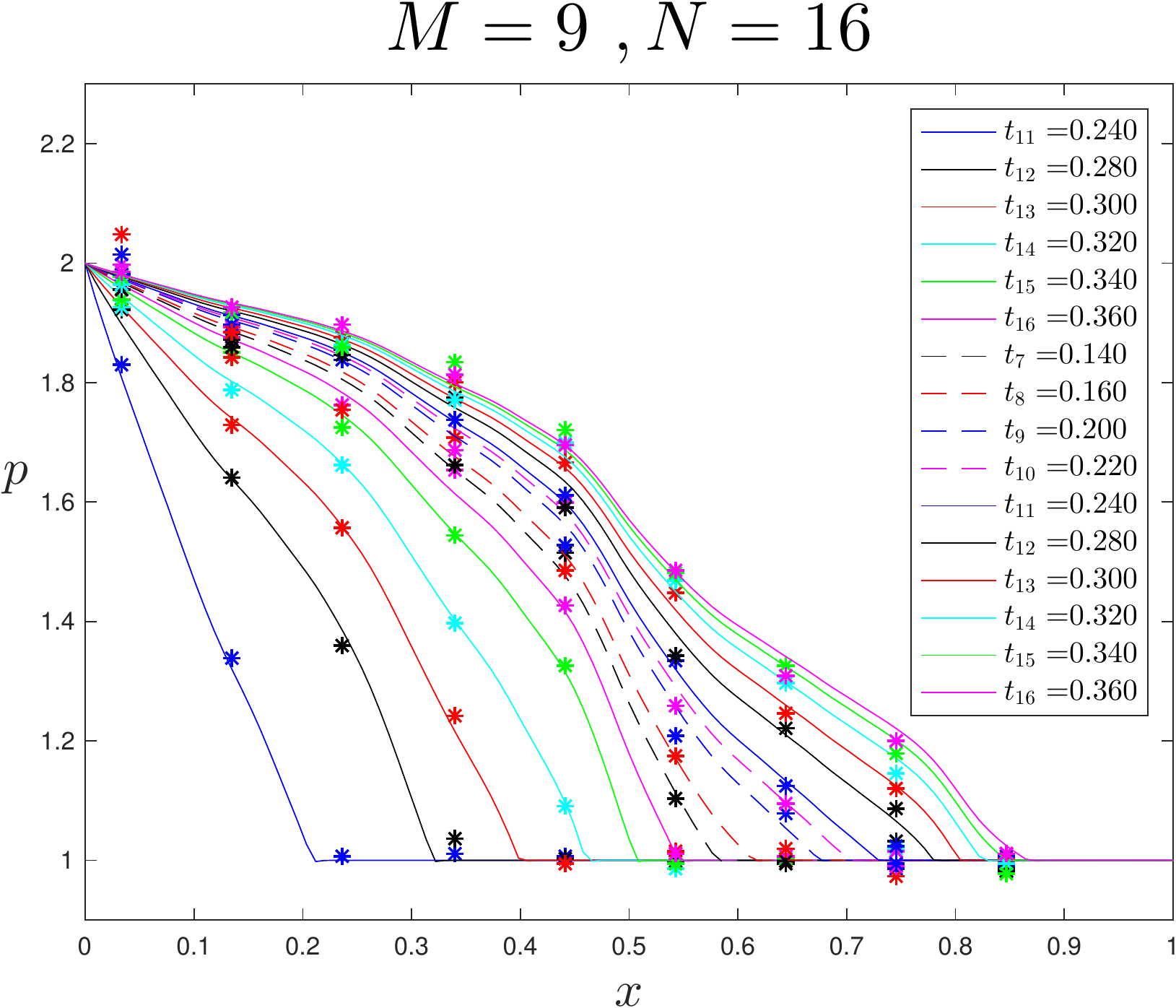}
 \caption{Top: Space-time measurement configuration for from left to right $N=1$, $N=3$, $N=8$ and $N=16$ observation times. Bottom: True pressure at observation times $\{p^{\dagger}(x,t_{n})\}_{n=1}^{N}$ together with synthetic pressure data $y_{n}^{p}$ for these measurement configurations.}  \label{Fig6BB}
\end{center}
\end{figure}

\begin{figure}[htbp]
\begin{center}

\includegraphics[scale=0.25]{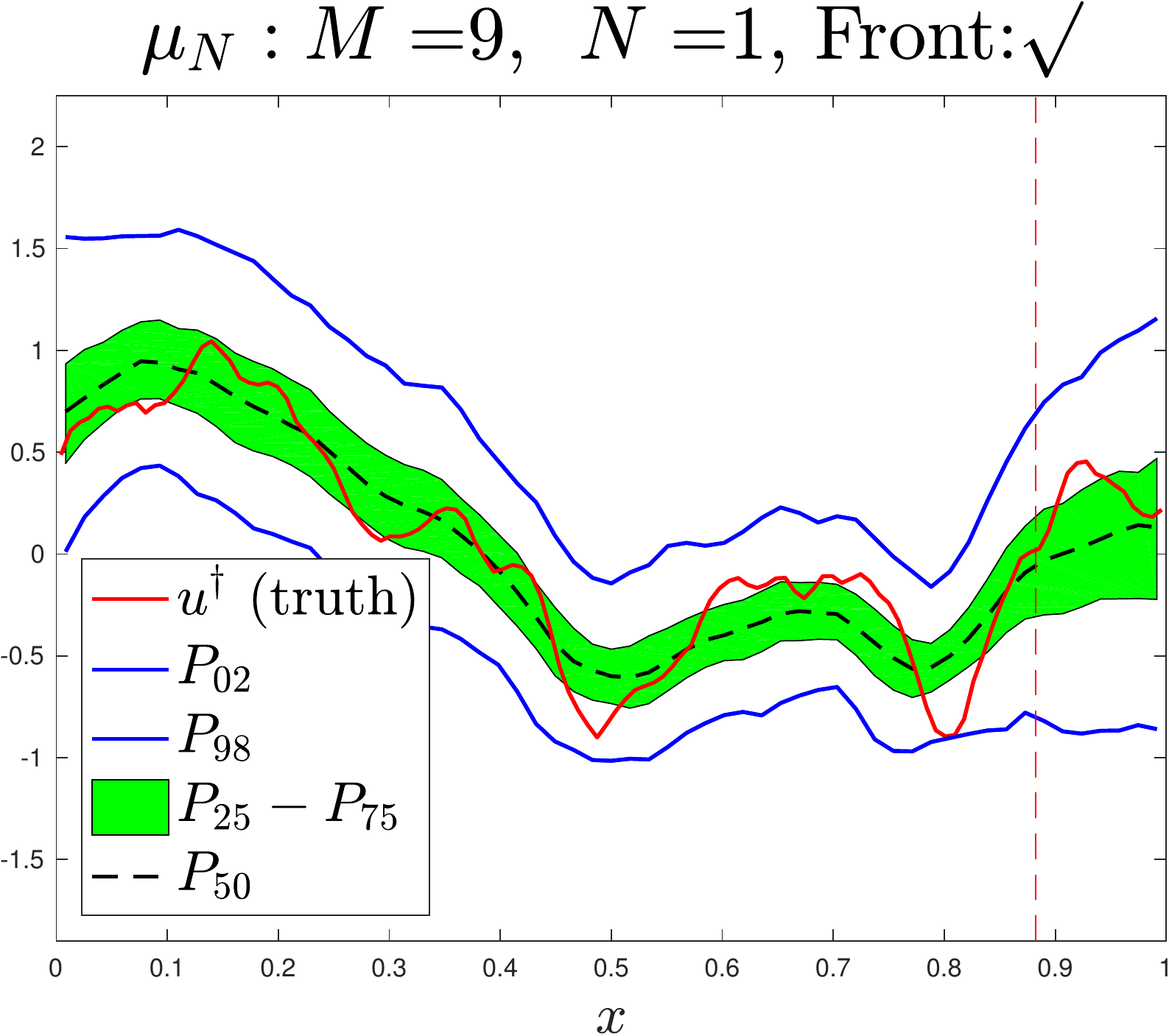}
\includegraphics[scale=0.25]{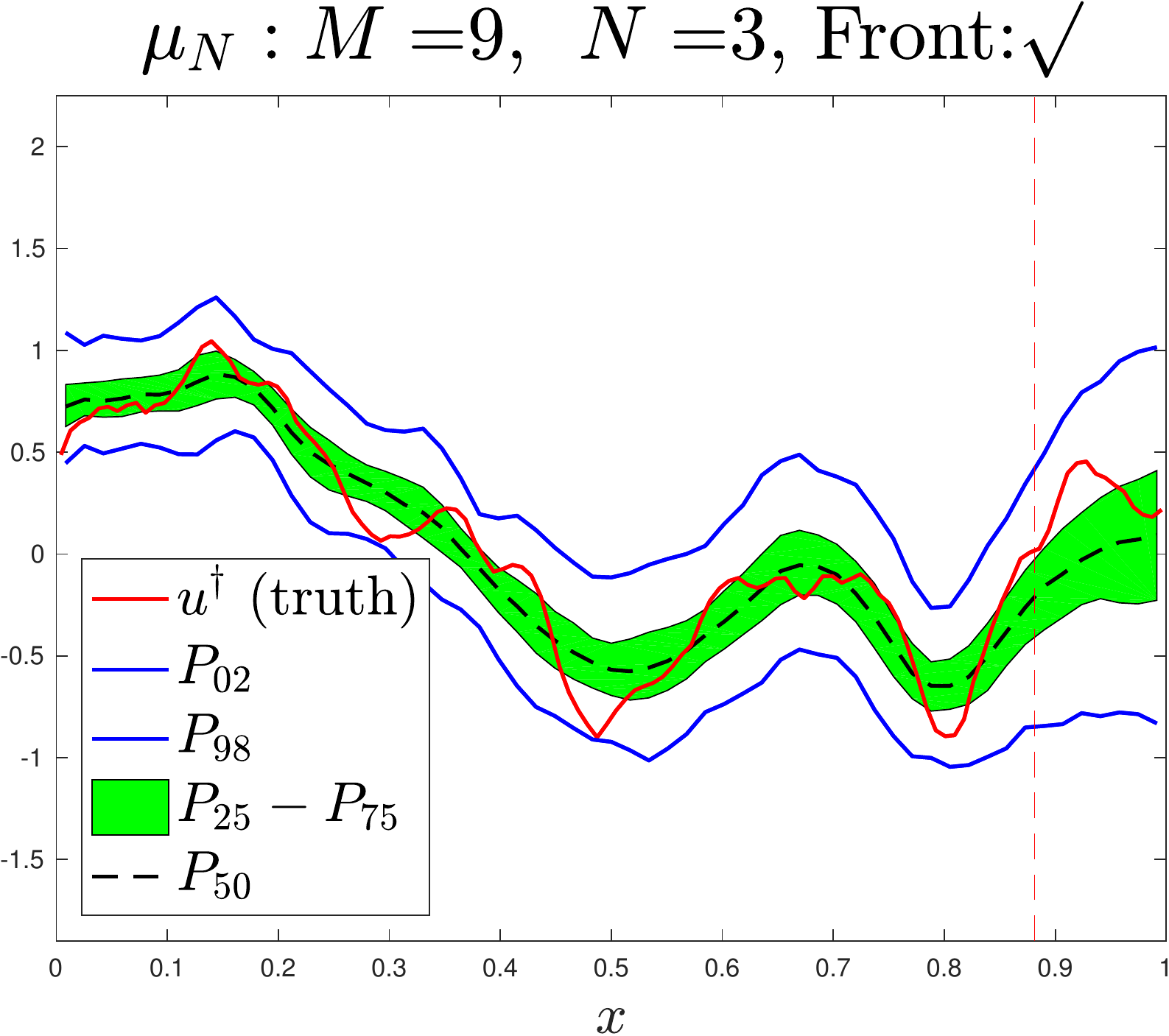}
\includegraphics[scale=0.25]{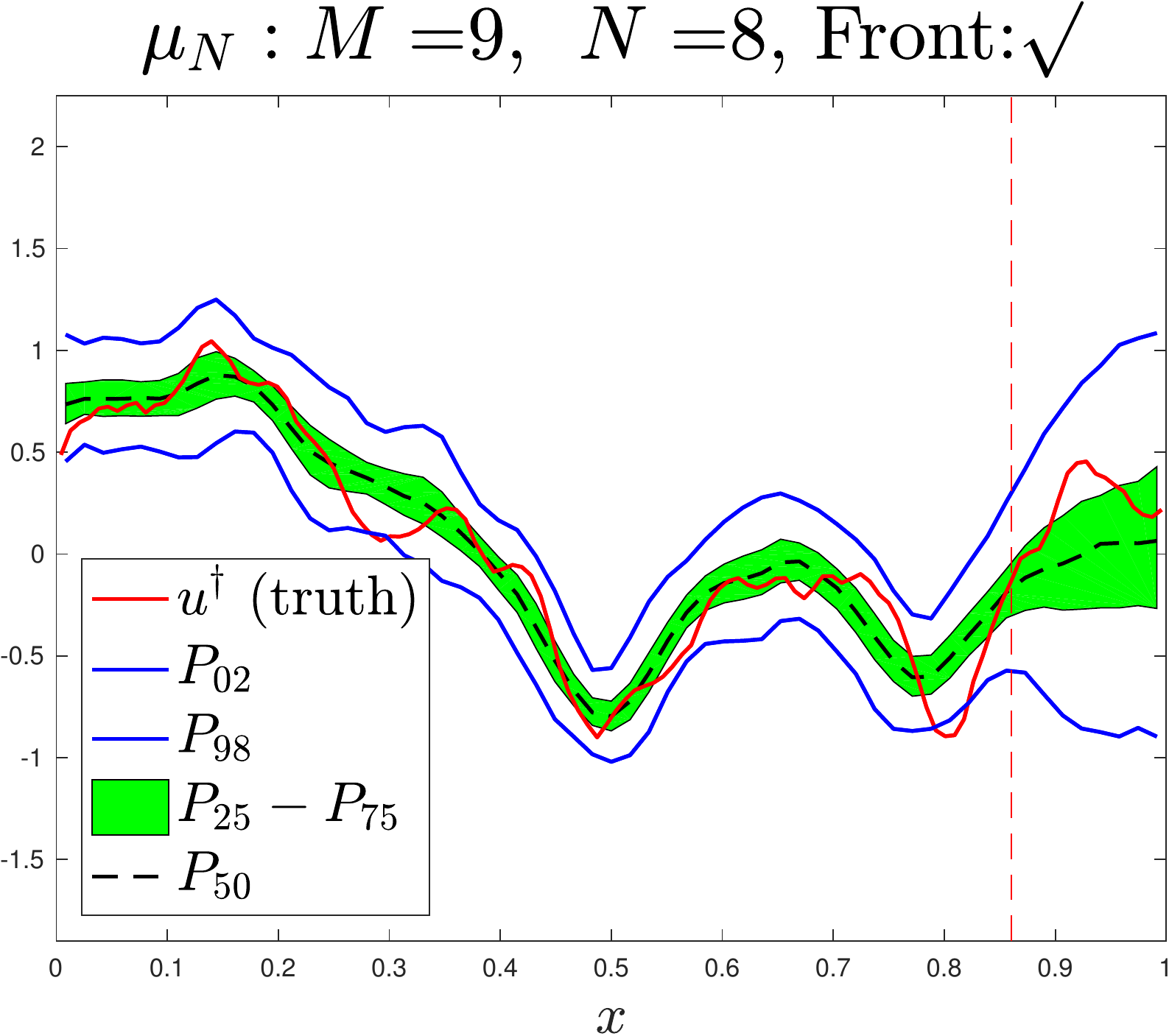}
\includegraphics[scale=0.25]{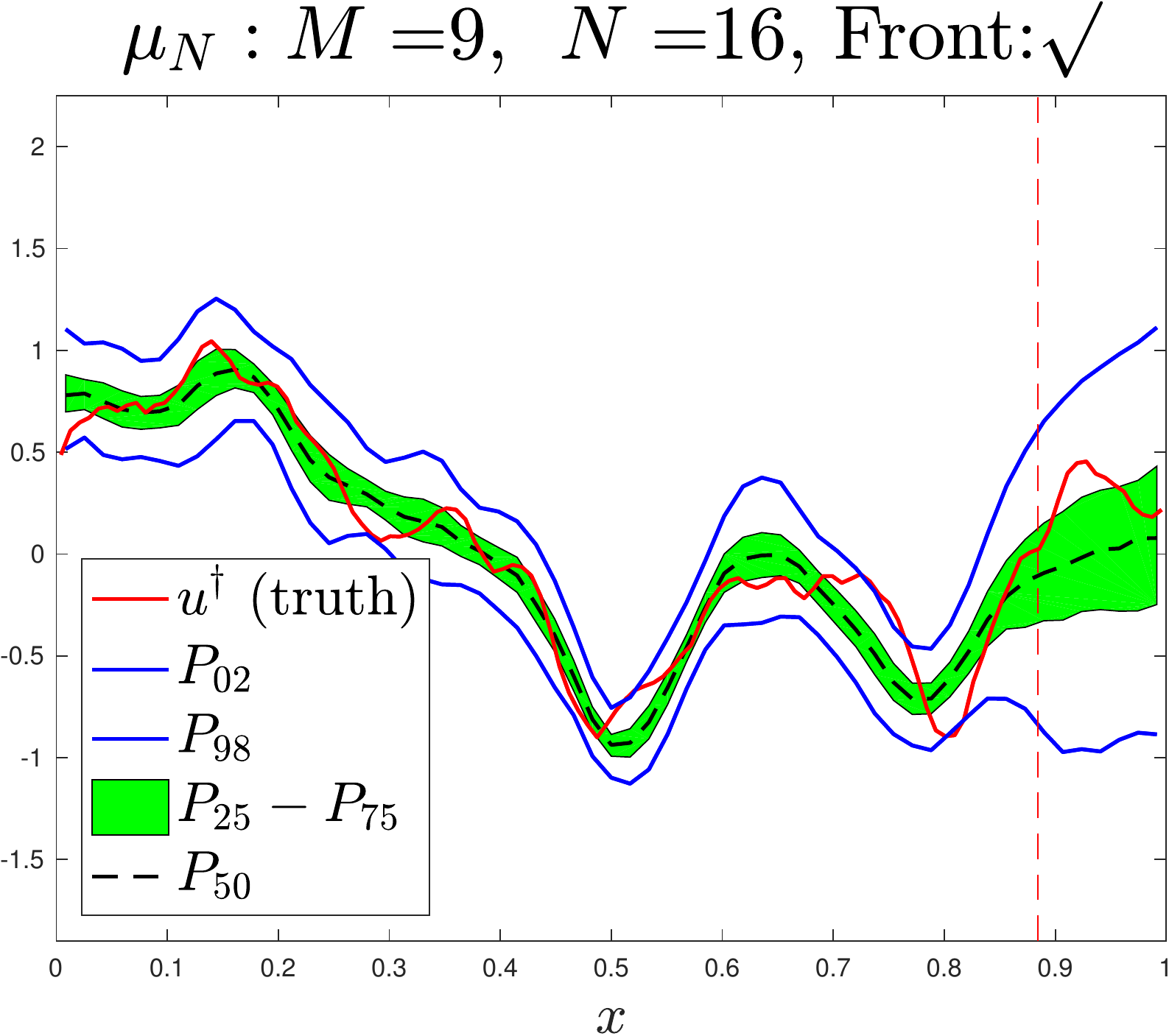}

 \caption{Percentiles of the final time posterior $\mu_{N}$ for different choices of total observation times $N$ and fixed $M=9$.} \label{Fig7}
\end{center}
\end{figure}

\begin{figure}[htbp]
\begin{center}
\includegraphics[scale=0.7]{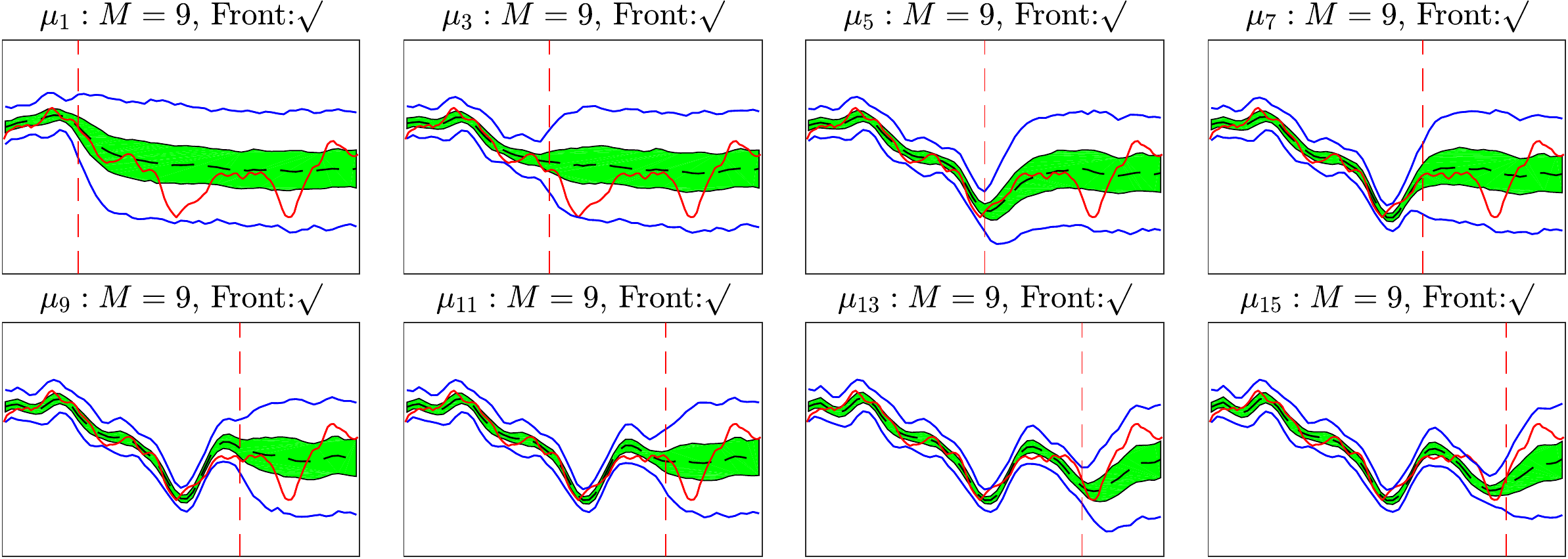}
\caption{Percentiles of the posteriors $\{\mu_{2n-1}\}_{n=1}^{8}$ obtained with $M=9$ pressure sensors and $N=16$ observation times.} \label{Fig8}
\end{center}
\end{figure}

The final error w.r.t the truth $\epsilon_{N}^{\Ga}$ as well as $\Sigma_{N}$ are both displayed in Figure \ref{Fig6C} (right). These numerical results indicate that increasing the number of observation times does not imply a decrease in either the error w.r.t truth or the uncertainty of the log-permeability at the final observation time. Although for $N>8$ we observe a smaller $\epsilon_{N}^{\Ga}$, the uncertainty in terms of $\Sigma_{n}$ is not substantially reduced. In fact, $\Sigma_{N}$ exhibits a slight increase for $N>8$, presumably due to the observational noise of these additional measurements. In Figure \ref{Fig7} we display the percentiles of the approximate posterior $\mu_{N}$ corresponding to the final observation time $t_{N}$ obtained for $N=1$, $N=3$, $N=8$ and $N=16$. While from Figure \ref{Fig6C} we notice that $\Sigma_{N}$ increases slightly for $N>8$, from Figure \ref{Fig7} we still observe a minor decrease in the uncertainty band defined by the posterior percentiles. It is essential to emphasize that even though our experiments suggest that assimilating measurements more frequently does not substantially improve our inference of the truth, the benefit of more frequent observations is the earlier reduction of the uncertainty which can be, in turn, used for the real-time optimization of the RTM process. This is illustrated in Figure \ref{Fig8}, where we show the percentiles of some of the log-permeability posteriors from one example with a total $N=16$ observation times. Indeed, the more frequent the observations, the better the characterization of the log-permeability field in the domain of the true moving front $D^{\dagger}(t_{n})$.

\subsubsection{The noise level}\label{noise}

\begin{figure}[htbp]
\begin{center}

\includegraphics[scale=0.7]{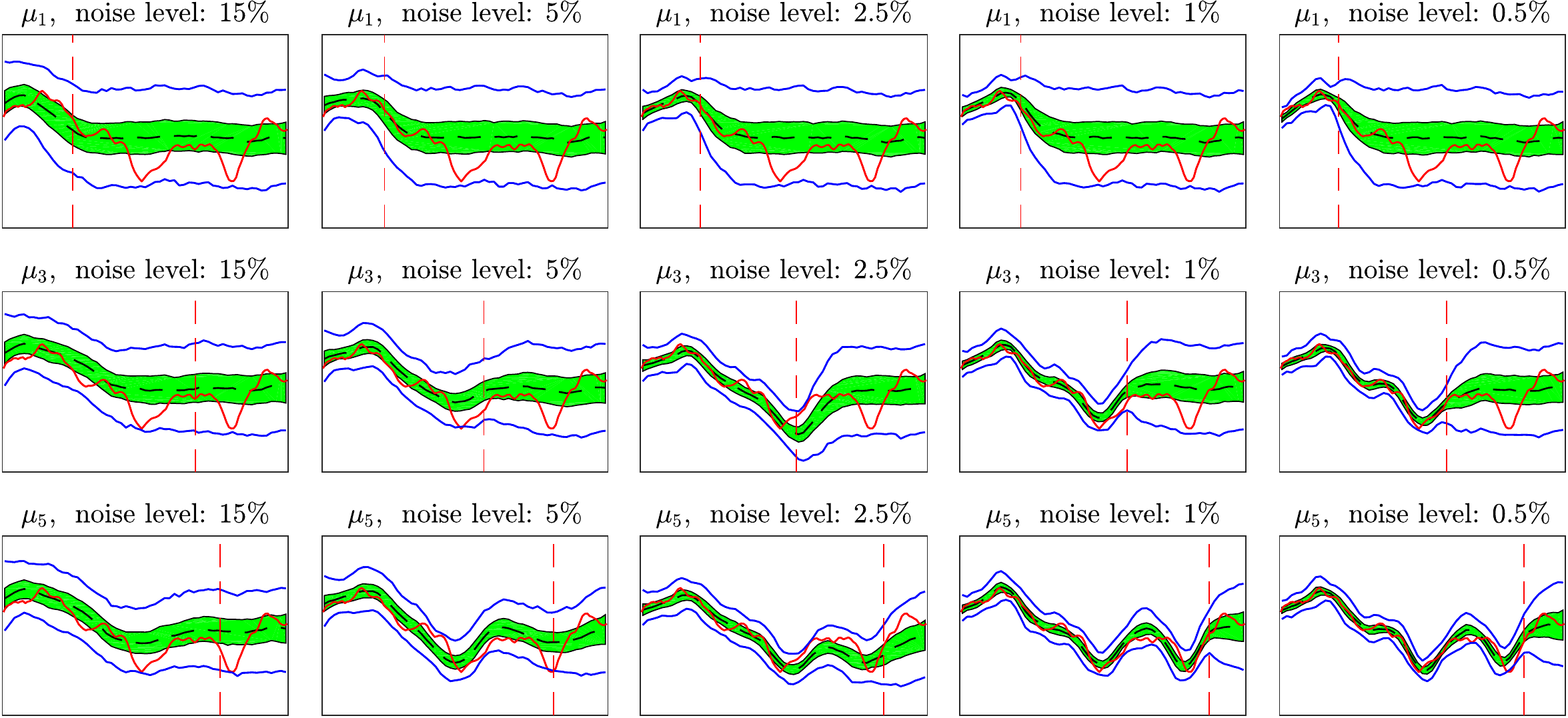}

 \caption{Percentiles of the posteriors $\{\mu_{n}\}_{n=1}^{5}$ obtained via REnKA with $J=1000$ from synthetic data with noise level of (from left to right) $15\%, 5\%, 2.5\%, 1\%, 0.5\%$.} \label{Fig9}
\end{center}
\end{figure}

The numerical investigation of the preceding section was conducted by means of numerical experiments with synthetic data with relatively small noise level ($1.5\%$). In this subsection we use REnKA to study the effect of the measurement noise level in the solution to the Bayesian inverse problem. Let us consider again the configuration of $M=9$ pressure sensors and $N=5$ observation times as those used in Section~\ref{set_up}. We generate both pressure and front location synthetic data as described in Section~\ref{set_up} with observational noise levels of $15\%$, $5\%$, $1\%$ and $0.5\%$. Percentiles of the posterior for different noise levels are displayed in Figure \ref{Fig9}. It comes as no surprise that for smaller noise levels, the posterior uncertainty bands defined by these percentiles are narrower and more concentrated around the true log-permeability. However, we find again that the reduction in the uncertainty is only observed in the region defined by the moving front $D^{\ast}(t_{n})$.

\begin{figure}[htbp]
\begin{center}

\includegraphics[scale=0.33]{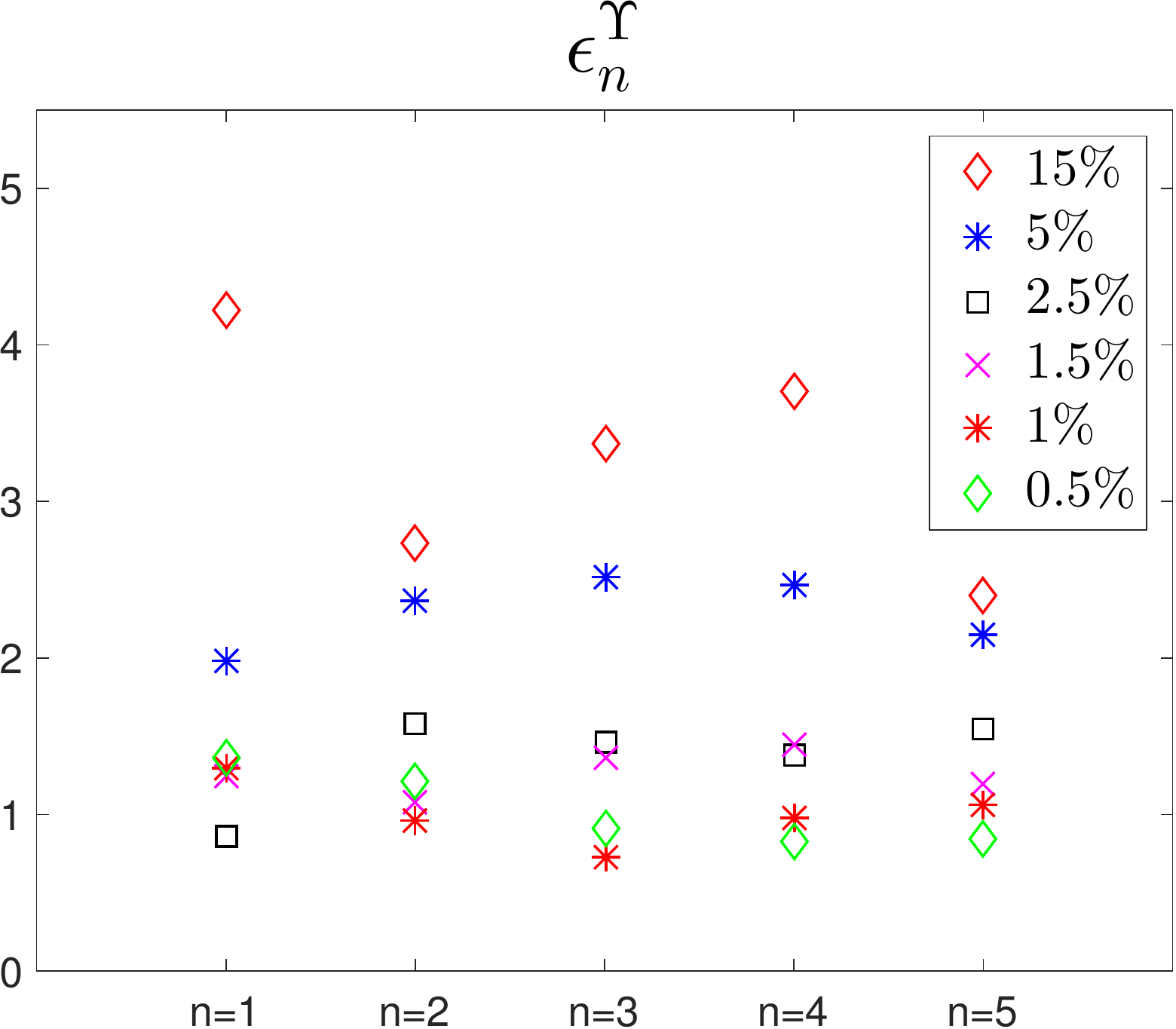}~~\includegraphics[scale=0.33]{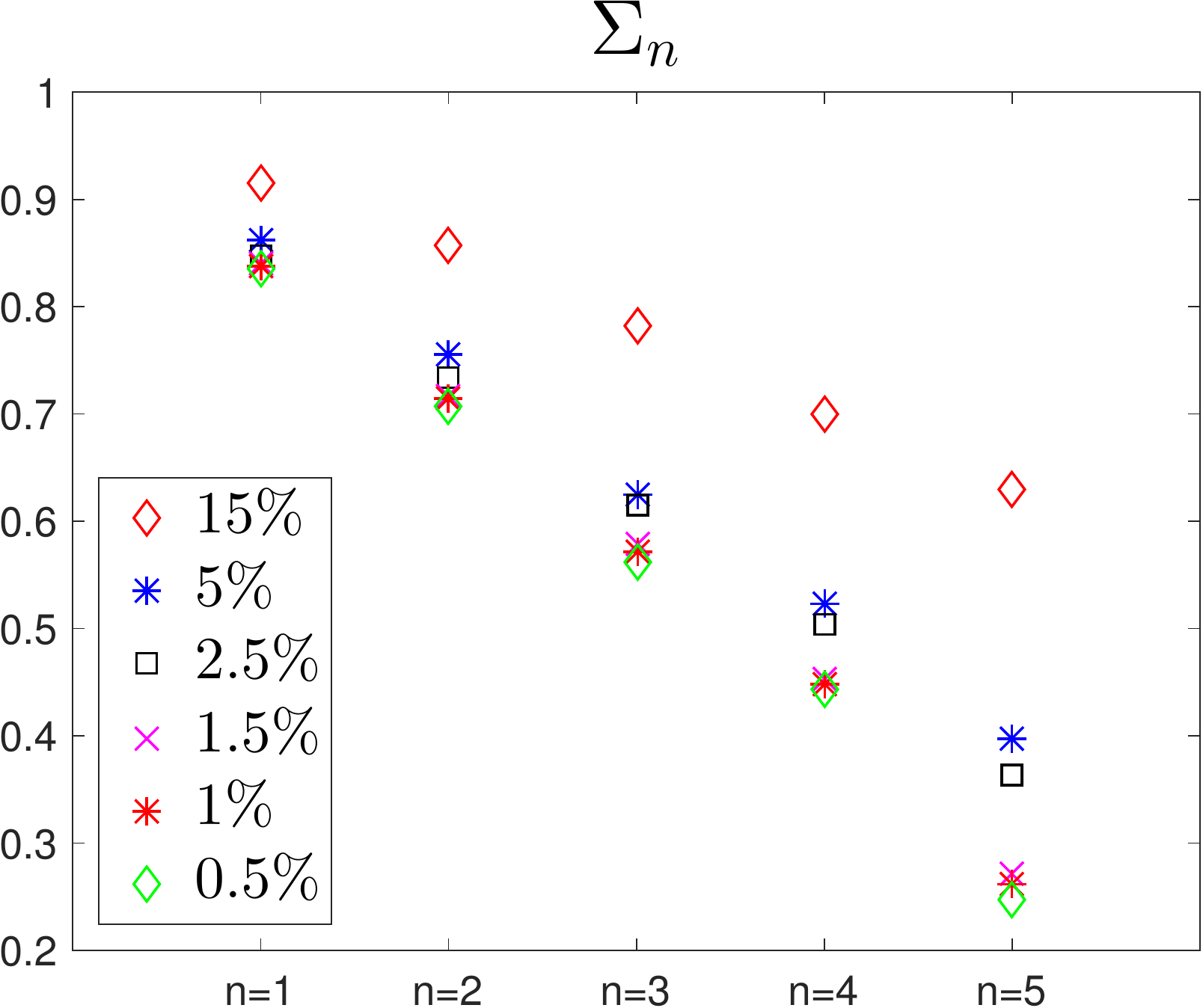}~~\includegraphics[scale=0.33]{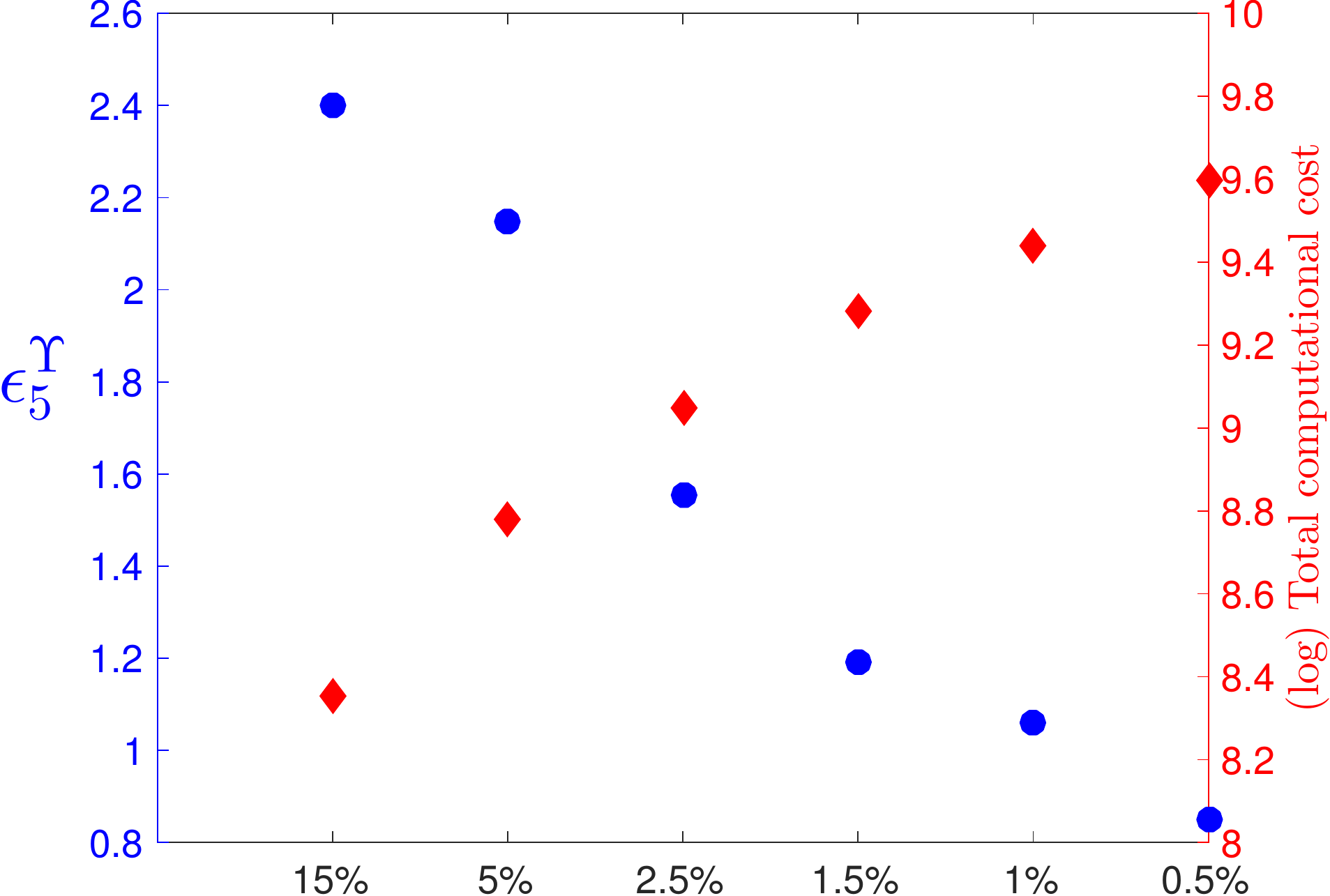}

 \caption{Left: Relative errors of the ensemble mean obtained via REnKA ($J=1000$) with respect to the truth $u^{\dagger}$ for synthetic data with different noise level.  Middle: norm of the variance of each posterior obtained with REnKA. Right: Number of tempering steps needed for the computation of each posterior measure $\mu_{n}$.} \label{Fig10}
\end{center}
\end{figure}

Similar to the results from the previous sections, the plot of $\epsilon_n$ (not shown) reveals that, at each observation time, inverting more accurate measurements of pressure and front does not necessarily implies that the log-permeability estimate is more accurate in the regions where the front has not yet arrived. In contrast, the plot of $\epsilon_{n}^{\Ga}$ displayed in Figure~\ref{Fig10} (left) allows us to clearly observe that smaller observational noise yield more accurate estimates of the truth.
In Figure \ref{Fig10} (middle) we display $\Sigma_{n}$ from which we can notice that smaller noise also results in lower uncertainty. Yet, measurements with noise levels below $1.5\%$ have little effect on both the reduction of the uncertainty and the accuracy of the ensemble mean at recovering the truth.

In Figure \ref{Fig9} (right) we show the final error with respect to the truth as well as the total computational cost involved in the approximation of the full sequence of posteriors. Smaller noise levels (i.e. more accurate measurements) require more tempering and thus larger number of intermediate measures. This, as discussed in Section~\ref{note}, is reflected in larger computational cost when more accurate measurements are assimilated.

\subsection{The 2D case}\label{REnKA_2D_invest}

In this section we apply REnKA for the Bayesian inversion of the 2D moving boundary problem described by (\ref{eq:2000})-(\ref{eq:2008}). In contrast to the 1D case, the numerical solution of the 2D moving boundary problem is much more computationally intensive. Therefore, the application of a fully Bayesian methodology such as the SMC sampler considered in Section~\ref{SMC_approach} is impractical for an online computation of the Bayesian posteriors in the 2D case. In this subsection, we exploit the capabilities of REnKA for providing an accurate yet computationally tractable approach for inferring preform (log) permeability alongside with its uncertainty.

\subsubsection{Formulation of the 2D Bayesian inverse problem}
Let us consider now the 2D moving boundary problem introduced (\ref{eq:2000})-(\ref{eq:2008}) from Section \ref{Intro}. We recall that we are interested in the inference of the log-permeability $u(x)=\log{\kappa(x)}$ given noisy measurements of the pressure field $\{p(x_{m},t_{n})\}_{m=1}^{M}$ from $M$ sensors located at points $\{x_{m}\}_{m=1}^{M}\subset D^{\ast}$ collected at a discrete observation times $\{t_{n}\}_{n=1}^{N}$. In addition, we wish to invert observations of the front location, or alternatively from the moving domain $D(t)$ that can be potentially obtained from CCT cameras such as in \cite{PC:PC23290,Andy_control07}. While in the 1D case we can trivially define observations of the (single point) front, in 2D \mt{the front} $\Ga(t_{n})$ is a curve which defines the moving domain $D(t)$. Therefore, rather than \mt{dealing with measurements of the front $\Ga(t)$ itself} we may assume pointwise measurements of $D(t)$ via its characteristic function defined by
\begin{equation}\label{moving}
\chi(x,t)\equiv  \left\{ \begin{array}{cc}
1& x\in D(t), \\
0 &x\notin D(t).    \end{array}\right.
\end{equation}
More precisely, we define observations of the form $\{\chi(x_{m}^{\Ga},t_{n})\}_{m=1}^{M_{\Ga}}$, where $\{x_{m}^{\Ga}\}_{m=1}^{M_{\Ga}}\subset D$ is an array of points \mt{for which} the  characteristic function of $D(t)$ is observed. In practice this array should be considered dense when a high resolution camera is used for capturing the moving domain.
Note that this mathematical description of front measurements is suitable as it enables its direct comparison with observations from digitalized images and also with the discrete formulation of  (\ref{eq:2001})-(\ref{eq:2008}) via the control volume finite element method (CV/FEM) in which the front location is characterized in terms of the filling factor (see details in \cite{SRTM16}) rather than using a parameterization of $\Ga$.

For specified (known) $p_{I}$, $p_{0}$, $\partial D_{I}$, $\partial D_{N}$, $\varphi$, \mt{the solution of} (\ref{eq:2001})-(\ref{eq:2008}) induces a map $u=\log{\kappa} \to [p(x,t), D(t)]$ which enables us to define the sequence of forward maps $\cG_{n}:C(\overline{D})\to \mathbb{R}^{M+M_{\Ga}}$ by means of
\begin{eqnarray}\label{eq:2010}
\cG_{n}(u)=\Big[\{p(x_{m},t_{n})\}_{m=1}^{M},\{\chi(x_{m}^{\Ga},t_{n})\}_{m=1}^{M_{\Ga}}\Big] .
\end{eqnarray}
To our best knowledge, uniqueness, existence and regularity theory for problem (\ref{eq:2001})-(\ref{eq:2008}) with non constant $\kappa(x)=e^{u(x)}$ is an open problem for $d>1$
(see a related discussion in \cite{SRTM16}). However, for the present work we assume that the following \mt{condition} holds.
 \begin{assumption}
The sequence of forward maps $\cG_{n}:C(\overline{D})\to \mathbb{R}^{M+M_{\Ga}}$ ($n=1,\dots, N$) are continuous.
\end{assumption}

We now follow the same formulation of the Bayesian inverse problem as the one we described in Section~\ref{BIP_1D}. At each observation time $t=t_{n}$, we collect noisy measurements of the front location $y_{n}^{\Ga}\in \mathbb{R}^{M_{\Ga}}$ as well as pressure measurements from sensors $y_{n}^{p}\in \mathbb{R}^{M}$. We assume that observations $y_{n}=[y_{n}^{\Ga},y_{p}]^{T}$ are related to the unknown via expressions (\ref{eq12C}) -(\ref{eq12D}) with $\cG_{n}$ defined in (\ref{eq:2010}). As before, both measurements of $D(t)$ (via its characteristic function) and pressures are assumed to be uncorrelated in time and independent from each other. Furthermore, we consider Gaussian priors $\mu_{0}=N(\overline{u},\cC)$ with a covariance operator $\cC$ that arises from the Whittle-Matern correlation function defined in (\ref{eq13}). The assumption of continuity of the forward maps as well as the fact that $\mu_{0}(C(\overline{D}^{\ast}))=1$ ensures existence of the sequence of posterior measures $\mu_{n}=\mathbb{P}(u\vert y_{1},\dots,y_{n})$ given by  Theorem \ref{Bayes} with the same definition of the likelihood functions introduced in (\ref{eq14}).  In the following section we use REnKA to to compute an ensemble approximation of $\{\mu_{n}\}_{n=1}^{N}$.

\subsubsection{2D Numerical Experiments}

\begin{figure}[htbp]
\begin{center}
\includegraphics[scale=0.25]{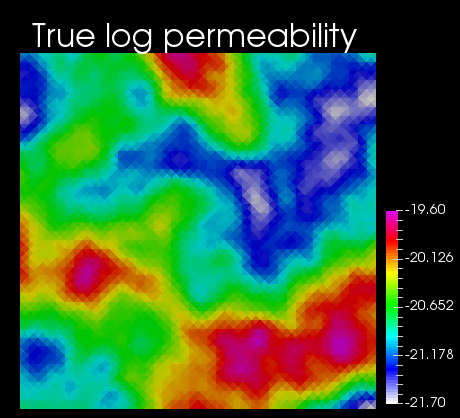}
\includegraphics[scale=0.26]{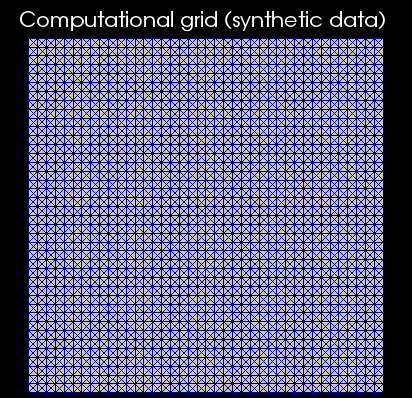}
\includegraphics[scale=0.26]{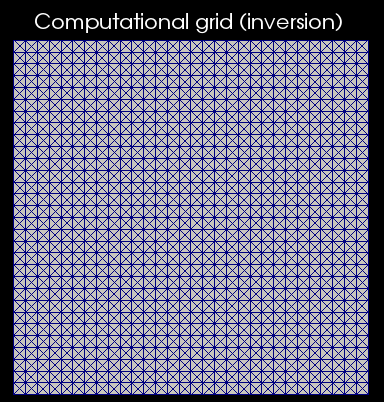}\\
\vspace{2.5mm}
\includegraphics[scale=0.2]{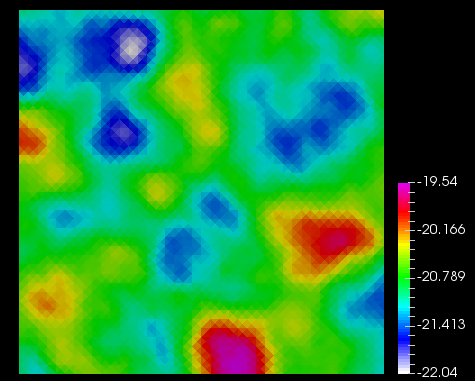}
\includegraphics[scale=0.2]{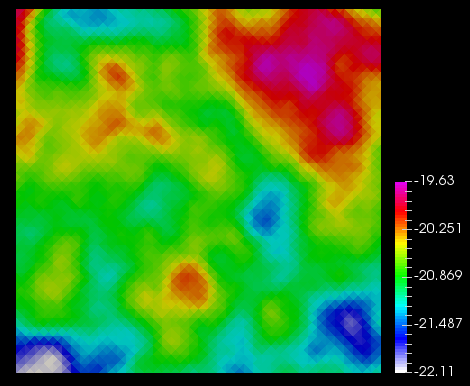}
\includegraphics[scale=0.2]{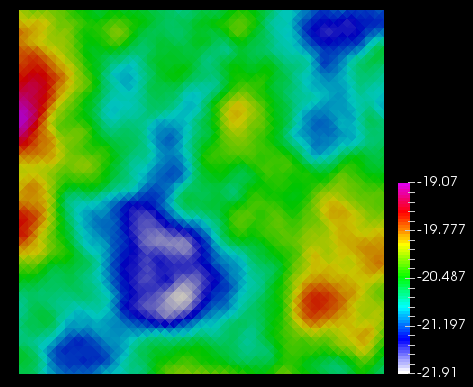}
\includegraphics[scale=0.2]{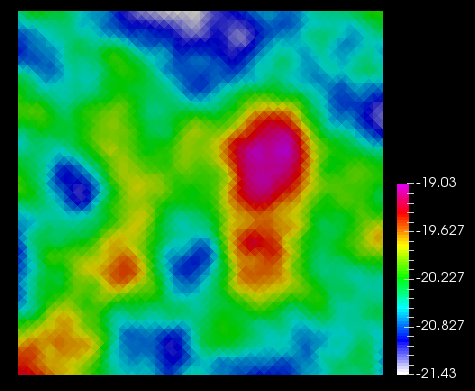}
\includegraphics[scale=0.2]{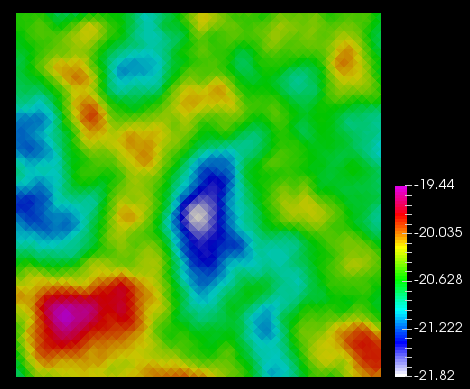}\\
\vspace{2.5mm}
\includegraphics[scale=0.31]{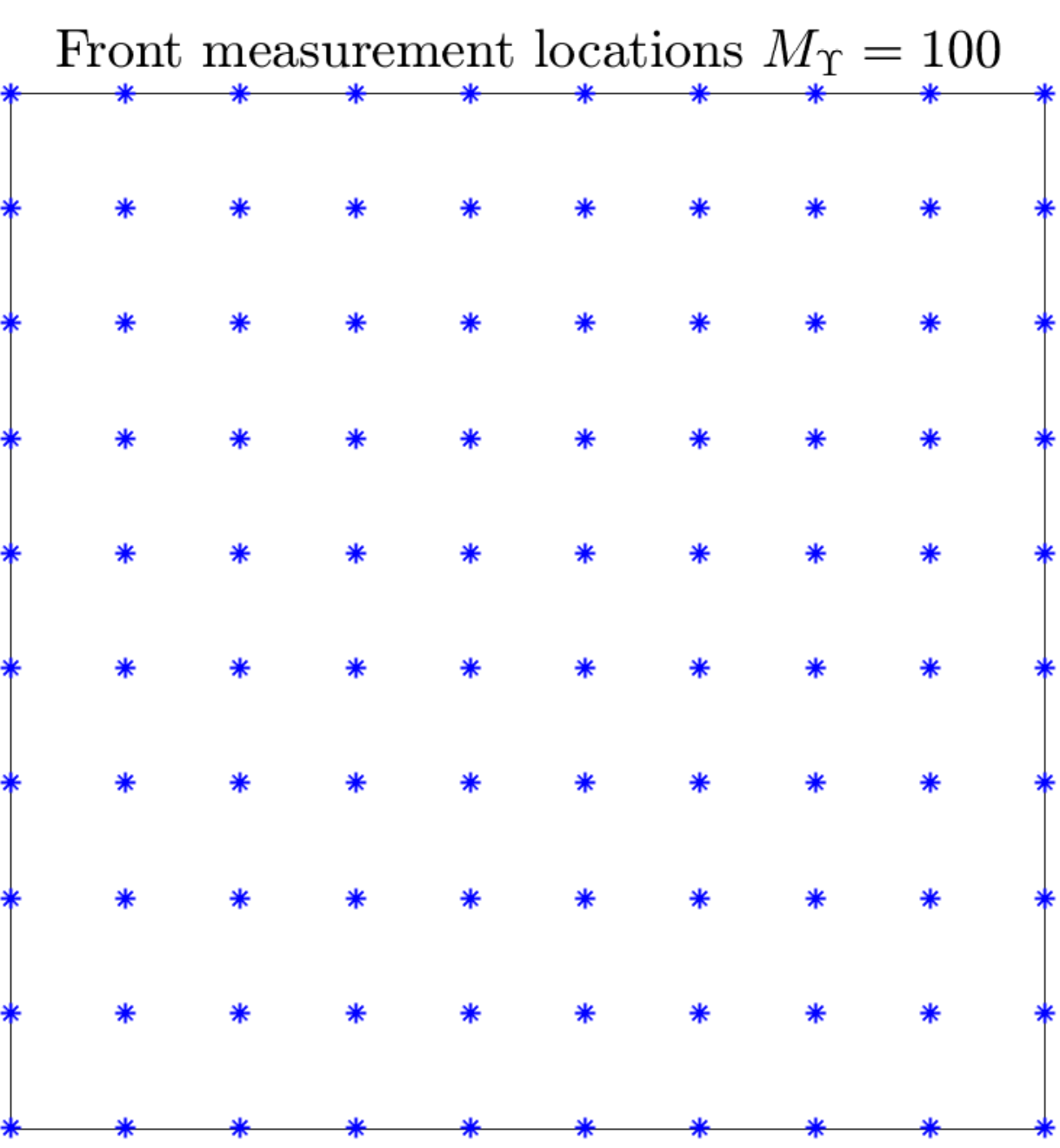}
\includegraphics[scale=0.31]{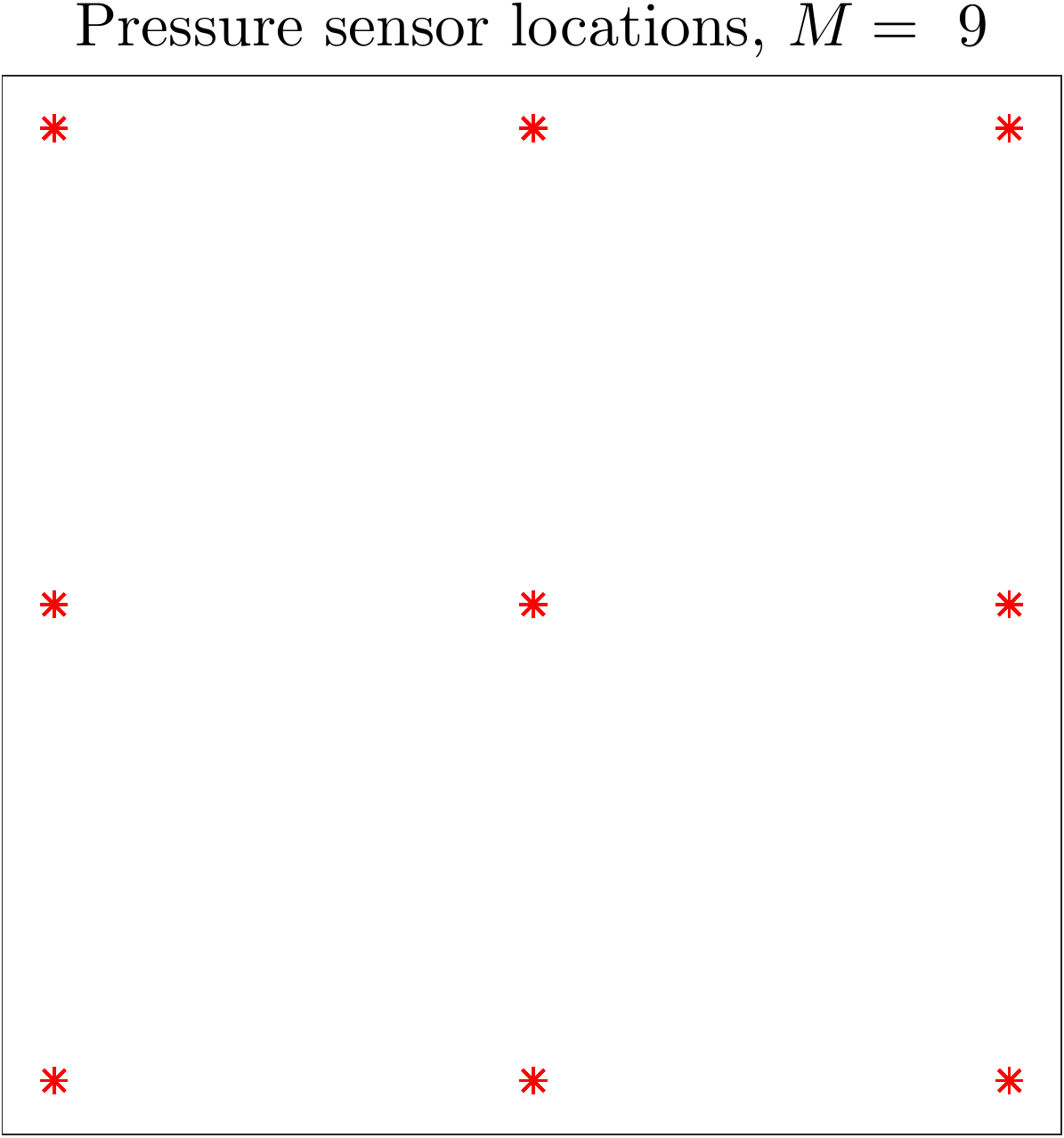}
\includegraphics[scale=0.31]{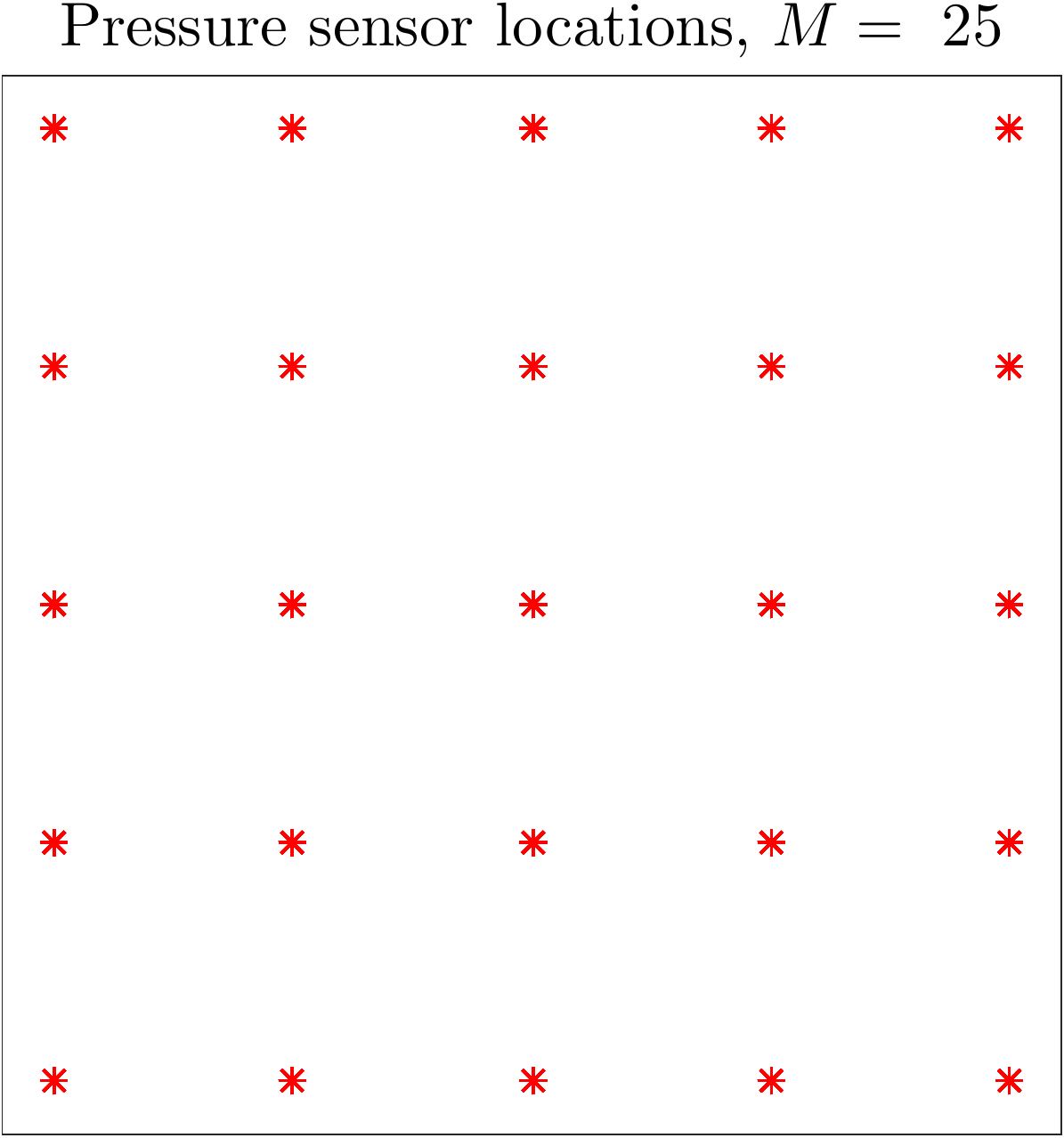}
\includegraphics[scale=0.31]{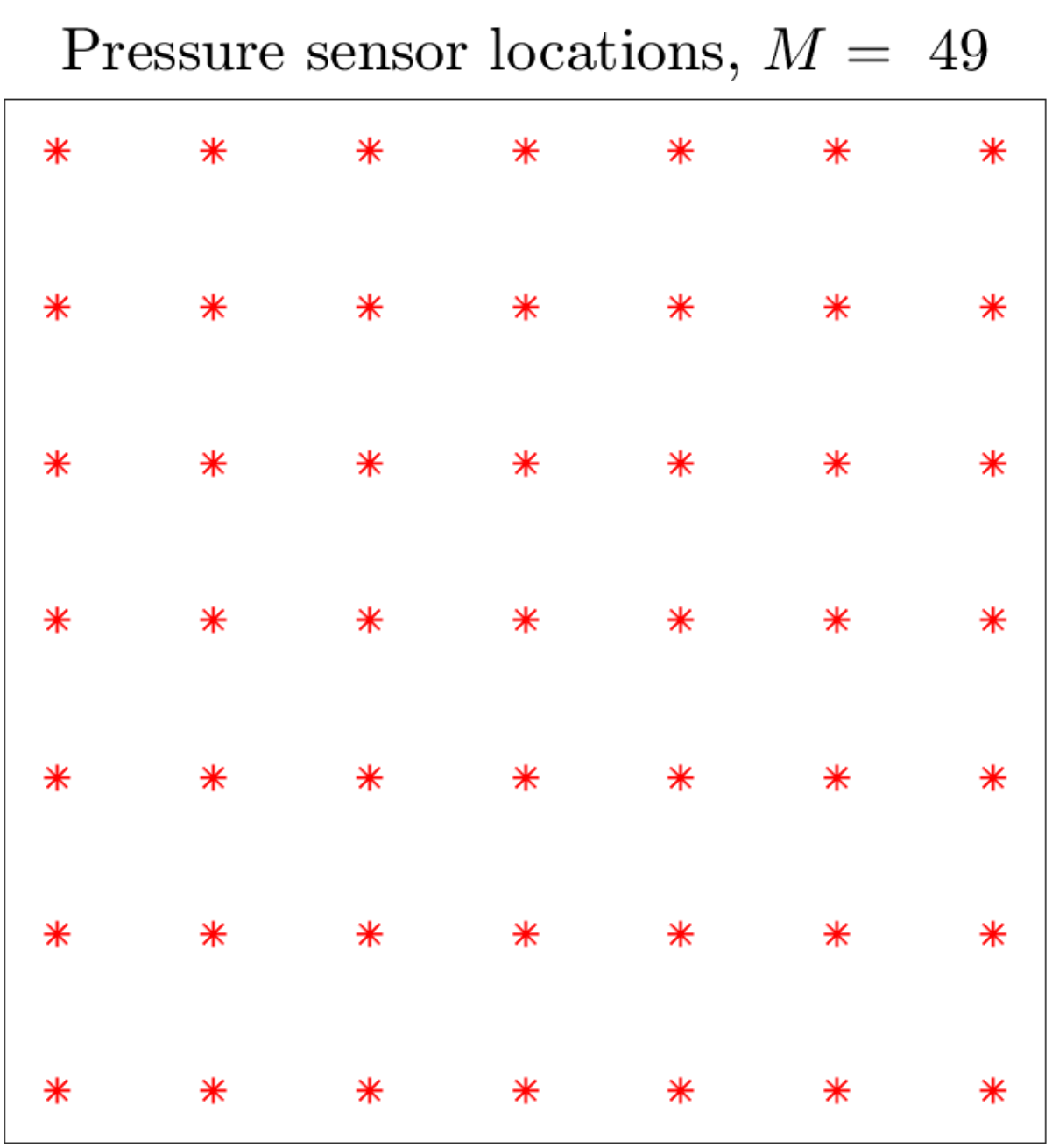}
 \caption{Top-left: true log permeability [$\log[\textrm{m}^{2}]$]. Top-middle: computational domain for the generation of synthetic data. Top-right: computational domain for the inversion (via REnKA). Middle row: Random draws from the Gaussian prior $\mu_{0}$ [$\log[\textrm{m}^{2}]$]
Bottom row: Measurement configuration for the moving front $M_{\Ga}=100$ (left); Pressure measurement configuration with (from left to right) $M=9$, $M=25$ and $M=49$ sensors. }  \label{Fig12B}
\end{center}
\end{figure}

In this subsection we apply REnKA for the solution of the 2D Bayesian inverse problem defined in the previous subsection. The forward model described by (\ref{eq:2000})-(\ref{eq:2008}) is solved numerically with the MATLAB code developed in \cite{SRTM16} and available from \newline \textit{https://github.com/parkmh/MATCVFEM}. This code is based on the interface-tracking control volume finite element method (CV/FEM) \cite{AS10,Hirt81,cvfem1}. For experiments of this subsection, we consider the following fixed values:
$$D^{\ast}=[0,1\textrm{m}]\times [0,1\textrm{m}],\quad  \mu= 0.1\textrm{Pa}\cdot \textrm{s},\quad \varphi=1,\quad p_{0}=1\textrm{MPa},\quad p_{I}=6\textrm{MPa}.$$
 Samples from $\mu_{0}$ are generated via KL parametrization as described in Section~\ref{Prior} with parameters $\sigma_{0}^{2}=0.25$, $\nu=1.5$, $l=0.1$ and $\overline{u}(x)= 0.0$ for all $x\in D^{\ast}$. Some draws from the prior are displayed in Figure \ref{Fig12B} (middle row). The log-permeability field is plotted in Figure \ref{Fig12B} (top-left) is a random draw from the prior that we use as the truth $u^{\dagger}$ for the present experiments.  

We use one dense configuration of $M_{\Ga}=100$ measurement locations $\{x_{m}^{\Ga}\}_{m=1}^{M_{\Ga}}$ for the observation of the moving domain given in terms of (\ref{moving}); these locations are displayed in Figure \ref{Fig12B} (bottom-left). We have selected a large number of measurements locations assuming that the moving domain can be densely observed with high-resolution cameras or dielectric sensors. In addition, we consider three possible configurations of $M=9$, $M=25$ and $M=49$ pressure sensors $\{x_{m}\}_{m=1}^{M}$, whose locations are shown in the left-middle to right panels of Figure \ref{Fig12B} (bottom). The summary of measurement configurations that we investigate are summarised below:
\small
\begin{eqnarray}\label{eq:final1}
(M=0, D(t):\checkmark),&(M=9, D(t):\checkmark),&\quad (M=25, D(t):\checkmark),\quad (M=49, D(t):\checkmark), \nonumber\\
(M=9, D(t): \mathcal{X}),& (M=25, D(t):\mathcal{X}),&\quad (M=49, D(t): \mathcal{X}), \label{combo2}
\end{eqnarray}
\normalsize
where $\checkmark$ (resp. $\mathcal{X}$) indicates whether the moving domain $D(t_{n})$ has been observed via (\ref{moving}).

\begin{figure}[htbp]
\begin{center}
\includegraphics[scale=0.275]{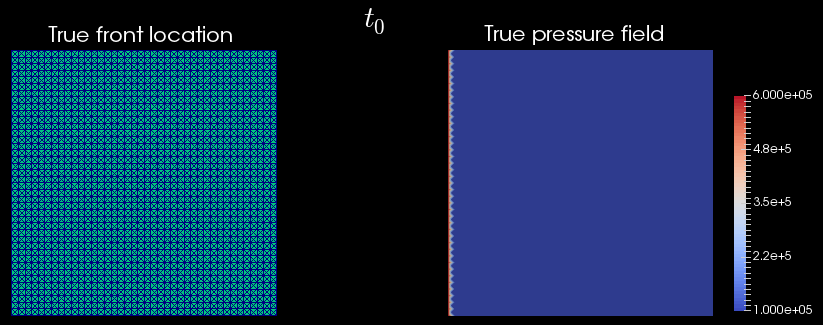}~~~\includegraphics[scale=0.275]{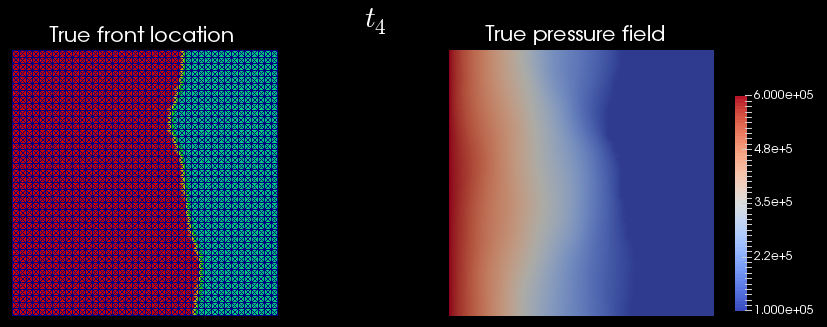}\\
\includegraphics[scale=0.275]{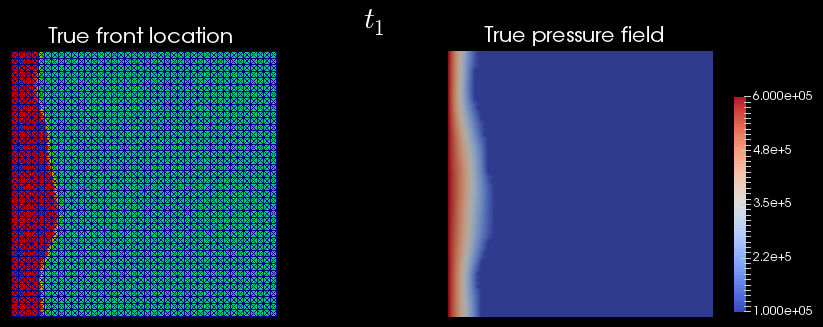}~~~\includegraphics[scale=0.275]{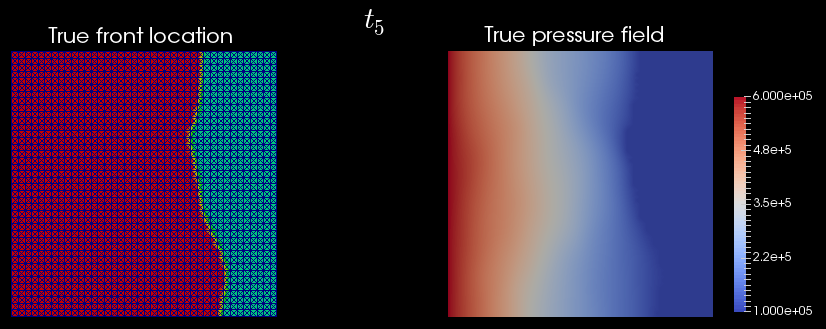}\\
\includegraphics[scale=0.275]{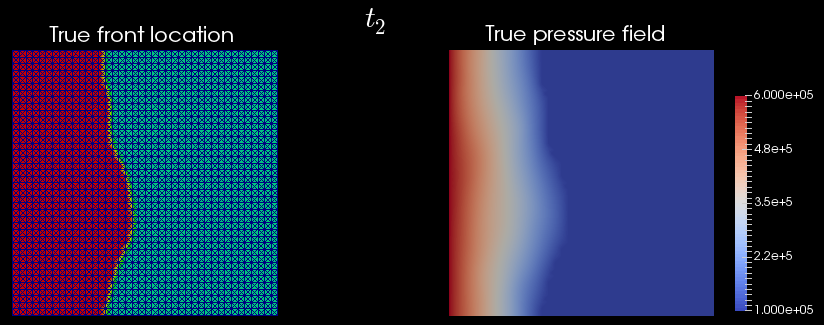}~~~\includegraphics[scale=0.275]{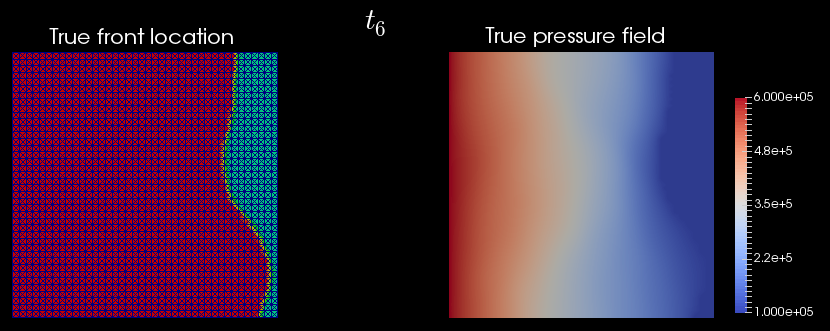}\\
\includegraphics[scale=0.275]{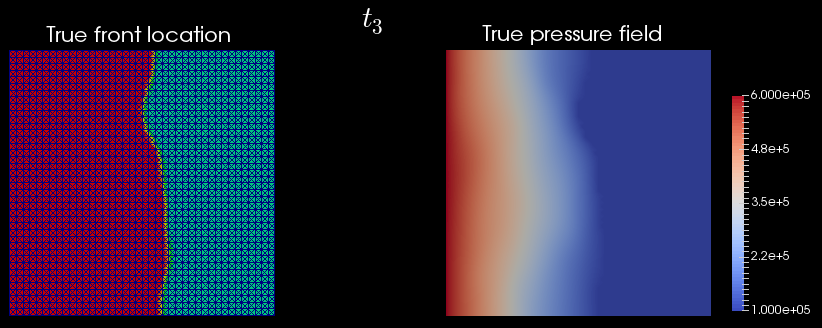}~~~\includegraphics[scale=0.275]{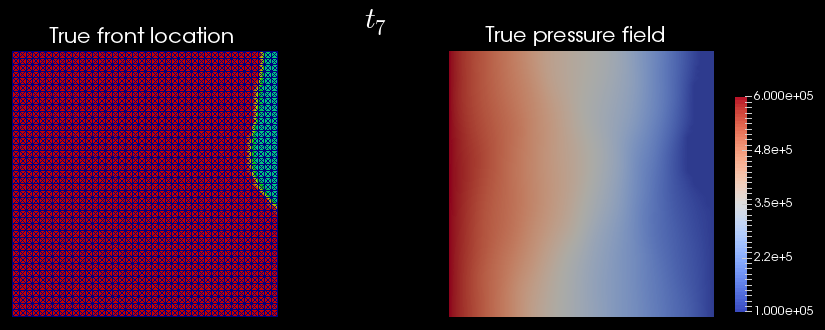}
 \caption{Snap shots of the true pressure field [Pa] and the true moving domain $D^{\dagger}(t_{n})$ for the initial time $t_{0}$ and the observation times $\{t_{n}\}_{n=1}^{7}$ [s].} \label{Fig13}
\end{center}
\end{figure}

We use the true log-permeability to numerically solve the forward model (\ref{eq:2000})-(\ref{eq:2008}) via the CV/FEM code described above. Then, for each of these measurement configurations, synthetic data with a realistic choice of $2.5\%$ Gaussian noise are generated in a similar manner to the one described in Section~\ref{set_up}. In order to avoid inverse crimes,
synthetic data are computed on a finer grid than the one we use for the Bayesian inversion via REnKA. These are shown in the middle and right panels of Figure \ref{Fig12B} (top). Snapshots of the true pressure field $p^{\dagger}$ at the initial time $t_{0}$ and observation times $\{t_{n}\}_{n=1}^{N}$ (in seconds) are displayed in Figure~\ref{Fig13} alongside with the corresponding true moving \mt{domain} $D^{\dagger}(t_{n})$.

\begin{figure}[htbp]
\begin{center}
\includegraphics[scale=0.28]{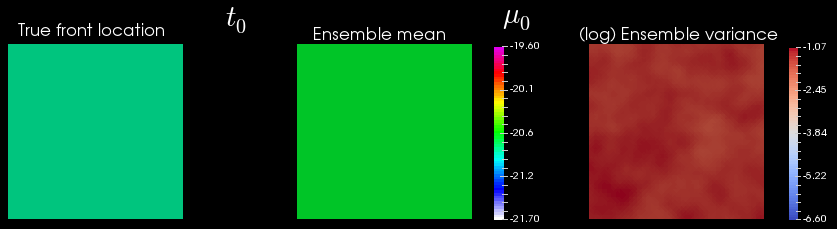}~~~~\includegraphics[scale=0.28]{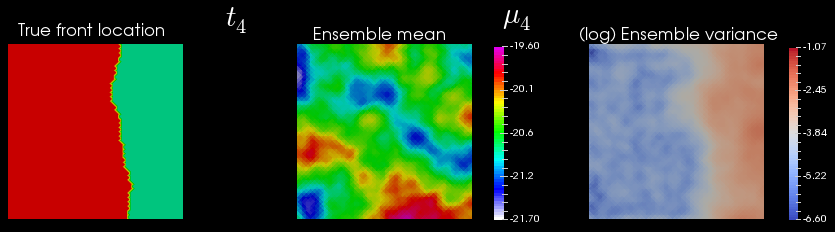}\\
\includegraphics[scale=0.28]{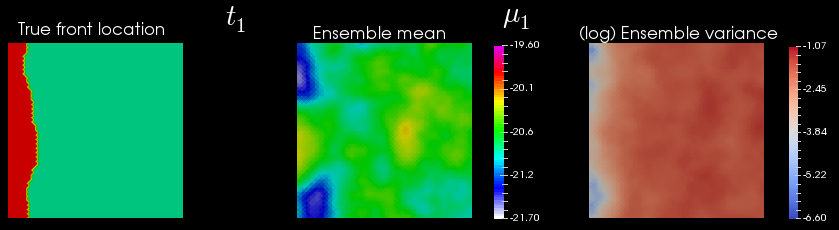}~~~~\includegraphics[scale=0.28]{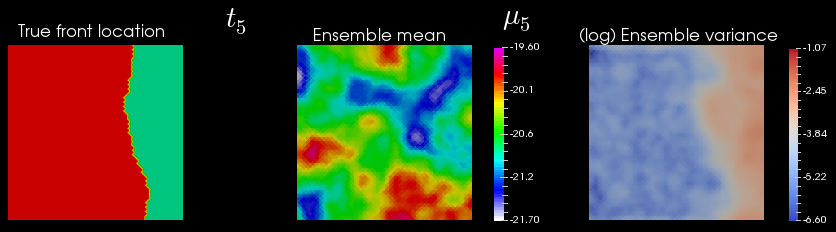}\\
\includegraphics[scale=0.28]{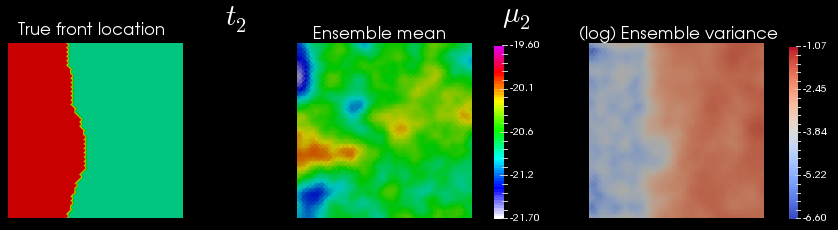}~~~~\includegraphics[scale=0.28]{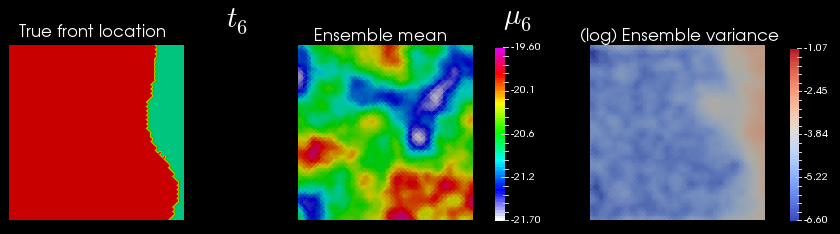}\\
\includegraphics[scale=0.28]{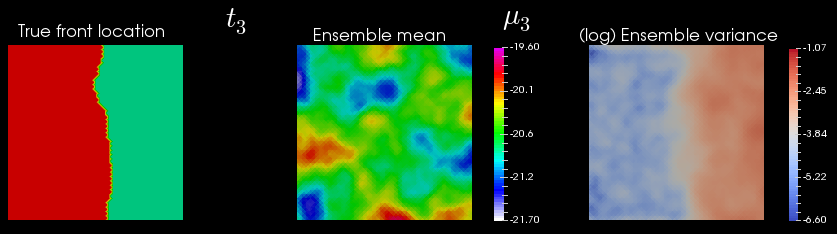}~~~~\includegraphics[scale=0.28]{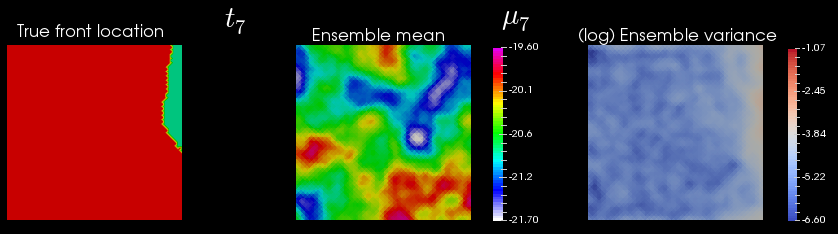}\\

 \caption{Ensemble mean and variance of $\mu_{n}$ obtained via REnKA at each observation time. The true moving boundary has been included at each of these observation times. }\label{Fig15}
\end{center}
\end{figure}

\begin{figure}[htbp]
\begin{center}
\includegraphics[scale=0.25]{./2D/True_log_perm.png}\\
\includegraphics[scale=0.275]{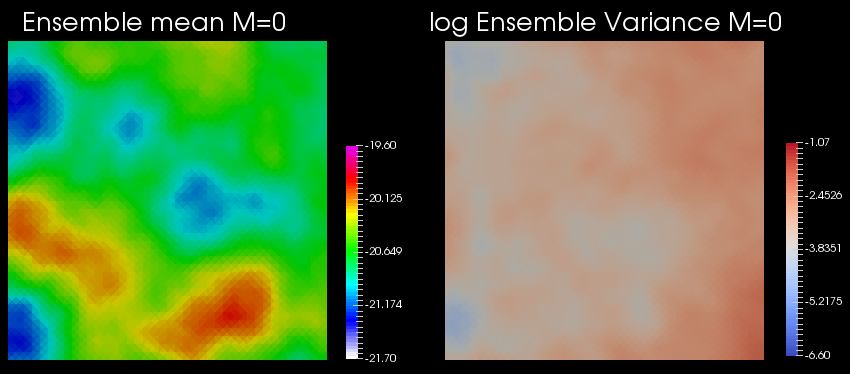}~~~\includegraphics[scale=0.275]{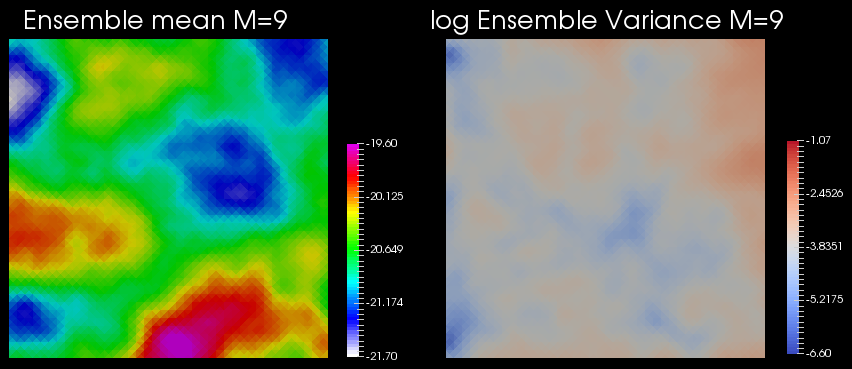}\\
\includegraphics[scale=0.275]{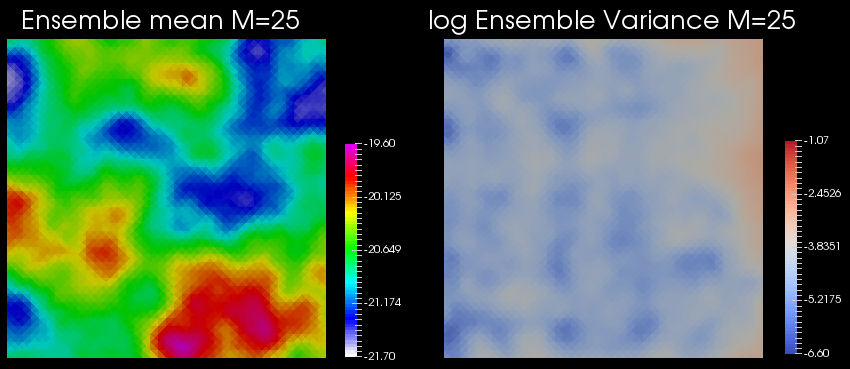}~~~\includegraphics[scale=0.275]{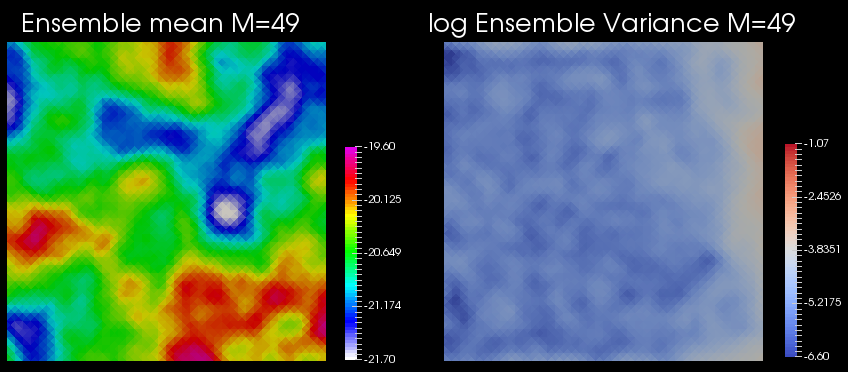}
 \caption{Top: True-log permeability: Bottom: Ensemble mean and variance of $\mu_{7}$ obtained via REnKA at the final observation time $t_{7}$.} \label{Fig16}
\end{center}
\end{figure}

For the configuration with ($M=49, D(t):\checkmark$) the ensemble mean and (log) variance of each posterior $\mu_{n}$ approximated with REnKA are displayed in Figure \ref{Fig15}. We note that as more observations (in time) are assimilated, the ensemble mean better captures the spatial features of the truth while the (variance) uncertainty is reduced in the region of the true moving domain $D^{\dagger}(t_n)$. In Figure \ref{Fig16} we show the ensemble mean and (log) variance of the final time posterior $\mu_{7}$ approximated via REnKA for different choices of the number of pressure sensors and with the inversion of the moving domain ($\checkmark$). Note that the pure inversion of the moving domain (i.e. $M=0$ pressure sensors) results in an informative measure of the log-permeability. It is clear that the accuracy in the estimation of the log-permeability improves with the number of pressure sensors.

\begin{figure}[htbp]
\begin{center}
\includegraphics[scale=0.35]{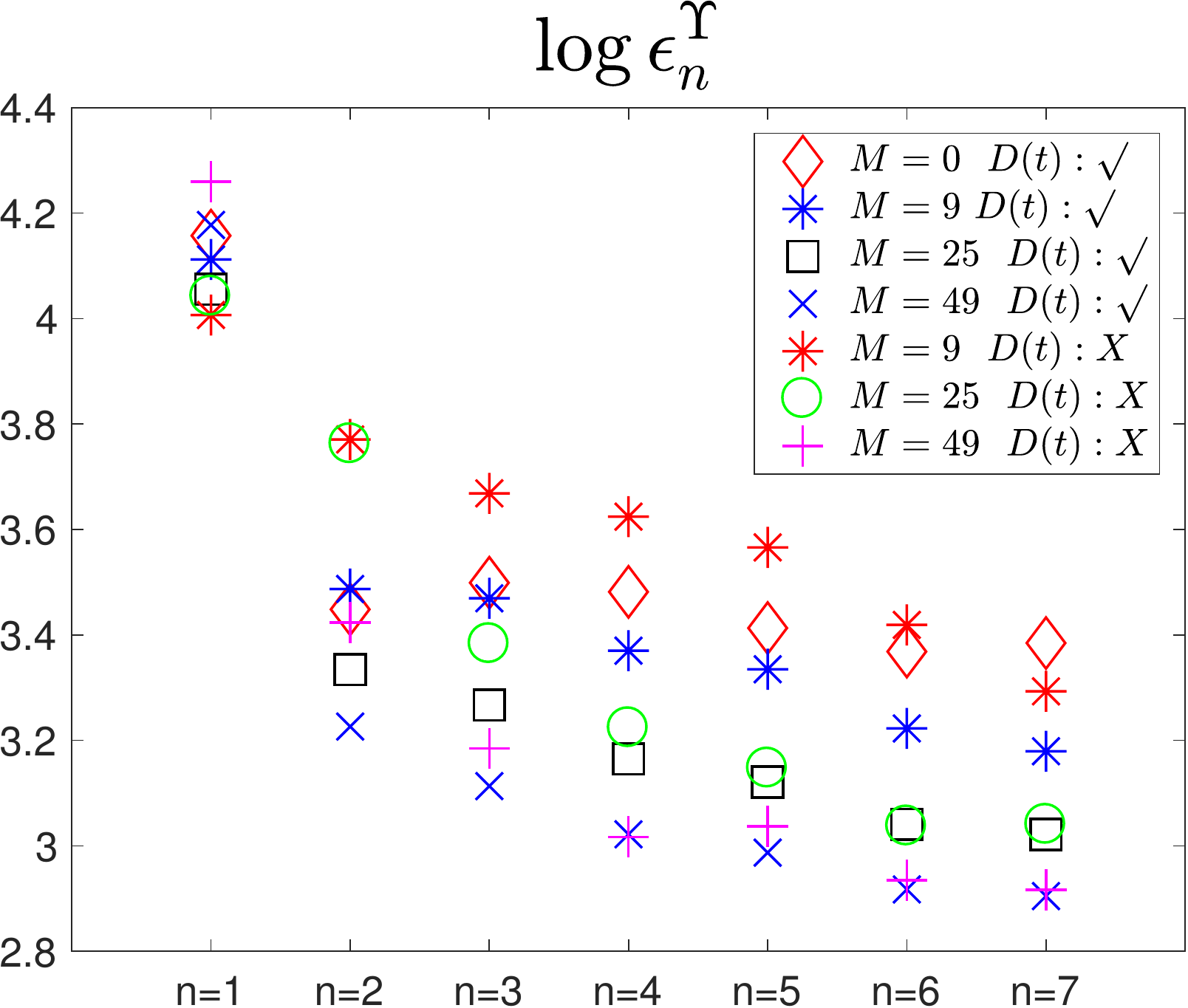}~~\includegraphics[scale=0.35]{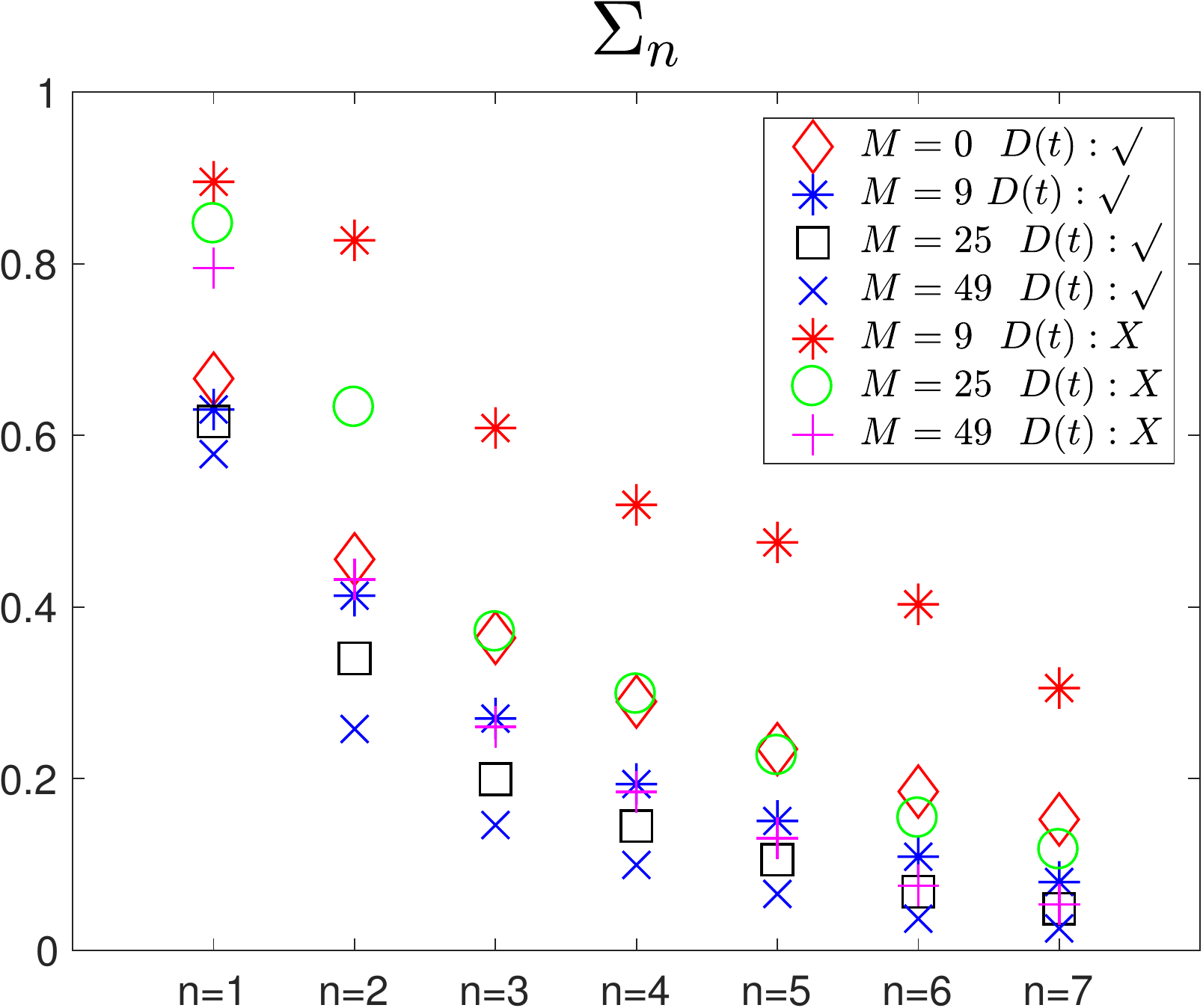}~~\includegraphics[scale=0.35]{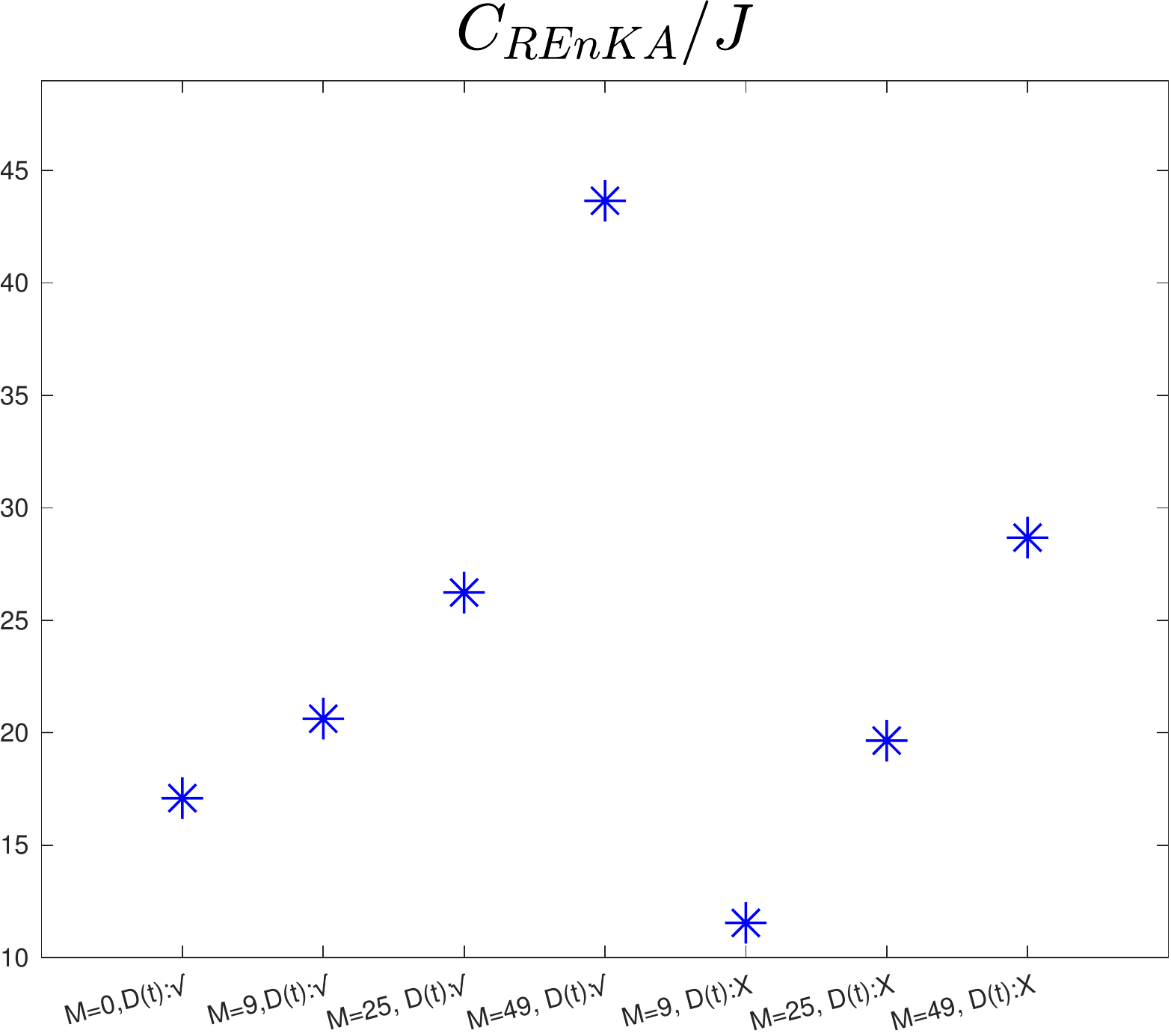}

 \caption{Performance of REnKA ($J=150$) for the the approximation of the posterior measures $\{\mu_{n}\}_{n=1}^{7}$ with measurement configurations from (\ref{eq:final1}). Left: $\log{\epsilon_{n}^{\Ga}}$ from (\ref{truth_err2}).  Middle: relative norm of the variance $\Sigma_{n}$ (\ref{truth_var0}). Right: Scalable (with respect to $J$) computational cost of REnKA.} \label{Fig17}
\end{center}
\end{figure}

From the plot of $\epsilon^{\Ga}$ displayed in Figure \ref{Fig17} (left), we note that, at the latest observation times ($n=6,7$), there is a clear improvement in the accuracy with increasing the number of pressure locations. Similar to the 1D case, we also observe that the benefit of inverting measurements from the moving domain is only noticeable when the number of pressure sensors is relatively small ($M=9$). This configuration of sensors is more realistic in practical settings. It is also worth noticing that, at the earliest observation times ($n=1,2$) when the front has not reached most pressure sensors, inverting measurements from the moving domain provides additional information of the log-permeability to the one provided only by pressure measurements. 

As more observations (in time) are assimilated, the reduction of the uncertainty in terms of the ensemble variance can be observed from the plot of $\Sigma_{n}$ displayed in Figure \ref{Fig17} (middle). From this plot we also note that the variance decreases as we increase the number of pressure sensors. The added value of measurements from the moving domain is also quite substantial and more noticeable for a small number of pressure sensors. In fact, note that smaller uncertainty has been achieved by inverting only the front ($M=0$) compared to the inversion of only pressure data from $M=9$ sensors. Here we also find that, at the earliest observation times, the additional inversion of measurements of the moving domain results in further reductions of the uncertainty in comparison to the inversion of only pressure data. While this investigation was conducted with a realistic choice of measurement noise (2.5\%), further studies should be conducted to understand the effect of the noise level on the uncertainty estimates of log-permeability in the 2D case.

Finally, the computational cost (see expression (\ref{cost_RENKA})) of approximating the sequence of posteriors $\{\mu_{n=1}^{7}\}$ via REnKA is displayed in Figure~\ref{Fig17} (right). This cost is expressed in terms of the number of $\cG_{7}$ forward model evaluations which, in turn, correspond to solving the moving boundary problem from $t=0$ to the last observation time $t_{7}$. Furthermore, this cost has been normalised by the number of particles $J$ used in REnKA. This normalisation enables us to provide a rough estimate of the scalable (with respect to $J$) computational cost of the REnKA (\ref{cost_RENKA}) if each evaluation of the forward map (see step 2(b) in Algorithm~\ref{EnKF_al}) is conducted in parallel. As discussed in Section~\ref{note}, increasing the number of measurements results in more tempering distributions (i.e. iterations) in the REnKA scheme. Therefore the computational cost increases with the number of measurements. However, for a realistic choice of pressure sensors $M=9$, we note that cost of inverting measurements of both front location and pressure sensor results (in average) in a scalable cost of 21 iterations. Since the number of particles that we use for REnKA is relatively low ($J=150$), such scalability with respect to the number of particles is reasonable with a high-end computer cluster and can be achieved within a few minutes.

\section{Summary and conclusions}\label{Conclusions}

In this work we studied the Bayesian inverse problem that arises from inferring physical properties in a setting for porous media flow with a moving boundary. Our investigation is focused on the inference of log-permeability from measurements of pressure and observation of the (moving) domain occupied by resin during the resin injection stage of RTM relevant to the fabrication of composite materials. We adopted the infinite-dimensional Bayesian approach to inverse problems where the aim is to characterise, at each observation time, the posteriors that arise from conditioning the log-permeability on pressure/front measurements. The simplicity of the 1D RTM model enabled us to show existence of the Bayesian posteriors in the aforementioned infinite-dimensional setting. These posteriors were then probed numerically with the dimension-independent SMC sampler for inverse problems from \cite{Kantas}. Our numerical experiments indicated that very large number of particles were needed to accurately approximate the Bayesian posteriors. This resulted in a high computational cost unfeasible for practical RTM settings.

In order to reduce computational cost of Bayesian inversions for practical RTM settings, we proposed a regularising ensemble Kalman algorithm (REnKA) that we motivated from the adaptive tempering SMC sampler of \cite{Kantas}. The proposed REnKA is based on Gaussian approximations of the sequence of Bayesian posteriors and thus, in general, asymptotic convergence of posterior expectations cannot be ensured. Nevertheless, our numerical results demonstrated that REnKA is robust with respect to tuneable parameters and provides reasonably accurate estimates of the posterior mean and variance with a computational cost affordable for practical RTM processes.

While measurements have been widely used with ad-hoc approaches to estimate permeability of preform in RTM, to the best of our knowledge, this work constitutes the first investigation that uses a Bayesian inverse modeling framework for the inference of preform permeability under the presence of uncertainty. From the numerical investigations that we conducted some conclusions and recommendations can be made with relevant implications to practical RTM settings. In particular, our synthetic experiments indicated that, when a small number of sensors (5 sensors in 1D and 9 sensors in 2D) are used to measure pressure, observing the front/moving domain can substantially reduce the uncertainty (variance) of the estimates of the log-permeability. This is particularly relevant in real experiments when the number of pressure sensors are \mt{usually} limited. However,  when the inversion is conducted with only measurements of the moving front, the reconstruction of the main spatial features of the true permeability are not recovered accurately.

Our results also display the benefit of the proposed sequential approach in updating our knowledge of the log-permeability as soon as measurements become available. Indeed, inverting measurements of pressure and front frequently in time, enabled us to reduce the uncertainty in the log-permeability. While the reduction of the uncertainty is mainly achieved within the region occupied by the resin at a given time, a decrease in the uncertainty (with respect to the prior) can also be observed in an unfilled region close to the front. Such a reduction of this uncertainty via the Bayesian posteriors can be valuable for decision-making purposes with the aim of an optimal control of RTM in real time. Finally, our numerical investigations show that the observation noise in pressure measurements and front location have a substantial effect \mt{on} the estimates of log-permeability and its uncertainties. Indeed, it comes as no surprise that more accurate measurements (e.g. $1\%$) result in estimates of log-permeability concentrated around the truth. Again, giving the limitation of deploying many pressure sensors within a real RTM scenario, it is then essential to use high precision pressure sensors to achieve enough confidence in the posterior uncertainties of the inferred permeability.

Although the context of this work is the resin injection in RTM processes, the Bayesian framework at the core of the proposed methodology, and the corresponding Gaussian approximations emerged from the proposed REnKA are generic, flexible, and thus transferable to a wide class of inverse problems constrained with PDEs with moving boundaries.

\section*{Acknowledgements}

This work was partially supported by the EPSRC grant EP/K031430/1. The authors are very grateful to Mikhail Matveev for useful discussions. MI wishes to thank Nikolas Kantas for helpful discussions on the computational aspects and implementation of SMC.


\appendix \label{Ape}

\section{Proof of Theorem \ref{Theo1}}\label{Ape_Theo}
For the proof of Theorem \ref{Theo1} we consider the dimensionless version of (\ref{eq8})-(\ref{eq10}) that we derive in the following subsection.
\subsection{Dimensional analysis}\label{dimension}
Let us consider the following change to dimensionless variables:
\begin{eqnarray}\label{ape1}
x\to x/x^{\ast}, \quad t\to t/t_{f},\quad p(x,t)\to p(x,t)/p_{0},\nonumber\\
u(x)\to u(x)-u_{0},\quad \Ga(t)\to \Ga(t)/x^{\ast},
\end{eqnarray}
where $t_{f}$ is a reference time and $u_{0}$ is a reference (constant) log-permeability. For simplicity we choose
$$\frac{p_{0}e^{u_{0}}t_{f}}{\mu (x^{\ast})^2}=1,\qquad \phi=1,\qquad p_{I}=2p_{0},$$
which enable us to transform (\ref{eq8})-(\ref{eq10}) into
\begin{eqnarray}\label{ape2}
\Ga(t) &=&W_{u}^{-1}(t), ~~~t>0
\end{eqnarray}
\begin{eqnarray}\label{ape3}
p(x,t) = \left\{ \begin{array}{cc}
2-\frac{F_u(x)}{F_u(\Ga(t))}, &\ t\geq 0,\  x\in D(t)\equiv (0,\Ga(t)),\\
1&  t\geq 0, \  x\in [0,1]\setminus D(t), \end{array}\right.
\end{eqnarray}
where, with a slight abuse in the notation, the variables $p$, $u$, $\Ga$, $x$ and $t$ are now the dimensionless variables. Similarly, the dimensionless filling time \eqref{eq10B} is given by
\begin{eqnarray}\label{ape4}
\tau^{\ast}= \int_{0}^{x^{*}}F_{u}(\xi)d\xi.
\end{eqnarray}
Let us note from \eqref{ape2} and \eqref{eq7} that
\begin{eqnarray}\label{ape5}
\frac{d\Ga}{dt}(t) =\frac{1}{F_{u}(\Ga(t))}.
\end{eqnarray}
Recall that $\{t_{n}\}_{n=1}^{N}$ is the collection of observation times with $0<t_{1}<t_{2}< \dots<  t_{N}$ and $\{x_{m}\}_{m=1}^{M}$ are the pressure measurement locations ($0<x_{1}<\dots<x_{M}\leq x^{\ast}$). We now prove two technical lemmas which are needed for the proof of Theorem~\ref{Theo1}.

\subsection{Technical lemmas}
For all $u\in X\equiv C[0,x^{\ast}]$, we define the norm
$$\vert\vert u\vert\vert\equiv \max_{x\in [0,x^{*}]}\vert u(x) \vert,$$
and denote by $(\Ga^{u},p^{u})$ the corresponding solutions of the dimensionless moving boundary problem (\ref{ape2})-(\ref{ape3}). Similarly, the filling time given by (\ref{ape4}) is denoted by $\tau^{\ast,u}$.

\begin{lemma}\label{lemma1}
For all $u\in X\equiv C[0,x^{\ast}]$, there exists a constant $\mathcal{A}_{u}$ such that
\begin{eqnarray}\label{ape6}
\vert \Ga^{u}(t) -\Ga^{u}(\hat{t})\vert  \leq  \mathcal{A}_{u}\vert t -\hat{t}\vert
\end{eqnarray}
for all $t,\hat{t}\in [t_{1},\tau^{\ast, u}]$. The constant $\mathcal{A}_{u}$ may depend only on $\vert\vert u\vert\vert,  t_{1}$ and $x^{\ast}$.
\end{lemma}
\textbf{Proof:}
Let $u\in X$ and $t\in [t_{1},\tau^{\ast, u}]$. From (\ref{ape5}) and \eqref{eq7} it is clear that $\Ga^{u}(t)$ is increasing and satisfies $\Ga^{u}(t)\leq \Ga^{u}(\tau^{*,u})=x^{*}$. Therefore,
\begin{eqnarray}\label{ape7}
F_{u}(\Ga^{u}(t)) =\int_{0}^{\Ga^{u}(t)}e^{-u(z)}dz\leq \int_{0}^{x^{\ast}}e^{-u(z)}dz\leq x^{\ast} e^{\vert\vert u\vert\vert}.
\end{eqnarray}
Then, from (\ref{ape5}) we have that
\begin{eqnarray}\label{ape8}
\frac{d\Ga^{u}}{dt}(t) =\frac{1}{F_{u}(\Ga^u(t))}\ge \frac{1}{x^{\ast}} e^{-\vert\vert u\vert\vert},
\end{eqnarray}
which implies
\begin{eqnarray}\label{ape9}
\Ga^{u}(t)  \ge \frac{t}{x^{\ast}} e^{-\vert\vert u\vert\vert}.
\end{eqnarray}
Similarly, note that
\begin{eqnarray}\label{ape10}
F_{u}(\Ga^{u}(t)) =\int_{0}^{\Ga^{u}(t)}e^{-u(z)}dz\ge \Ga^{u}(t)e^{-\vert\vert u\vert\vert}\ge  \frac{t}{x^{\ast}} e^{-2\vert\vert u\vert\vert},
\end{eqnarray}
and so
\begin{eqnarray}\label{ape11}
\frac{d\Ga^{u}}{dt}(t) =\frac{1}{F_{u}(\Ga^{u}(t))}\leq  \frac{x^{\ast}}{t} e^{2\vert\vert u\vert\vert}\leq  \frac{x^{\ast}}{t_{1}} e^{2\vert\vert u\vert\vert},
\end{eqnarray}
for all $t\in[t_{1},\tau^{*,u}]$. The Mean Value Theorem combined with (\ref{ape11}) yields (\ref{ape6}).  $\Box$

\begin{lemma}\label{lemma2}
For all $u,v\in X$, there exists a constant $\mathcal{B}_{u,v}$,  such that
\begin{eqnarray}\label{ape12}
\vert \Ga^{u}(t)-\Ga^{v}(t)\vert &\leq \mathcal{B}_{u,v} \vert\vert u-v\vert\vert,\\
\vert p^{u}(x_{m},t)-p^{v}(x_{m},t)\vert& \leq \mathcal{B}_{u,v} \vert\vert u-v\vert\vert, \label{ape12B}
\end{eqnarray}
for all $t\in [t_{1},\min\{\tau^{\ast,u},\tau^{\ast,v}\}]$ and all $m=1,\dots,M$. Moreover,
\begin{eqnarray}\label{ape13}
\vert \tau^{*,u}-\tau^{*,v}\vert  \leq \mathcal{B}_{u,v} \vert\vert u-v\vert\vert.
\end{eqnarray}
The constant $\mathcal{B}_{u,v}$ may depend only on $\vert\vert u\vert\vert, \vert\vert v\vert\vert, t_{1}$ and $x^{\ast}$.
\end{lemma}
\textbf{Proof:} From the Mean Value Theorem it is not difficult to see that
\begin{eqnarray}\label{ape14}
\vert F_{u}(x)-F_{v}(x) \vert \leq  x e^{\max\{\vert\vert u\vert\vert,~ \vert\vert v\vert\vert\}}\vert\vert u-v\vert\vert\leq  x~ \mathcal{M}_{u,v}\vert\vert u-v\vert\vert,
\end{eqnarray}
where
$$\mathcal{M}_{u,v}\equiv e^{\max\{\vert\vert u\vert\vert,~ \vert\vert v\vert\vert\}}.$$
It is also not difficult to see that
\begin{eqnarray}\label{ape15}
\vert F_{v}(\Ga^{u}(t))-F_{v}(\Ga^{v}(t)) \vert\leq \Big\vert  \int_{\Ga^{u}(t)}^{\Ga^{v}(t)}e^{-v(z)}dz\vert\leq  \vert \Ga^{u}(t)-\Ga^{v}(t)\vert e^{\vert\vert v\vert\vert}\leq  \mathcal{M}_{u,v}\vert \Ga^{u}(t)-\Ga^{v}(t)\vert .
\end{eqnarray}
From (\ref{ape14})-(\ref{ape15}), and the fact that $\Ga^{u}(t)\leq x^{*}$, we have
\begin{eqnarray}\label{ape16}
\vert F_{u}(\Ga^{u}(t))-F_{v}(\Ga^{v}(t)) \vert \leq \vert F_{u}(\Ga^{u}(t))-F_{v}(\Ga^{u}(t))\vert+\vert F_{v}(\Ga^{u}(t)-F_{v}(\Ga^{v}(t)) \vert \\
\leq \mathcal{M}_{u,v}\Big[ x^{\ast}\vert\vert u-v\vert\vert+ \vert \Ga^{u}(t)-\Ga^{v}(t)\vert \Big] .\nonumber
\end{eqnarray}
From \eqref{ape11} we get
\begin{eqnarray}\label{ape17}
\frac{1}{F_{u}(\Ga^{u}(t))}\frac{1}{F_{v}(\Ga^{v}(t))}\leq  \Big[\frac{x^{\ast}}{t_{1}}\Big]^2 e^{2\vert\vert u\vert\vert}e^{2\vert\vert v\vert\vert}\leq \Big[\frac{x^{\ast}}{t_{1}}\Big]^2 \mathcal{M}_{u,v}^4. 
\end{eqnarray}
Therefore
\begin{eqnarray}\label{ape18}
\Big \vert \frac{d\Ga^{u}}{dt}(t)-\frac{d\Ga^{v}}{dt}(t)\Big\vert \leq  \frac{\vert F_{u}(\Ga^{u}(t))-F_{v}(\Ga^{v}(t)) \vert }{  F_{u}(\Ga^{u}(t))F_{v}(\Ga^{v}(t))} \nonumber \\
\leq   \Big[\frac{x^{\ast}}{t_{1}}\Big]^2\mathcal{M}_{u,v}^5\Bigg[ x^{\ast}\vert\vert u-v\vert\vert+ \vert \Ga^{u}(t)-\Ga^{v}(t)\vert \Bigg] .\nonumber
\end{eqnarray}
We recall that $\Ga^{u}(0)=\Ga^{v}(0)=0$ and use Gronwall's inequality to conclude that
\begin{eqnarray}\label{ape19}
\vert \Ga^{u}(t)-\Ga^{v}(t)\vert \leq \exp\Bigg[\int_{0}^{t}\Big[\frac{x^{\ast}}{t_{1}}\Big]^2\mathcal{M}_{u,v}^5ds   \Bigg]~\int_{0}^{t}\frac{(x^{\ast})^3}{(t_{1})^2}\mathcal{M}_{u,v}^5 \vert\vert u-v\vert\vert ds\nonumber\\
=\exp\Bigg[t\Big[\frac{x^{\ast}}{t_{1}}\Big]^2\mathcal{M}_{u,v}^5  \Bigg]~t\frac{(x^{\ast})^3}{(t_{1})^2}\mathcal{M}_{u,v}^5 \vert\vert u-v\vert\vert .
\end{eqnarray}
From (\ref{ape4}) we see that
\begin{eqnarray}\label{ape20}
t\leq \min\{\tau^{\ast,u},\tau^{\ast,v}\}=\min\Bigg\{  \int_{0}^{x^{*}} F_{u}(\xi) d\xi, \int_{0}^{x^{*}} F_{v}(\xi) d\xi \Bigg\} \leq  (x^{\ast})^2 \mathcal{M}_{u,v}
\end{eqnarray}
which we combine with (\ref{ape19}) to obtain
\begin{eqnarray}\label{ape21}
\vert \Ga^{u}(t)-\Ga^{v}(t)\vert \leq
\mathcal{C}_{u,v} \vert\vert u-v\vert\vert ,
\end{eqnarray}
where
\begin{eqnarray}\label{ape22}
\mathcal{C}_{u,v}\equiv \exp\Bigg[\frac{(x^{\ast})^4}{t_{1}^2}\mathcal{M}_{u,v}^6  \Bigg]~\frac{(x^{\ast})^5}{(t_{1})^2}\mathcal{M}_{u,v}^6 .
\end{eqnarray}
Hence, (\ref{ape11}) is proved.

Let us now consider the case $x_{m}> \Ga^{u}(t)$ and $x_{m}> \Ga^{v}(t)$, then
\begin{eqnarray}\label{ape23}
\vert p^{u}(x_{m},t)-p^{v}(x_{m},t) \vert =0
\end{eqnarray}
for all $t\in [0,\min\{\tau^{*,u},\tau^{*,v}\}]$. Assume now that $x_{m}\leq \Ga^{u}(t)$, $x_{m}\leq \Ga^{v}(t)$, and let us note that
\begin{eqnarray}\label{ape24}
F_{u}(\Ga^{u}(t)) =\int_{0}^{\Ga^{u}(t)}e^{-u(z)}dz \ge \int_{0}^{x_{m}}e^{-u(z)}dz = F_{u}(x_{m}) .
\end{eqnarray}
Therefore, from (\ref{ape3}), (\ref{ape11}), (\ref{ape24}), (\ref{ape14}), (\ref{ape16}) and (\ref{ape21}),  we find
\begin{eqnarray}\label{ape25}
\vert p^{u}(x_{m},t)-p^{v}(x_{m},t) \vert \leq \frac{1}{ F_{v}(\Ga^{v}(t)) }\Bigg\vert F_{v}(x_m)-F_{u}(x) +\frac{F_{u}(x_m)}{  F_{u}(\Ga^{u}(t))}(F_{u}(\Ga^{u}(t))-F_{v}(\Ga^{v}(t)) \Bigg\vert \nonumber\\
 \leq \frac{x^{*}}{t_{1} }e^{2\vert\vert v\vert\vert}\Bigg[ \vert  F_{v}(x_{m})-F_{u}(x_{m}) \vert +\vert F_{u}(\Ga^{u}(t))-F_{v}(\Ga^{v}(t))\vert  \Bigg]\nonumber\\
 \leq \frac{x^{*}}{t_{1} }\mathcal{M}_{u,v}^3\Bigg[  x_{m} \vert\vert u-v\vert\vert 
+x^{\ast} \vert\vert u-v\vert\vert+ \vert \Ga^{u}(t)-\Ga^{v}(t)\vert  \Bigg]
 \leq \frac{x^{*}}{t_{1} }\mathcal{M}_{u,v}^3
(2x^{\ast} +\mathcal{C}_{u,v})\vert\vert u-v\vert\vert.
\end{eqnarray}
Consider now the case $x_{m}\leq \Ga^{u}(t)$, $x_{m}> \Ga^{v}(t)$. From (\ref{ape3}) and (\ref{ape24}) that $p^{u}(x_{m},t)\ge 1$. Therefore,
\begin{eqnarray}
\vert p^{u}(x_{m},t)-p^{v}(x_{m},t) \vert =p^{u}(x_{m},t)-1=
 \frac{1}{ F_{u}(\Ga^{u}(t)) }(  F_{u}(\Ga^{u}(t)) - F_{u}(x_m)) \nonumber\\
= \frac{1}{ F_{u}(\Ga^{u}(t)) }(  F_{u}(\Ga^{u}(t))- F_{u}(\Ga^{v}(t)) +F_{u}(\Ga^{v}(t))- F_{u}(x_m)).
\end{eqnarray}
Since $F_{u}(\Ga^{v}(t))- F_{u}(x_{m})<0$ (recall $\Ga^{v}(t)< x_{m}$) and $F_{u}(\Ga^{u}(t))\ge F_{u}(\Ga^{v}(t)) $, it then follows from \eqref{ape15}, \eqref{ape11} and \eqref{ape21} that
\begin{eqnarray}\label{ape27}
\vert p^{u}(x_{m},t)-p^{v}(x_{m},t) \vert < \frac{1}{ F_{u}(\Ga^{u}(t)) }(  F_{u}(\Ga^{u}(t))- F_{u}(\Ga^{v}(t)) )\leq   \frac{x^{\ast}}{t_{1}} \mathcal{M}_{u,v}^3\vert \Ga^{u}(t))- \Ga^{v}(t)\vert\nonumber\\
\leq   \frac{x^{\ast}}{t_{1}} \mathcal{M}_{u,v}^3 \mathcal{C}_{u,v} \vert\vert u-v\vert\vert
\leq    \frac{x^{*}}{t_{1} }\mathcal{M}_{u,v}^3
(2x^{\ast} +\mathcal{C}_{u,v})\vert\vert u-v\vert\vert.
\nonumber
\end{eqnarray}
Hence, (\ref{ape12}) is proved.

Finally, from (\ref{ape4}) and (\ref{ape14}), it is easy to see that
\begin{eqnarray}\label{ape28}
\vert \tau^{\ast,u}-\tau^{\ast,v}\vert \leq  \int_{0}^{x^{*}} \vert F_{u}(\xi) -F_{v}(\xi)\vert  d\xi\leq \frac{1}{2}(x^{*})^2 \mathcal{M}_{u,v}\vert\vert u-v\vert\vert .
\end{eqnarray}
We combine \eqref{ape21}, \eqref{ape23}, \eqref{ape25}-\eqref{ape28} and then \eqref{ape12}-\eqref{ape13} follows with
$$\mathcal{B}_{u,v}\equiv \min\Big\{ \mathcal{C}_{u,v},\frac{x^{*}}{t_{1} }\mathcal{M}_{u,v}^3
(2x^{\ast} +\mathcal{C}_{u,v}),\frac{1}{2}(x^{*})^2 \mathcal{M}_{u,v}\Big\} .$$
$\Box$

\subsection{Proof of Theorem \ref{Theo1}}
Given $t\ge t_{1}$ fixed, we first establish continuity of the following map $u\to \Ga^{u}(t\wedge\tau^{\ast,u} )$. Let $u,v\in X$ and without loss of generality assume that $\tau^{\ast,v}\leq \tau^{\ast,u}$. Note that
\begin{eqnarray}\label{ape29}
\vert \Ga^{u}(t\wedge\tau^{\ast,u} )-\Ga^{v}(t\wedge\tau^{\ast,v} )\vert\leq   \vert \Ga^{u}(t )-\Ga^{v}(t )\vert \mathbb{I}_{t<\tau^{\ast,v}}+  \vert \Ga^{u}(t )-\Ga^{v}(\tau^{\ast,v} )\vert \mathbb{I}_{t\in [\tau^{\ast,v},\tau^{\ast,u}]}\nonumber\\
+  \vert \Ga^{u}(\tau^{\ast,u} )-\Ga^{v}(\tau^{\ast,v} )\vert \mathbb{I}_{t>\tau^{\ast,u}}.
\end{eqnarray}
Since $\Ga^{u}(\tau^{\ast,u} )=\Ga^{v}(\tau^{\ast,v} )=x^{\ast}$, the previous expression reduces to
\begin{eqnarray}\label{ape30}
\vert \Ga^{u}(t\wedge\tau^{\ast,u} )-\Ga^{v}(t\wedge\tau^{\ast,v} )\vert\leq   \vert \Ga^{u}(t )-\Ga^{v}(t )\vert \mathbb{I}_{t<\tau^{\ast,v}}+  \vert \Ga^{u}(t )-\Ga^{v}(\tau^{\ast,v} )\vert \mathbb{I}_{t\in [\tau^{\ast,v},\tau^{\ast,u}]}.
\end{eqnarray}
We observe that we can write
\begin{eqnarray}\label{ape31}
\Ga^{u}(t )-\Ga^{v}(\tau^{\ast,v} ) =   \frac{1}{2}[\Ga^{u}(t )-\Ga^{u}(\tau^{\ast,u} )] + \frac{1}{2}[ \Ga^{u}(t )-\Ga^{u}(\tau^{\ast,v} )] .
\end{eqnarray}
Then by Lemmas~\ref{lemma1} and~\ref{lemma2}, we obtain
\begin{eqnarray}\label{ape33}
\vert\Ga^{u}(t )-\Ga^{v}(\tau^{\ast,v} ) \vert\leq  \mathcal{A}_{u} \vert t-\tau^{\ast,u}\vert +\mathcal{A}_{u}\vert t-\tau^{\ast,v}\vert  \nonumber\\
=\mathcal{A}_{u} (\tau^{\ast,u}-t) +\mathcal{A}_{u} (t-\tau^{\ast,v})=
\mathcal{A}_{u}( \tau^{\ast,u}-\tau^{\ast,v}) \leq  \mathcal{A}_{u}\mathcal{B}_{u,v}\vert\vert u-v\vert\vert
\end{eqnarray}
for all $t\in [\tau^{*,v},\tau^{*,u}]$. We combine (\ref{ape33}) with (\ref{ape29}) and use Lemma \ref{lemma1} again to obtain
\begin{eqnarray}\label{ape34}
\vert \Ga^{u}(t\wedge\tau^{\ast,u} )-\Ga^{v}(t\wedge\tau^{\ast,v} )\vert\leq  \mathcal{B}_{u,v}( 1+ \mathcal{A}_{u})\vert\vert u-v\vert\vert
\end{eqnarray}
for all $t\geq t_{1}$ which establishes the continuity of $u\to \Ga^{u}(t\wedge\tau^{\ast,u} )$ and, in particular, of $\cG_{n}^{\Ga}(u)=\Ga^{u}(t_{n}\wedge\tau^{\ast,u} )$ for all $n=1,\dots,N$.

Similarly, we now prove the continuity of $u\to p^{u}(x_{m},t\wedge\tau^{\ast,u} )$. We note that
\begin{eqnarray}\label{ape35}
\vert p^{u}(x_{m},t\wedge\tau^{\ast,u} )-p^{v}(x_{m},t\wedge\tau^{\ast,v} )\vert\leq   \vert p^{u}(x_{m},t )-p^{v}(x_{m},t )\vert \mathbb{I}_{t<\tau^{\ast,v}}+  \nonumber\\
\vert p^{u}(x_{m},t )-p^{v}(x_{m},\tau^{\ast,v} )\vert \mathbb{I}_{t\in [\tau^{\ast,v},\tau^{\ast,u}]}
+  \vert p^{u}(x_{m},\tau^{\ast,u} )-p^{v}(x_{m},\tau^{\ast,v} )\vert \mathbb{I}_{t>\tau^{\ast,u}} \, .
\end{eqnarray}
From (\ref{ape3}) it follows
\begin{eqnarray}\label{ape36}
\vert p^{u}(x_{m},t )-p^{v}(x_{m},\tau^{\ast,v} )\vert =\Big\vert\frac{F_{u}(x_{m})}{F_{u}(\Ga^{u}(t))}-\frac{F_{v}(x_{m})}{F_{v}(\Ga^{v}(\tau^{\ast,v}))} \Big\vert\nonumber\\
\leq \frac{1}{F_{u}(\Ga^{u}(t)) }\Big[\Big\vert F_{u}(x_{m})-F_{v}(x_{m}) \Big\vert+  \frac{F_{v}(x_{m})}{F_{v}(x^{*})}\vert  F_{v}(\Ga^{v}(\tau^{\ast,v}))-  F_{u}(\Ga^{u}(t)))   \vert\Big]
\end{eqnarray}
for all $t\in [\tau^{\ast,v},\tau^{\ast,u}]$. Using (\ref{ape33}) as well as similar arguments to the ones used before, we obtain
\begin{eqnarray}\label{ape37}
\vert    F_{u}(\Ga^{u}(t))-F_{v}(\Ga^{v}(\tau^{\ast,v})) \vert \leq \vert    F_{u}(\Ga^{u}(t))-F_{u}(\Ga^{v}(\tau^{*,v}))\vert  +\vert F_{u}(\Ga^{v}(\tau^{*,v}))-F_{v}(\Ga^{v}(\tau^{\ast,v})) \vert  \nonumber\\
\leq \mathcal{M}_{u,v}\vert \Ga^{u}(t) -\Ga^{v}(\tau^{*,v})\vert +\vert F_{u}(x^{*})-F_{v}(x^{*}) \vert\nonumber\\
\leq\mathcal{M}_{u,v}\vert \Ga^{u}(t) -\Ga^{v}(\tau^{*,v})\vert +x^{\ast}~ \mathcal{M}_{u,v}\vert\vert u-v\vert\vert
\leq  \mathcal{M}_{u,v}(\mathcal{A}_{u}\mathcal{B}_{u,v} +x^{\ast})\vert\vert u-v\vert\vert .
\end{eqnarray}
We use (\ref{ape37}), (\ref{ape11}) and the fact that $F_{v}(x_{m})\leq F_{v}(x^{*})$ to rewrite (\ref{ape36}) as follows
\begin{eqnarray}\label{ape38}
\vert p^{u}(x_{m},t )-p^{v}(x_{m},\tau^{\ast,v} )\vert  \leq  \frac{x^{\ast}}{t_{1}} \mathcal{M}_{u,v}^{2} \Big[ x_{m}\mathcal{M}_{u,v}+  \mathcal{M}_{u,v}(\mathcal{A}_{u}\mathcal{B}_{u,v} +x^{\ast})\Big]\vert\vert u-v\vert\vert\nonumber\\
 \leq  2\frac{x^{\ast}}{t_{1}} \mathcal{M}_{u,v}^{3} \Big[ x^{*}+  \mathcal{A}_{u}\mathcal{B}_{u,v} \Big]\vert\vert u-v\vert\vert .
\end{eqnarray}
From similar arguments it is easy to see that
\begin{eqnarray}\label{ape39}
\ \vert p^{u}(x_{m},\tau^{\ast,u} )-p^{v}(x_{m},\tau^{\ast,v} )\vert=\Big\vert\frac{F_{u}(x_{m})}{F_{u}(x^{*})}-\frac{F_{v}(x_{m})}{F_{v}(x^{*})} \Big\vert\nonumber\\
\leq \frac{1}{F_{u}(x^*) }\Big[\Big\vert F_{u}(x_{m})-F_{v}(x_{m}) \Big\vert+  \frac{F_{v}(x_{m})}{F_{v}(x^{*})}\vert  F_{v}(x^*)-  F_{u}(x^*)   \vert\Big]\nonumber\\
 \leq  \frac{x^{\ast}}{t_{1}} \mathcal{M}_{u,v}^{2}\Big[ x_{m}\mathcal{M}_{u,v} +  x^{*}\mathcal{M}_{u,v} \vert\Big]\vert\vert u-v\vert\vert \leq  2\frac{(x^{\ast})^2}{t_{1}} \mathcal{M}_{u,v}^{3} \vert\vert u-v\vert\vert .
\end{eqnarray}
We use (\ref{ape38})-(\ref{ape39}) and Lemma \ref{lemma2} to conclude that
\begin{eqnarray}
\vert p^{u}(x_{m},t\wedge\tau^{\ast,u} )-p^{v}(x_{m},(t\wedge\tau^{\ast,v} )\vert\leq   \Big[\mathcal{B}_{u,v}+4\frac{x^{\ast}}{t_{1}} \mathcal{M}_{u,v}^{3} ( x^{*}+  \mathcal{A}_{u}\mathcal{B}_{u,v} )\Big]\vert\vert u-v\vert\vert
\end{eqnarray}
which proves the continuity of $u\to p^{u}(x_{m},t\wedge\tau^{\ast,u} )$ for all $t\ge t_{1}$. The continuity of $\cG_{n}^{p}(u)$ then follows. $\Box$

\section{SMC sampler and pcn-MCMC algorithm}\label{Ape_SMC}

In Algorithm \ref{MCMC_al} we display the pcn-MCMC method that we use for the mutation step in the SMC sampler of \cite{Kantas} discussed in Section~\ref{Kantas_SMC} and summarized in Algorithm~\ref{SMC_al} below.

\begin{algorithm}[H]
\caption{\small pcn-MCMC to generate samples from a $\mu_{n,r}$-invariant Markov kernel}\label{MCMC_al}
\begin{algorithmic}
\STATE Select $\beta\in (0,1)$ and an integer $N_{\mu}$.
\FOR{ $j=1,\dots, J$}
\STATE Initialize $\nu^{(j)}(0)=  \hat{u}_{n,r}^{(j)}$
  \WHILE{$\alpha\leq N_{\mu}$}
\STATE{(1) \textbf{pcN proposal}. Propose $u_{prop}$ from
\begin{eqnarray}\label{prop_ape}
u_{prop}=\sqrt{1-\beta^2}\nu^{(j)}(\alpha)+(1-\sqrt{1-\beta^2})\overline{u}+\beta\xi, \qquad \textrm{with}~~ \xi\sim N(0,\mathcal{C}) \nonumber
\end{eqnarray}}
\STATE{(2) Set $\nu^{(j)}(\alpha+1)=u_{prop}$ with probability $a(\nu^{(j)}(\alpha),u)$ and  $\nu^{(j)}(\alpha+1)=\nu^{(j)}(\alpha)$ with probability $1-a(\nu^{(j)}(\alpha),u)$, where
\begin{eqnarray}\label{eq:3.6}
a(u,v)=\min\Big\{1,  \frac{ l_{n,r}(u_{prop},y_{n} )}{l_{n,r}(v,y_{n})  }\Big\} \nonumber
\end{eqnarray}
with $l_{n,r}$ defined in (\ref{eq34C})}
\STATE{(3) $\alpha \gets \alpha+1$}
  \ENDWHILE
\ENDFOR
\end{algorithmic}
\end{algorithm}

\begin{algorithm}[H]
\caption{SMC algorithm for High-Dimensional Inverse Problems}\label{SMC_al}
\begin{algorithmic}
\STATE Let $\{u_{0,0}^{(j)}\}_{j=1}^{J}\sim \mu_{0}$ be the initial ensemble of $J$ particles.
\STATE Define the tunable parameters $J_{thresh}$ and $N_{\mu}$.
\FOR{ $n=1,\dots,N$}
\STATE Set $r=0$ and $\phi_{n,0}=0$
\WHILE{ $\phi_{n,r}<1$}
\STATE $r\to r+1$
\STATE  \textbf{Compute the $n$th likelihood} (\ref{eq14}) $l_{n}(u_{n,r-1}^{(j)},y_{n})$ (for $j=1,\dots, J$)
\STATE   \textbf{Compute \mt{the tempering parameter} $\phi_{n,r}$}:
\IF{ $\min_{\phi\in (\phi_{n,r-1},1)}\textrm{ESS}_{n,r}(\phi)>J_{thresh}$}
\STATE set $\phi_{n,r}=1$.
\ELSE
\STATE compute $\phi_{n,r}$ such that $\textrm{ESS}_{n,r}(\phi)\approx J_{thresh}$ using a bisection algorithm on $(\phi_{n,r-1},1]$.
\ENDIF

\STATE \textbf{Computing weights} from expression (\ref{eq36}) $W_{n,r}^{(j)}\equiv \mathcal{W}_{n,r-1}^{(j)}[\phi_{n,r}]$
\STATE \textbf{Resample}. Let $(p_{n}^{(1)},\dots,p_{n}^{(N_{p})})\in \mathcal{R}(W_{n,r}^{(1)},\dots,W_{n,r}^{(N_{p})})$, Set $\hat{u}_{n,r}^{(j)}\equiv u_{n,r-1}^{(p_{n}^{(j)})}$ and $W_{n,r}^{(j)}=\frac{1}{J}$
\STATE\textbf{Mutation}. Sample $u_{n,r}^{(j)}\sim \mathcal{K}_{n,r}(\hat{u}_{n,r}^{(j)},\cdot )$ via Algorithm \ref{MCMC_al}.
\ENDWHILE
\STATE Set $u_{n+1,0}^{(j)}\equiv u_{n,r}^{(j)}$. Approximate $\mu_{n}$ by $\mu_{n}^{J}\equiv \frac{1}{J}\sum_{j=1}^{J} \delta_{u_{n,r}^{(j)}}$
\ENDFOR
\end{algorithmic}
\end{algorithm}

\section{Numerical implementation of the 1D forward model}\label{num_imple}
In this section we discuss the key aspects of the numerical implementation of the dimensionless version of the 1D RTM forward model derived in Section~\ref{dimension}.

Note that $F_{u}$ defined in (\ref{eq7}) can we written as
\begin{eqnarray}\label{eq:ape1:2}
F_{u}(x)\equiv \int_{0}^{x}e^{-u(z)}dz =\int_{0}^{1}e^{-u(z)}H(x-z)dz ,
\end{eqnarray}
where
$$H(x)=\left\{\begin{array}{cc}
0, &\textrm{if}~x<0,\\
1,& \textrm{if}~x\ge 0,\end{array}\right.
$$
is the Heaviside function. In order to approximate (\ref{ape2})-(\ref{ape3}), we discretize the domain $[0,1]$ with $S$ subintervals with end points defined by $x_{s+1/2}=[1/2+s]\Delta x$ $(s=0,\dots S)$, where $\Delta x= x^{\ast}/S$ and the centers of the cells are $x_{s}=\frac{x_{s-1/2}+x_{s+1/2}}{2}$. Let us consider a piecewise constant approximation of the unknown $u$ defined on the centers of the cells, i.e.
$$u(x)\approx \sum_{s=1}u_{s}\chi_{[x_{s-1/2},x_{s+1/2}]}, $$
where $u_{s}=u(x_{s})$. Therefore, (\ref{eq:ape1:2}) can be approximated by
\begin{eqnarray}\label{eq:ape1:3}
F_{u}(x) \approx \sum_{s=1}^{S} e^{-u_{s} }\Bigg[\frac{1}{2}+\frac{1}{2}\tanh{r (x-x_{s})}\Bigg]\Delta x,
\end{eqnarray}
where we have replaced $H(x)$ with its smooth approximation $\hat{H}(x)=\frac{1}{2}+\frac{1}{2}\tanh{r x}$ (with $r=300$) \cite{bracewell1978fourier}.

We consider a temporal domain $[0,0.4]$ discretized with $K$ points $t_{k}=k \Delta t$, where $\Delta t \equiv \ t_{f}/K$. An implicit backward Euler scheme applied to the dimensionless version of (\ref{eq5}) yields
\begin{eqnarray}\label{eq:ape1:4}
\frac{1}{\Delta t}(\Ga_{k+1}-\Ga_{k})-\frac{1}{F_{u}(\Ga_{k+1})}=0,\qquad \Ga_{0}=0,
\end{eqnarray}
where $\Ga_{k}\equiv \Ga(t_{k})$. For the approximation of (\ref{eq:ape1:4}), we use (\ref{eq:ape1:3}). The solution of the resulting nonlinear equation is implemented in MATLAB by means of the routine \texttt{fzero}. Once $\Ga_{k}$ is computed, we evaluate the pressure field from
\begin{eqnarray}\label{eq:ape1:5}
p(x,t_{k}) = 2-\frac{F_u(x)}{F_u(\Ga_{k})}
\end{eqnarray}
at the mesh points $x_{s+1/2}$ defined earlier.
\section{Performance of the SMC sampler}

\begin{figure}[htbp]
\begin{center}

\includegraphics[scale=0.33]{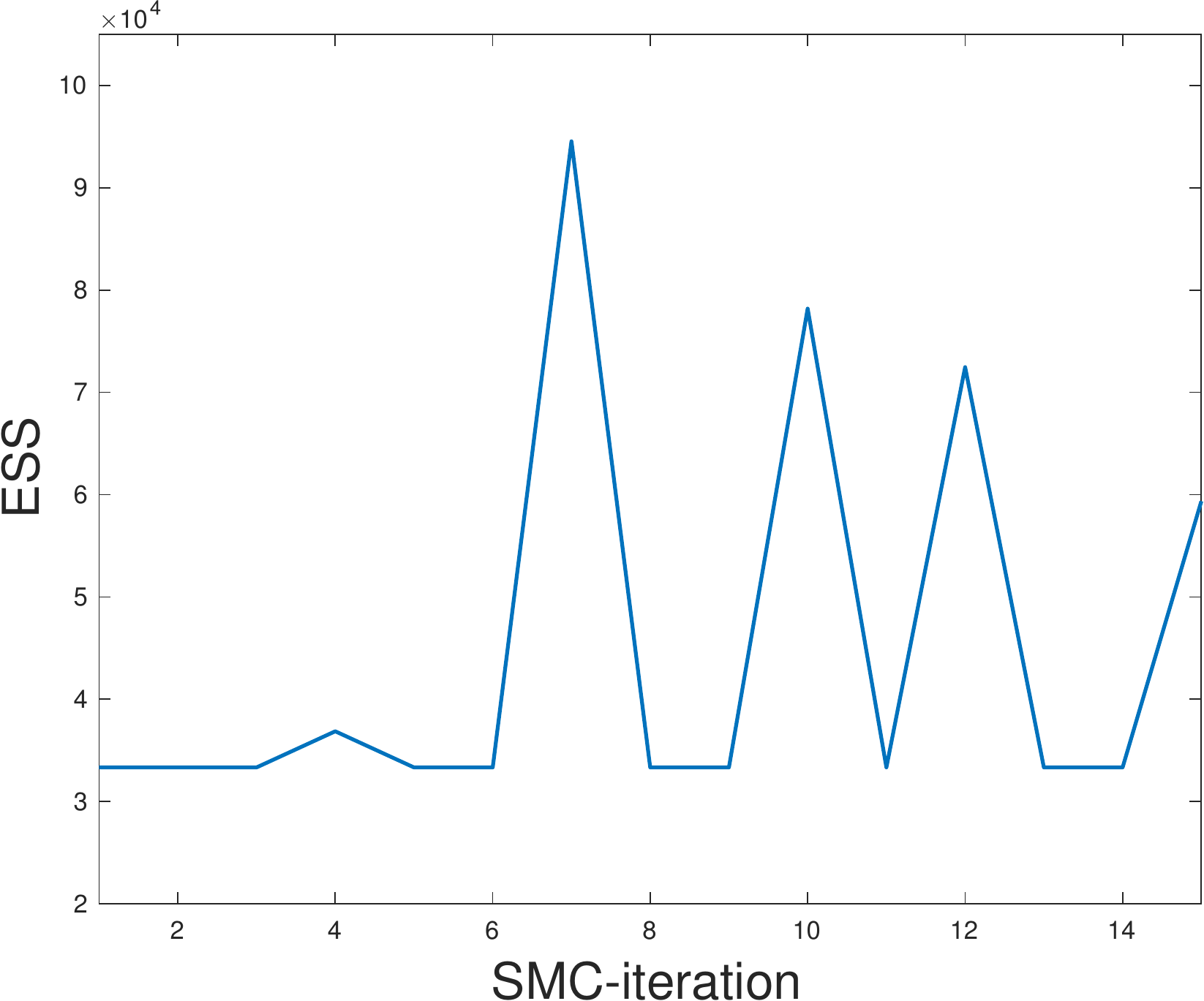}
\includegraphics[scale=0.33]{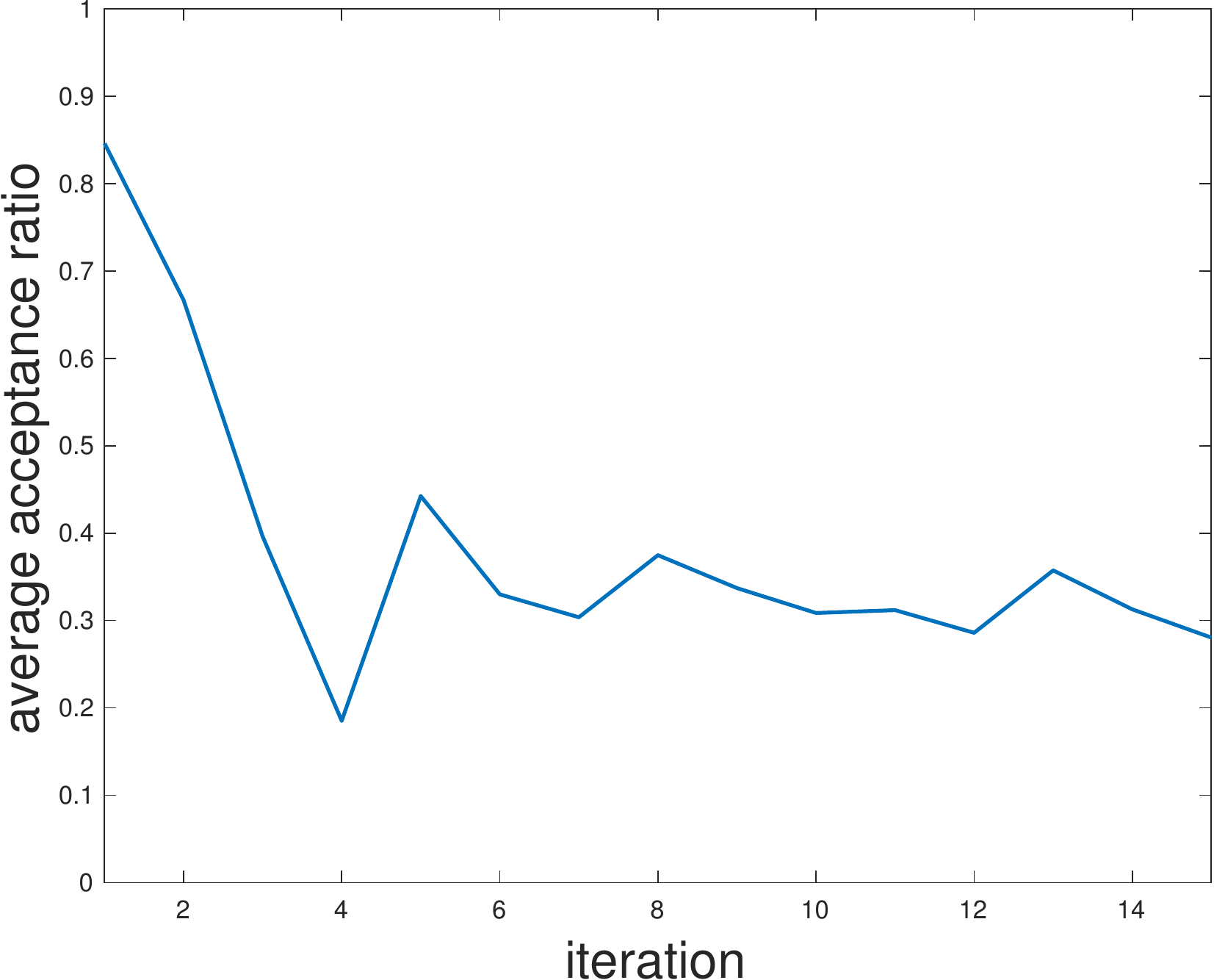}
\includegraphics[scale=0.33]{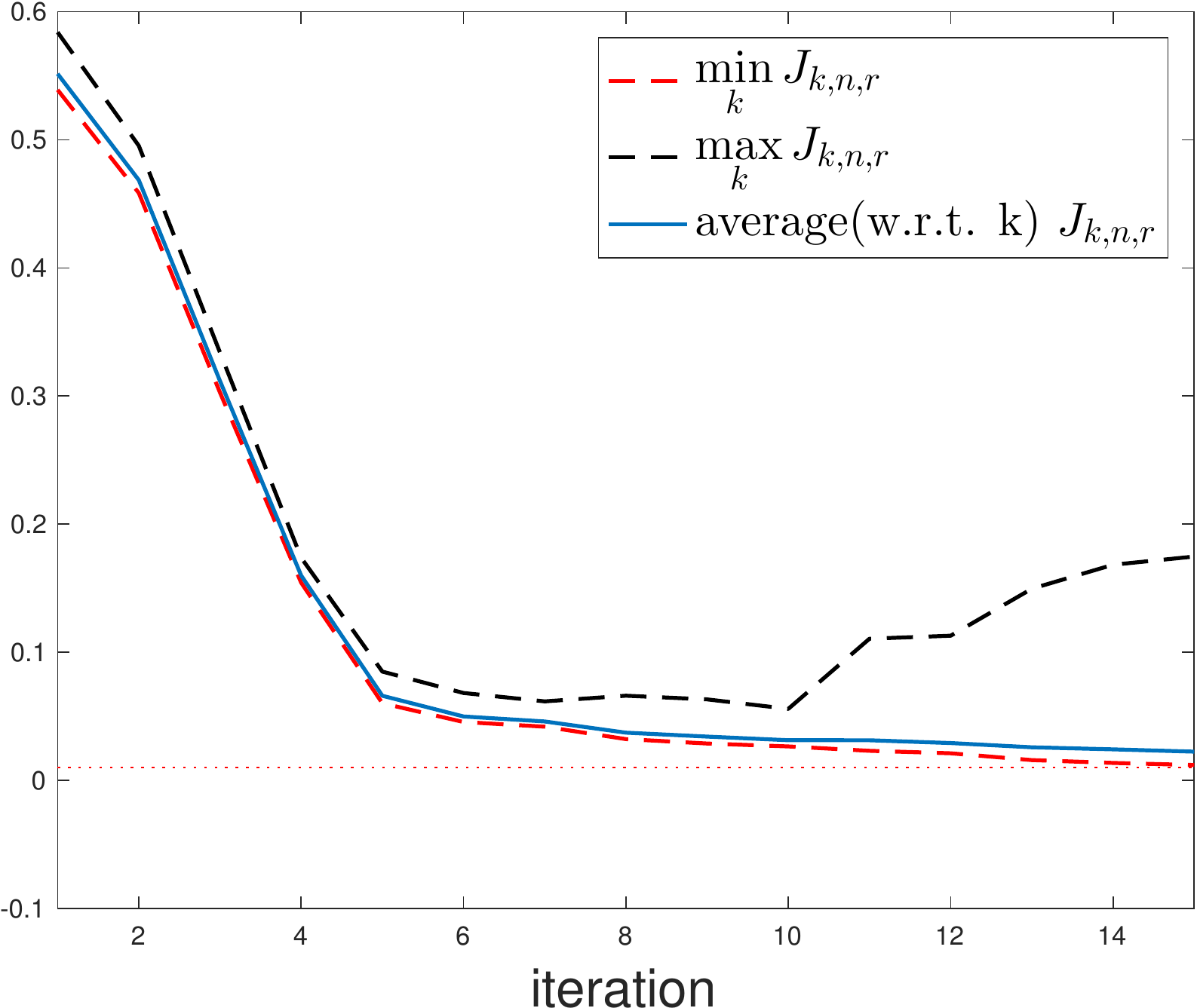}
 \caption{SMC diagnostics. Left: Effective Sample Size. Middle: Acceptance ratio of pcn-MCMC. Right: Plots of $\max_{k}J_{k,n,r}$, $\min_{k}J_{k,n,r}$ and the average with respect to $k$ of $J_{k,n,r}$.} \label{Fig_per1}
\end{center}
\end{figure}

In this section we discuss the performance of the SMC sampler (Algorithm \ref{SMC_al}) applied in Section~\ref{benchmark}. The first measure of performance is ESS defined in (\ref{eq37}) as a function of iterations given by $\sum_{p=1}^{n} q_{p}$, with $q_{p}$ given in (\ref{it})). This is displayed in Figure~\ref{Fig_per1} (left). By design of $\phi_{n,r}$, ESS oscillates between the tunable parameter $J_{thresh}$ (when $\phi_{n,r}<1$) and larger values when $\phi_{n,r}>1$. Note that more iterations (i.e. tempering) is needed for computing $\mu_{1}$. This comes as no surprise due the substantial difference between $\mu_{0}$ and $\mu_{1}$, which arises from the assimilation of data at the first observation time $t_{1}$. By ensuring that the ESS does not fall below the threshold $J_{thresh}$, we avoid a sharp failure in the SMC sampler as discussed in Section~\ref{Kantas_SMC}.

We now monitor performance of the mutation step (Step (e) of Algorithm~\ref{SMC_al}) of the SMC sampler that controls the diversity added to the population of particles. As suggested in \cite{Kantas}, this can be monitored by the average acceptance ratio with respect to the number of particles in Algorithm \ref{MCMC_al}. We plot this ratio in Figure \ref{Fig_per1} (middle). For the present work, we have tuned $\beta$ in the proposal of Algorithm \ref{MCMC_al} so that the average acceptance ratio for the measures is no less than $30\%$. Additionally, it is suggested in \cite{Kantas} to monitor the quality of the efficacy of the mutation step by means of
\begin{eqnarray}\label{perfo}
J_{k,n,r}=\frac{1}{2}\frac{\sum_{j=1}^{J} \vert u_{k,n,r}^{(j)}(M) -u_{k,n,r}^{(j)} (0) \vert } { \vert u_{k,n,r}^{(j)}(0) -\mathbb{E}(\mu_{n,r}^{J} (0) )\vert } ,
\end{eqnarray}
where $u_{k,n,r}^{(j)}(0)$ (resp. $u_{k,n,r}^{(j)}(M)$) denotes the $k$ KL coefficient of the $j$th particle $u_{n,r}^{(j)}$ before the mutation step (resp. after $M$ MCMC iterations). In Figure \ref{Fig_per1} (right) we display the plots of $\max_{k}J_{k,n,r}$, $\min_{k}J_{k,n,r}$ and the average with respect to $k$ of $J_{k,n,r}$. As suggested in \cite{Kantas}, this quantity should not be less than 0.09, which is indicated by a horizontal dotted line in Figure \ref{Fig_per1} (right). For further discussions of the role of (\ref{perfo}) in assessing the mutation step in SMC we refer the reader to  \cite{Kantas}.

Finally, in order to assess stability of the scheme with respect to the prior ensemble, we consider 4 different runs of the SMC sampler with a different set of initial $J=10^5$ particles. The errors (\ref{truth_err}) with respect to the truth  for all these cases are displayed in Table~\ref{TableAp1}. These results indicate that there is indeed quite a good agreement in the estimator of SMC provided by the mean of the particles.

\begin{table}
\caption{Error with respect to the truth for different runs of SMC.}
\centering
\begin{tabular}{|c|c|c|c|c|}
\hline
 &run  1& run 2& run 3& run 4\\
\hline
$\epsilon_{1}$ & 0.679 & 0.681 & 0.681 & 0.682 \\
$\epsilon_{2}$ & 0.656 & 0.652 & 0.653 & 0.656 \\
$\epsilon_{3}$ & 0.485 & 0.488 & 0.476 & 0.485 \\
$\epsilon_{4}$ & 0.522 & 0.520 & 0.519 & 0.522 \\
$\epsilon_{5}$ & 0.339 & 0.338 & 0.344 & 0.339 \\
\hline
\end{tabular}
\label{TableAp1}
\end{table}

\bibliographystyle{siamplain}
 \bibliography{Ensemble_bib}

\end{document}